\documentclass[preprint,12pt]{elsarticle}
\usepackage{psfrag}
\usepackage{pstricks}
\usepackage{pstricks-add}
\usepackage{graphics} % for pdf, bitmapped graphics files
\usepackage{graphicx}          
\usepackage{ulem}
\usepackage{epsfig} % for postscript graphics files
\usepackage{mathptmx} % assumes new font selection scheme installed
\usepackage{times} % assumes new font selection scheme installed
\usepackage{amsmath} % assumes amsmath package installed
\usepackage{amssymb}  % assumes amsmath package installed
\usepackage{amsfonts}
\usepackage{pgfplots}
\usepackage{epstopdf} %converting to PDF % Comment for arXiv

\usepackage{float} % To put figure exactly in the place specified
\usepackage[utf8]{inputenc}
\usepackage[T1]{fontenc}

\usepackage{url}
\usepackage{hyperref}

\newtheorem{assumption}{Assumption}

\def\C19{Covid-19}
\def\ARA{Auvergne-Rhône-Alpes }
\pgfplotsset{compat=1.14}

\journal{Annual Reviews in Control}

\begin{document}

\begin{frontmatter}

\title{The Ockham's razor applied to \\ 
COVID-19 model fitting French data}

\author[label1]{Mirko Fiacchini and Mazen Alamir}
\address[label1]{Univ. Grenoble Alpes, CNRS, Grenoble INP, GIPSA-lab, 38000 Grenoble, France. {\tt \{mirko.fiacchini, mazen@alamir\}@gipsa-lab.fr}}

\begin{abstract}

This paper presents a data-based simple model for fitting the available data of the \C19 pandemic evolution in France. The time series concerning the 13 regions of mainland France have been considered for fitting and validating the model. An extremely simple, two-dimensional model with only two parameters demonstrated to be able to reproduce the time series concerning the number of daily demises caused by \C19, the hospitalizations, intensive care and emergency accesses, the daily number of positive test and other indicators, for the different French regions. These results might contribute to stimulate a debate on the suitability of much more complex models for reproducing and forecasting the pandemic evolution since, although relevant from a mechanistic point of view, they could lead to nonidentifiability issues.

\end{abstract}


\end{frontmatter}

\section{Introduction}

The \C19 pandemic spread affecting the whole world, causing several millions of infected cases and thousands of deaths, attracted huge efforts of researchers from different scientific fields aiming at a better understanding of the virus evolution features. Besides the epidemiologists and the medical researchers, also the data scientists and the automatic control community mobilize, one of the major issue being the inference and identification of the dynamical system ruling the pandemic evolution. The virus evolution dynamics knowledge, in fact, would allow to estimate the pandemic state and predict its future evolution and thus serve as modelling tool to infer the effects of the containment measures and support the health institutions choices. A recent survey, \cite{alamo2020data} offers a detailed overview on the recent scientific publications regarding modelling, forecasting and controlling the evolution of \C19 pandemic based on data science and control theory.\\

Among the simpler and more popular epidemic compartmental models, recently used for fitting the data of the spread of \C19, there are the SIR model, first formulated in the seminar work \cite{kermack1927contribution}, and the SEIR model \cite{fang2020transmission,kucharski2020early,wu2020nowcasting}. The SIR model consists in simple differential equations representing the interaction between the populations of susceptible, infected, and recovered individuals, whereas, the SEIR models, considers also the population of exposed individuals, for modelling the incubation period. Variations are presented in the literature to take into account other populations involved in the \C19 epidemic disease evolution, for instance in \cite{crokidakis2020data,gevertz2020novel,morato2020optimal} that included also the asymptomatic individuals, and \cite{giordano2020sidarthe,giordano2020modelling} that introduces the SIDARTHE model, compartmental \C19 model composed by 8 populations and 16 parameters. Also spatial models \cite{brockmann2013hidden}, to model the geographical interconnection between regions, and network models \cite{el2012social} to represent the social interaction between individuals and populations, have been formulated to reproduce the epidemic evolution. Also data-based, model-free models have been proposed to predict the \C19 spread evolution, \cite{huang2020data}.\\

Besides non-parametric approaches for model inference, \cite{ferretti2020quantifying,roda2020difficult} for instance, parametric model identification has been widely used. The problem of parametric identification for fitting the model with the available data is strongly affected by several issues, such that the time variation of the parameters \cite{cori2013new}, the lack of full access to the data concerning the populations \cite{mckinley2009inference}, the presence of uncertainties affecting the model \cite{qian2020sensitivity,fang2020transmission}. These sources of uncertainties and lack of exact knowledge may lead to the issue of nonidentifiability \cite{roda2020difficult}, that consists in practice in the existence of different sets of parameters that permits to reproduce the available data. When this occurs, it is not possible to univocally identify the real set of parameters, and then to discriminate them from all the other potential solutions of the identification problem. The consequence might be a low accuracy and a poor forecasting power of the identified model. The use of simple models \cite{burnham2011aic,roda2020difficult,postnikov2020estimation} and the validation of the identified system on several sets of data \cite{portet2020primer}, from different regions for instance, are possible solutions to the nonidentifiability issue.\\

This paper is guided by the Ockham's razor, or law of parsimony, criterion, which substantially claims that the simplest solution is most likely the right one since based on less hypotheses. As stated above for the nonidentifiability problem, indeed, the simpler is the model able to fit and reproduce the available data, the higher is its reliability in terms of accuracy and forecasting power. Starting from the available data concerning the \C19 pandemic evolution in the 13 regions of mainland France, an extremely simple dynamical model has been formulated with only two states and two parameters, common to all the regions. The model has been identified on the measures of deaths caused by the \C19 and then validated on several other time series, concerning hospitalizations, intensive care accesses, emergency admissions etc. for all the 13 regions.\\ 
 
The paper is organized as follows: Section\ref{sec:data} presents the time series concerning the excess of deaths due to \C19, used to identify the model; Section~\ref{sec:fit} deals with the model and its fitting to deaths excess data; Section~\ref{sec:valid} regards the validation of the identified model on other times series. The paper terminates with some conclusions.

\section{Deaths excess data for fitting}\label{sec:data}

Several datasets concerning the evolution of the \C19 in France are available that can be used to identify the dynamics parameters. The data are organized by gender, by department, by age and by region. In this work focuses on the dynamics of the \C19 evolution within the different regions of mainland France, aiming at identifying a single model, as simple as possible, that permits to fit and reproduce the regional evolution of \C19 pandemic.  \\

\begin{figure}[H]
 \centering
\includegraphics[width=13.6cm]{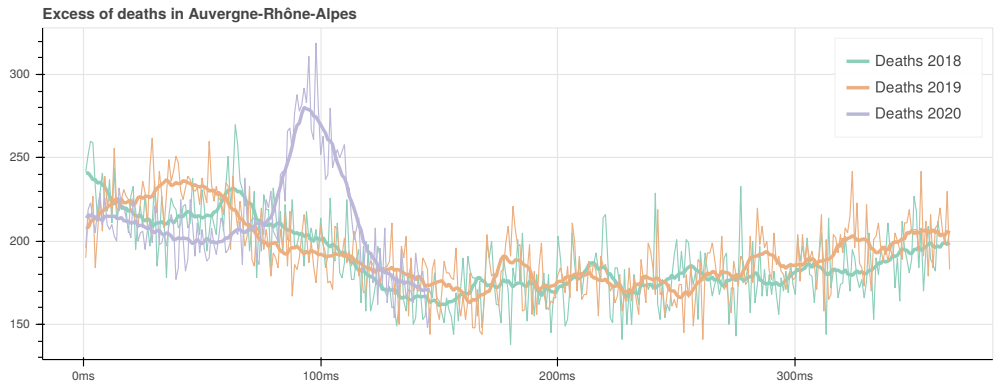}
 \caption{Evolution of daily demises in 2018, 2019 and 2020, in \ARA region. In bold, the average over a 14-days-long window.}\label{fig4}
\end{figure}

Concerning the number of demises due to the \C19, in particular, three time series are available. Firstly, the INSEE, the French National Institute of Statistics and Economic Studies, provides the data of all the daily demises occurred since January 1, 2018, in a single file, organized by region, in \cite{insee}. In Fig~\ref{fig4} the evolution of deaths in \ARA region, concerning the years 2018, 2019 and 2020 are represented. Besides the time series, the smoothed series, computed by averaging over a 14-days-long window, are drown in bold, together with the average of values of 2018 and 2019. The data concerning the 13 regions of mainland France are shown in \ref{app:excess}. \\

A measure of the daily deaths for \C19 is the difference between the daily demises occurred in 2020 and the average of those of 2018 and 2019. This difference, in fact, is supposed to represent (via some correction that are mentioned in the sequel) the excess of deaths occurred in the period of \C19 pandemic spread in France and can be supposed to be the effect of the epidemic. 

Other two time series are available that represent measures of the demises caused by the \C19, in particular the number of deaths registered at the hospitals, {\tt incid\_dc} and the number of demises that have been electronically registered and transmitted to the Inserm\footnote{French National Institute of Health and Medical Research.} institute. More details on these two time series can be found in Section~\ref{sec:valid}. For obtaining a reasonably reliable guess of the number of deaths caused by \C19, these two signals have been smoothed by averaging them over a 14-days-long time window.\\

\begin{figure}[H]
 \centering
\includegraphics[width=13.6cm]{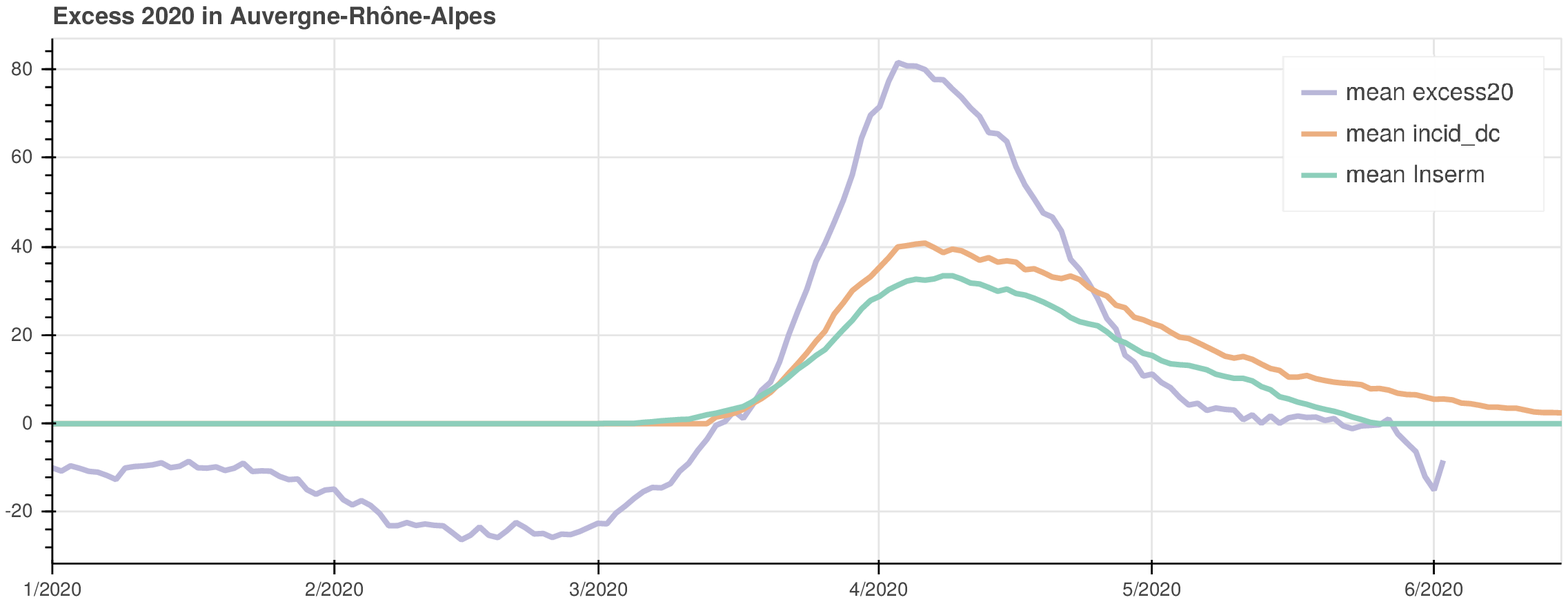}
 \caption{Excess of death in 2020 with respect to the average of 2018/2019, {\tt mean excess20}; deaths certified in hospitals {\tt mean incid\_dc} and deaths electronically certified to Inserm {\tt mean Inserm}, averaged over a 14-days-long window, in \ARA region.} \label{fig5}
\end{figure}

The time series considered for model fitting, depicted in Fig.~\ref{fig5} for \ARA region (see \ref{app:excessfit} for the other regions), are:\\

\begin{itemize}
 \item the excess of deaths of 2020 with respect to the average of 2018 and 2019, averaged over a 14-days-long window, denoted {\tt mean excess20};\\
 \item the average over a 14-days-long window of hospital demises time series {\tt incid\_dc}, denoted {\tt mean incid\_dc};\\
 \item the average over a 14-days-long window of demises registered by Inserm {\tt inserm\_dc}, denoted {\tt mean Inserm}.\\
\end{itemize}

From the inspection of Fig.~\ref{fig5}, and considering the excess of deaths of 2020 as a reliable measure of the real amount of demises caused by \C19, it can be inferred that:\\

\begin{itemize}
 \item the cases registered in the hospitals and those certified to Inserm represent a sub-estimation of the real number of demises due to \C19 in the first weeks after the lockdown of March 17;\\
 
 \item the pandemic in \ARA might have been starting at the beginning of March, as the curve, although at negative values (i.e. deaths in 2020 below the average of 2018/2019), presents an exponentially increasing behavior since the first days of March;\\
 
 \item the lockdown might have affected also the other causes of mortality (for instance due to the traffic reduction, working activity partial stop, etc.) leading to values of excess around $0$ in the last week of April and in May, when the certified demises, both in hospitals and through electronic certification, are higher than the mortality excess.\\
 
\end{itemize}

The last consideration, in particular, seems to indicate that the mortality excess of 2020 represents an underestimation of cases of demises due to \C19, particularly in the last part of the period taken into account. To compensate this effect, that can be observed in the data concerning many of the mainland France regions, see the \ref{app:excess20corr}, an heuristic criterion has been employed, to avoid that the time series considered as a reliable measure of the deaths caused by \C19 is lower than the number of observed cases.\\

\begin{figure}[H]
 \centering
\includegraphics[width=13.6cm]{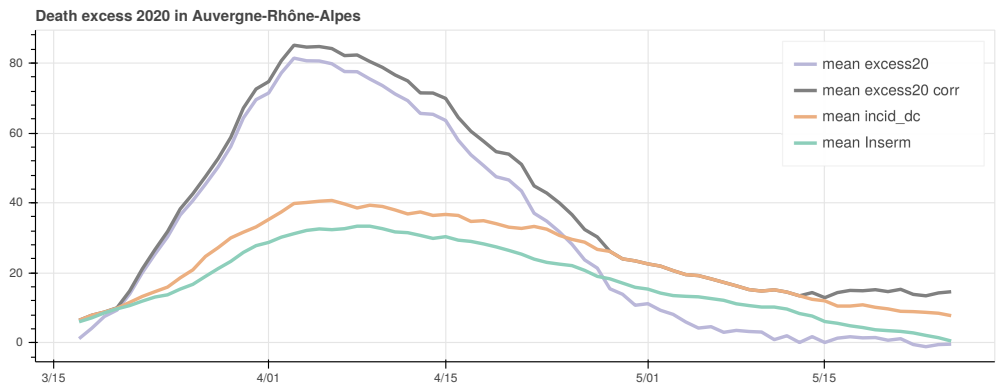}
 \caption{Excess of death in 2020 with respect to the average of 2018/2019, {\tt mean excess20}; deaths certified in hospitals {\tt mean incid\_dc} and deaths electronically certified to Inserm {\tt mean Inserm}, averaged over a 14-days-long window. Corrected average of excess {\tt mean excess20 corr}, in \ARA region.} \label{fig5b}
\end{figure}

In particular, a linear function of the time has been added to the time series {\tt mean excess20} and the maximum between it and the signal {\tt mean incid\_dc} is taken. The resulting time series {\tt mean excess20 corr} in Fig~\ref{fig5b} has been employed to identify the system.\\

\section{Model fitting the deaths excess}\label{sec:fit}

A first hypothesis is formulated to express the fact that the demises caused by \C19 is proportional to the number of infected individuals.\\

\begin{assumption}\label{ass:1}
The number of daily deaths due to \C19 is proportional to the number of infected individuals.\\
\end{assumption}

Consider the dynamics relating the number of \C19 deaths and the number of infected cases employed in several epidemiological compartmental models:

\begin{equation}\label{eq:dDalphaI}
 \frac{d D(t)}{dt} = \alpha I(t).
\end{equation}

Assumption~\ref{ass:1} is equivalent to suppose satisfaction of (\ref{eq:dDalphaI}), often employed in the compartmental epidemic models, and then that the number of new demises for \C19 is proportional to the value of the infected cases, with proportionality coefficient $\alpha$. Thus, a reliable measure of the daily demises would provide a scaled image of the \C19 infected cases. \\

Moreover, as a second hypothesis, the resulting signal {\tt mean excess20 corr} in Fig.~\ref{fig5} is considered hereafter as a reasonably reliable guess of the amount of daily demises due to \C19 infection.\\

\begin{assumption}
  The time series {\tt mean excess20 corr}, denoted $f(t)$, is a reliable measure of the number of daily demises caused by \C19.\\
\end{assumption}

Under these assumptions, then, $f(t)$ is a measured approximation of the value of the function $\alpha I(t)$, from (\ref{eq:dDalphaI}), in fact:

\begin{equation}\label{eq:ft}
 \frac{d D(t)}{dt} = \alpha I(t) \simeq f(t).
\end{equation}

Given the time series $f(t)$ that represents the evolution of the population of \C19 infected individuals, multiplied by an known constant $\alpha$, its dynamics can be identified. \\

The simpler epidemic compartmental models, yet considered among the most robust from the information theory point of view, see \cite{burnham2011aic,alamo2020data}, is the SIR model. The SIR model, and its variation SEIR, have been recently used for fitting the spread of \C19 \cite{fang2020transmission,kucharski2020early,wu2020nowcasting}. In SIR (or SIRD) models, the populations of susceptible ($S$), infected ($I$), recovered ($R$) and dead ($D$) individuals (often the recovered and dead individuals are considered in a single compartment $R$) are modeled by the differential equations:

%\begin{equation}
\begin{align}\label{SIRD}
& \frac{d S}{dt} = - \beta I S & \nonumber \\
&\frac{d I}{dt} = \beta I S - \gamma I - \alpha I & \nonumber \\
&\frac{d R}{dt} = \gamma I & \nonumber \\
&\frac{d D}{dt} = \alpha I & \nonumber \\
\end{align}\\
%\end{equation}\\

To pursue the aim of fitting the data with simple model, though, consider the evolution of the state variable defined as $f(t)$ in (\ref{eq:ft}). From $f(t) = \alpha I(t)$ and the dynamics of $I(t)$ in (\ref{SIRD}), then\\

\begin{equation}\label{eq:fdelta}
 \frac{d f (t)}{dt} = \beta(t) f(t) S(t) - \gamma f(t) - \alpha f(t) = \delta(t) f(t)  
\end{equation}\\
with $\delta(t) = \beta(t) S(t) - \gamma - \alpha$. Note that $\beta(t)$, that is the coefficient of the rate of infection, has been considered time-varying. In fact, the variation of susceptible individuals, i.e. the whole region population (more than 8 million people in \ARA) at the beginning of the pandemic, might not be the cause of the excess of deaths drastic decreasing. This demises drop should most likely be caused by the containment measures, the lockdown in particular, that reduced the rate of infection.\\ 

Concerning, the evolution of $\delta(t)$, related to the reproduction number, a first analysis of the data showed a behavior typical of first order non-autonomous linear systems, evolving exponentially from positive values at the beginning of the pandemic expansion to negative ones during its decreasing. Thus, denoting with\\

\begin{itemize}
 \item $F$ the set of indices of the 13 regions of mainland France\footnote{Île-de-France (11), Grand Est (44), Auvergne-Rhône-Alpes (84), Hauts-de-France (32), Provence-Alpes-Côte d'Azur (93), Bourgogne-Franche-Comté (27), Occitanie  (76), Pays de la Loire (52), Centre-Val de Loire (24), Bretagne (53), Normandie (28), Nouvelle-Aquitaine (75) and Corse (94)};\\

 \item $f_i(t)$ the time series of the corrected excess of deaths in 2020, obtained as illustrated above, for the $i$-th region with $i \in F$;\\

 \item $\delta_i(t)$ the parameter of equation (\ref{eq:fdelta}), for the $i$-th region with $i \in F$;\\
\end{itemize}

we shall prove that the following simple two-parameter model can quite nicely accounts for the measurements concerning the 13 regions in the mainland France:\\

\begin{align}\label{eq:ID}
 & \frac{d f_i (t)}{dt} = \delta_i(t) f_i(t) & \nonumber \\  
 &\frac{d \delta_i(t)}{dt} = a \delta_i(t) + u & \nonumber \\
\end{align}\\

for every region $i$, with two common parameters, $a$ and $u$. The objective of the identification is to find the optimal values of $a$ and $u$, common to the dynamical systems of all the 13 regions of mainland France, and the initial conditions $f_i(t_0)$ and $\delta_i(t_0)$, for every $i$-th region, such that the simulation of (\ref{eq:ID}) fits the signal {\tt mean excess20 corr}. \\

Hence, denoting:\\

\begin{itemize}
  \item $k_0$ the initial day of the identification interval, i.e. the lockdown date of March 17th;\\
 \item $k_f$ the final day of the identification interval, April 28th;\\
\end{itemize}
the following problem is posed:\\

\begin{align}\label{eq:optprobl}
\min_{a, u, \hat{f}_{i, 0}, \delta_{i, 0}} & \quad \sum_{i \in F} \sum_{k = k_0}^{k_f} q_i  (\hat{f}_i(k) - f_i(k))^2 \nonumber \\
\text{s.t.} \quad & \hat{f}_i(k+1) = (1 + \delta_i(k)) \hat{f}_i(k), \qquad \forall k_0 \leq k \leq k_f, \ \forall i \in F,\nonumber \\
& \delta_i(k+1) = (1 + a) \delta_i(k) + u, \qquad  \forall k_0 \leq k \leq k_f, \  \forall i \in F, \nonumber \\
& \hat{f}_i(k_0) = \hat{f}_{i,0}, \qquad  \forall i \in F, \nonumber \\
& \delta_i(k_0) = \delta_{i,0}, \qquad  \forall i \in F.\nonumber \\
\end{align}\\

Problem (\ref{eq:optprobl}) is a least-squared minimization problem with nonconvex constraints and 28 variables, that are:\\

\begin{itemize}
 \item the two parameters $a$ and $u$;\\
 
 \item the 26 initial conditions $\hat{f}_{i,0}$ and $\delta_{i,0}$, with $i \in F$, for the 13 regions.\\
\end{itemize}

Parameters $q_i$ are chosen to appropriately weight the errors of regions with different values of excess of deaths in 2020. \\

To solve the nonconvex problem (\ref{eq:optprobl}), first an initial suboptimal solution has been obtained by solving a convex problem involving only the dynamics of $\delta_i$, with variables $a, u$ and $\delta_{i,0}$. This solution, together with an {\it ad-hoc} selection of $\hat{f}_{i,0}$, has been used as initial condition of the nonconvex optimization problem. In particular, the Torczon algorithm has been employed, that is a derivative free nonconvex optimization algorithm and whose python implementation can be found at \cite{torczon}.\\

The optimal values for the parameters  $a$ and $u$ in (\ref{eq:ID}) obtained are

\begin{equation}\label{eq:optpar}
 a = -7.1139 \cdot 10^{-2}, \qquad u = -6.2489 \cdot 10^{-3},
\end{equation}

and the optimal initial conditions are given in Table~\ref{tab:table1}. Notice the initial condition of $f_i$ for the two regions most affected by the \C19 spread, i.e. Île-de-France (11) and Grand Est (44), are around 40 units, much higher than those of the other regions.\\

\begin{table}[h!]
  \begin{center}
    \caption{Optimal initial conditions for (\ref{eq:ID}) with region $i \in F$}
    \label{tab:table1}
    \begin{tabular}{|c|c|c|c|c|c|c|c|} 
      \hline
      {$\delta_{84,0}$} & {$\delta_{27,0}$} & {$\delta_{24,0}$} & {$\delta_{44,0}$} & {$\delta_{32,0}$} & {$\delta_{11,0}$} & {$\delta_{75,0}$} \\
      \hline
        $0.3221$ &  $0.2667$ & $0.3598$ &  $0.2054$ &  $0.3032$ &  $0.3211$ &
        $0.3167$\\
      \hline
      \hline
    {$f_{84,0}$} & {$f_{27,0}$} & {$f_{24,0}$} & {$f_{44,0}$} & {$f_{32,0}$} & {$f_{11,0}$} & {$f_{75,0}$}\\
      \hline
        $7.0714$ & $6.277$ & $1.7143$ & $42.8393$ & $7.9161$ & $37.0938$ & $1$\\
      \hline
    \end{tabular}\\
    \vspace*{0.2cm}
    \begin{tabular}{|c|c|c|c|c|c|} 
      \hline
{$\delta_{76,0}$} & {$\delta_{52,0}$} & {$\delta_{53,0}$} & {$\delta_{94,0}$} & {$\delta_{28,0}$} & {$\delta_{93,0}$}\\
      \hline
$0.295$ &  $0.3088$ &  $0.2284$ & $0.1204$ &  $0.3737$ &  $0.2923$\\
      \hline
      \hline
{$f_{76,0}$} & {$f_{52,0}$} & {$f_{53,0}$} & {$f_{94,0}$} & {$f_{28,0}$} & {$f_{93,0}$}\\
      \hline
$3.4643$ & $2.3189$ & $1.8857$ & $1.3255$ & $1.2857$ & $5.2883$\\
      \hline
    \end{tabular}
  \end{center}
\end{table}

\begin{figure}[H]
 \centering
\includegraphics[width=13.6cm]{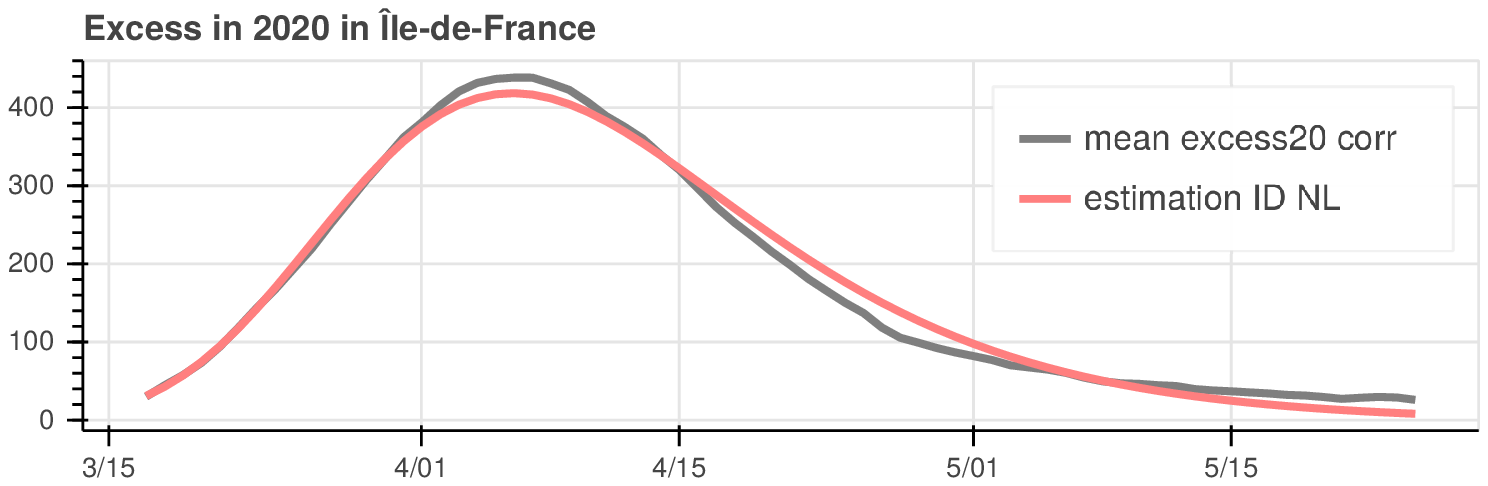}
\includegraphics[width=13.6cm]{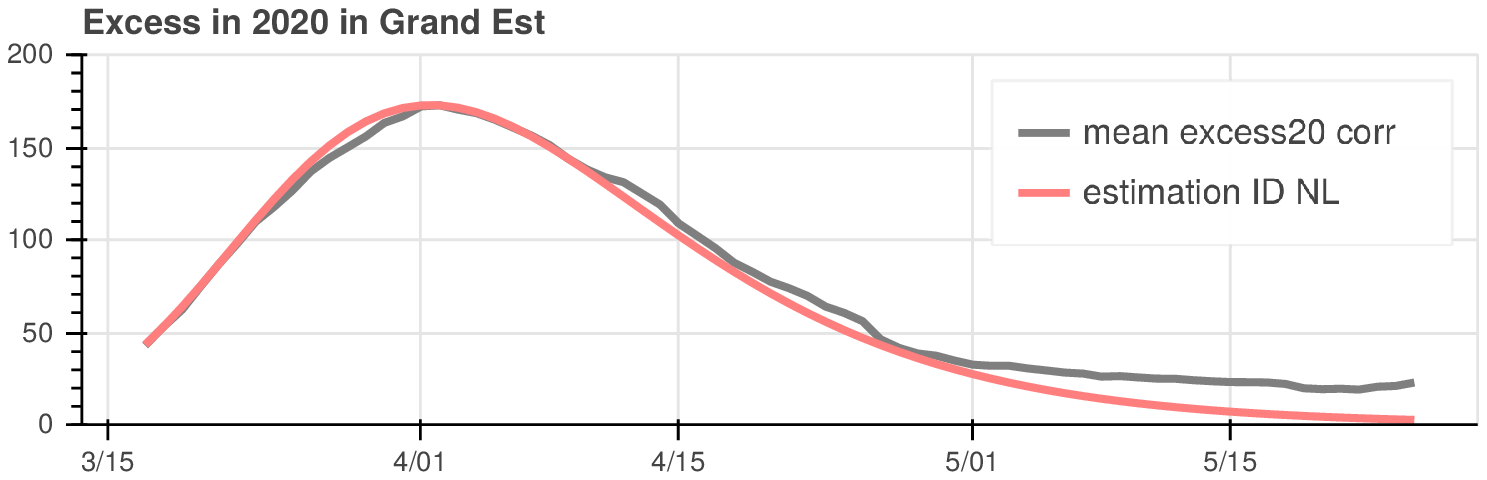}
\includegraphics[width=13.6cm]{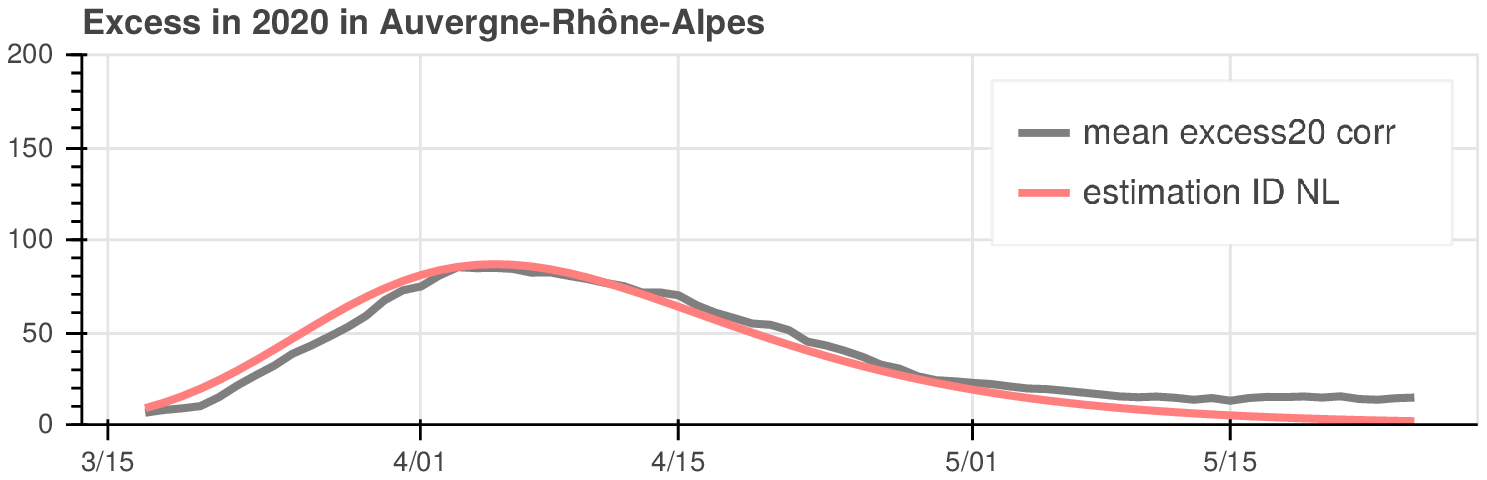}
\includegraphics[width=13.6cm]{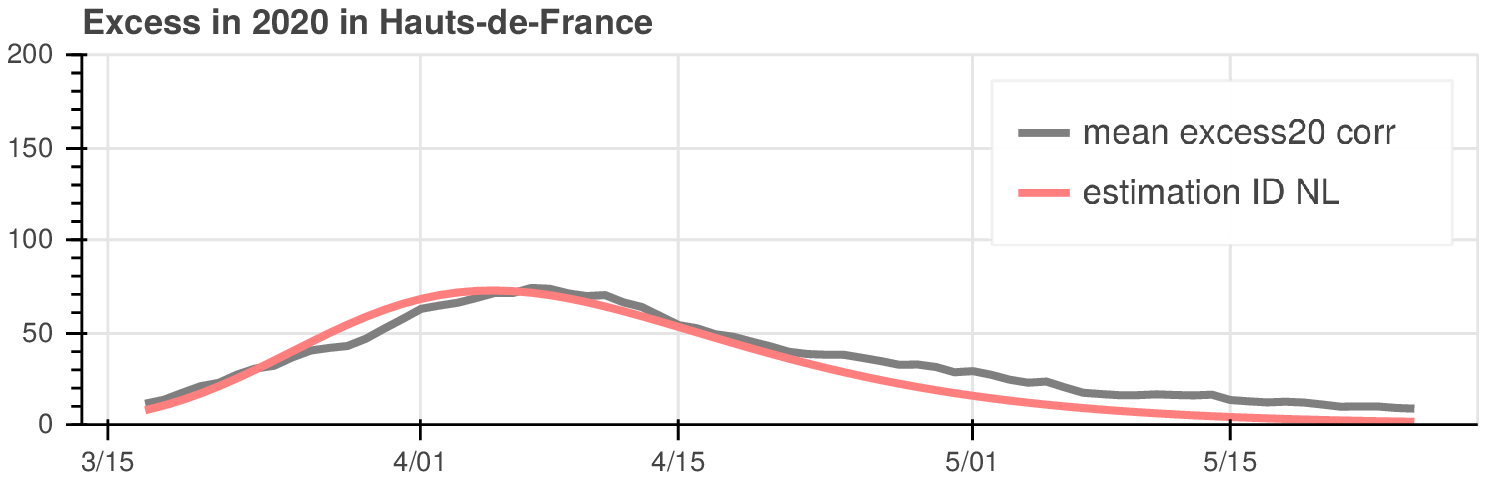}
 \caption{Comparison of time series {\tt mean excess20 corr} and trajectory of  (\ref{eq:ID}) with optimal parameters, denoted {\tt estimation ID NL}, for Île-de-France (11) and Grand Est (44), Auvergne-Rhône-Alpes (84) and Hauts-de-France (32) regions.}\label{figI2}
\end{figure}

\begin{figure}[H]
 \centering
\includegraphics[width=13.6cm]{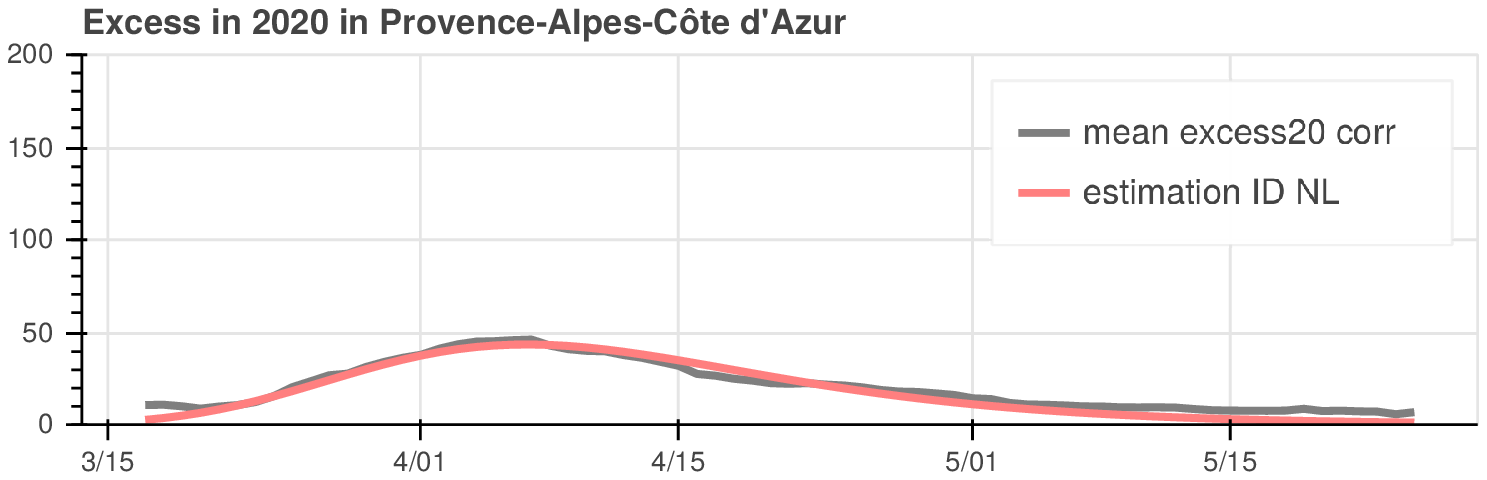}
\includegraphics[width=13.6cm]{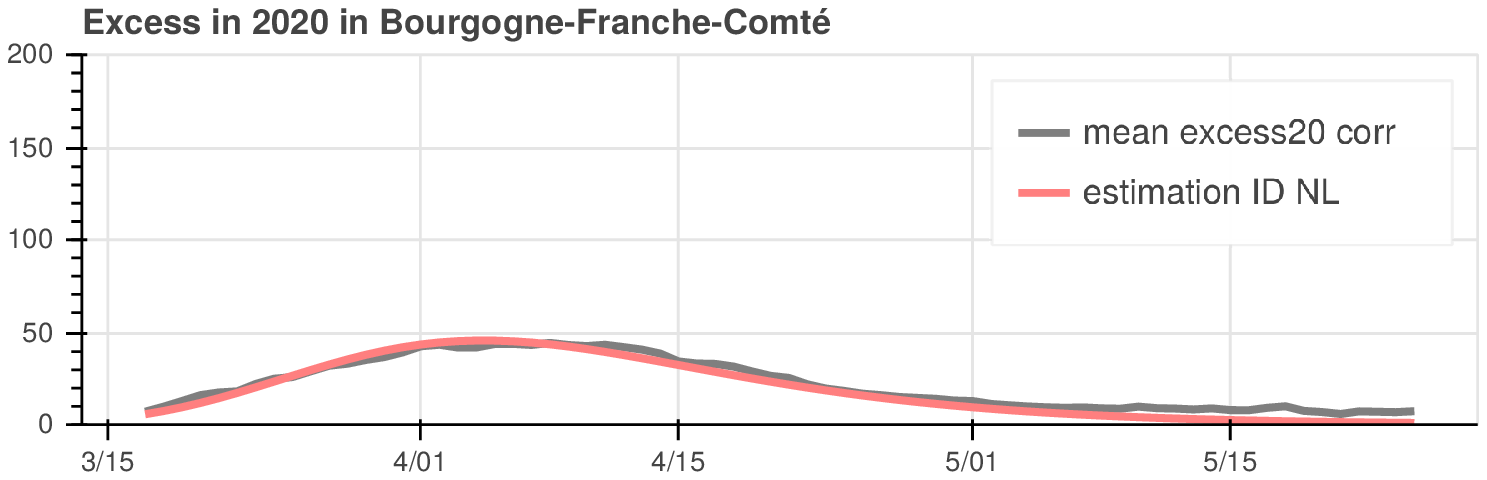}
\includegraphics[width=13.6cm]{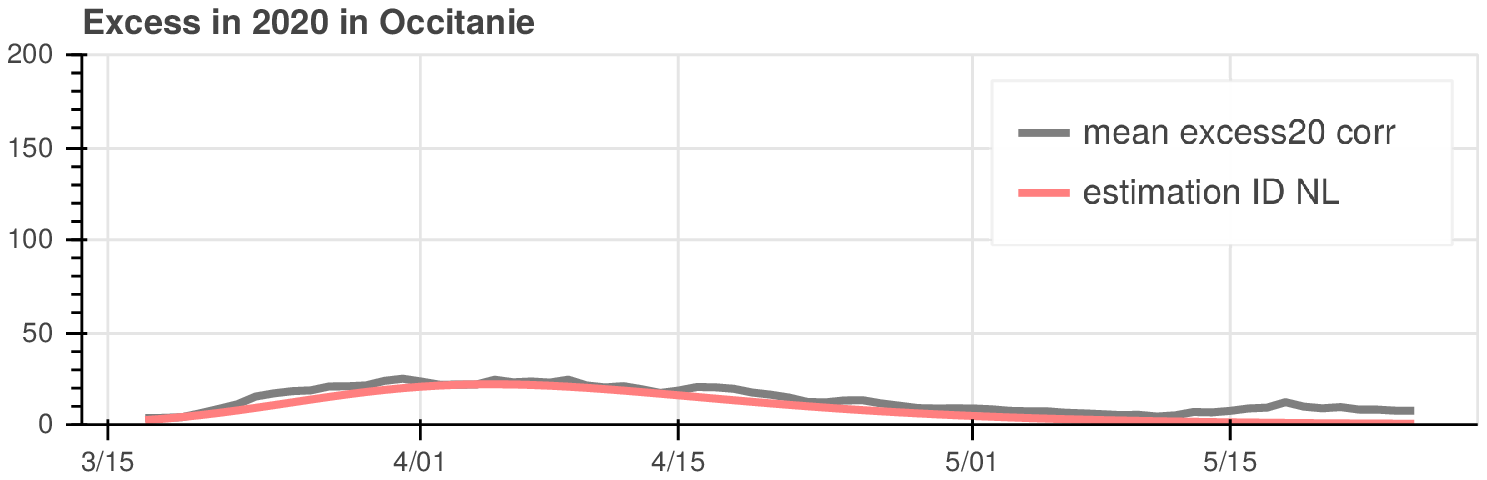}
\includegraphics[width=13.6cm]{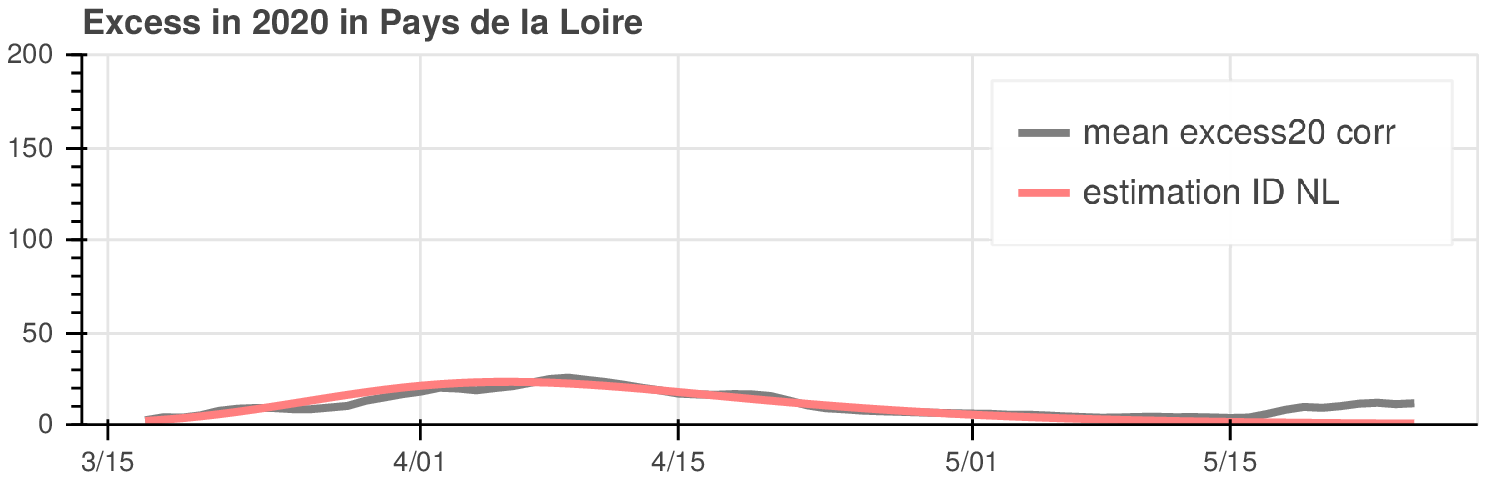}
 \caption{Comparison of time series {\tt mean excess20 corr} and trajectory of  (\ref{eq:ID}) with optimal parameters, denoted {\tt estimation ID NL}, for Provence-Alpes-Côte d'Azur (93) and Bourgogne-Franche-Comté (27), Occitanie  (76) and Pays de la Loire (52) regions.}\label{figI4}
\end{figure}

\begin{figure}[H]
 \centering
\includegraphics[width=13.6cm]{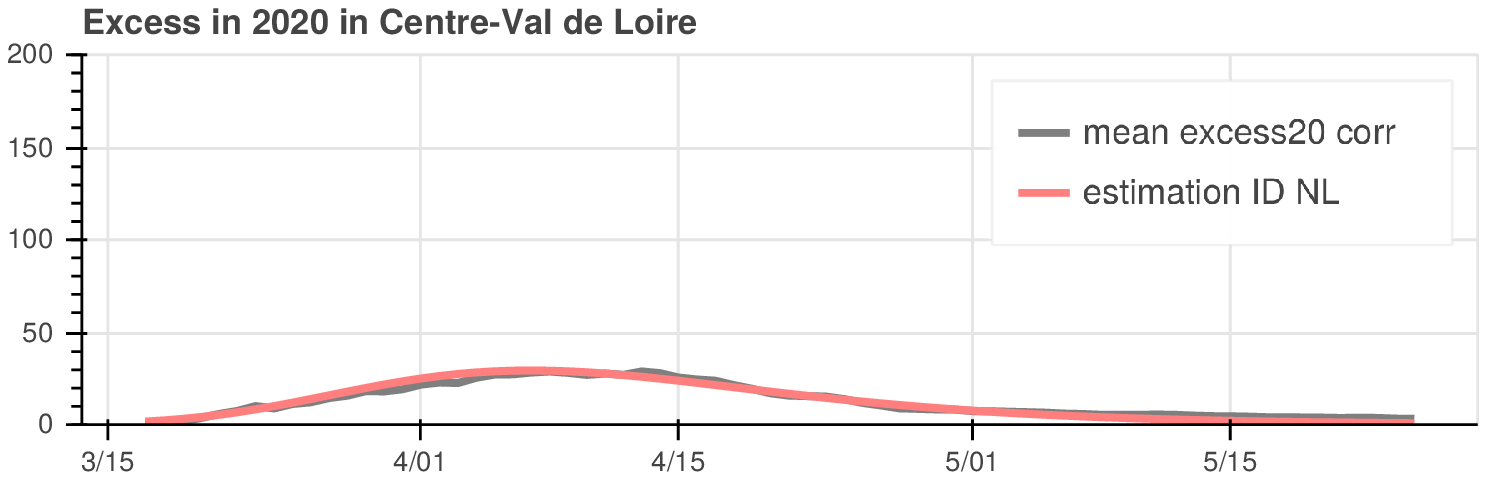}
\includegraphics[width=13.6cm]{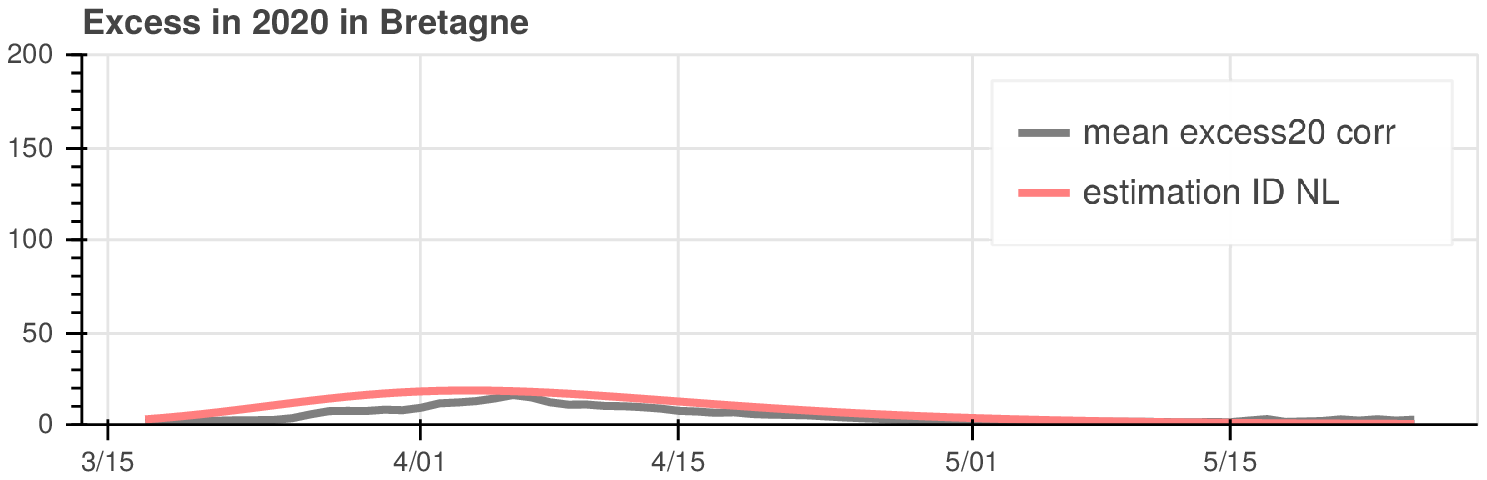}
\includegraphics[width=13.6cm]{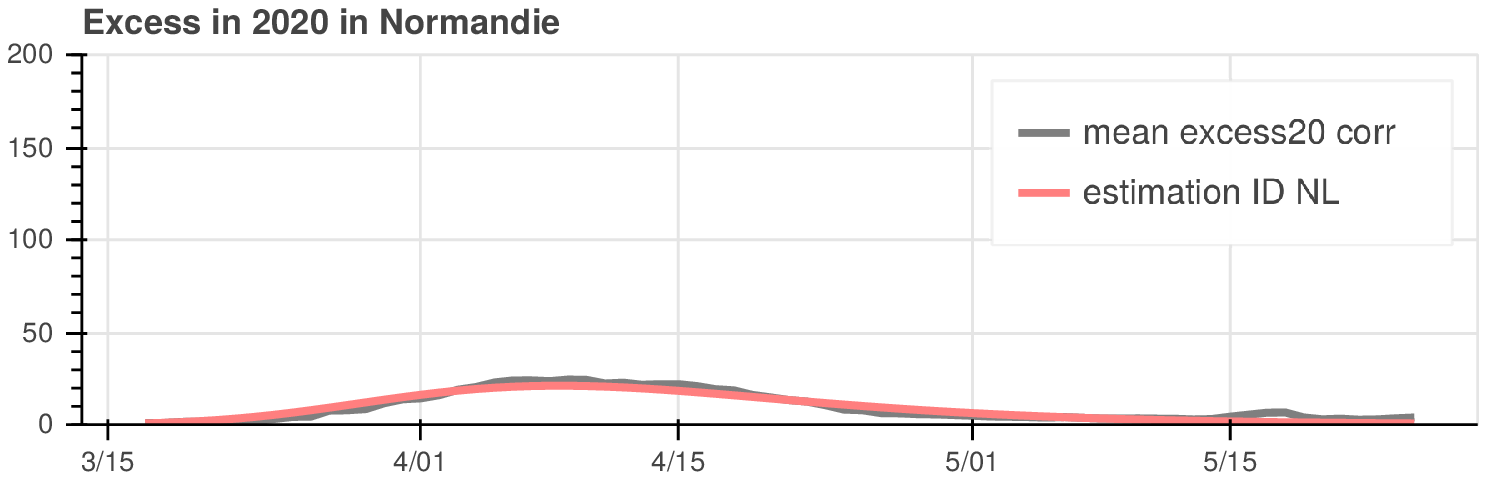}
\includegraphics[width=13.6cm]{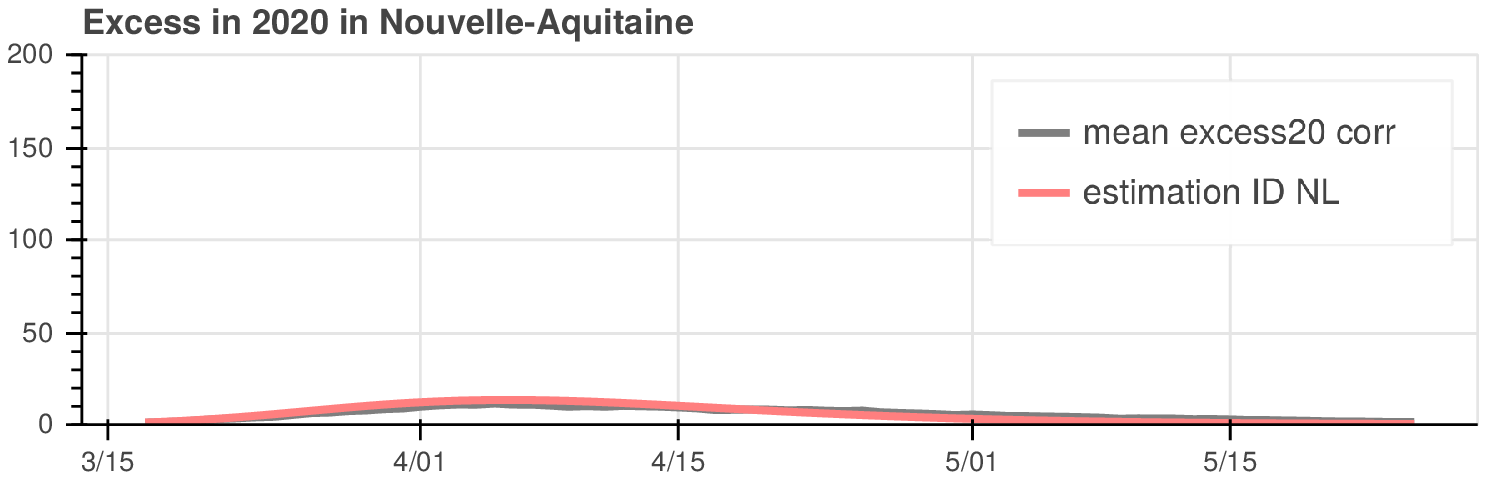}
 \caption{Comparison of time series {\tt mean excess20 corr} and trajectory of  (\ref{eq:ID}) with optimal parameters, denoted {\tt estimation ID NL}, for Centre-Val de Loire (24) and Bretagne (53), Normandie (28) and Nouvelle-Aquitaine (75) regions.}\label{figI6}
\end{figure}

\begin{figure}[H]
 \centering
\includegraphics[width=13.6cm]{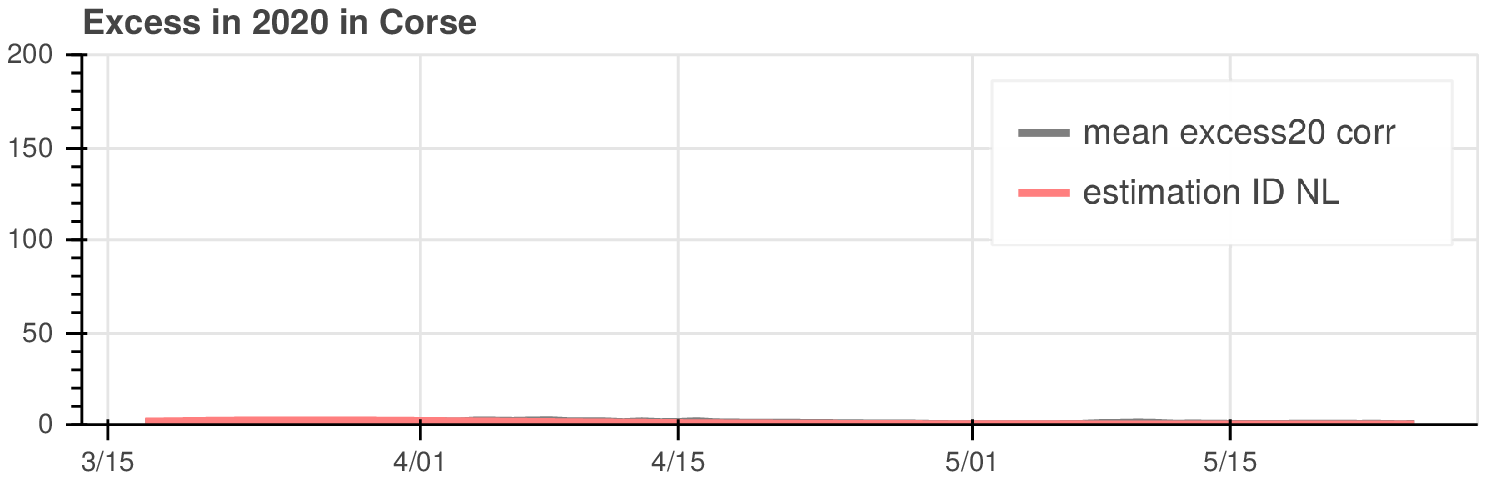}
\includegraphics[width=13.6cm]{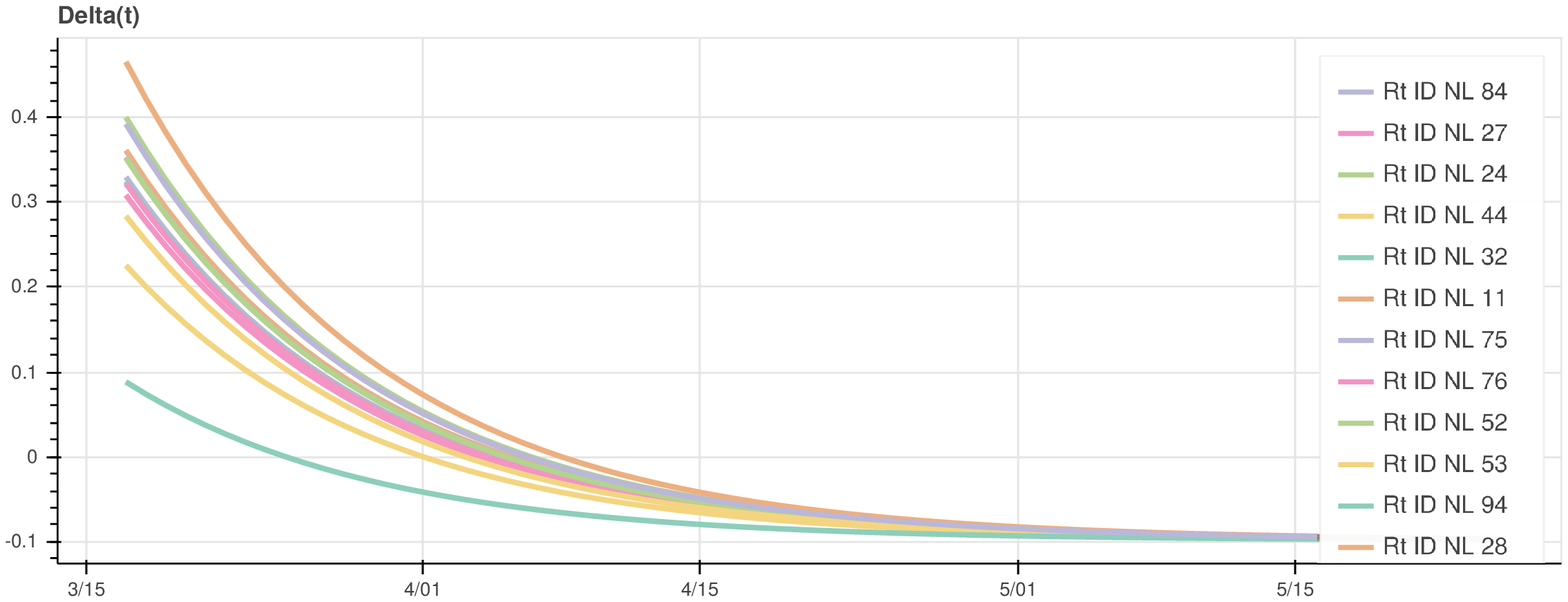}
 \caption{Comparison of time series {\tt mean excess20 corr} and trajectory of  (\ref{eq:ID}) with optimal parameters, denoted {\tt estimation ID NL}, for Corse (94) region, and evolutions of $\delta_i(t)$.}\label{figI7}
\end{figure}

The results of simulating system (\ref{eq:ID}) with parameters (\ref{eq:optpar}) and initial conditions as in Table~\ref{tab:table1}, denoted {\tt estimation ID NL}, are illustrated in Figs.~\ref{figI2} -\ref{figI7}. Notice that the $y$ axis in all the figures ranges between $0$ and $200$, except Île-de-France that reaches $450$. The fitting, obtained with a two-parameters two dimensional system, is remarkable.\\

\subsection{Discussion on identifiability and forecast}

Although the model that proved to reproduce the evolution of the \C19 spread in the 13 French regions is extremely simple, with only two parameters and two states, the system involving the infected cases is nonidentifiable. In fact, it is easy to see that every pair of time series $I_i(t)$ and parameter $\alpha$ such that \\

\begin{equation*}
f_i(t) = \alpha I_i(t) 
\end{equation*}\\

for the $i \in F$ region, leads to the same trajectory of system (\ref{eq:ID}). Then, from the observation of the 2020 deaths access only it is not possible to identify the real value of $\alpha$ and, then, at the same time the time series of the infected population.\\

The perspective is even worse for what concerns the forecast power of system (\ref{eq:ID}). Indeed, the state evolution depends on the value of $\delta_i(t)$, that can be seen as a time-varying parameter. In the case under analysis, its evolution, caused most likely by the containment measures, the lockdown application and the people habits adaptation, has been identified with satisfactory precision. This would imply, then, that the effect of complete lockdown seems to be modelled by the specific values of $a$ and $u$, resulting from the optimization process. On the other hand, it is not clear how to predict the future evolution of $\delta_i(t)$ for $i \in F$, whose knowledge is unavoidable to forecast the evolution of the number of demises due to the pandemic. Further knowledge, on historic data, epidemiology and clinical data, as done for influenza pandemic for instance \cite{nishiura2009early}, might help to achieve this crucial task.\\

\section{Model validation on further data}\label{sec:valid}
In the previous sections, the model has been fitted with the data regarding the daily deaths that can be considered as consequences of the \C19 pandemic. As several other datasets are available from gouvernamental sources, though, that might be related to the pandemic evolution, we aim at check whether simple relations between these data and the identified model exist. In the affirmative case, this would validate the fitted model.\\

As said above, many time series regarding the \C19 daily evolution in France are available. In particular, in \cite{datagouv}, the daily data on the \C19 can be found, between mid-March and mid-June 2020, regarding: \\
\begin{itemize}
 \item the number of new hospitalizations, signal {\tt incid\_hosp}; \\
 \item the number of patients entering the intensive care services, denoted {\tt incid\_rea}; \\
 \item the number of patient demises, {\tt incid\_dc};\\
 \item and the number of recovered patients, {\tt incid\_rad}. \\
\end{itemize}

\begin{figure}[H]
 \centering
\includegraphics[width=13.6cm]{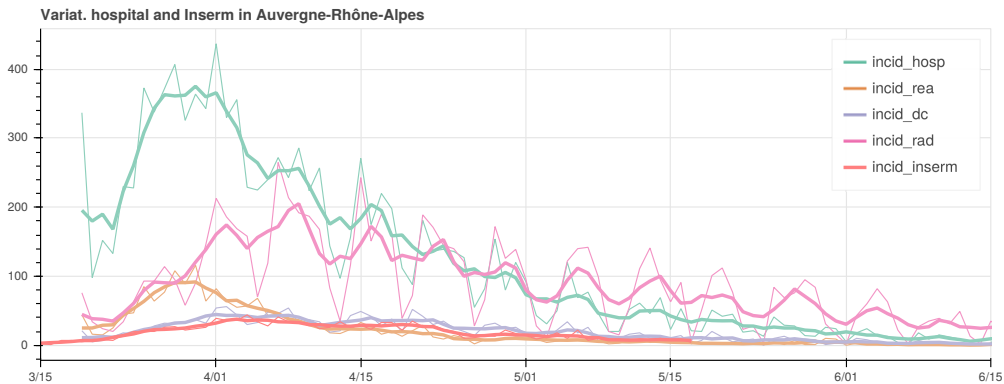}
 \caption{Daily evolution of: new hospitalizations ({\tt incid\_hosp}); patients entering in intensive care ({\tt incid\_rea}); patient demises ({\tt incid\_dc}); recovered patients ({\tt incid\_rad}) and electronically certified demises, ({\tt incid\_inserm}), in Auvergne-Rhône-Alpes. In bold, the average over a 5-days-long window.}\label{fig1}
\end{figure}

Fig.~\ref{fig1} shows the evolution of the time series in the region Auvergne-Rhône-Alpes, in thin lines the data from the files are depicted, while in bold the signals smoothed by averaging the values over 5 days are drown. \\

Moreover, Fig.\ref{fig1} shows also the time series, \cite{datagouv_inserm}, of the deaths for  \C19 that have been electronically certified by the CepiDc laboratory, which is the Inserm laboratory in charge of elaborating the statistics regarding the medical demises. This time series, denoted {\tt incid\_inserm} in Fig.~\ref{fig1}, concerns the deaths occurred in public or private hospitals, nursing homes, or other places, that have been electronically registered. This time series considers also part of the cases included in {\tt incid\_dc}. Recall that the average of time series {\tt incid\_dc} and {\tt incid\_inserm} have been employed in the definition of the signal {\tt mean excess20 corr}, on which the model has been fitted, also see Fig.~\ref{fig5} and Fig.~\ref{fig5b}.\\

\begin{figure}[H]
 \centering
\includegraphics[width=13.6cm]{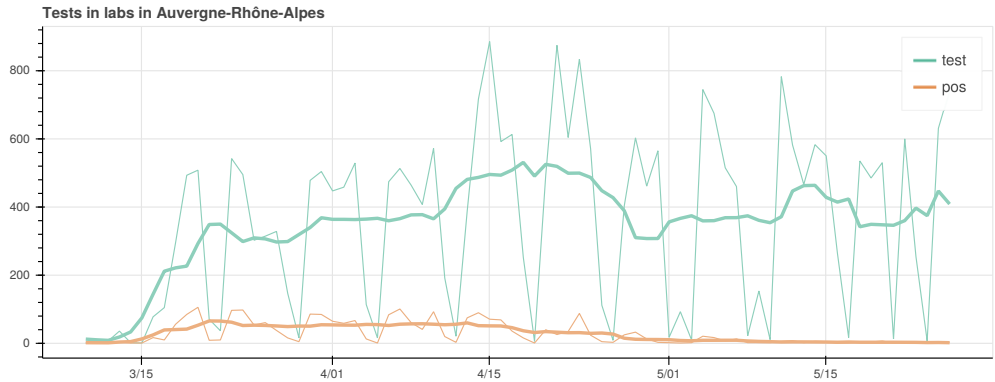}
 \caption{Daily evolution of \C19 test ({\tt test}) and tests resulted positive ({\tt pos}), in Auvergne-Rhône-Alpes. In bold, the average over a 5-days-long window.}\label{fig2}
\end{figure}

The site of the French government gives also to access to the data concerning the daily \C19 tests realized in every region and to the number of positive tests, see \cite{datagouv_test}. The evolution of the time series concerning the \ARA region, smoothed by averaging it over a 5-days-long window, are depicted in Fig.~\ref{fig2}.\\

At the site \cite{datagouv_sosurg}, files are available reporting, for every French region and every day since mid-February, the data concerning the accesses to the hospital emergency services and the activity of SOS Médecins, an association of doctors providing emergency assistance. In particular, the data series regard:\\

\begin{itemize}
 \item the overall number of admissions to hospital emergency services,
 
 {\tt nbre\_pass\_tot};\\
 
 \item the number of admissions to emergency for presumed \C19 cases, 
 
 {\tt nbre\_pass\_corona};\\
 
 \item the number of hospitalizations among the admissions to emergency for presumed \C19 cases, {\tt nbre\_hospit\_corona};\\
 \item the overall number of medical acts of SOS Médecins, {\tt nbre\_act\_tot};\\
 \item the number of medical acts of SOS Médecins of presumed \C19 cases, {\tt nbre\_act\_corona}.\\
\end{itemize}

\begin{figure}[H]
 \centering
\includegraphics[width=13.6cm]{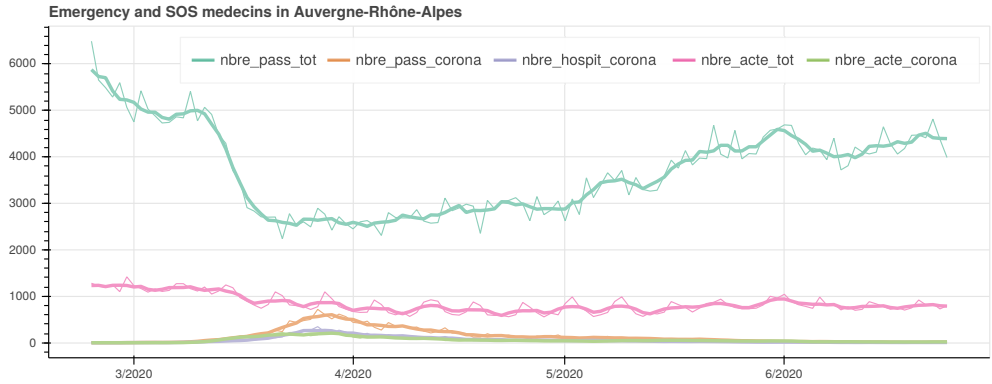}
 \caption{Daily evolution of: admissions to emergency, {\tt nbre\_pass\_tot};
admissions to emergency for \C19, {\tt nbre\_pass\_corona}; hospitalizations among the admissions to emergency for \C19, {\tt nbre\_hospit\_corona}; medical acts of SOS Médecins, {\tt nbre\_act\_tot};
medical acts of SOS Médecins of \C19 cases, {\tt nbre\_act\_corona}. In bold, the average over a 5-days-long window.}\label{fig3}
\end{figure}

As can be noticed in Fig~\ref{fig3}, the number of admissions to emergency and the SOS Médecins, in \ARA region, dropped around the middle of March, when the \C19 epidemic spread started in France and the lockdown was imposed by the French health authorities, on March 17. Nevertheless, as can be remarked from Fig~\ref{fig3}, the number of presumed or overt \C19 cases started consistently growing at the same time in \ARA region. Analogous data evolution can be observed in several French regions, where the \C19 pandemic spread relevantly.  \\

Consider here the $i$-th region, the index is omitted for notation simplicity. Suppose that the number of hospitalizations, that is the time series {\tt incid\_hosp} in Fig.~\ref{fig1} denoted here as $H(t)$, is proportional to the number of infected individuals at a previous instant:\\

\begin{equation*}
H(t) = \mu f(t - \eta).\\
\end{equation*}\\

The time series $f(t)$ being known, parameters $\mu$ and $\eta$ can be identified by simple least-squared minimization, repeated for different values of $\eta$. The same can be done for the other time series illustrated in Section~\ref{sec:data}, that is \\

\begin{itemize}
 \item new hospitalizations, {\tt incid\_hosp}; \\
 \item new intensive cares patients, {\tt incid\_rea}; \\
 \item daily patient demises, {\tt incid\_dc};\\
 \item recovered patients, {\tt incid\_rad}, \\
 \item positive tested cases, {\tt pos}, \\
 \item daily demises certified to Inserm, {\tt incid\_inserm}, \\
 \item admissions to emergency for presumed \C19 cases, {\tt nbre\_pass\_corona} (denoted {\tt pass\_corona} in the tables);\\
 \item new hospitalizations among the admissions to emergency for presumed \C19 cases, {\tt nbre\_hospit\_corona} (denoted {\tt hospit\_corona} in the tables);\\
 \item medical acts of SOS Médecins of presumed \C19 cases, {\tt nbre\_act\_corona} (denoted {\tt act\_corona} in the tables)\\
\end{itemize}

The results of the fitting for the three regions that experienced the most relevant pandemic spread, i.e. Île-de-France (11), Grand Est (44) and Auvergne-Rhône-Alpes (84) are shown in Sections~\ref{ssec:IdF}, \ref{ssec:GE}, and \ref{ssec:ARA}, respectively, those concerning the other regions can be found in  \ref{app:valid}.\\

The optimal values of parameters $\eta$ and $\mu$ for the three regions are given in Tables~\ref{tab:table2} and \ref{tab:table3}. It can be noticed that the curves of the excess of deaths are reproduced, scaled and shifted in time, in the measured time series. A coherent sequence of event could be inferred from the values of $\eta$. It seems, in fact, that the peak of SOS Médecins acts occurred first, followed three days later by the peaks at the emergency services; then the peaks at intensive care units happened followed by the peaks of hospitalizations. Afterward, the peak of the excess of deaths, i.e. of $f_i$, occurred, while the peaks of the deaths registered at hospitals and by Inserm came, few days later. Finally, the curve of recovered patients arrived.\\

It is also interesting to note that, while the temporal sequence are substantially the same for Île-de-France (11) and Auvergne-Rhône-Alpes (84), in Grand Est (44) region, where the most important and early \C19 cluster revealed, the events appear to be compressed in time, the peaks at the emergency units, for SOS Médecins and for positive test occurred around 5 days before than the other two regions. \\

Moreover, from the comparison between the values of $\mu$ parameters, that are the proportionality coefficients between the evolution of excess of deaths and the considered time series, interesting informations can be recovered, in particular on the degree of stress and saturation of the regional health systems. Considering for instance the acts of SOS Médecins, while there have been almost 2 acts for every demise of \C19 in \ARA, in Grand Est the number of deaths have been higher than the capability of the institution diagnosis, while in Île-de-France the number of demises seems to be more than the double of the SOS Médecins acts. Analogous conclusions can be drawn by other indicators: the activity of the emergency units ({\tt pass\_corona} and {\tt hospit\_corona}); the number of positive test; the hospitalization and intensive care accesses, in the three regions. It appear evident the degree of overwhelming saturation of the health systems of Grand Est and, even more dramatically, of Île-de-France region.
% Île-de-France (11) and Grand Est (44) regions
% Auvergne-Rhône-Alpes (84) and Hauts-de-France (32)
% Provence-Alpes-Côte d'Azur (93) and Bourgogne-Franche-Comté (27)
% Occitanie  (76) and Pays de la Loire (52)
% Centre-Val de Loire (24) and Bretagne (53)
% Normandie (28) and Nouvelle-Aquitaine (75)
% Corse (94) region
% 

\begin{table}[h!]
  \begin{center}
    \caption{Optimal initial conditions for (\ref{eq:ID}) with region $i \in F$}
    \label{tab:table2}
    \begin{tabular}{|c|c|c|c|c|c|} 
      \hline
& {\tt incid\_hosp} & {\tt incid\_rea} & {\tt incid\_dc} & {\tt incid\_rad} & {\tt incid\_inserm}\\
      \hline
      \hline
 $\eta_{11} $ & $4$ &  $6$ & $-2$ &  $-5$ &  $-1$\\
      \hline  
 $\mu_{11} $  & $2.890469$ & $0.530296$ & $0.474127$ & $0.951683$ & $0.275693$\\
      \hline
      \hline
 $\eta_{44} $ & $3$ &  $5$ & $-1$ &  $-5$ &  $0$\\
      \hline  
$\mu_{44}$ & $2.846007$ & $0.524773$ & $0.505064$ & $0.966889$ & $0.331267$\\
      \hline
      \hline
 $\eta_{84} $ & $5$ &  $6$ & $-3$ &  $-3$ &  $-2$\\
      \hline  
$\mu_{84}$ & $3.430858$ & $0.806820$ & $0.399725$ & $1.261361$ & $0.343933$\\ 
      \hline
    \end{tabular}
  \end{center}
\end{table}

\begin{table}[h!]
  \begin{center}
    \caption{Optimal initial conditions for (\ref{eq:ID}) with region $i \in F$}
    \label{tab:table3}
    \begin{tabular}{|c|c|c|c|c|} 
      \hline
& {\tt pos} & {\tt pass\_corona} & {\tt hospit\_corona} & {\tt act\_corona}\\
      \hline
      \hline
 $\eta_{11} $ & $5$ & $7$ & $ 7 $ & $11$\\
      \hline  
 $\mu_{11} $  & $1.496454$ & $1.053246$ & $0.470787$ & $0.385416$\\
      \hline
      \hline
 $\eta_{44} $ & $-1$ & $2$ & $ 3 $ & $6$\\
      \hline  
$\mu_{44}$ & $0.853961$ & $2.487286$ & $1.116551$ & $0.859851$\\
      \hline
      \hline
 $\eta_{84}$ & $5$ & $6$ & $ 7 $ & $10$\\
      \hline  
$\mu_{84}$ & $0.693771$ & $5.060418$ & $2.269344$ & $1.831710$\\ 
      \hline
    \end{tabular}
  \end{center}
\end{table}

\vspace*{10cm}

\newpage
\subsection{Île-de-France}\label{ssec:IdF}

\vspace{-0.2cm}

\begin{figure}[H]
 \centering
\includegraphics[width=6.8cm]{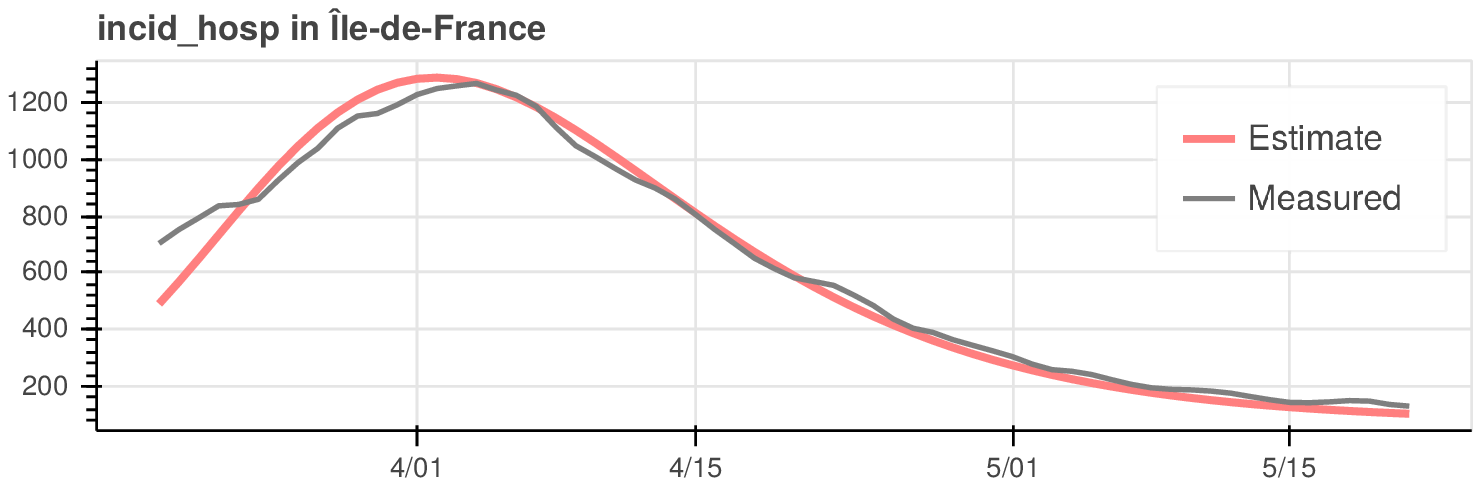}
\includegraphics[width=6.8cm]{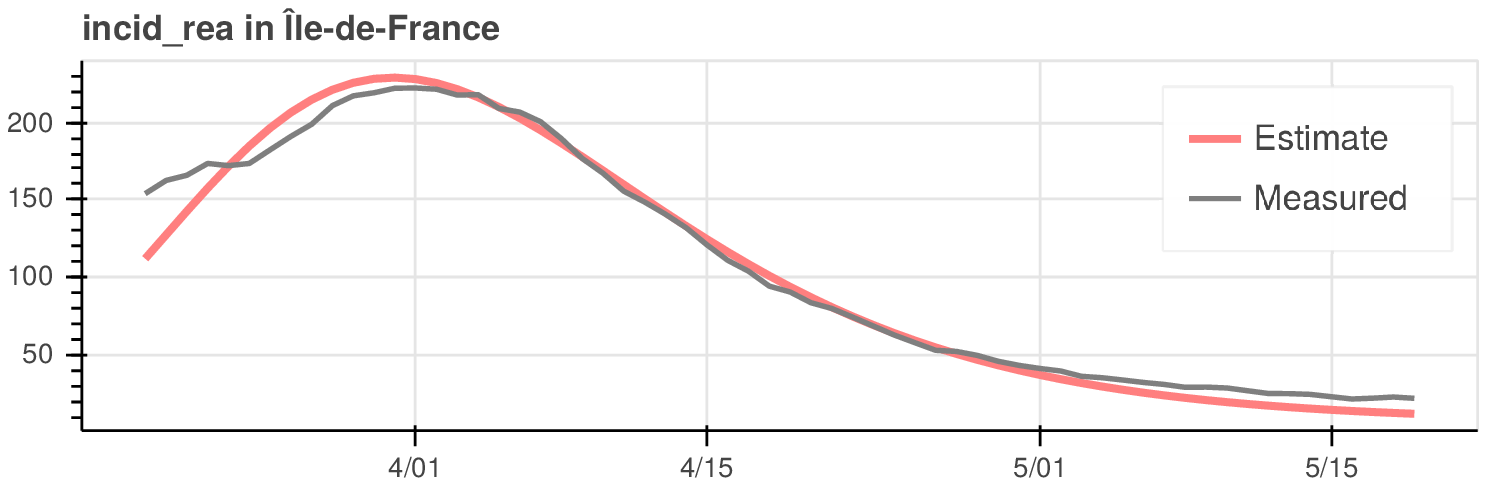}
 \vspace*{-0.7cm} \caption{Comparison of time series {\tt incid\_hosp} and {\tt incid\_rea} averaged over a 14-days-long window, with the fitted outputs, for Île-de-France (11) region.}
\end{figure}

\vspace{-0.7cm}

\begin{figure}[H]
 \centering
\includegraphics[width=6.8cm]{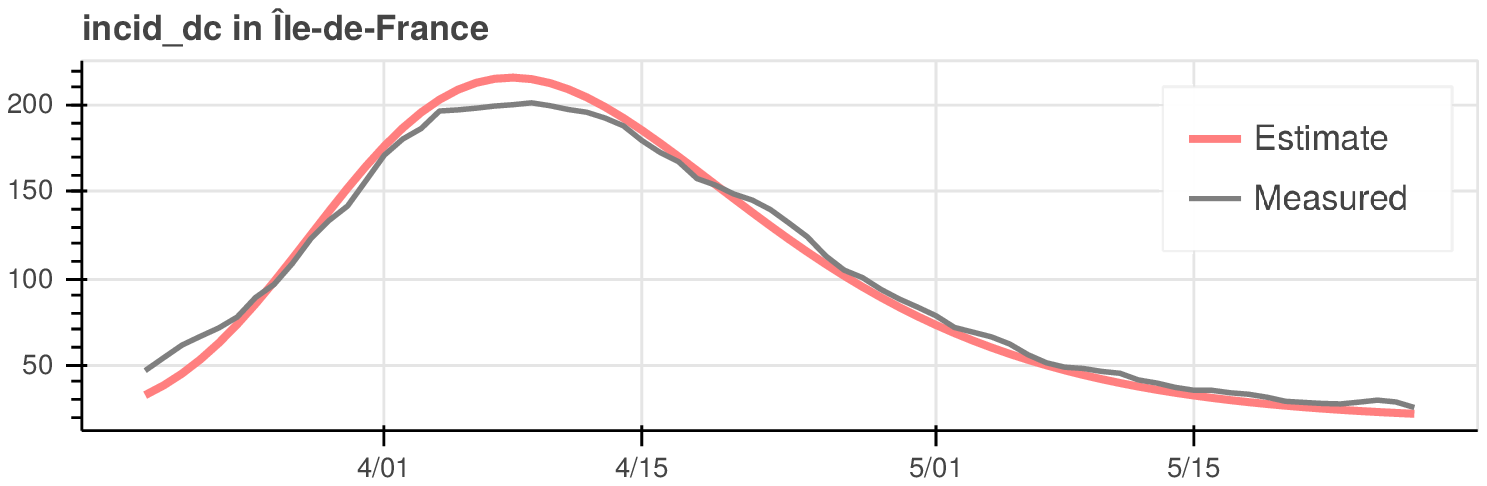}
\includegraphics[width=6.8cm]{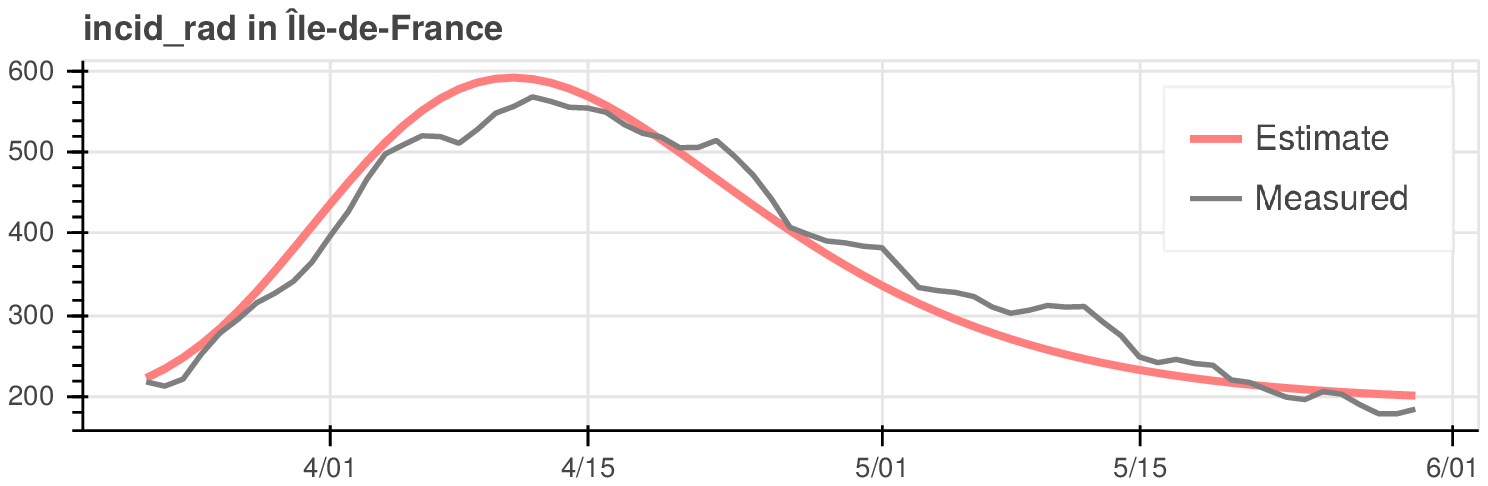}
\vspace{-0.7cm}
 \caption{Comparison of time series {\tt incid\_dc} and {\tt incid\_rad} averaged over a 14-days-long window, with the fitted outputs, for Île-de-France (11) region.}
\end{figure}

\vspace{-0.7cm}

\begin{figure}[H]
 \centering
\includegraphics[width=6.8cm]{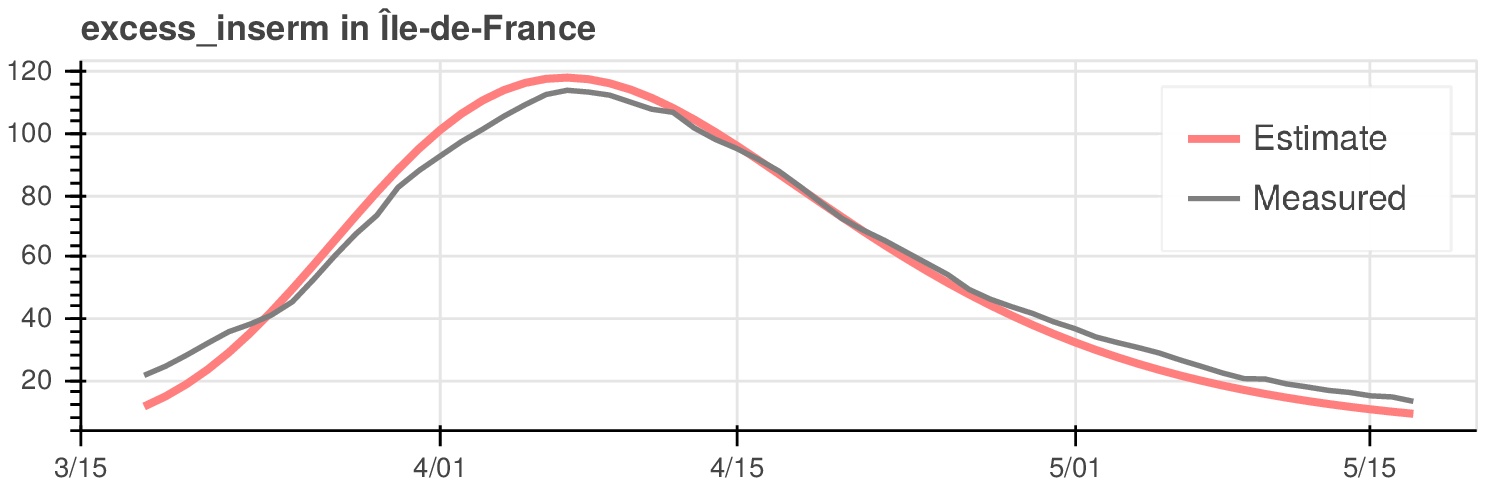}
\includegraphics[width=6.8cm]{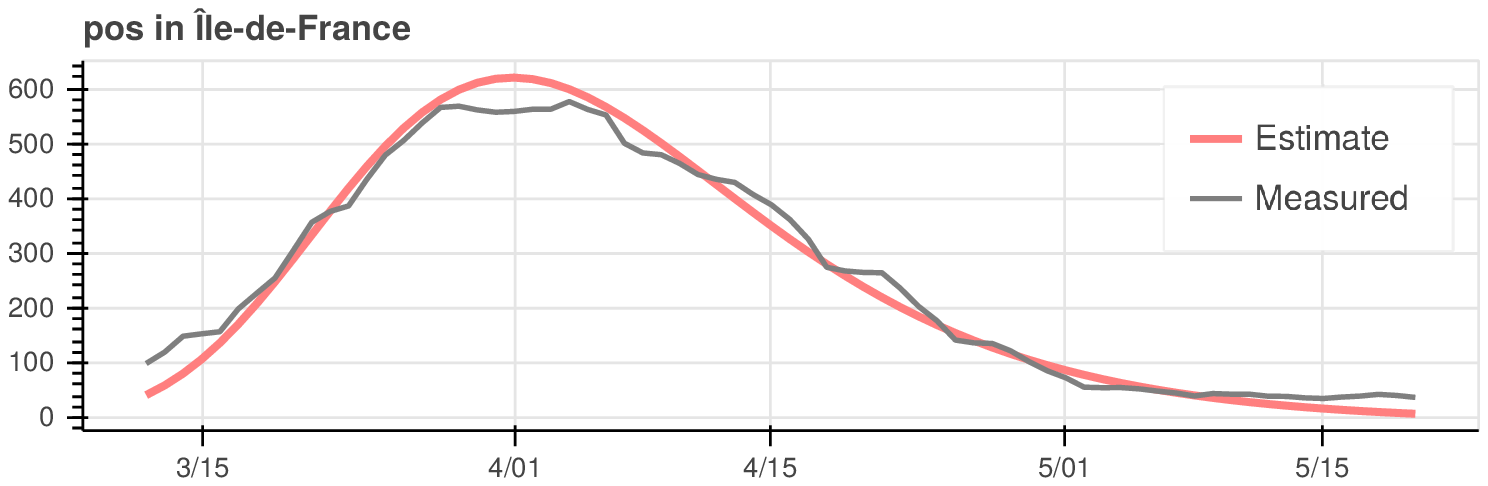}
\vspace{-0.7cm}
 \caption{Comparison of time series {\tt incid\_inserm} and {\tt pos} averaged over a 14-days-long window, with the fitted outputs, for Île-de-France (11) region.}
\end{figure}

\vspace{-0.7cm}

\begin{figure}[H]
 \centering
\includegraphics[width=6.8cm]{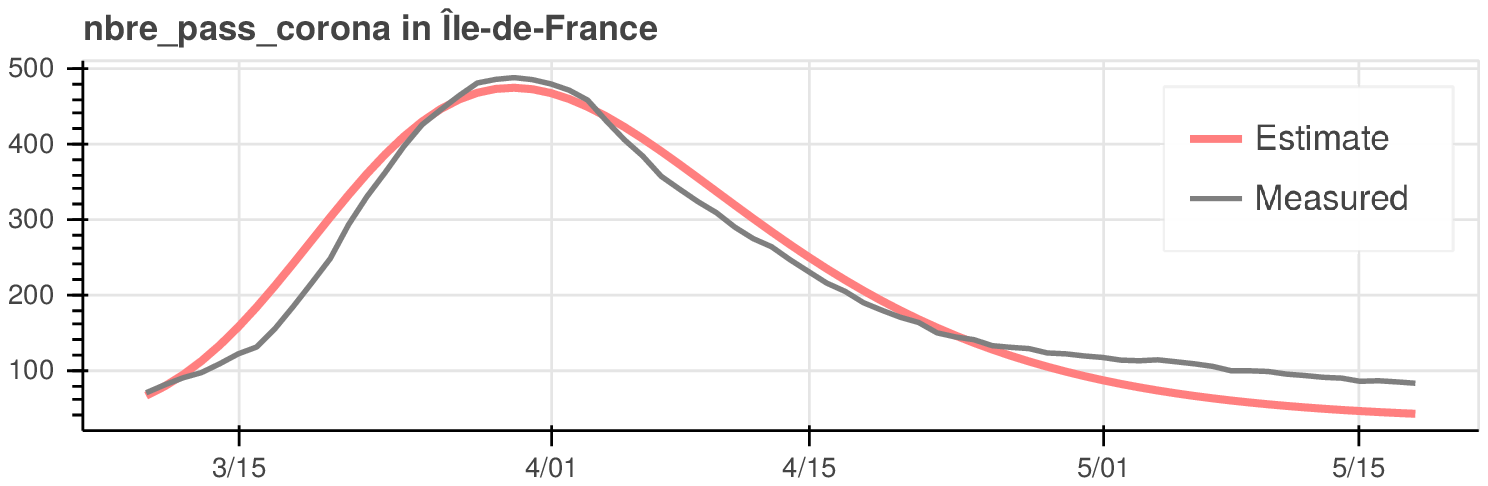}
\includegraphics[width=6.8cm]{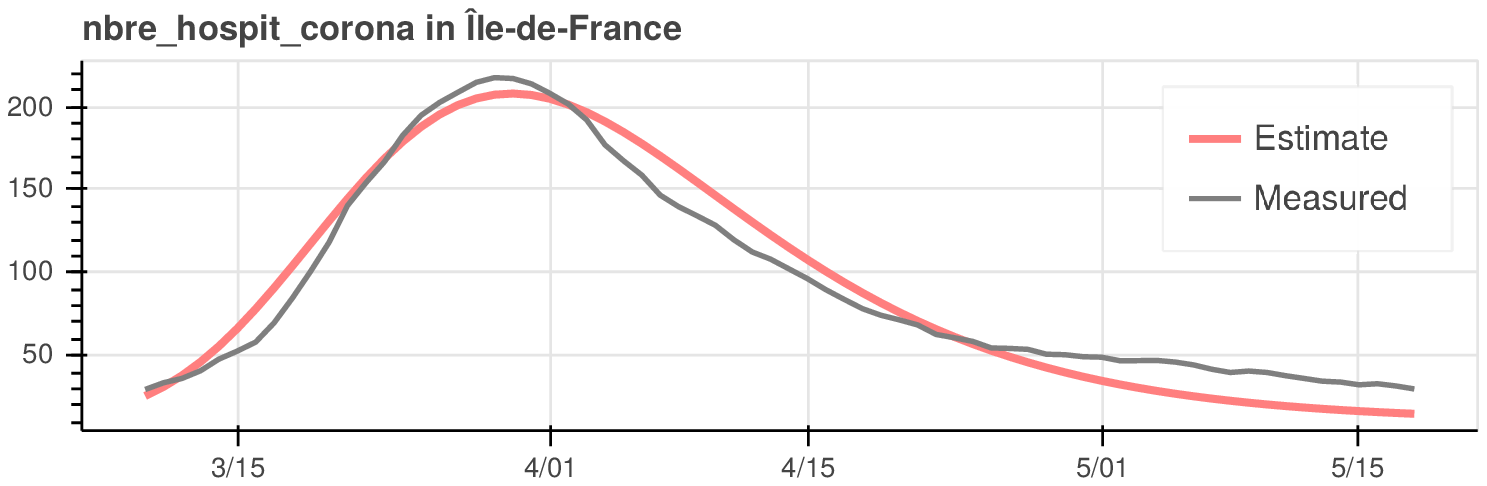}
\vspace{-0.7cm}
 \caption{Comparison of time series {\tt nbre\_pass\_corona} and 
 {\tt nbre\_hospit\_corona} averaged over a 14-days-long window, with the fitted outputs, for Île-de-France (11) region.}
\end{figure}
\vspace{-0.7cm}

\begin{figure}[H]
 \centering
\includegraphics[width=6.8cm]{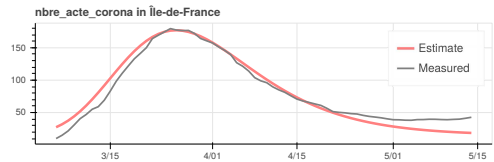}
\vspace{-0.3cm}
 \caption{Comparison of time series {\tt nbre\_acte\_corona} averaged over a 14-days-long window, with the fitted output, for Île-de-France (11) region.}
\end{figure}

\subsection{Grand Est}\label{ssec:GE}

\vspace{-0.2cm}

\begin{figure}[H]
 \centering
\includegraphics[width=6.8cm]{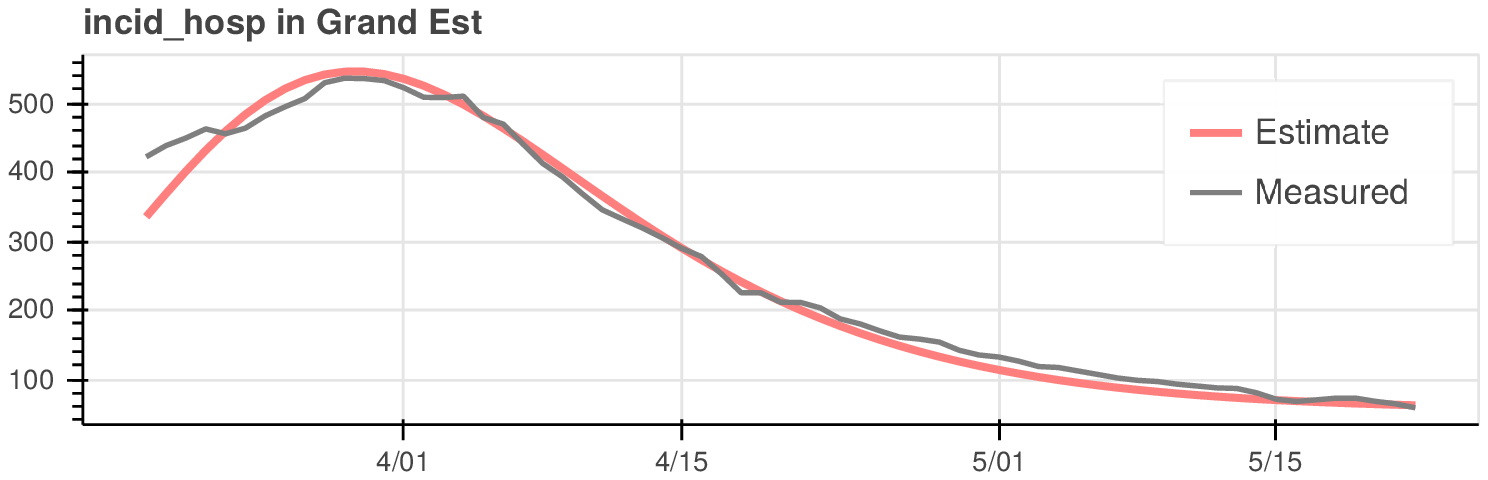}
\includegraphics[width=6.8cm]{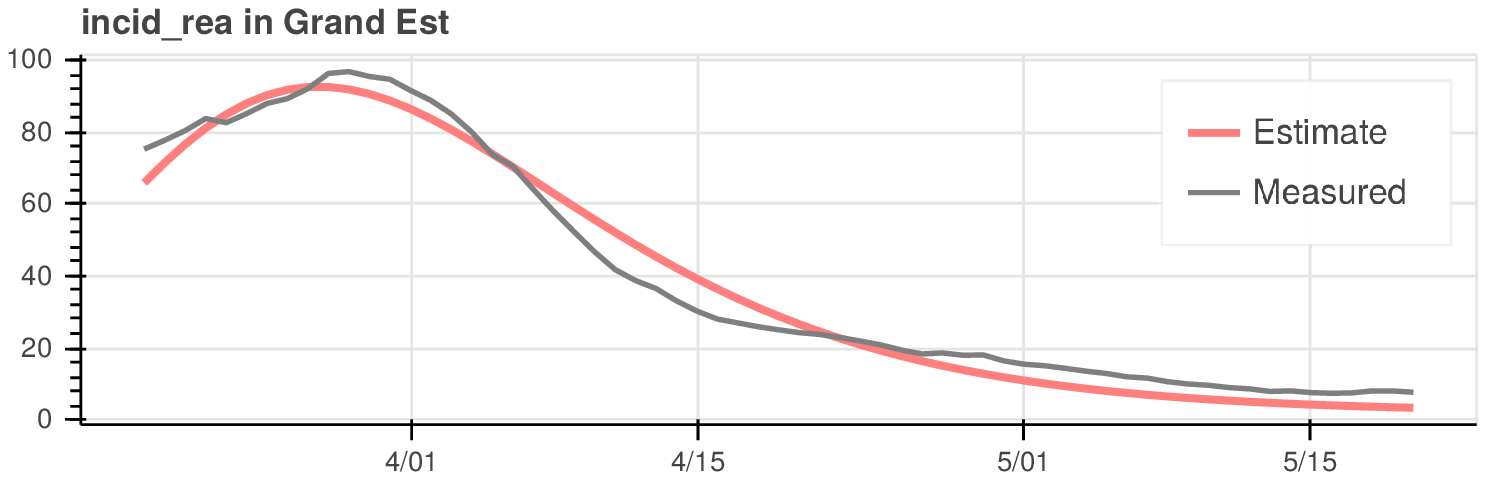}
\vspace{-0.7cm}
 \caption{Comparison of time series {\tt incid\_hosp} and {\tt incid\_rea} averaged over a 14-days-long window, with the fitted outputs, for Grand Est (44) region.}
\end{figure}

\vspace{-0.7cm}
\begin{figure}[H]
 \centering
\includegraphics[width=6.8cm]{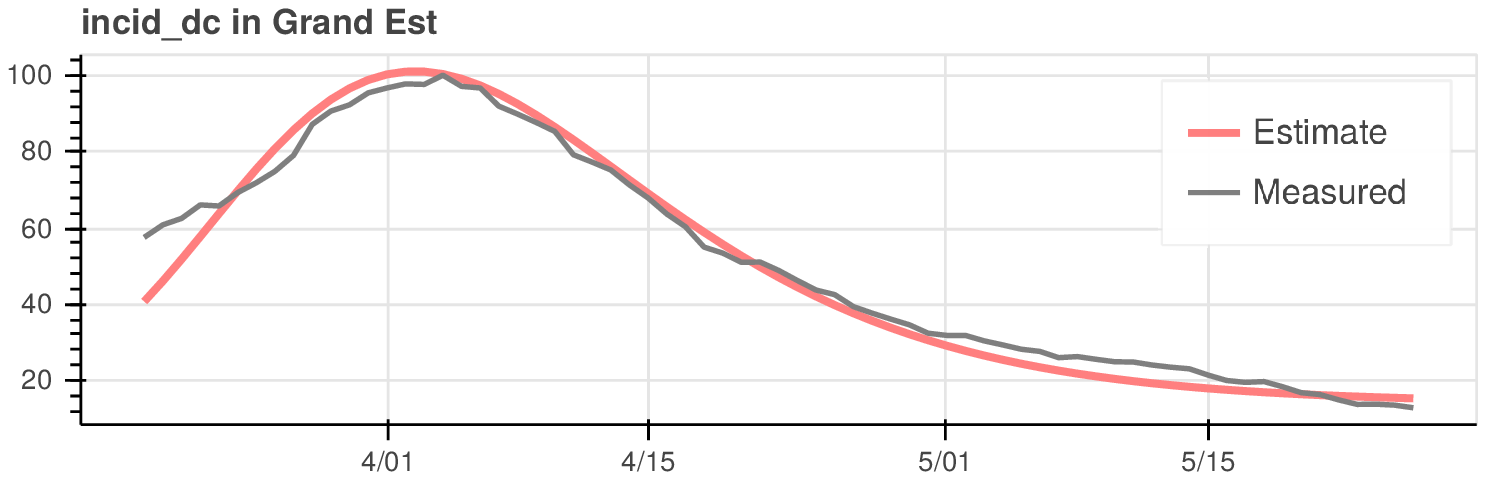}
\includegraphics[width=6.8cm]{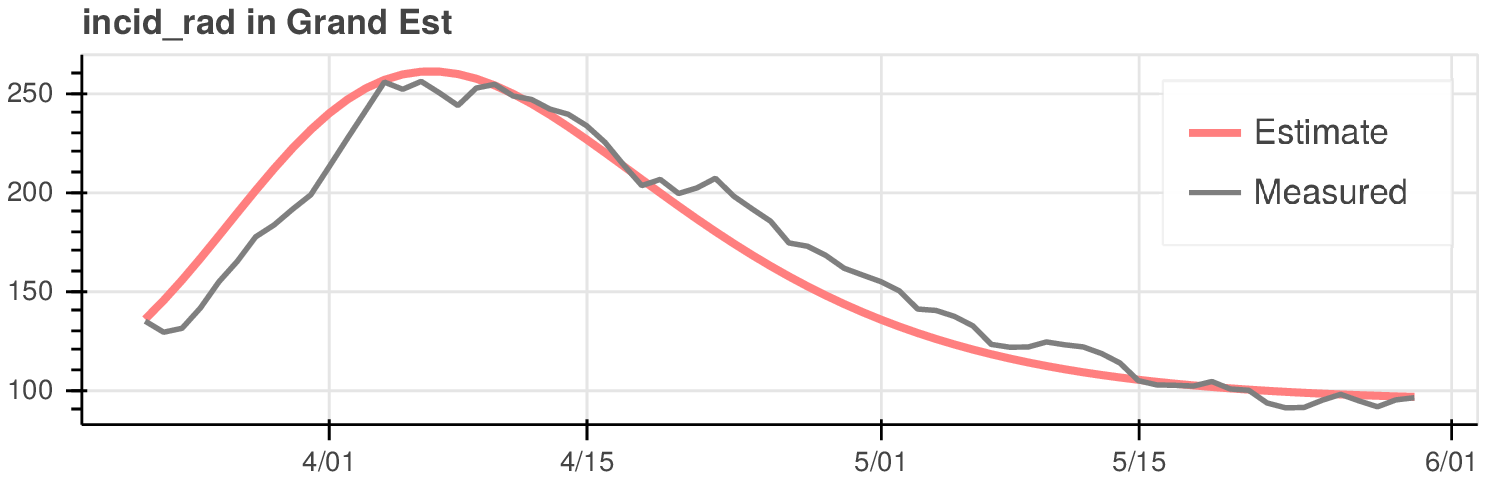}
\vspace{-0.7cm}
 \caption{Comparison of time series {\tt incid\_dc} and {\tt incid\_rad} averaged over a 14-days-long window, with the fitted outputs, for Grand Est (44) region.}
\end{figure}

\vspace{-0.7cm}
\begin{figure}[H]
 \centering
\includegraphics[width=6.8cm]{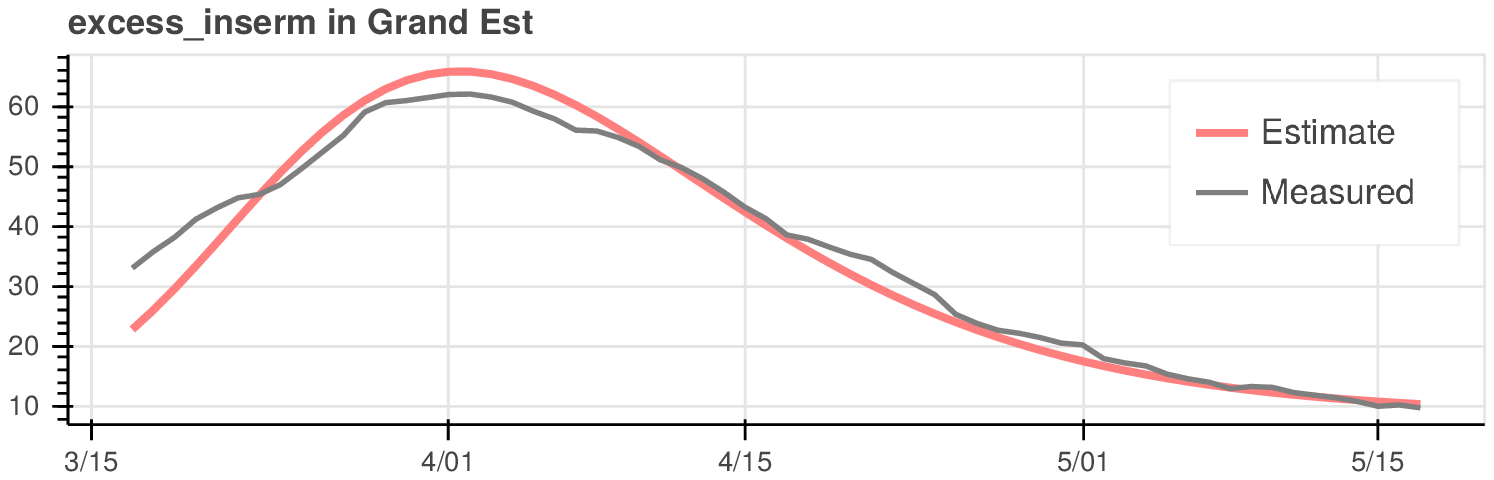}
\includegraphics[width=6.8cm]{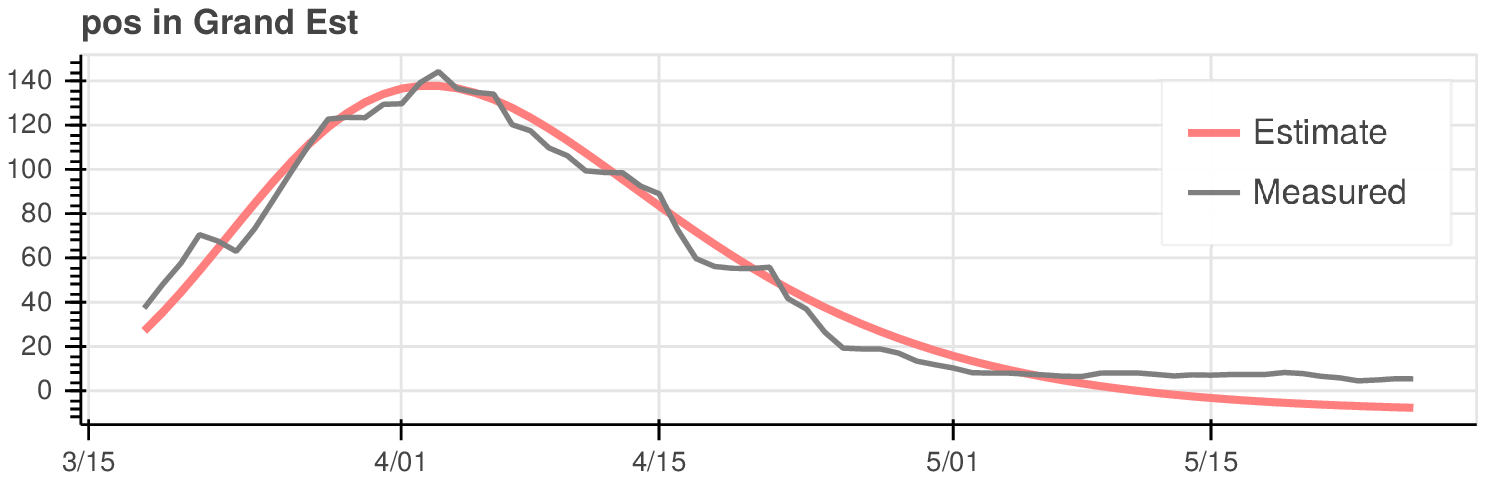}
\vspace{-0.7cm}
 \caption{Comparison of time series {\tt incid\_inserm} and {\tt pos} averaged over a 14-days-long window, with the fitted outputs, for Grand Est (44) region.}
\end{figure}

\vspace{-0.7cm}
\begin{figure}[H]
 \centering
\includegraphics[width=6.8cm]{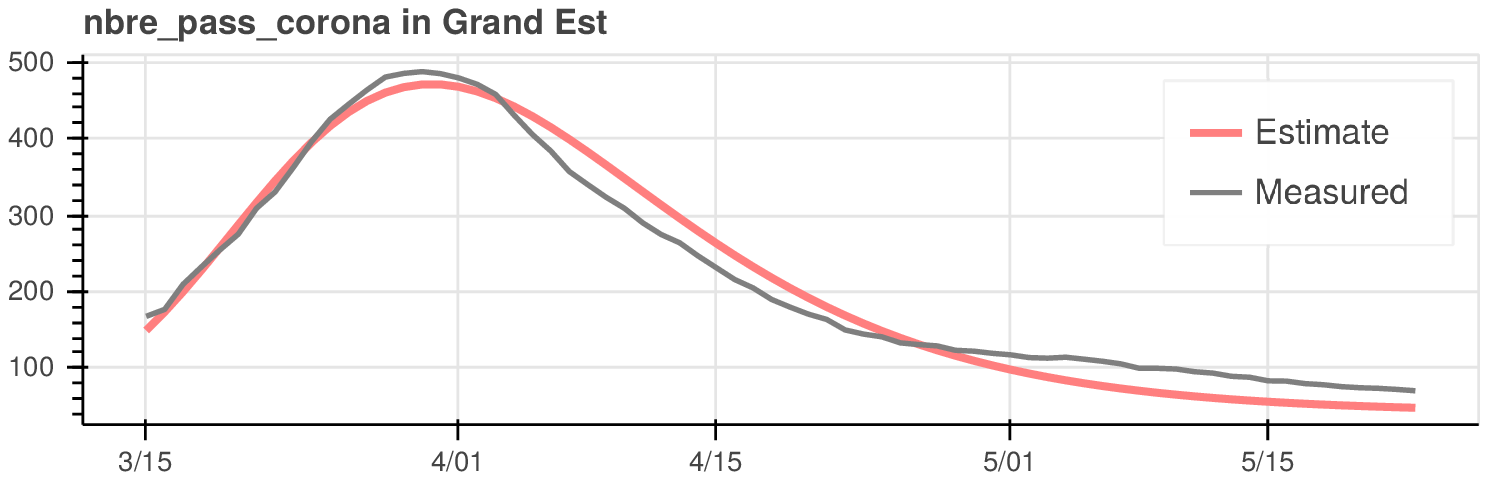}
\includegraphics[width=6.8cm]{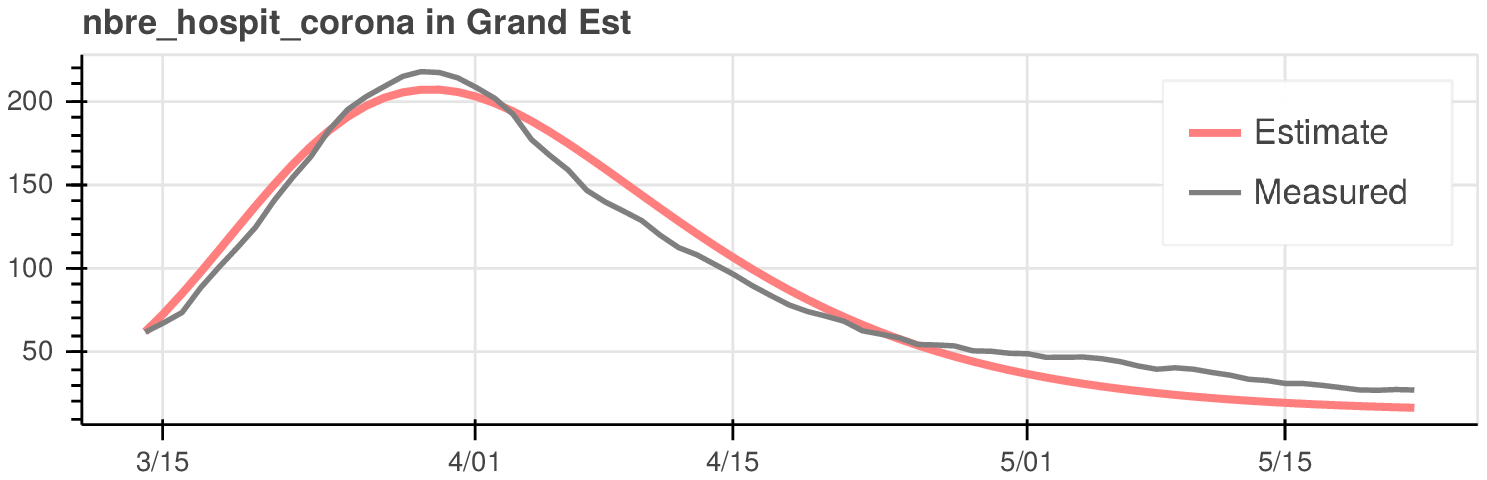}
\vspace{-0.7cm}
 \caption{Comparison of time series {\tt nbre\_pass\_corona} and 
 {\tt nbre\_hospit\_corona} averaged over a 14-days-long window, with the fitted outputs, for Grand Est (44) region.}
\end{figure}
\vspace{-0.7cm}
\begin{figure}[H]
 \centering
\includegraphics[width=6.8cm]{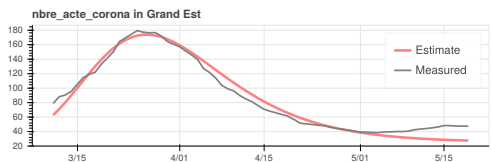}
\vspace{-0.3cm}
 \caption{Comparison of time series {\tt nbre\_acte\_corona} averaged over a 14-days-long window, with the fitted output, for Grand Est (44) region.}
\end{figure}

\subsection{\ARA}\label{ssec:ARA}

\vspace{-0.2cm}

\begin{figure}[H]
 \centering
\includegraphics[width=6.8cm]{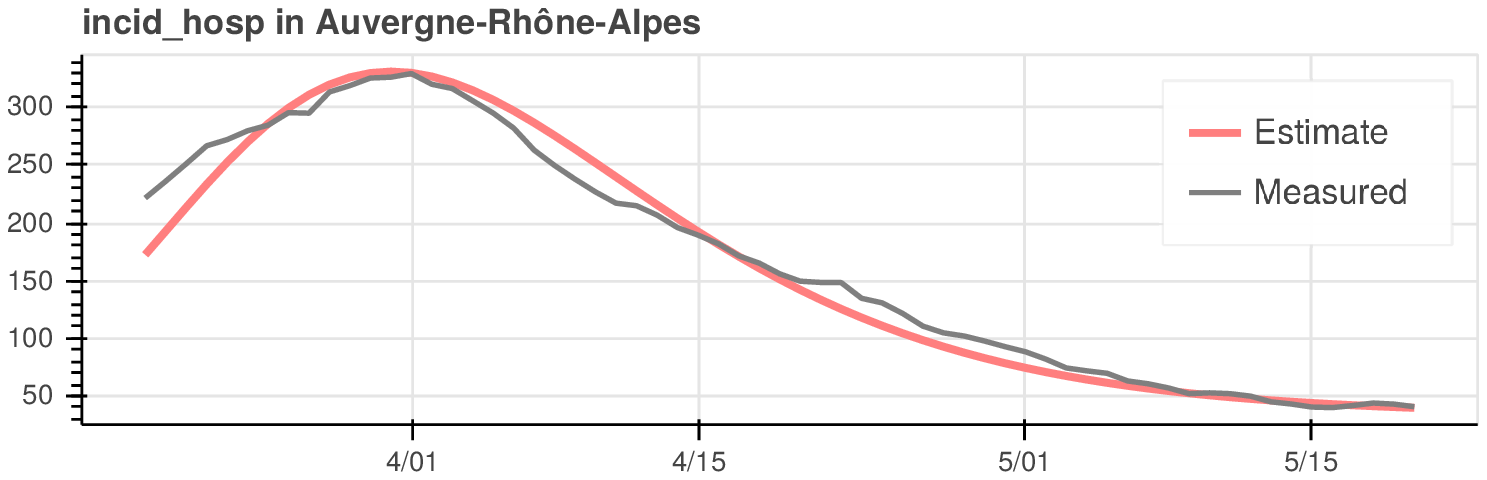}
\includegraphics[width=6.8cm]{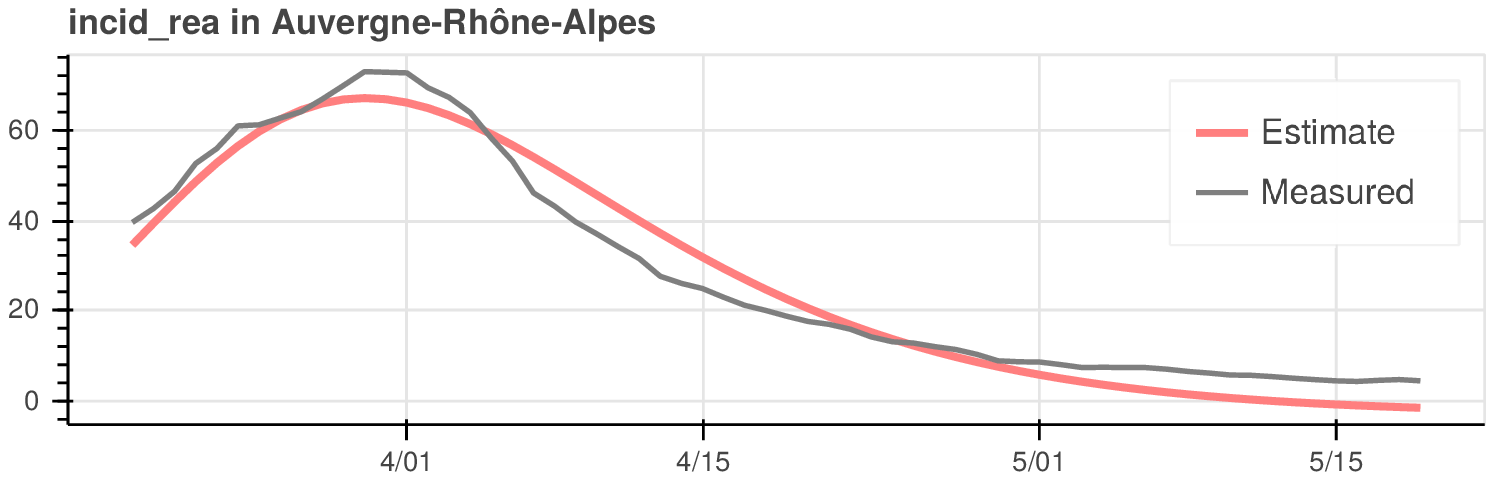}
 \vspace{-0.7cm}
\caption{Comparison of time series {\tt incid\_hosp} and {\tt incid\_rea} averaged over a 14-days-long window, with the fitted outputs, for \ARA (84) region.}
\end{figure}

\vspace{-0.7cm}
\begin{figure}[H]
 \centering
\includegraphics[width=6.8cm]{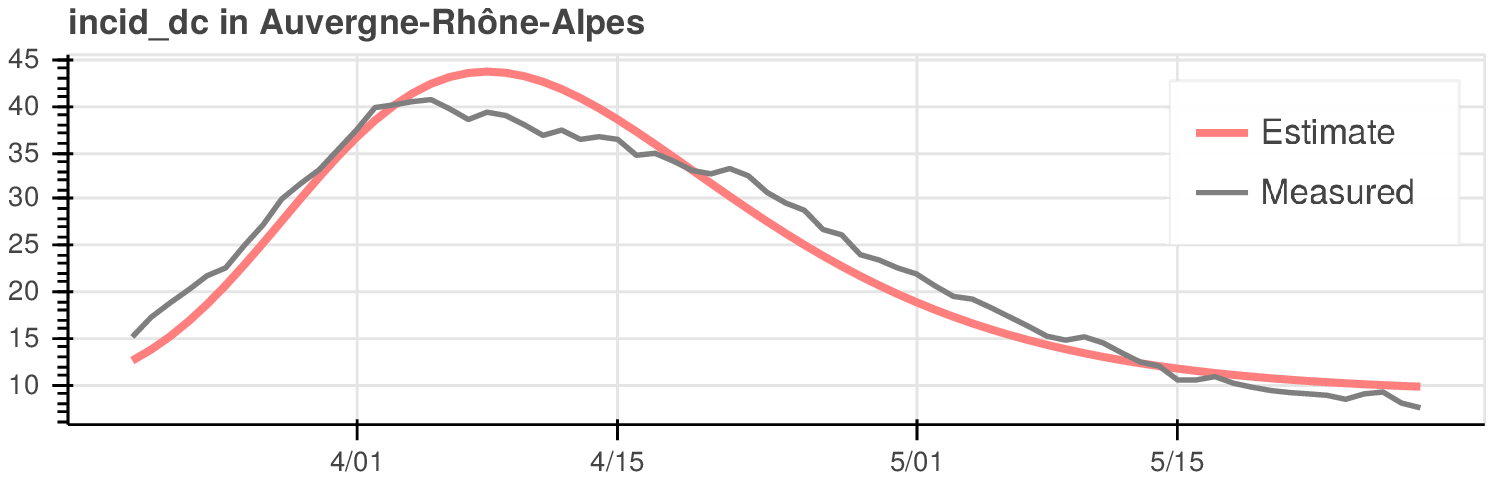}
\includegraphics[width=6.8cm]{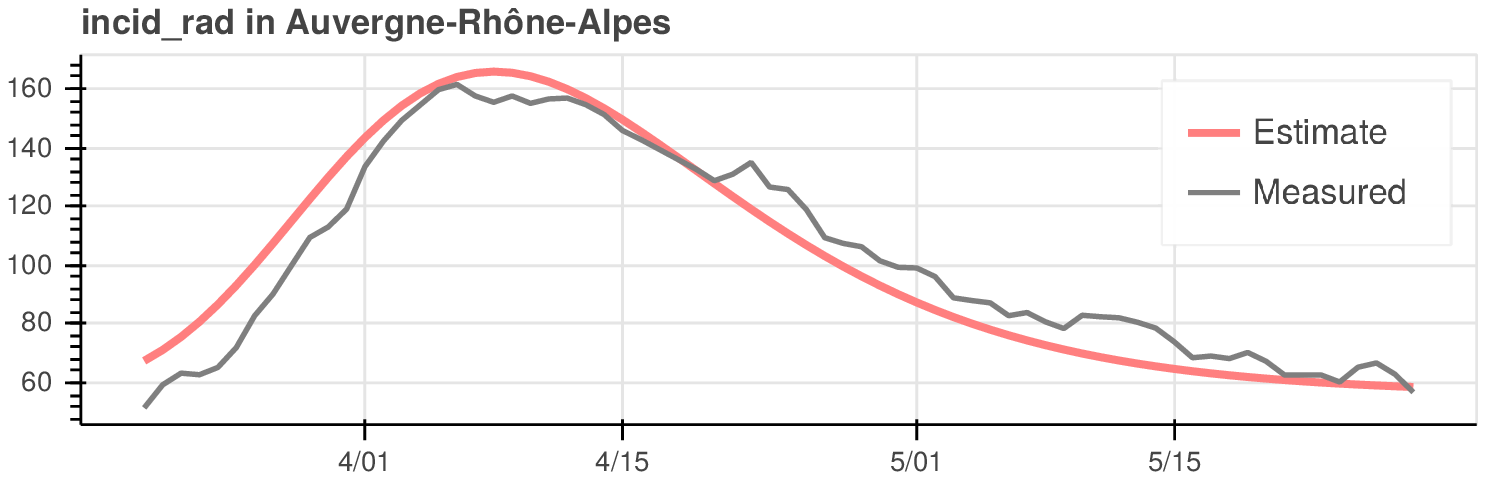}
\vspace{-0.7cm}
 \caption{Comparison of time series {\tt incid\_dc} and {\tt incid\_rad} averaged over a 14-days-long window, with the fitted outputs, for \ARA (84) region.}
\end{figure}

\vspace{-0.7cm}
\begin{figure}[H]
 \centering
\includegraphics[width=6.8cm]{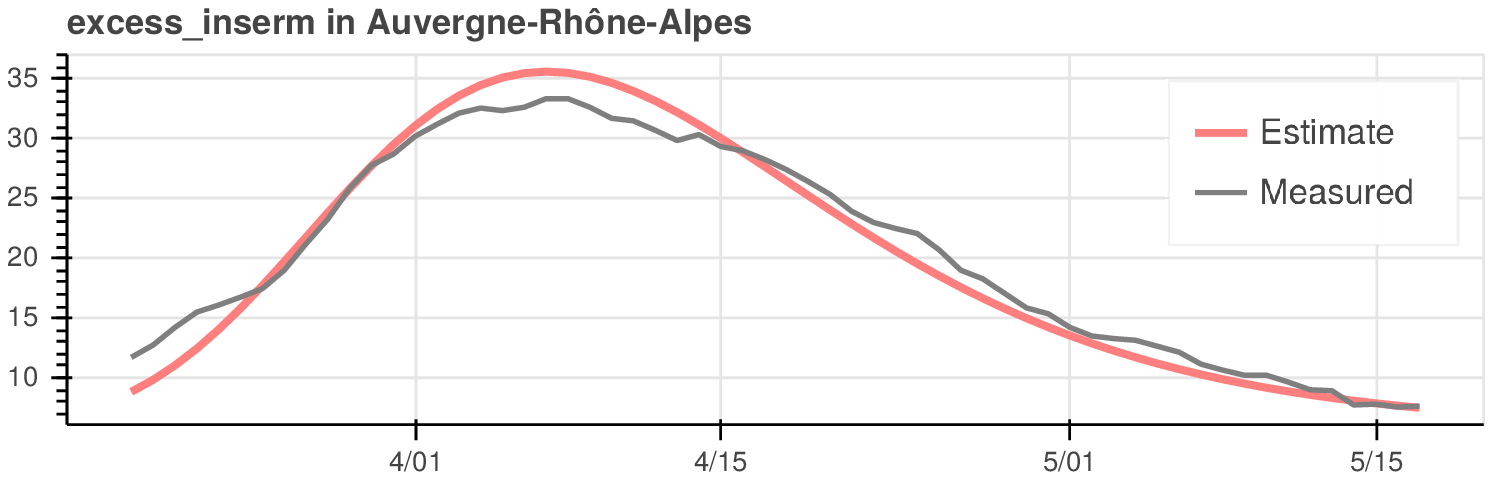}
\includegraphics[width=6.8cm]{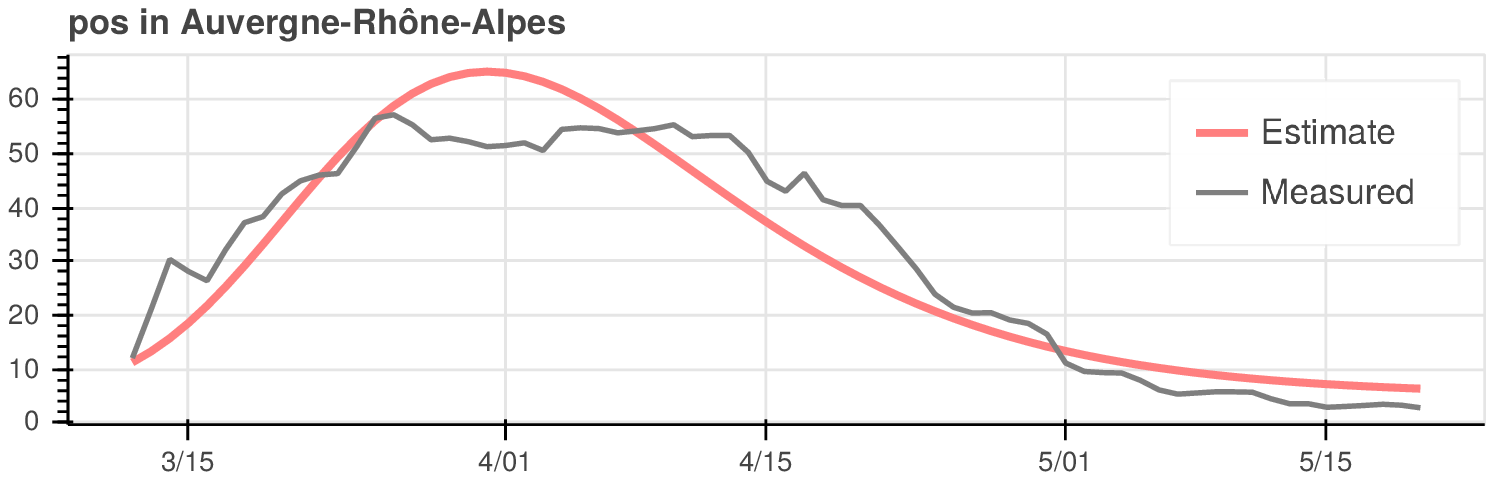}
\vspace{-0.7cm}
 \caption{Comparison of time series {\tt incid\_inserm} and {\tt pos} averaged over a 14-days-long window, with the fitted outputs, for \ARA (84) region.}
\end{figure}

\vspace{-0.7cm}
\begin{figure}[H]
 \centering
\includegraphics[width=6.8cm]{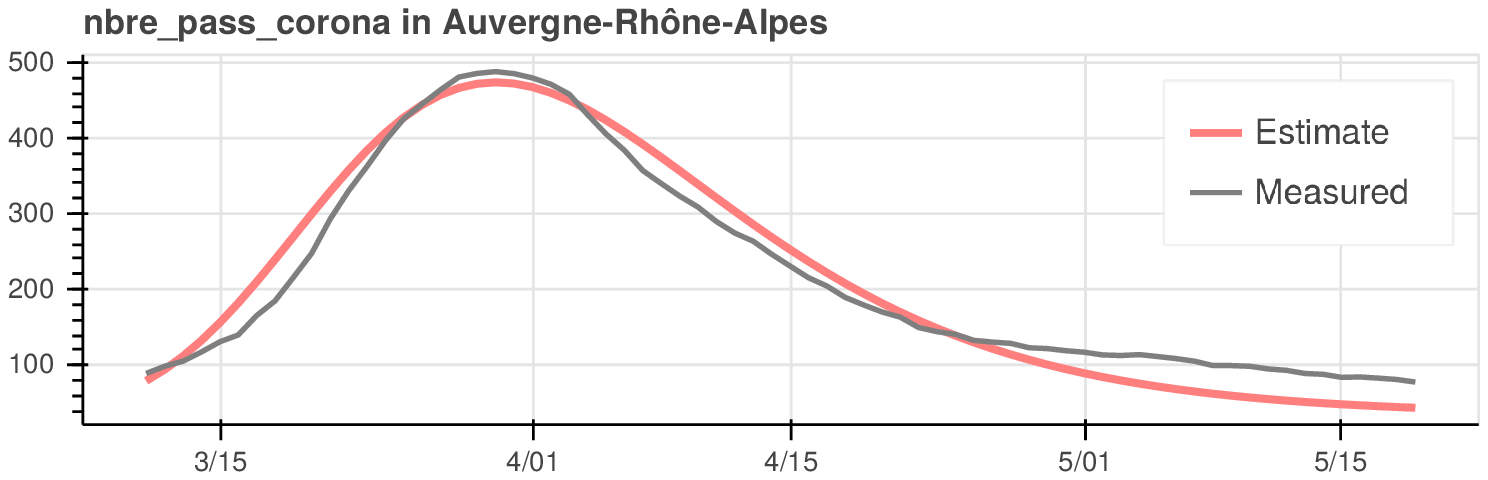}
\includegraphics[width=6.8cm]{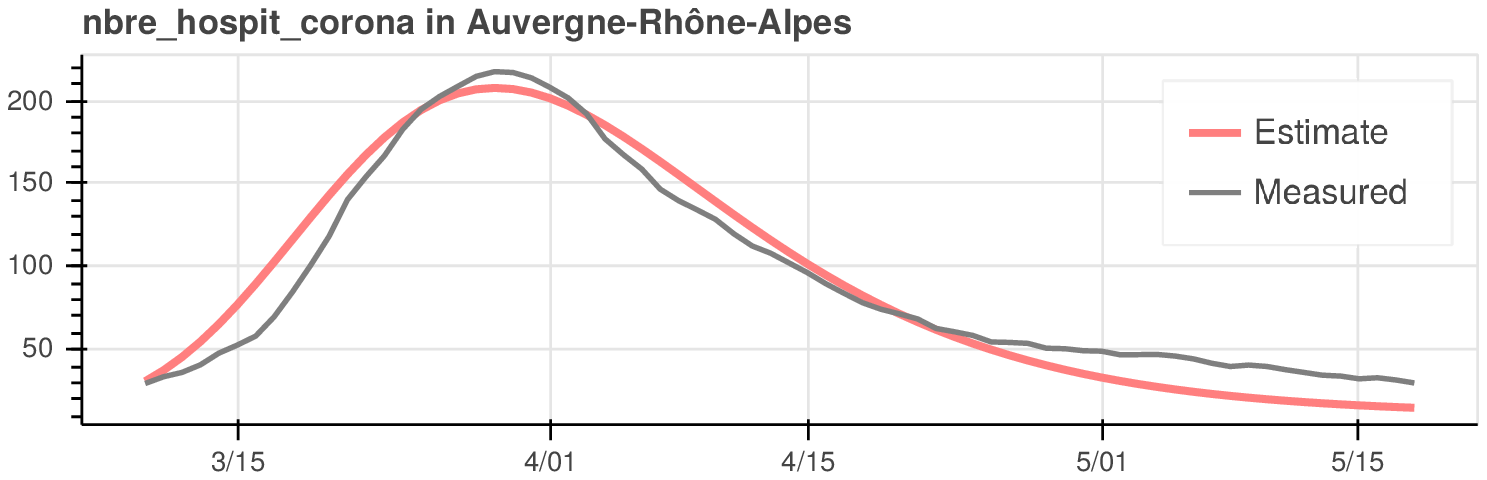}
\vspace{-0.7cm}
 \caption{Comparison of time series {\tt nbre\_pass\_corona} and 
 {\tt nbre\_hospit\_corona} averaged over a 14-days-long window, with the fitted outputs, for \ARA (84) region.}
\end{figure}
\vspace{-0.7cm}
\begin{figure}[H]
 \centering
\includegraphics[width=6.8cm]{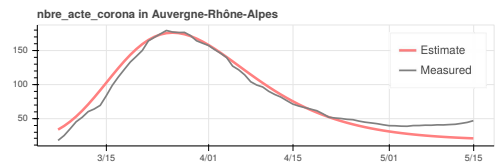}
\vspace{-0.3cm}
 \caption{Comparison of time series {\tt nbre\_acte\_corona} averaged over a 14-days-long window, with the fitted output, for \ARA (84) region.}
\end{figure}

\section{Conclusions}

In this paper, the available data concerning the \C19 spread in the 13 different regions of mainland France have been considered to build and identify a dynamical model. The aim was to fit the data with a model as simple as possible, to prevent and overcome the possible problem of nonidentifiability. An extremely simple dynamical system, with two states and only two common parameters for all the regions, permitted to reproduce the considered time series, that are the data concerning hospitalizations, intensive cares accesses, emergency and SOS Médecins activity, \C19 testing. These results seem to indicate that, with the available data only, more complex epidemiological models might incur in nonidentifiability and raise issues on their forecasting power. Exploiting additional data and further knowledge could be a solution to this problem.

 \bibliographystyle{elsarticle-num} 
 \bibliography{Covid19}

% \end{document}

\newpage 

\appendix
\section*{Appendix}

% Île-de-France (11) and Grand Est (44) regions
% Auvergne-Rhône-Alpes (84) and Hauts-de-France (32)
% Provence-Alpes-Côte d'Azur (93) and Bourgogne-Franche-Comté (27)
% Occitanie  (76) and Pays de la Loire (52)
% Centre-Val de Loire (24) and Bretagne (53)
% Normandie (28) and Nouvelle-Aquitaine (75)
% Corse (94) region
% 

\section{Deaths excess in French regions}\label{app:excess}

\begin{figure}[H]
 \centering
\includegraphics[width=6.8cm]{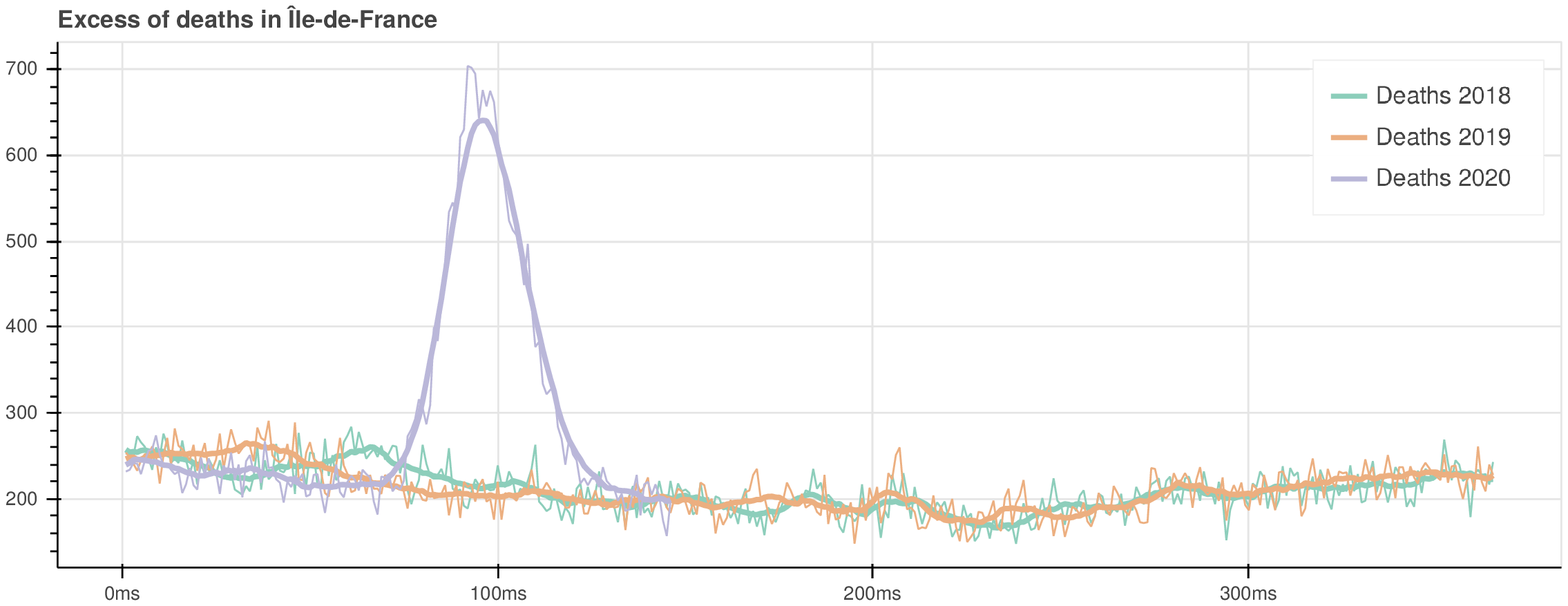}
\includegraphics[width=6.8cm]{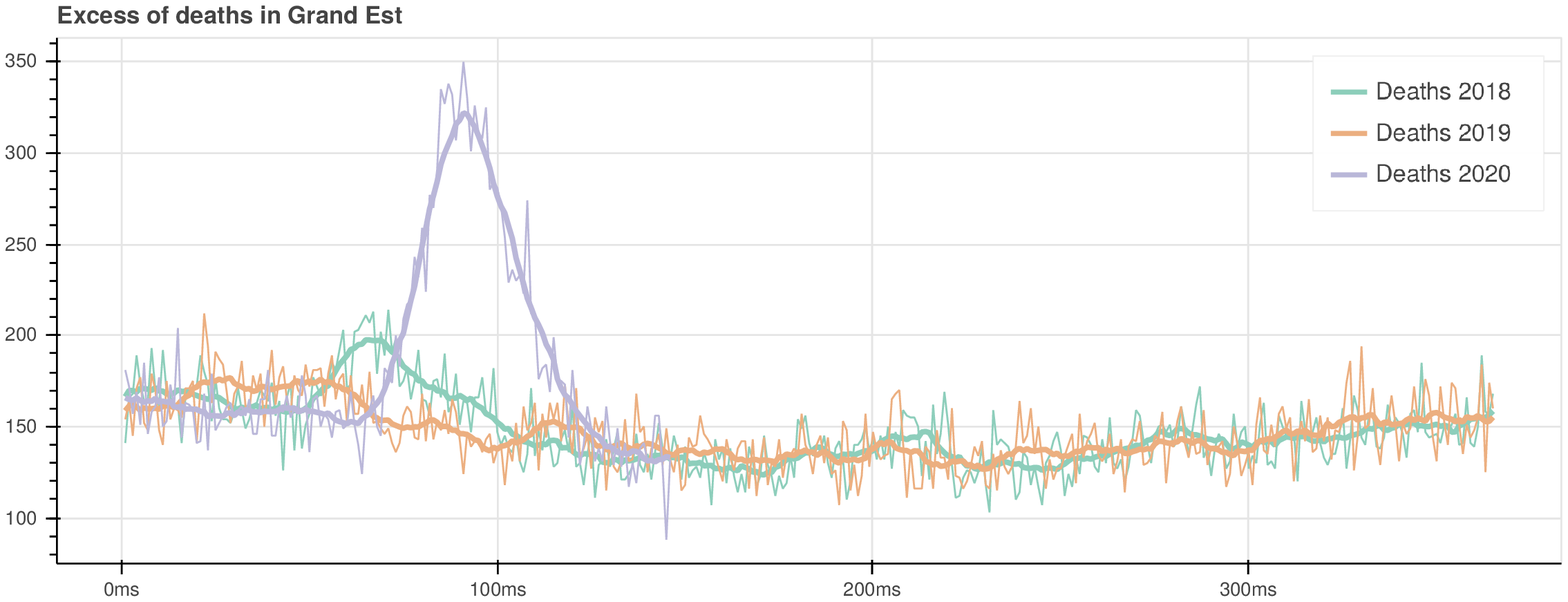}
 \caption{Evolution of daily demises in 2018, 2019 and 2020, in Île-de-France (11) and Grand Est (44) regions. In bold, the average over a 14-days-long window.}
\end{figure}

\begin{figure}[H]
 \centering
\includegraphics[width=6.8cm]{1Covid/1_5mort_reg84.eps}
\includegraphics[width=6.8cm]{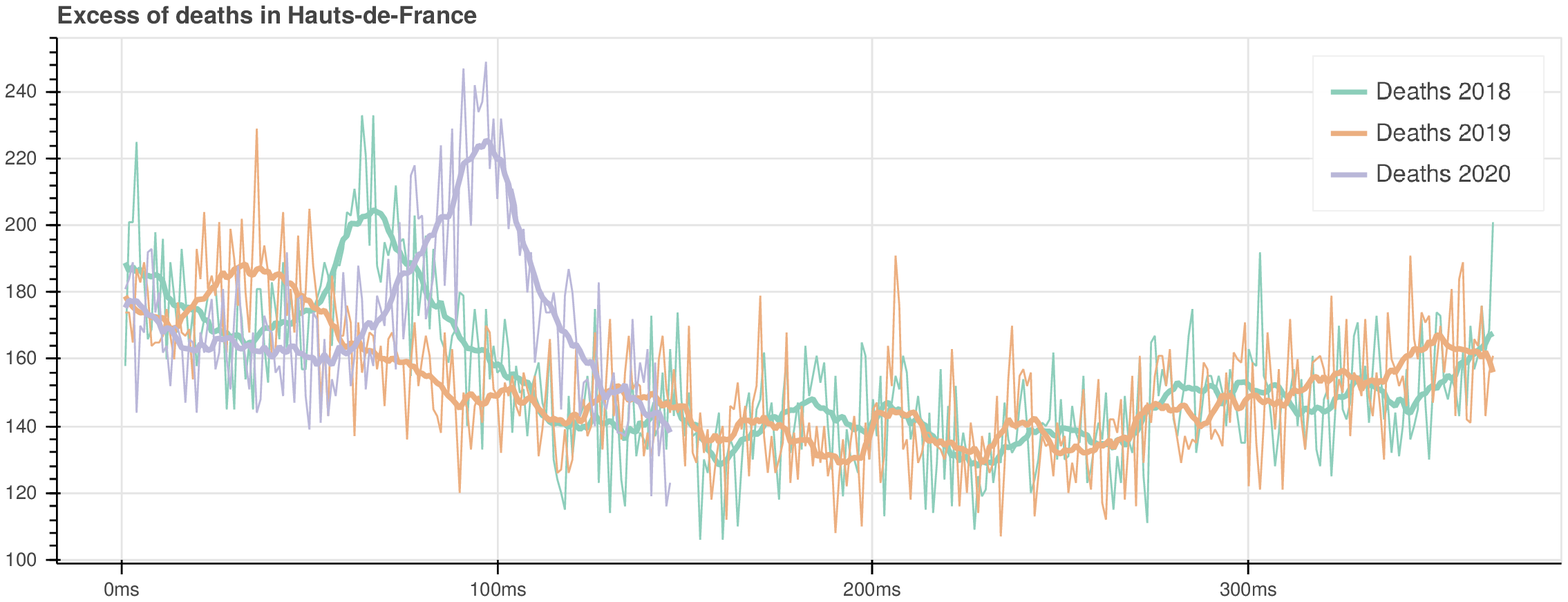}
 \caption{Evolution of daily demises in 2018, 2019 and 2020, in Auvergne-Rhône-Alpes (84) and Hauts-de-France (32) regions. In bold, the average over a 14-days-long window.}
\end{figure}

\begin{figure}[H]
 \centering
\includegraphics[width=6.8cm]{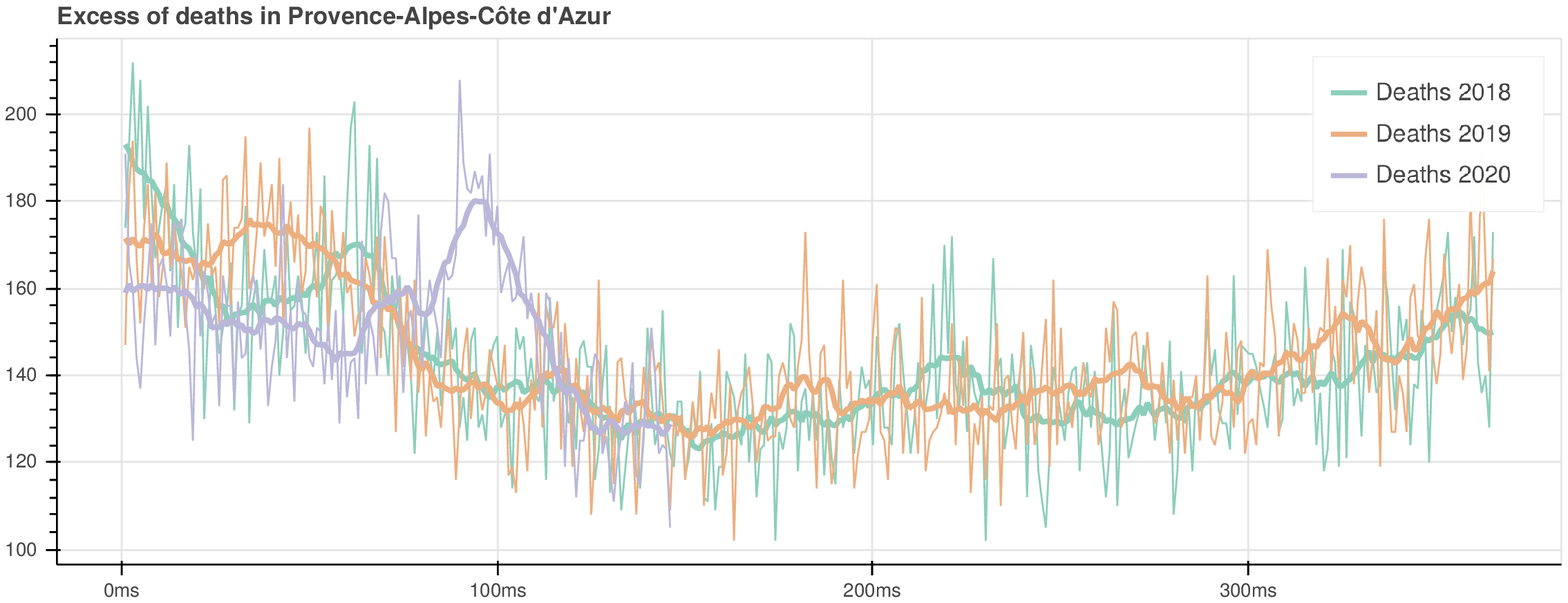}
\includegraphics[width=6.8cm]{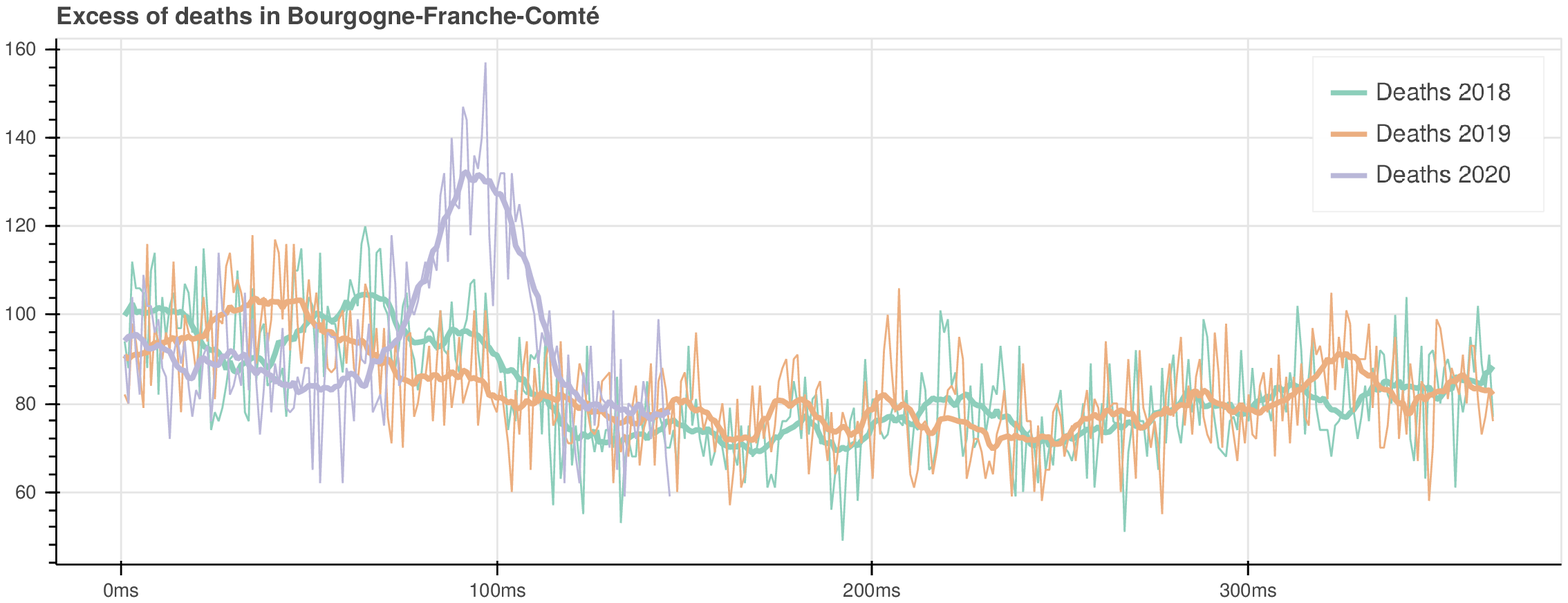}
 \caption{Evolution of daily demises in 2018, 2019 and 2020, in Provence-Alpes-Côte d'Azur (93) and Bourgogne-Franche-Comté (27) regions. In bold, the average over a 14-days-long window.}
\end{figure}

\begin{figure}[H]
 \centering
\includegraphics[width=6.8cm]{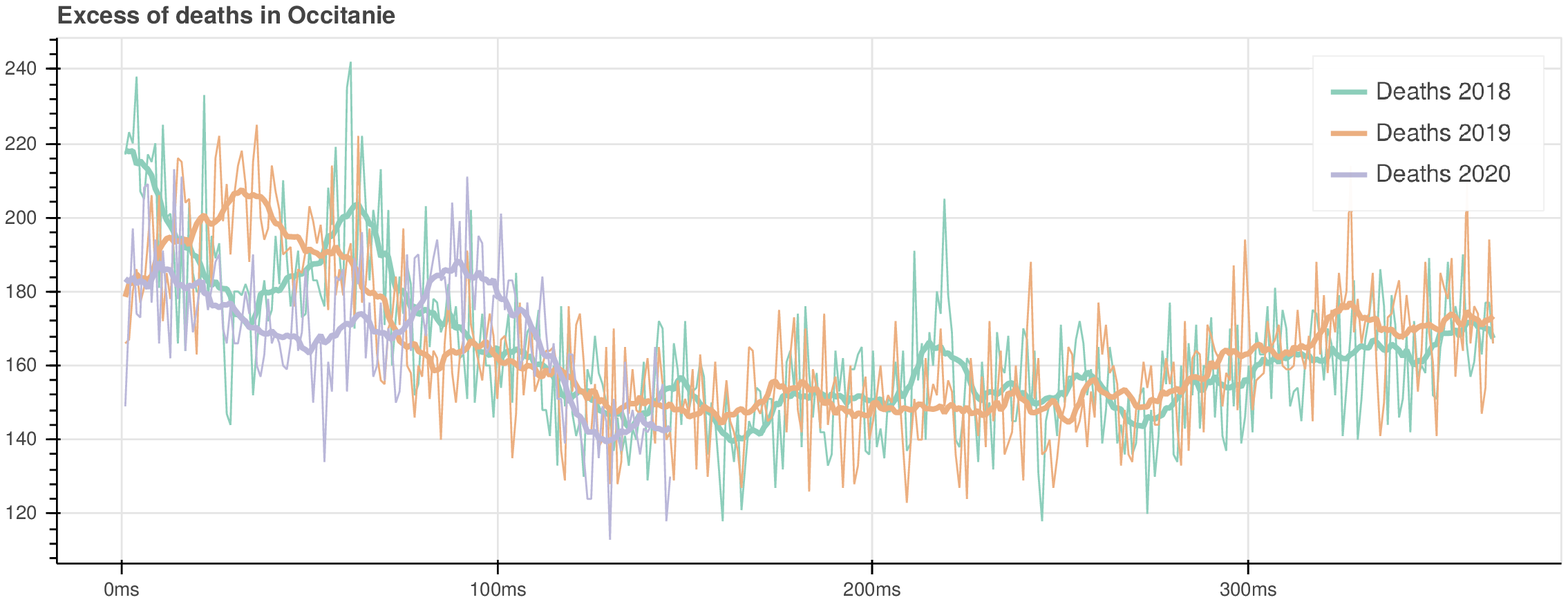}
\includegraphics[width=6.8cm]{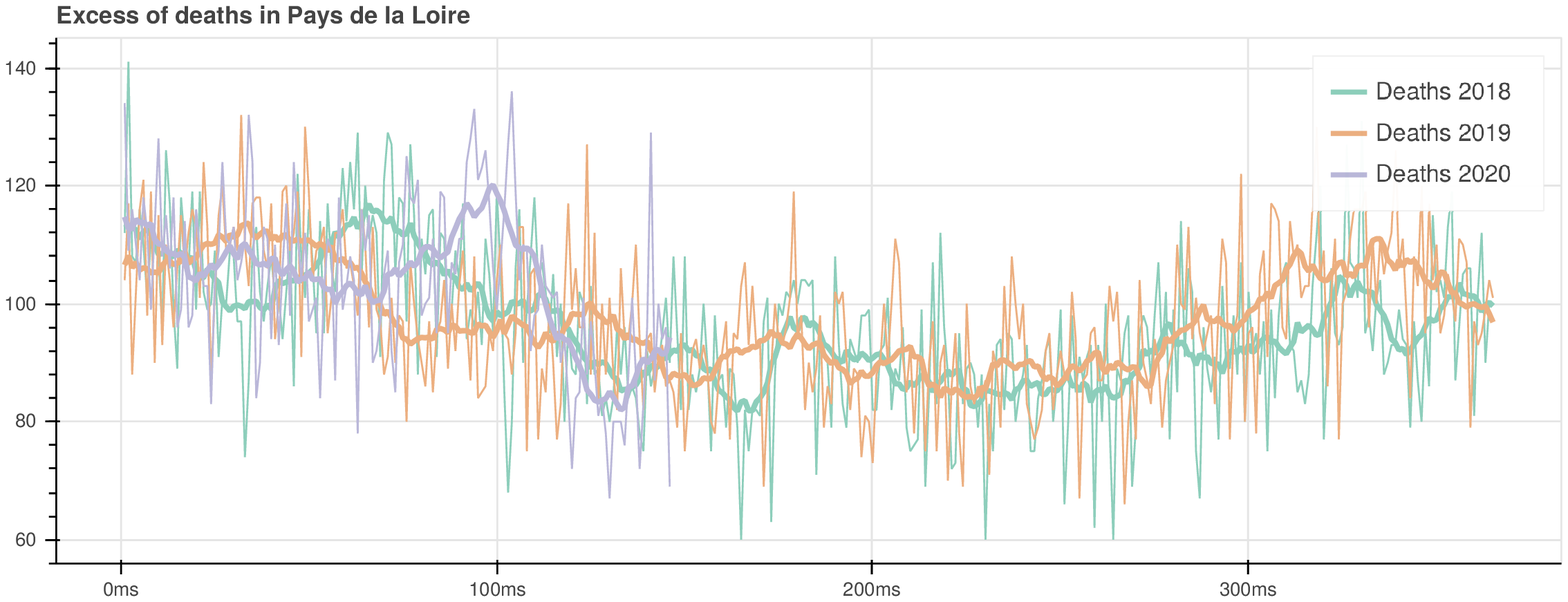}
 \caption{Evolution of daily demises in 2018, 2019 and 2020, in Occitanie  (76) and Pays de la Loire (52) regions. In bold, the average over a 14-days-long window.}
\end{figure}

\begin{figure}[H]
 \centering
\includegraphics[width=6.8cm]{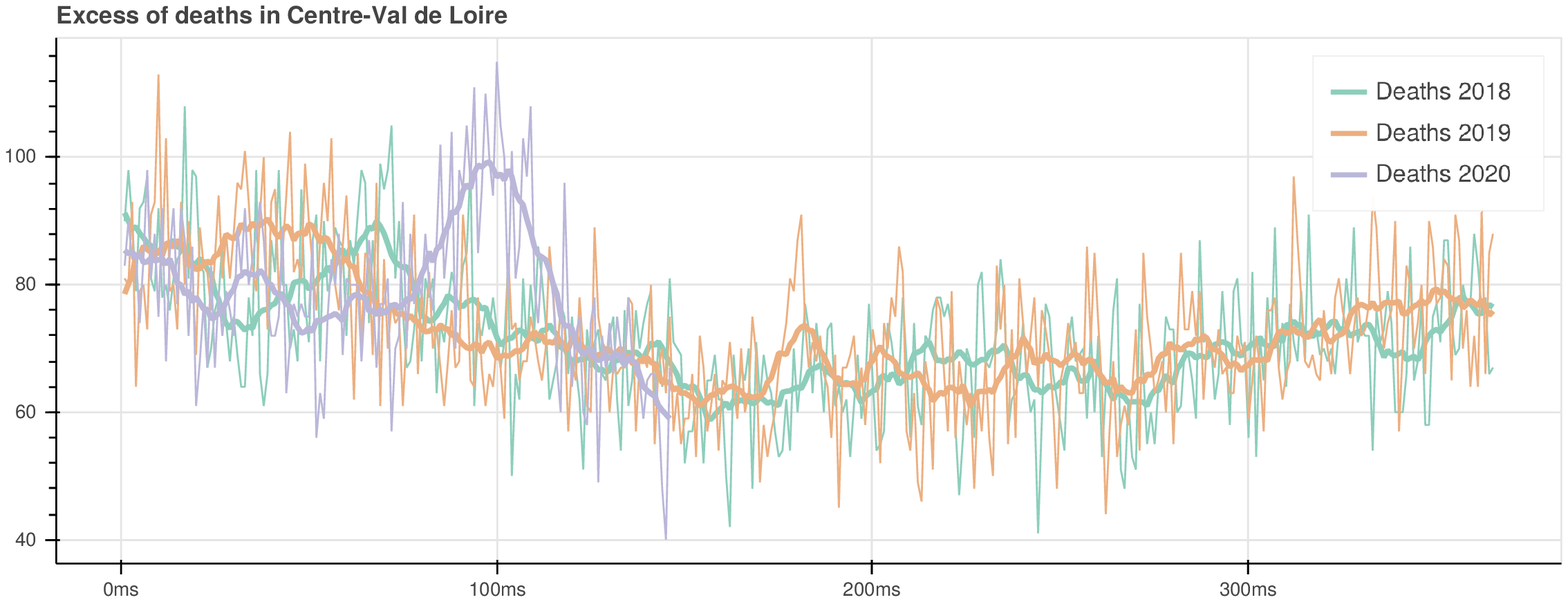}
\includegraphics[width=6.8cm]{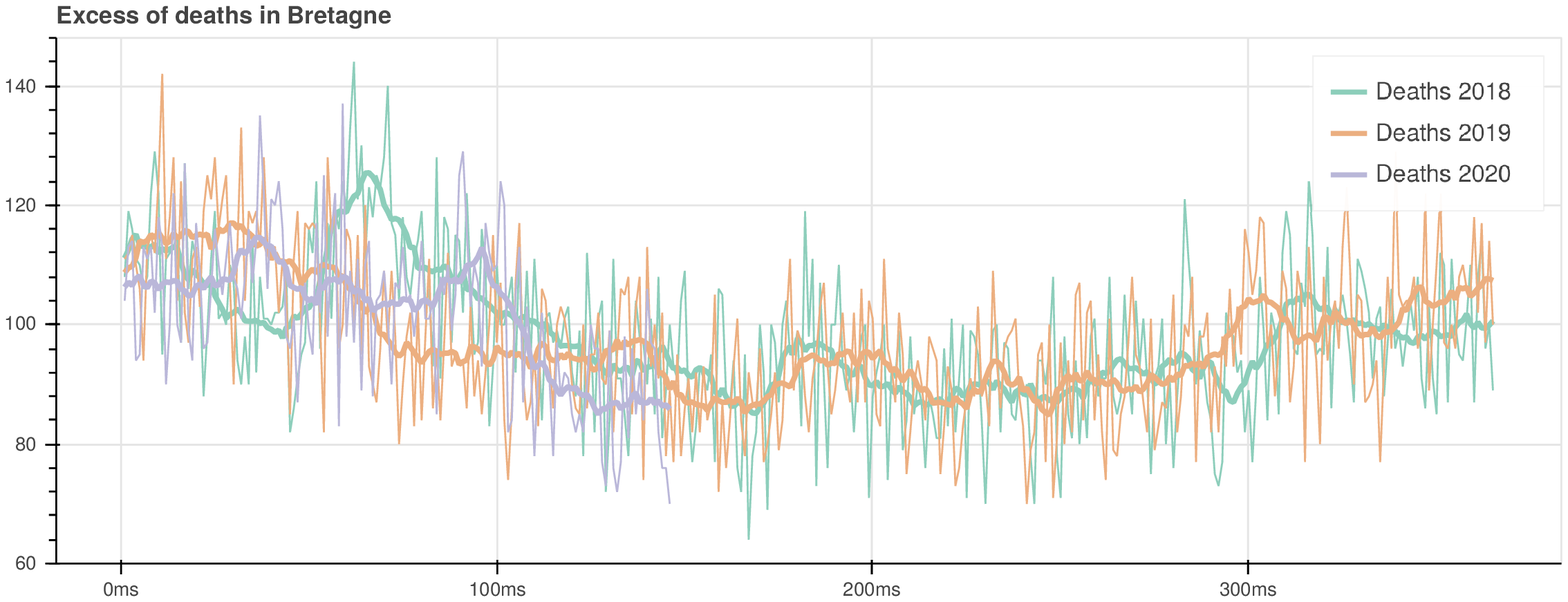}
 \caption{Evolution of daily demises in 2018, 2019 and 2020, in Centre-Val de Loire (24) and Bretagne (53) regions. In bold, the average over a 14-days-long window.}
\end{figure}

\begin{figure}[H]
 \centering
\includegraphics[width=6.8cm]{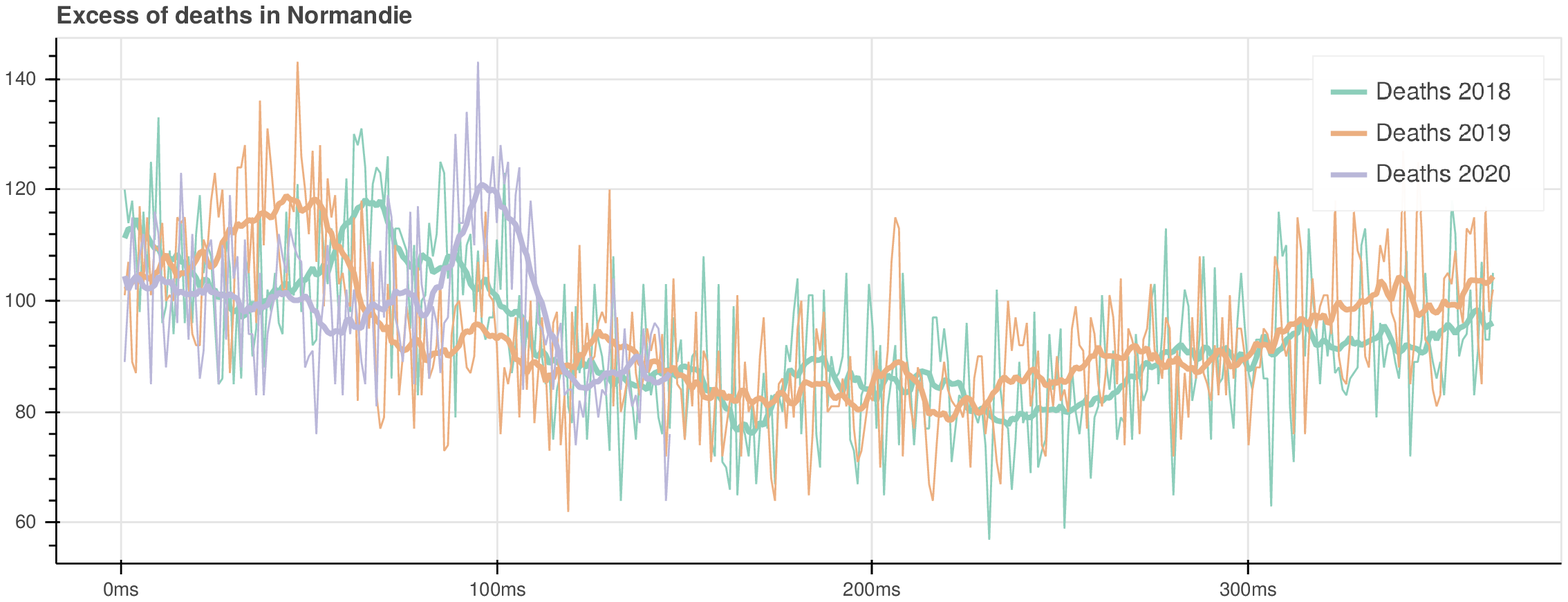}
\includegraphics[width=6.8cm]{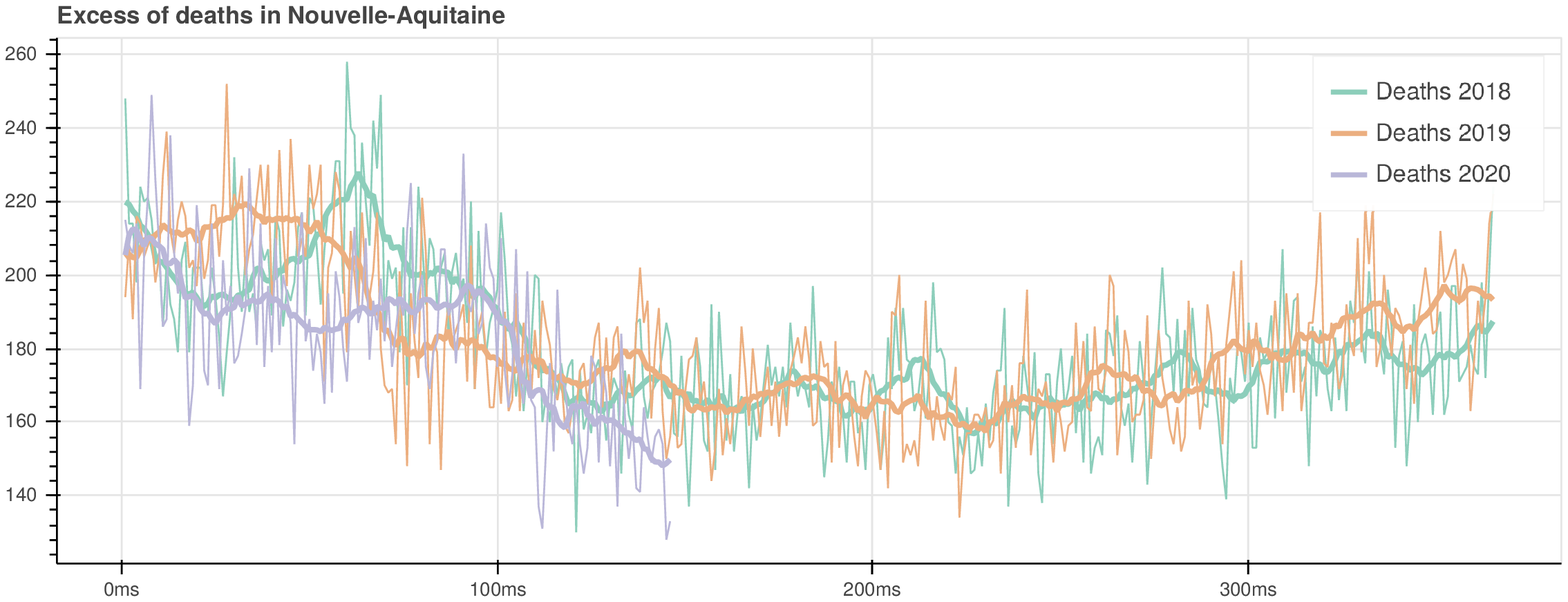}
 \caption{Evolution of daily demises in 2018, 2019 and 2020, in Normandie (28) and Nouvelle-Aquitaine (75) regions. In bold, the average over a 14-days-long window.}
\end{figure}

\begin{figure}[H]
 \centering
\includegraphics[width=6.8cm]{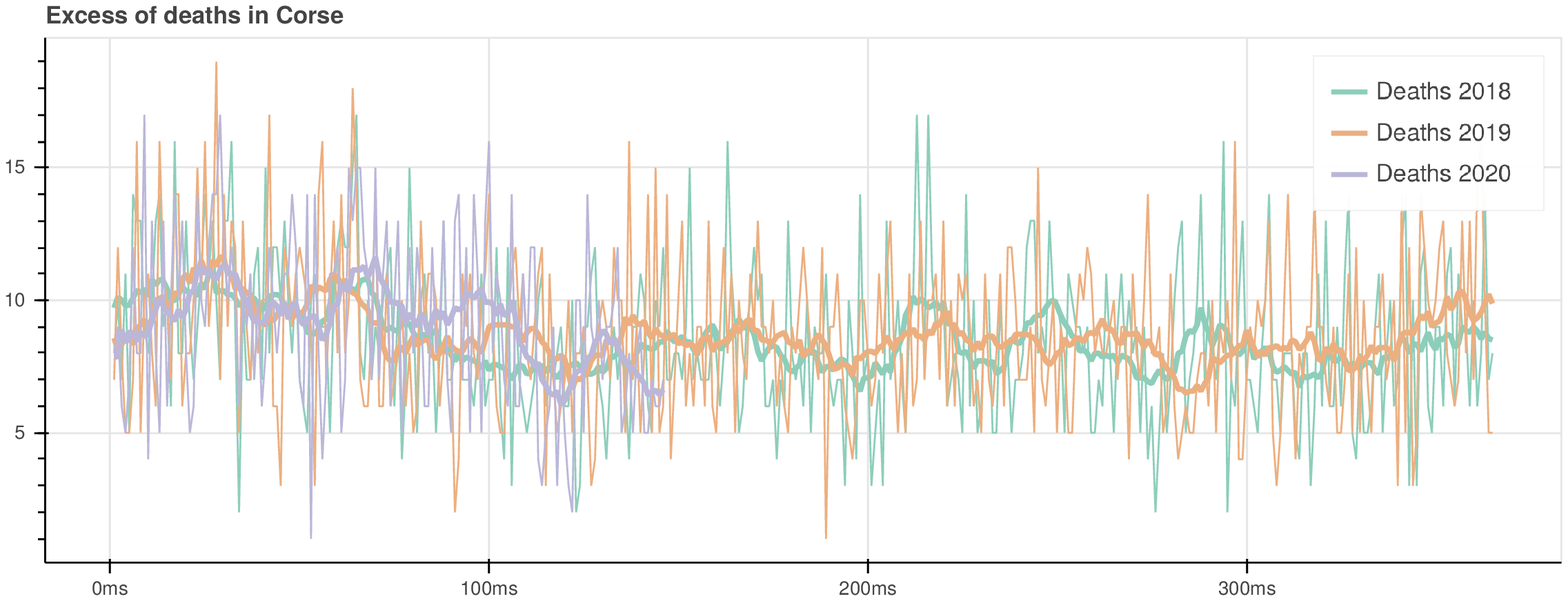}
 \caption{Evolution of daily demises in 2018, 2019 and 2020, in Corse (94) region. In bold, the average over a 14-days-long window.}
\end{figure}

\newpage

\section{Deaths time series in French regions}\label{app:excessfit}

\begin{figure}[H]
 \centering
\includegraphics[width=6.8cm]{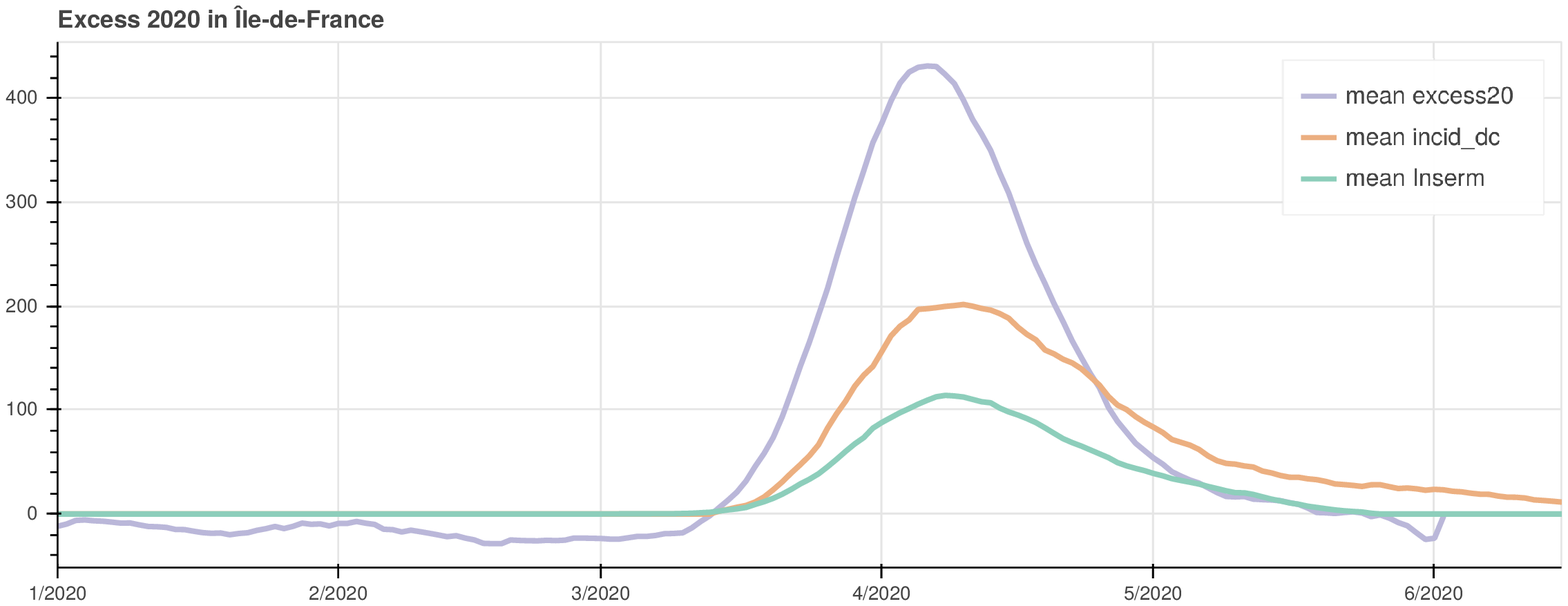}
\includegraphics[width=6.8cm]{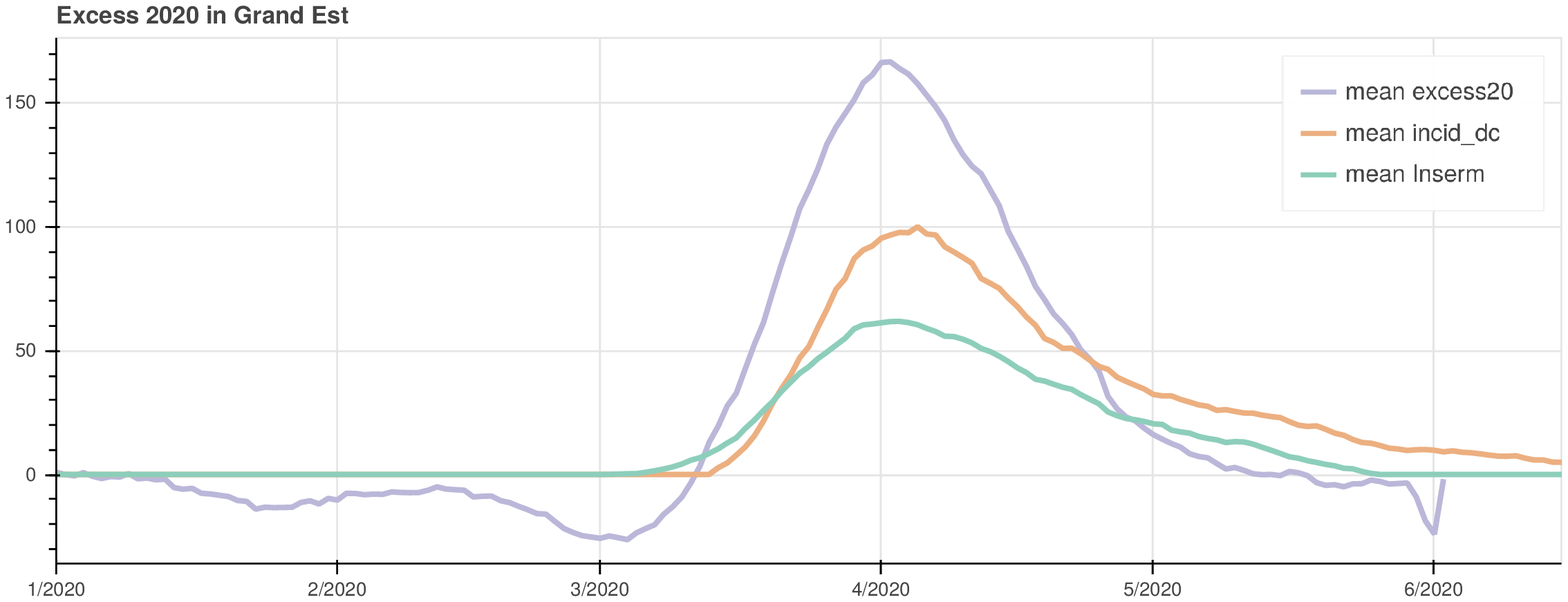}
 \caption{Excess of death in 2020 with respect to the average of 2018/2019, {\tt mean excess20}; deaths certified in hospitals {\tt mean incid\_dc} and deaths electronically certified to Inserm {\tt mean Inserm}, averaged over a 14-days-long window from January to June 2020, in Île-de-France (11) and Grand Est (44) regions.} 
\end{figure}

\begin{figure}[H]
 \centering
\includegraphics[width=6.8cm]{2Covid/2_2_reg84.eps}
\includegraphics[width=6.8cm]{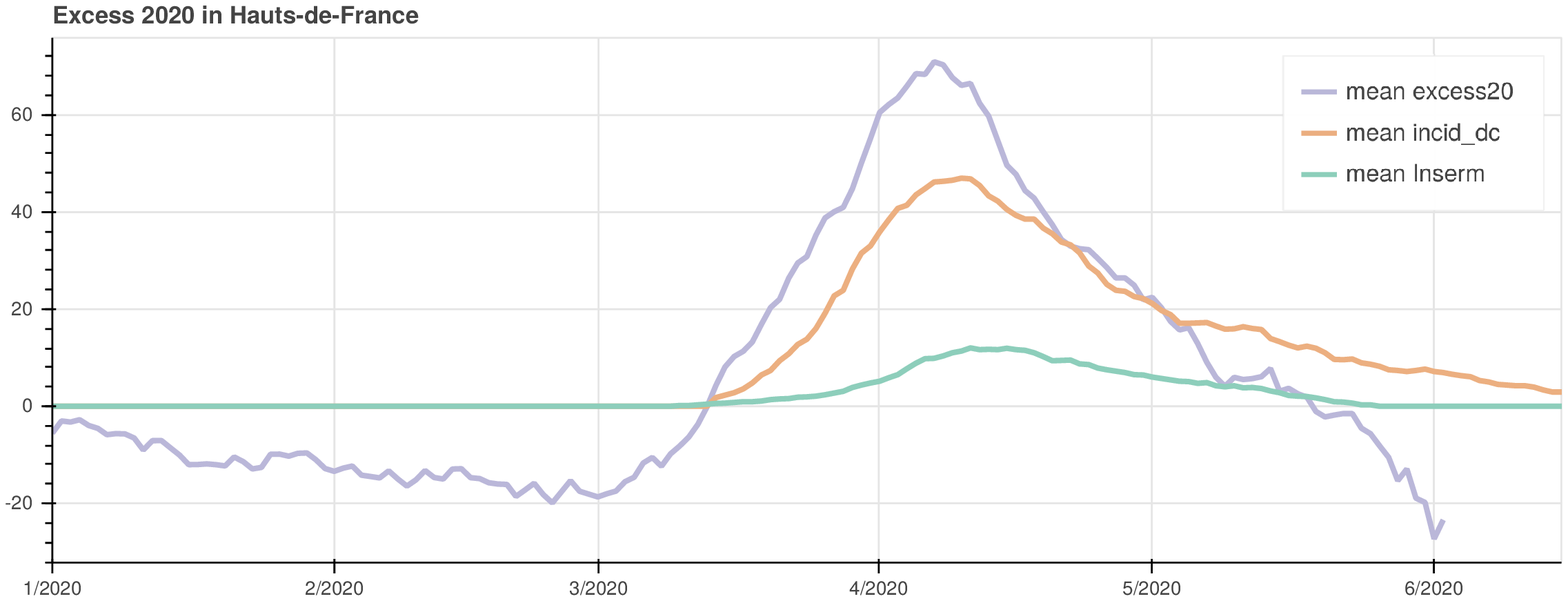}
 \caption{Excess of death in 2020 with respect to the average of 2018/2019, {\tt mean excess20}; deaths certified in hospitals {\tt mean incid\_dc} and deaths electronically certified to Inserm {\tt mean Inserm}, averaged over a 14-days-long window, in Auvergne-Rhône-Alpes (84) and Hauts-de-France (32) regions.} 
\end{figure}

\begin{figure}[H]
 \centering
\includegraphics[width=6.8cm]{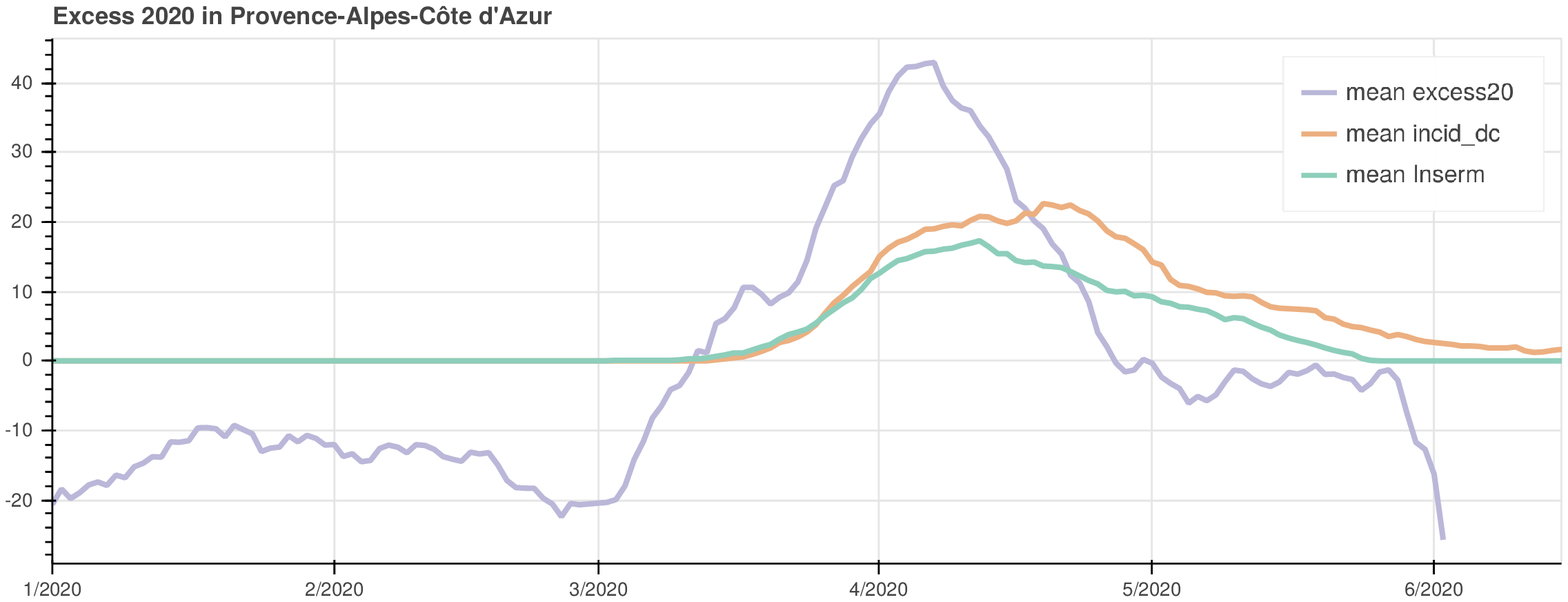}
\includegraphics[width=6.8cm]{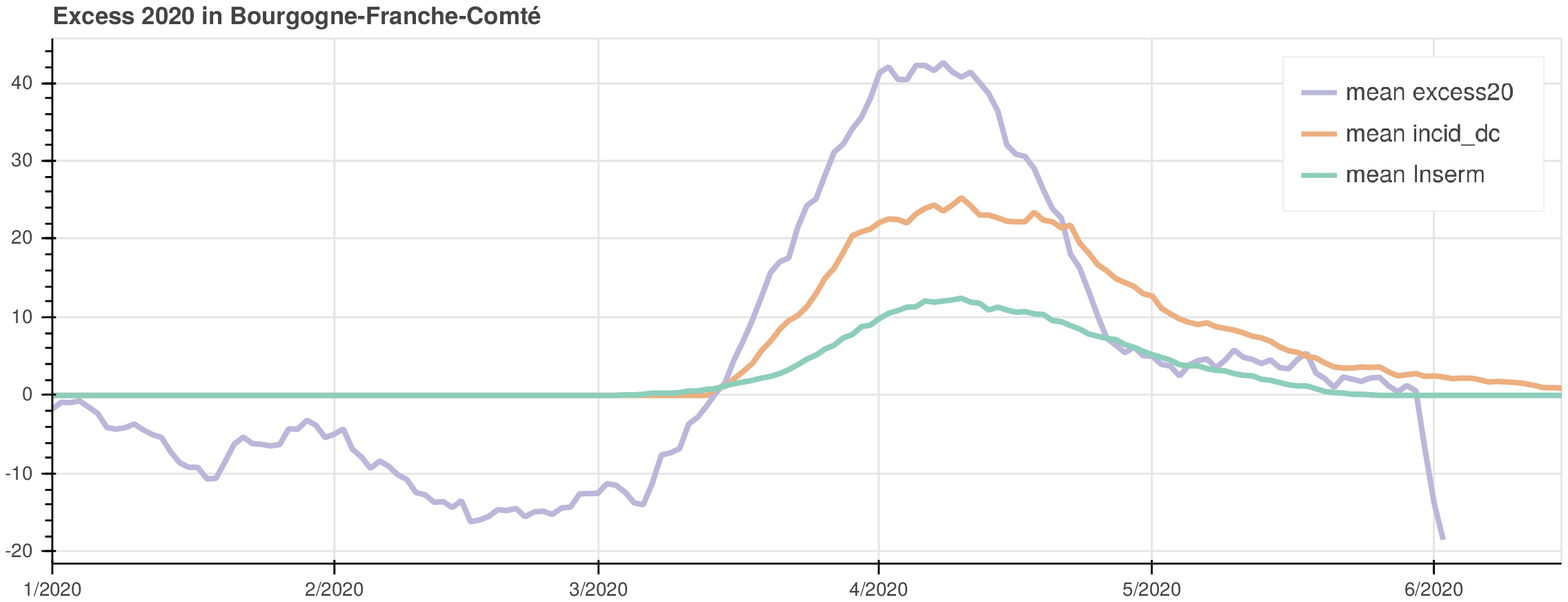}
 \caption{Excess of death in 2020 with respect to the average of 2018/2019, {\tt mean excess20}; deaths certified in hospitals {\tt mean incid\_dc} and deaths electronically certified to Inserm {\tt mean Inserm}, averaged over a 14-days-long window from January to June 2020, in Provence-Alpes-Côte d'Azur (93) and Bourgogne-Franche-Comté (27) regions.} 
\end{figure}

\begin{figure}[H]
 \centering
\includegraphics[width=6.8cm]{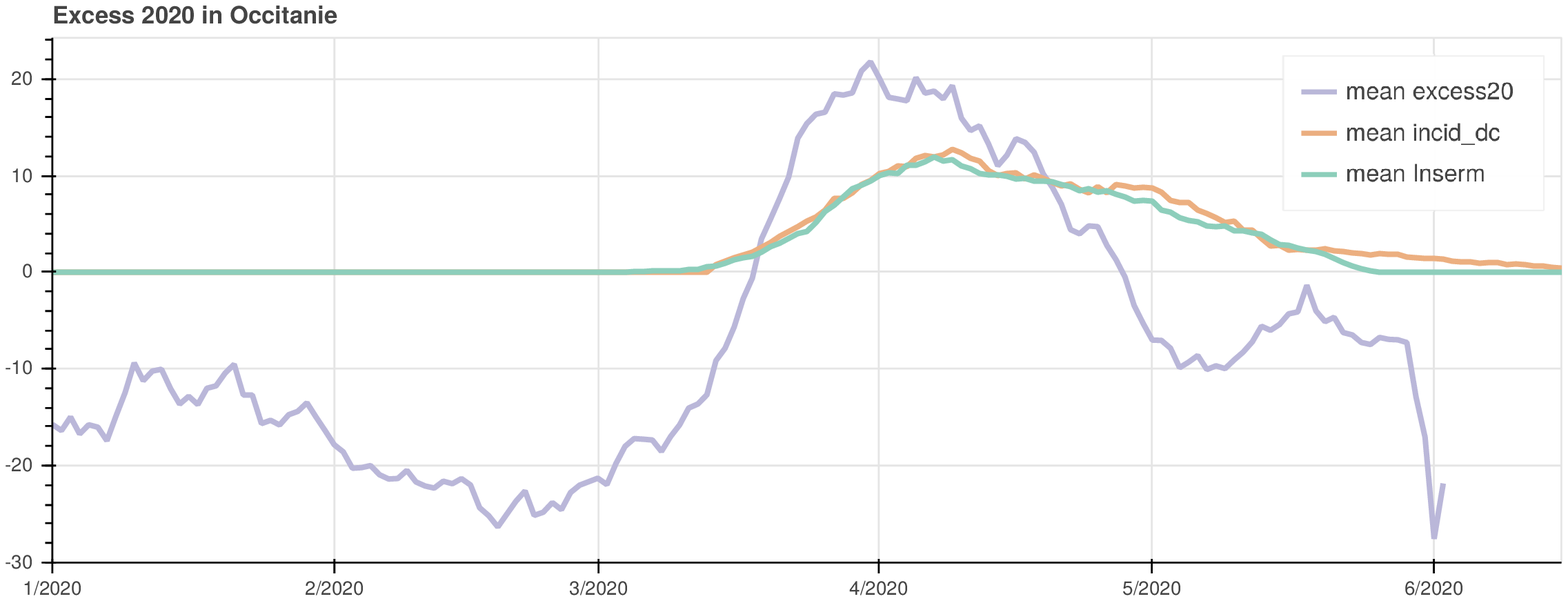}
\includegraphics[width=6.8cm]{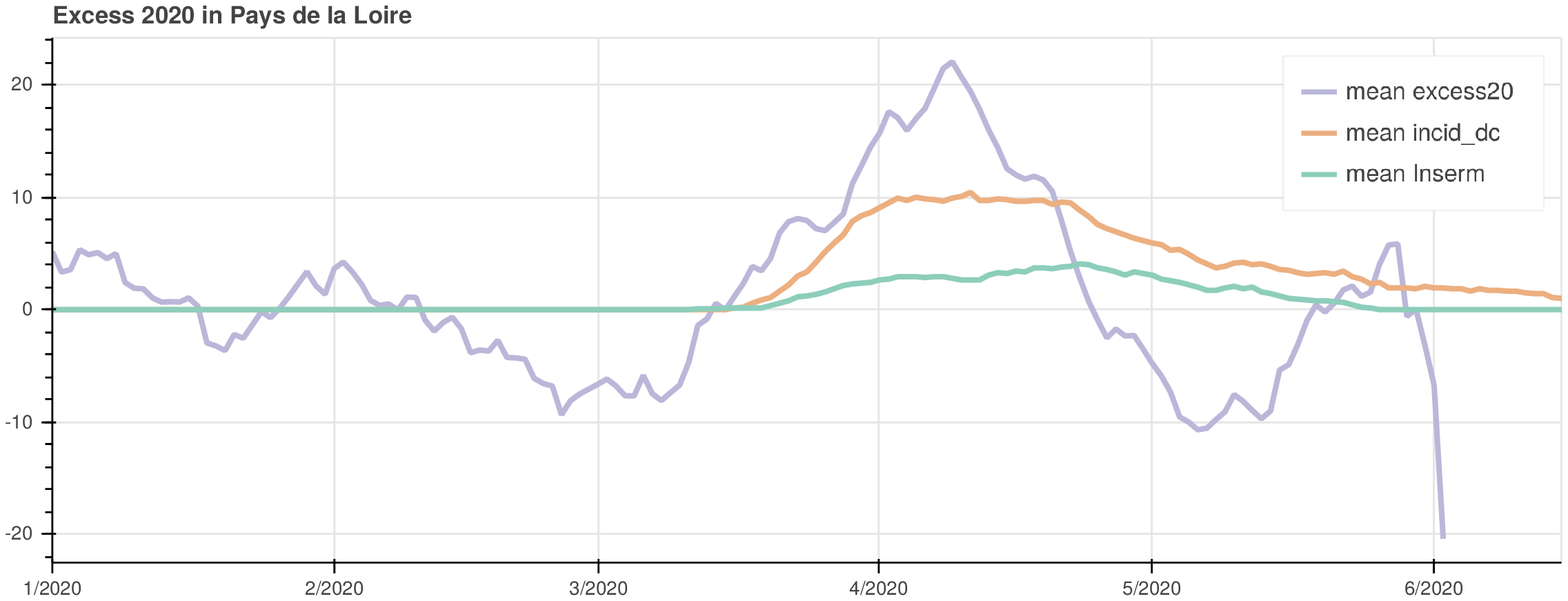}
 \caption{Excess of death in 2020 with respect to the average of 2018/2019, {\tt mean excess20}; deaths certified in hospitals {\tt mean incid\_dc} and deaths electronically certified to Inserm {\tt mean Inserm}, averaged over a 14-days-long window from January to June 2020, in Occitanie  (76) and Pays de la Loire (52) regions.} 
\end{figure}

\begin{figure}[H]
 \centering
\includegraphics[width=6.8cm]{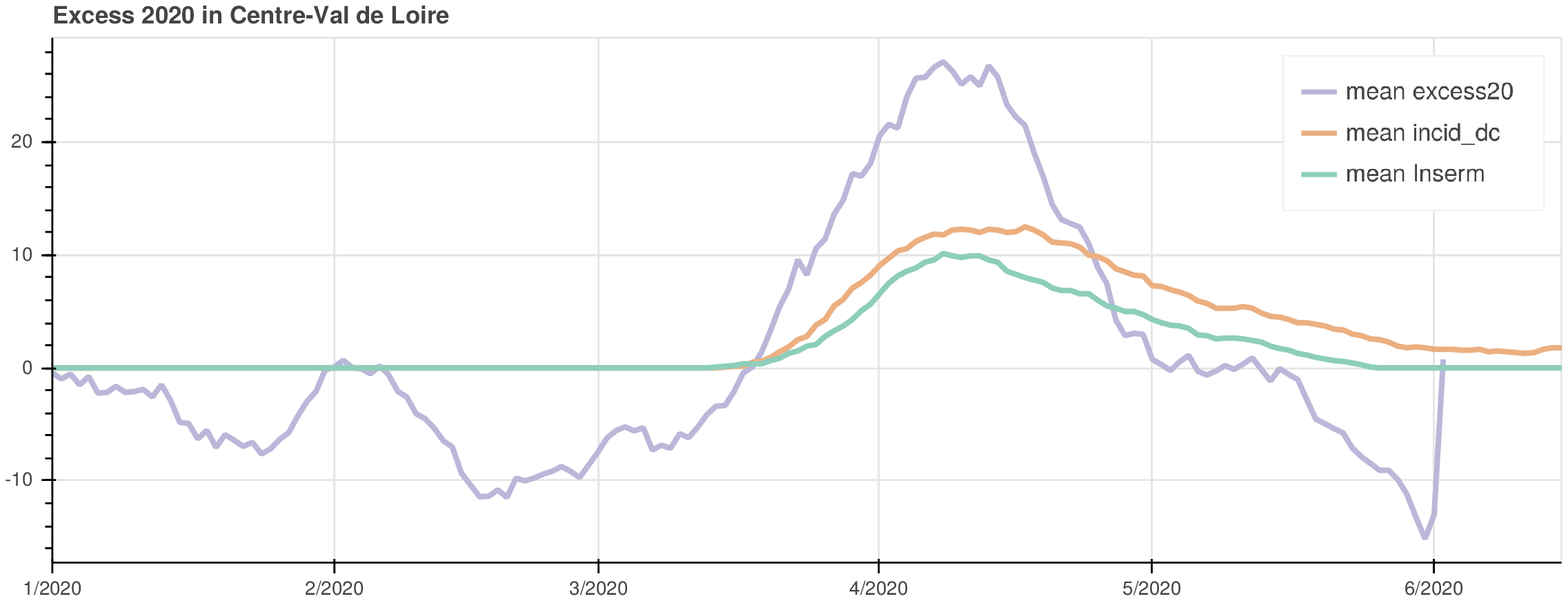}
\includegraphics[width=6.8cm]{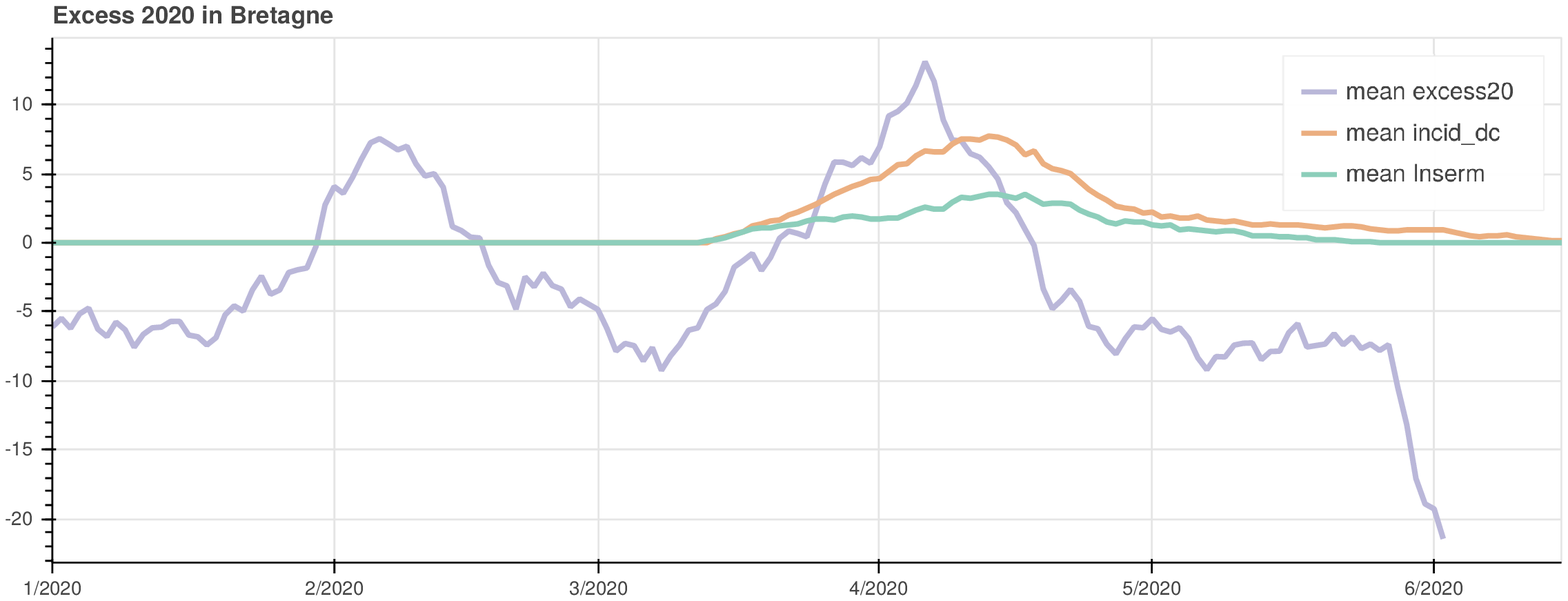}
 \caption{Excess of death in 2020 with respect to the average of 2018/2019, {\tt mean excess20}; deaths certified in hospitals {\tt mean incid\_dc} and deaths electronically certified to Inserm {\tt mean Inserm}, averaged over a 14-days-long window from January to June 2020, in Centre-Val de Loire (24) and Bretagne (53) regions.} 
\end{figure}

\begin{figure}[H]
 \centering
\includegraphics[width=6.8cm]{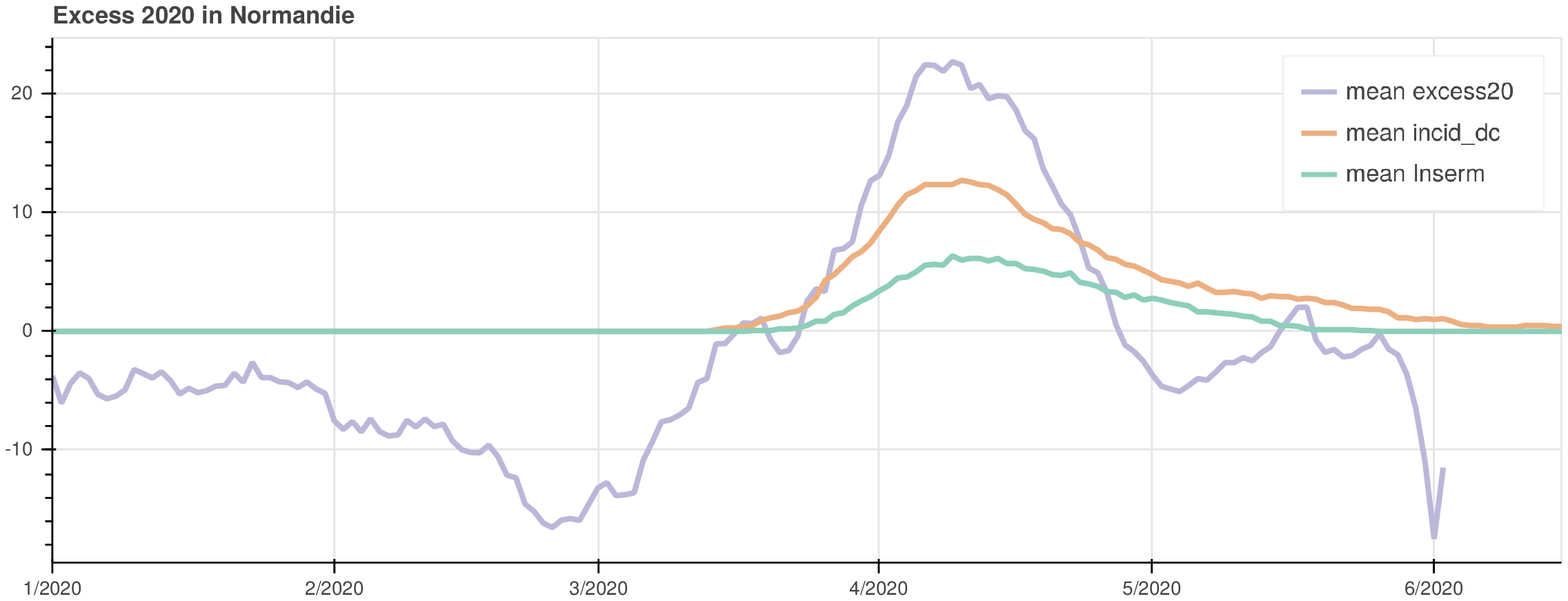}
\includegraphics[width=6.8cm]{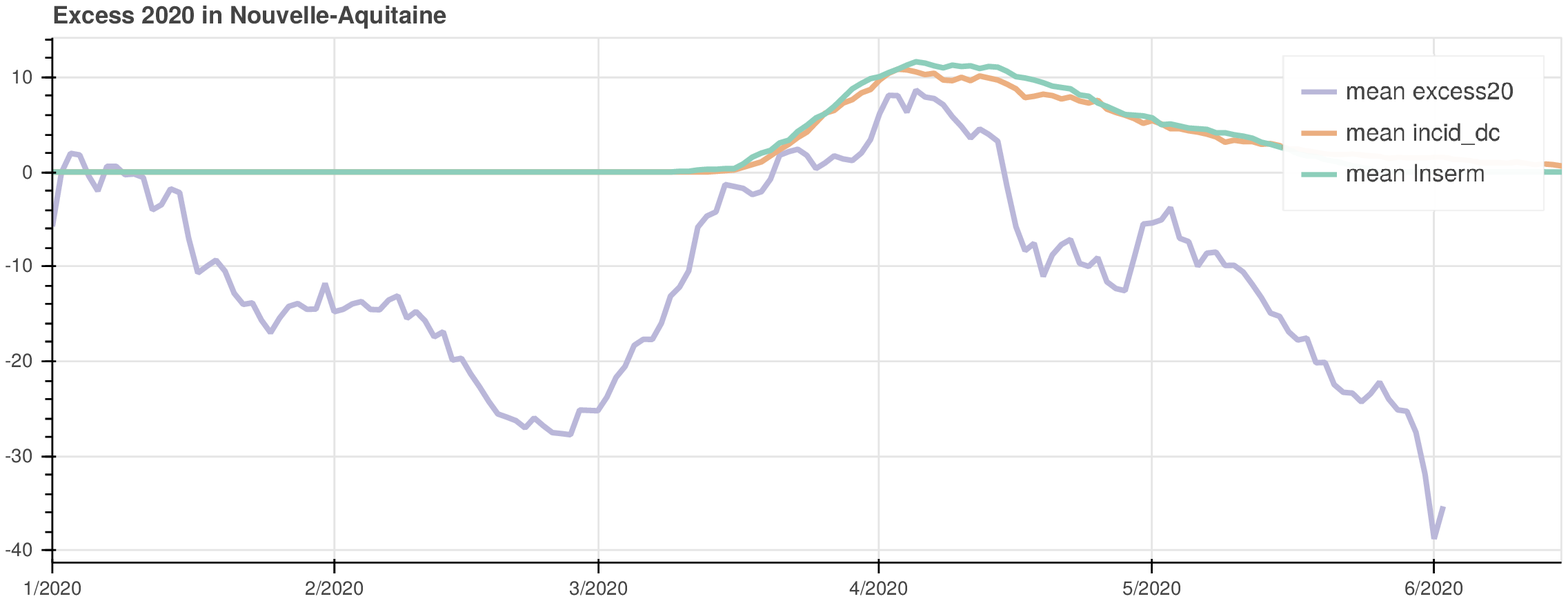}
 \caption{Excess of death in 2020 with respect to the average of 2018/2019, {\tt mean excess20}; deaths certified in hospitals {\tt mean incid\_dc} and deaths electronically certified to Inserm {\tt mean Inserm}, averaged over a 14-days-long window from January to June 2020, in Normandie (28) and Nouvelle-Aquitaine (75) regions.} 
\end{figure}

\begin{figure}[H]
 \centering
\includegraphics[width=6.8cm]{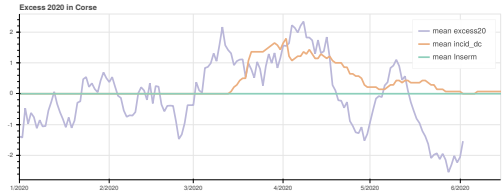}
 \caption{Excess of death in 2020 with respect to the average of 2018/2019, {\tt mean excess20}; deaths certified in hospitals {\tt mean incid\_dc} and deaths electronically certified to Inserm {\tt mean Inserm}, averaged over a 14-days-long window from January to June 2020, in Corse (94) region.} 
\end{figure}

%%%%%%%%%%%%%%%%%%%%%%%%%%%%%%%%%%%%%%%%%%%%%%%%%%%%%%%%%%%%%%%%%%%%%
\newpage

\section{Deaths time series for fitting in French regions}\label{app:excess20corr}

\begin{figure}[H]
 \centering
\includegraphics[width=6.8cm]{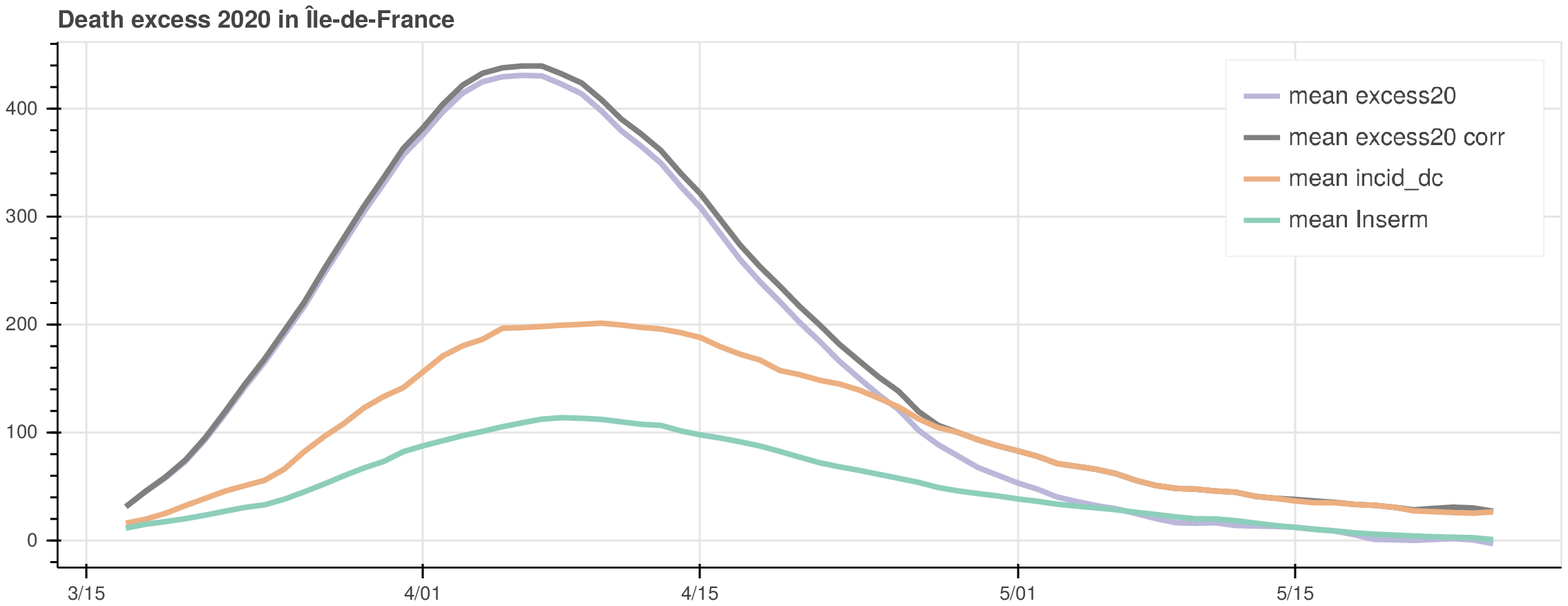}
\includegraphics[width=6.8cm]{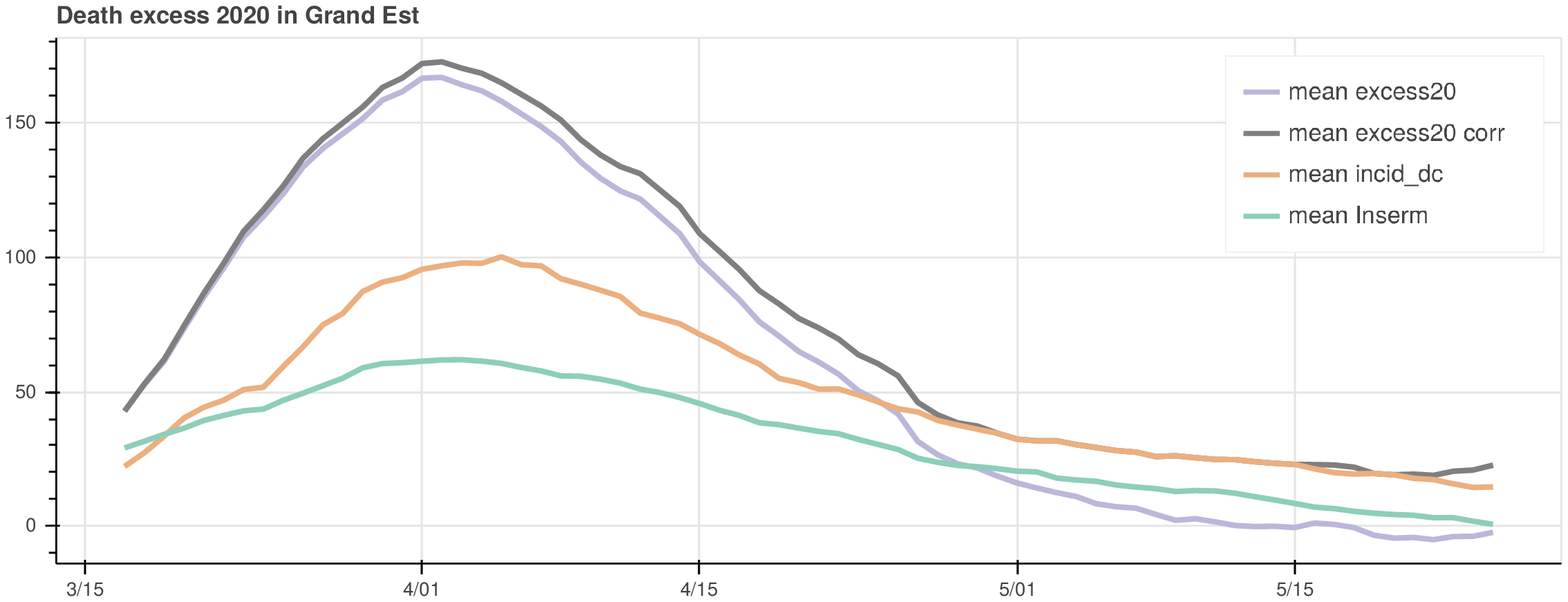}
 \caption{Excess of death in 2020 with respect to the average of 2018/2019, {\tt mean excess20}; deaths certified in hospitals {\tt mean incid\_dc} and deaths electronically certified to Inserm {\tt mean Inserm}, averaged over a 14-days-long window. Corrected average of excess {\tt mean excess20 corr}, in the identification interval March 17th - April 28th, in Île-de-France (11) and Grand Est (44) regions.} 
\end{figure}

\begin{figure}[H]
 \centering
\includegraphics[width=6.8cm]{2Covid/2_3_reg84.eps}
\includegraphics[width=6.8cm]{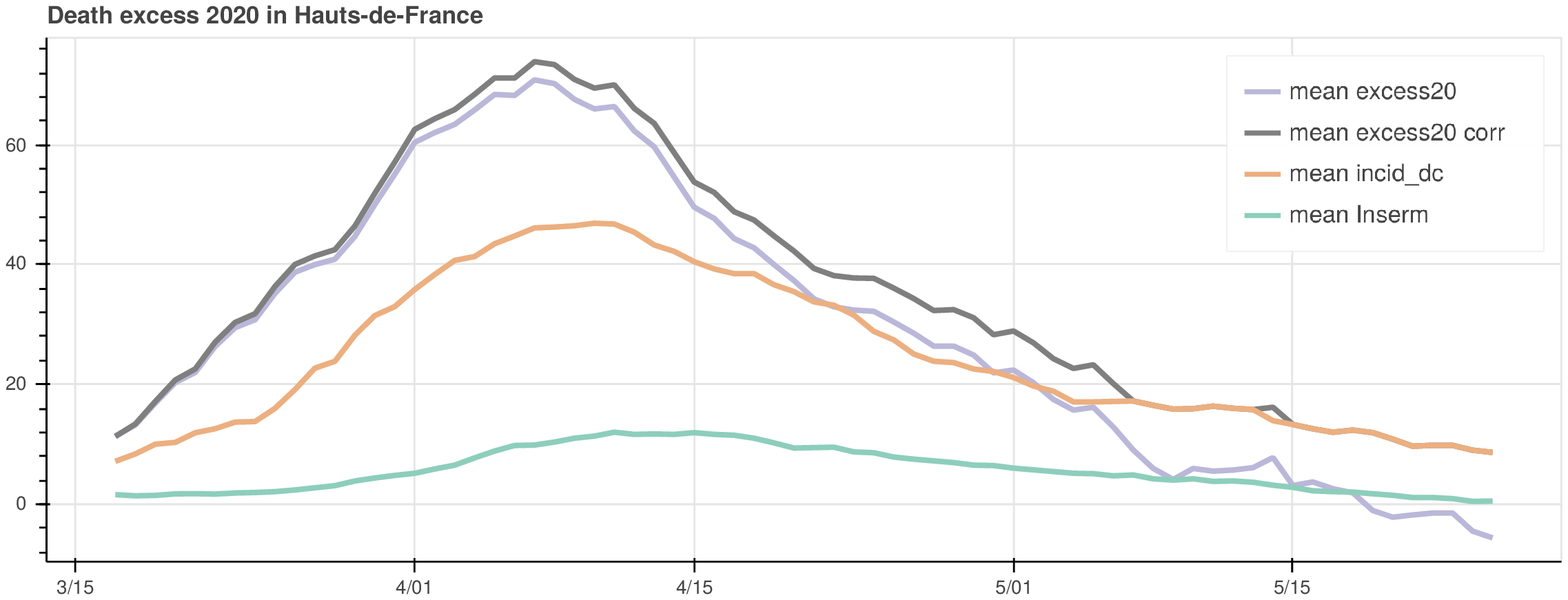}
 \caption{Excess of death in 2020 with respect to the average of 2018/2019, {\tt mean excess20}; deaths certified in hospitals {\tt mean incid\_dc} and deaths electronically certified to Inserm {\tt mean Inserm}, averaged over a 14-days-long window. Corrected average of excess {\tt mean excess20 corr}, in the identification interval March 17th - April 28th, in Auvergne-Rhône-Alpes (84) and Hauts-de-France (32) regions.} 
\end{figure}

\begin{figure}[H]
 \centering
\includegraphics[width=6.8cm]{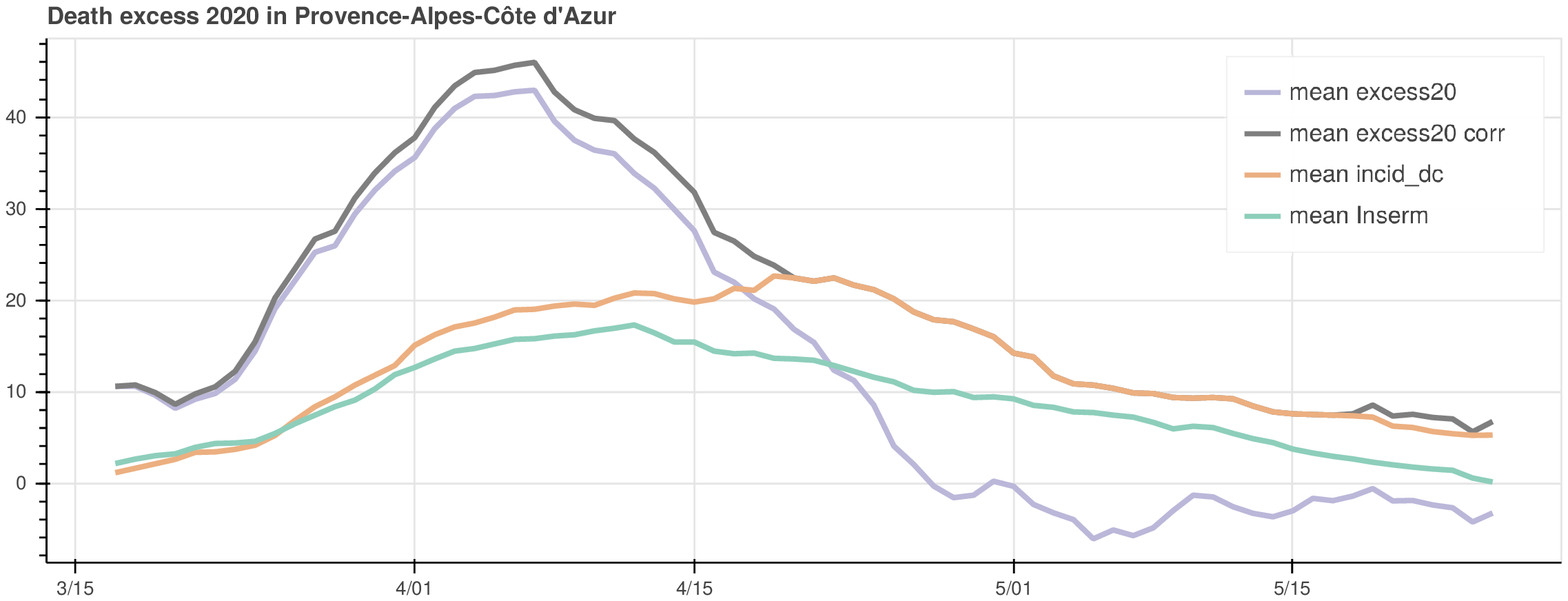}
\includegraphics[width=6.8cm]{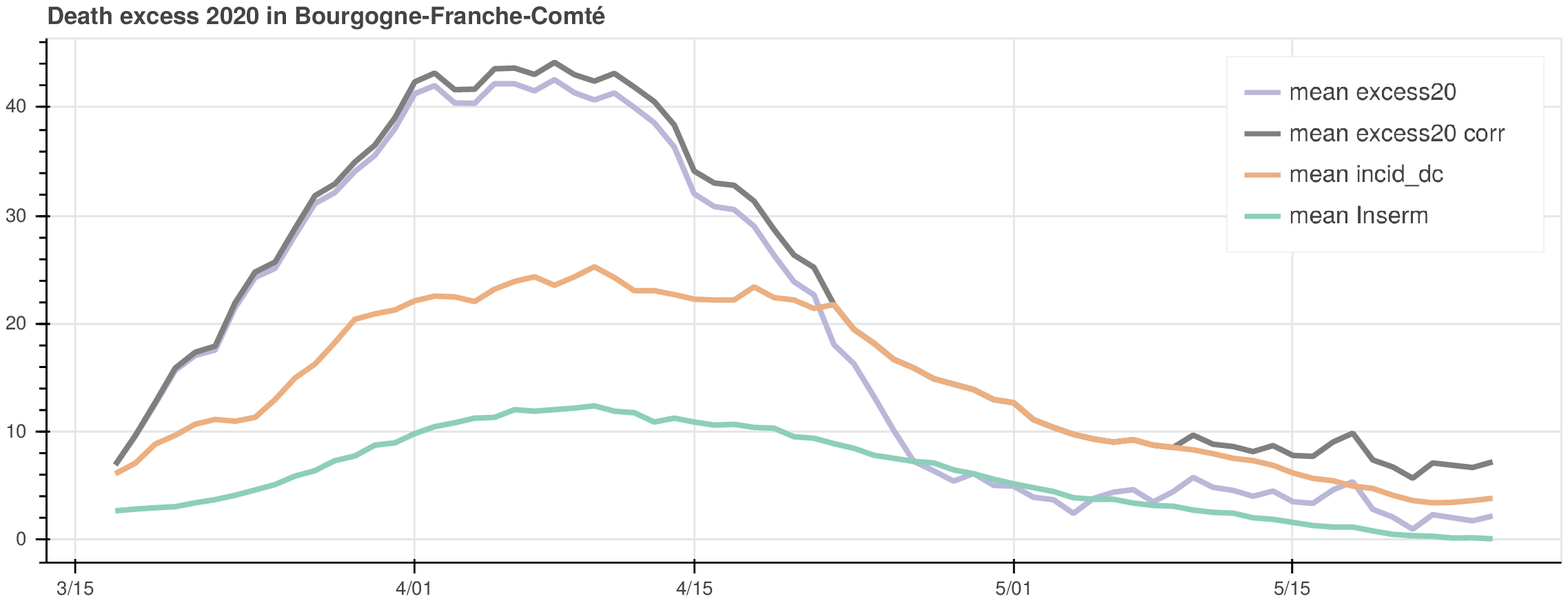}
 \caption{Excess of death in 2020 with respect to the average of 2018/2019, {\tt mean excess20}; deaths certified in hospitals {\tt mean incid\_dc} and deaths electronically certified to Inserm {\tt mean Inserm}, averaged over a 14-days-long window. Corrected average of excess {\tt mean excess20 corr}, in the identification interval March 17th - April 28th, in Provence-Alpes-Côte d'Azur (93) and Bourgogne-Franche-Comté (27)
regions.} 
\end{figure}

\begin{figure}[H]
 \centering
\includegraphics[width=6.8cm]{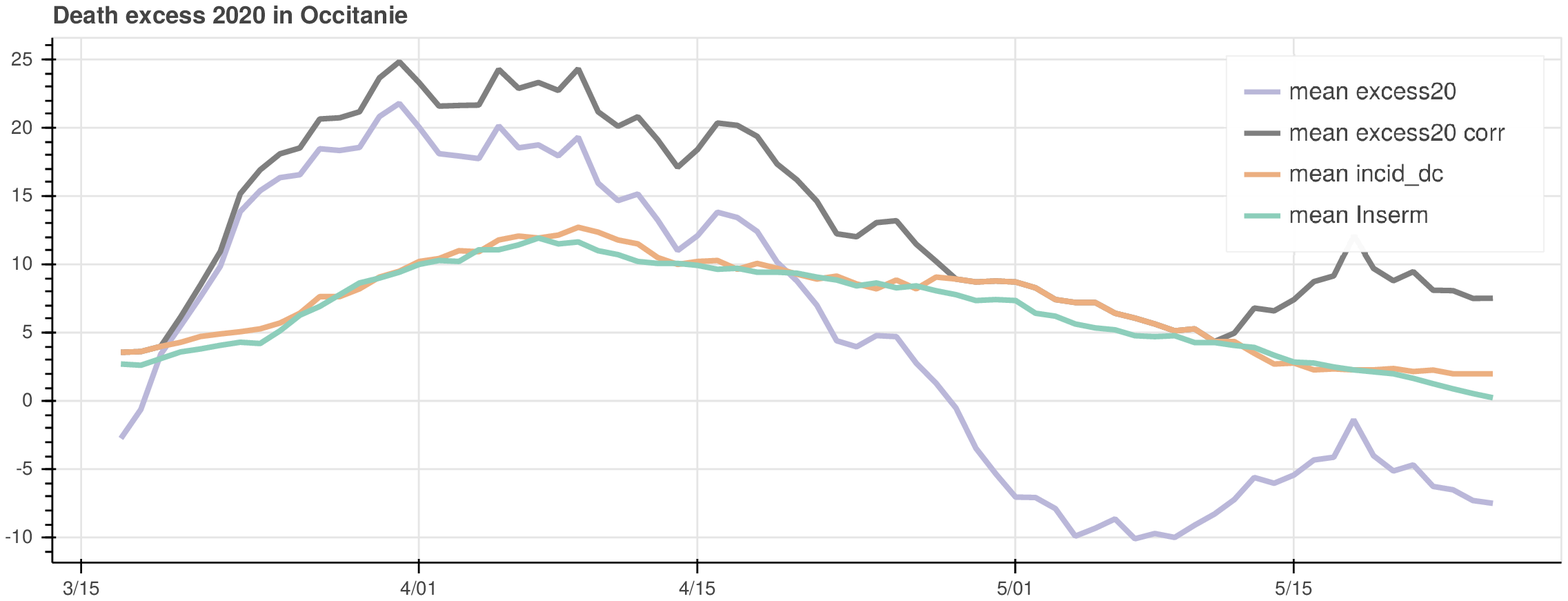}
\includegraphics[width=6.8cm]{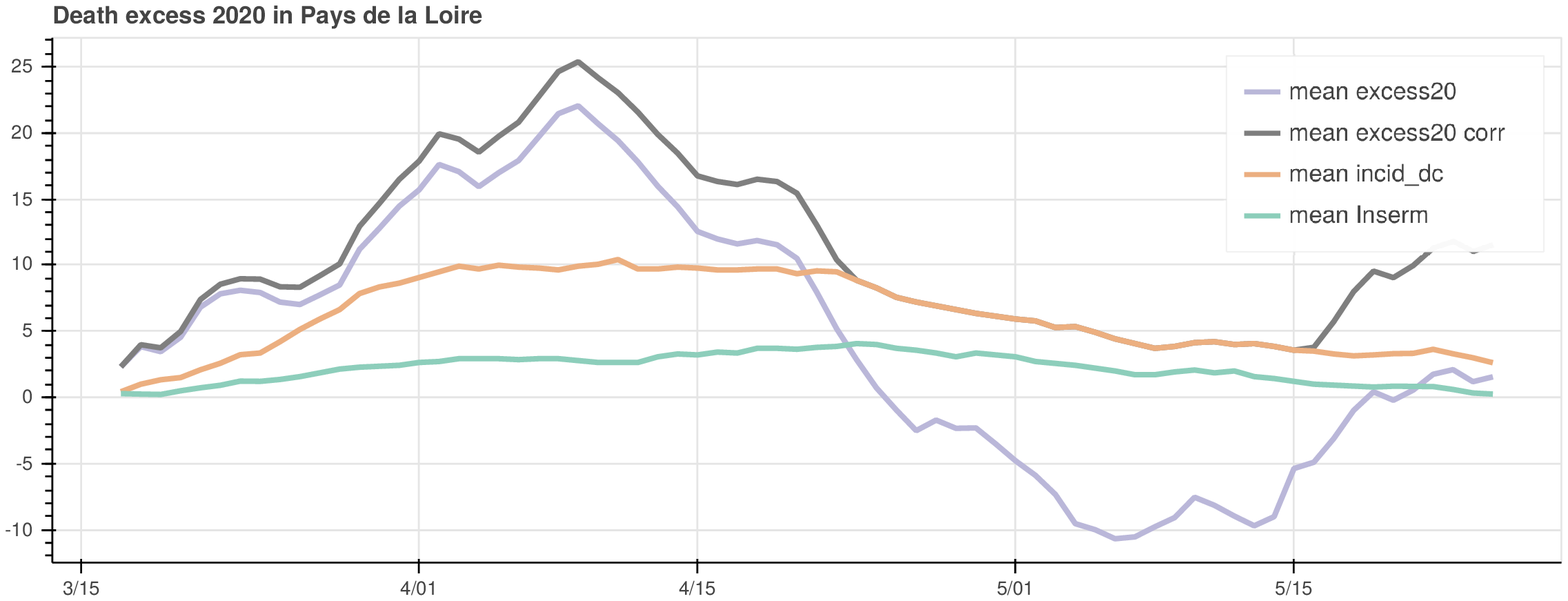}
 \caption{Excess of death in 2020 with respect to the average of 2018/2019, {\tt mean excess20}; deaths certified in hospitals {\tt mean incid\_dc} and deaths electronically certified to Inserm {\tt mean Inserm}, averaged over a 14-days-long window. Corrected average of excess {\tt mean excess20 corr}, in the identification interval March 17th - April 28th, in Occitanie  (76) and Pays de la Loire (52) regions.} 
\end{figure}

\begin{figure}[H]
 \centering
\includegraphics[width=6.8cm]{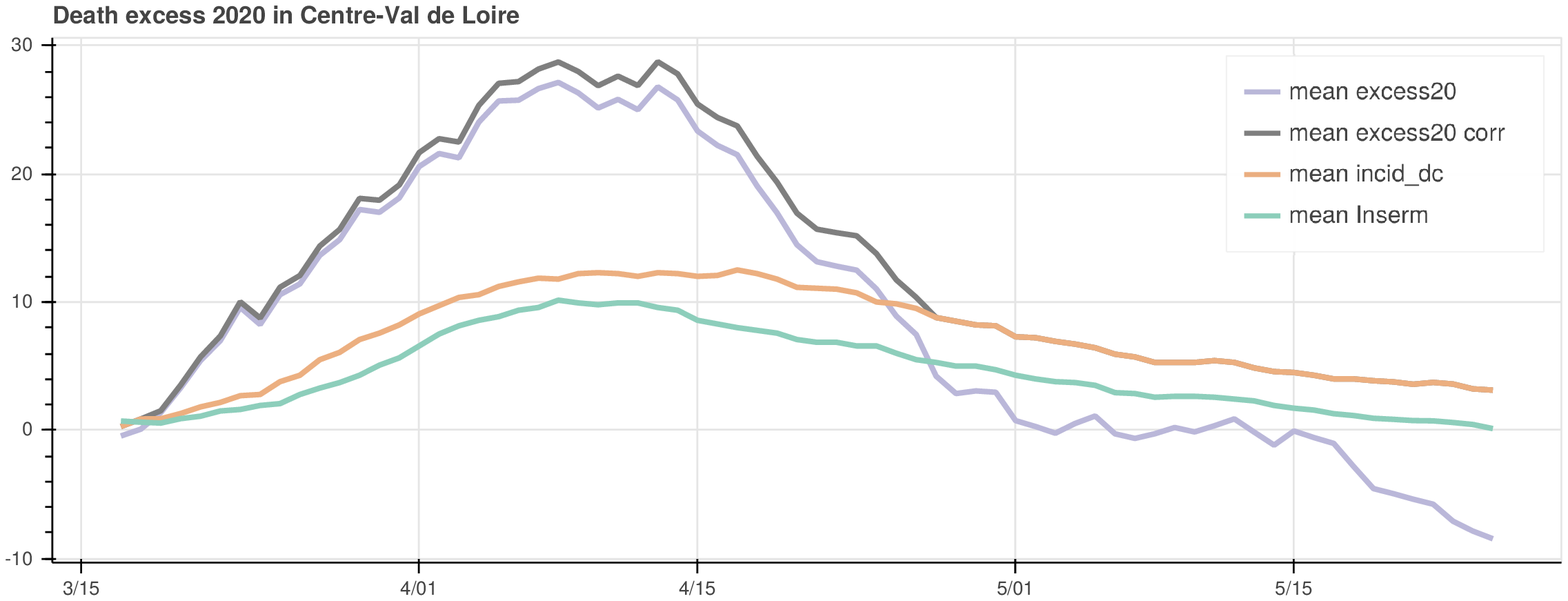}
\includegraphics[width=6.8cm]{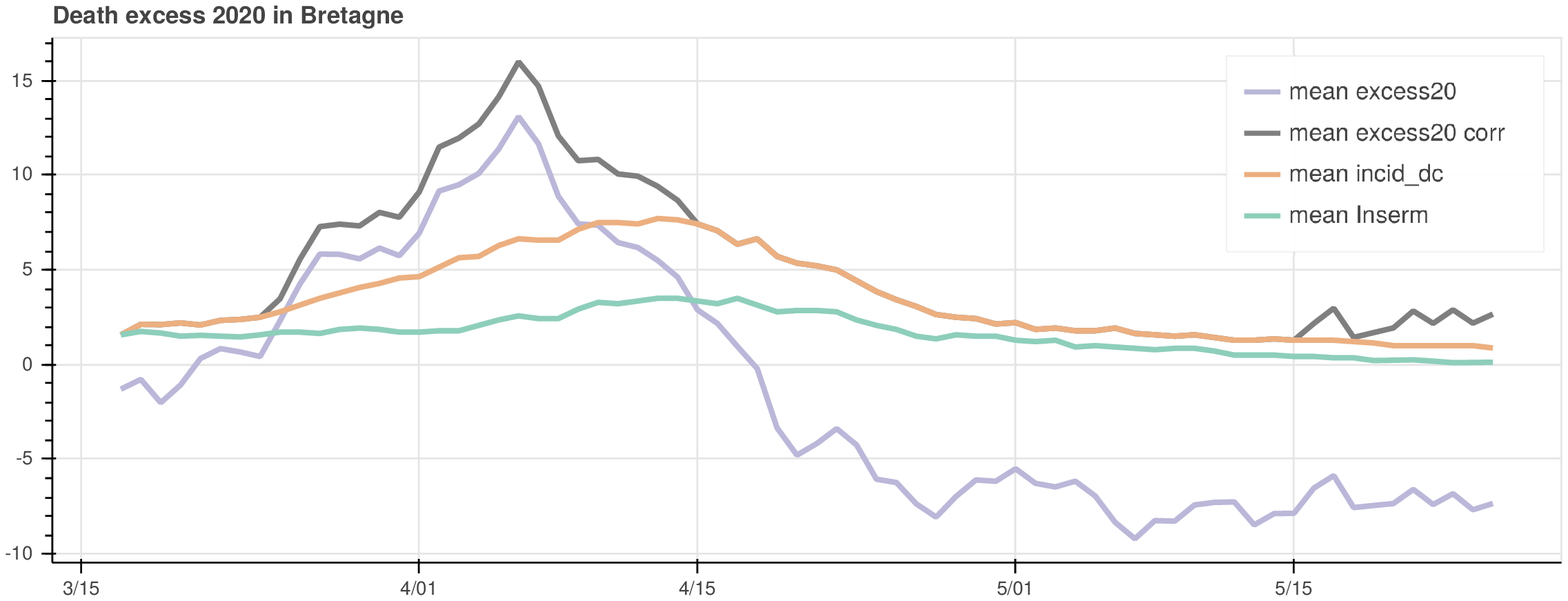}
 \caption{Excess of death in 2020 with respect to the average of 2018/2019, {\tt mean excess20}; deaths certified in hospitals {\tt mean incid\_dc} and deaths electronically certified to Inserm {\tt mean Inserm}, averaged over a 14-days-long window. Corrected average of excess {\tt mean excess20 corr}, in the identification interval March 17th - April 28th, in Centre-Val de Loire (24) and Bretagne (53) regions.} 
\end{figure}

\begin{figure}[H]
 \centering
\includegraphics[width=6.8cm]{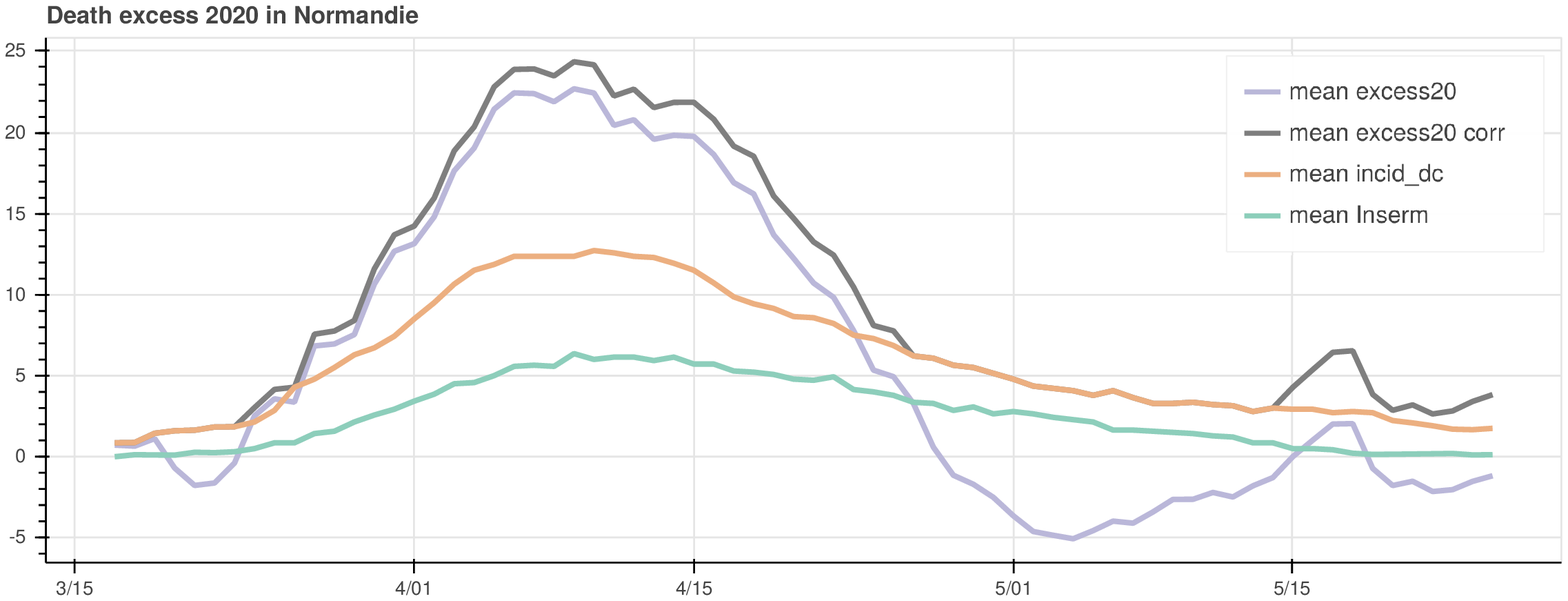}
\includegraphics[width=6.8cm]{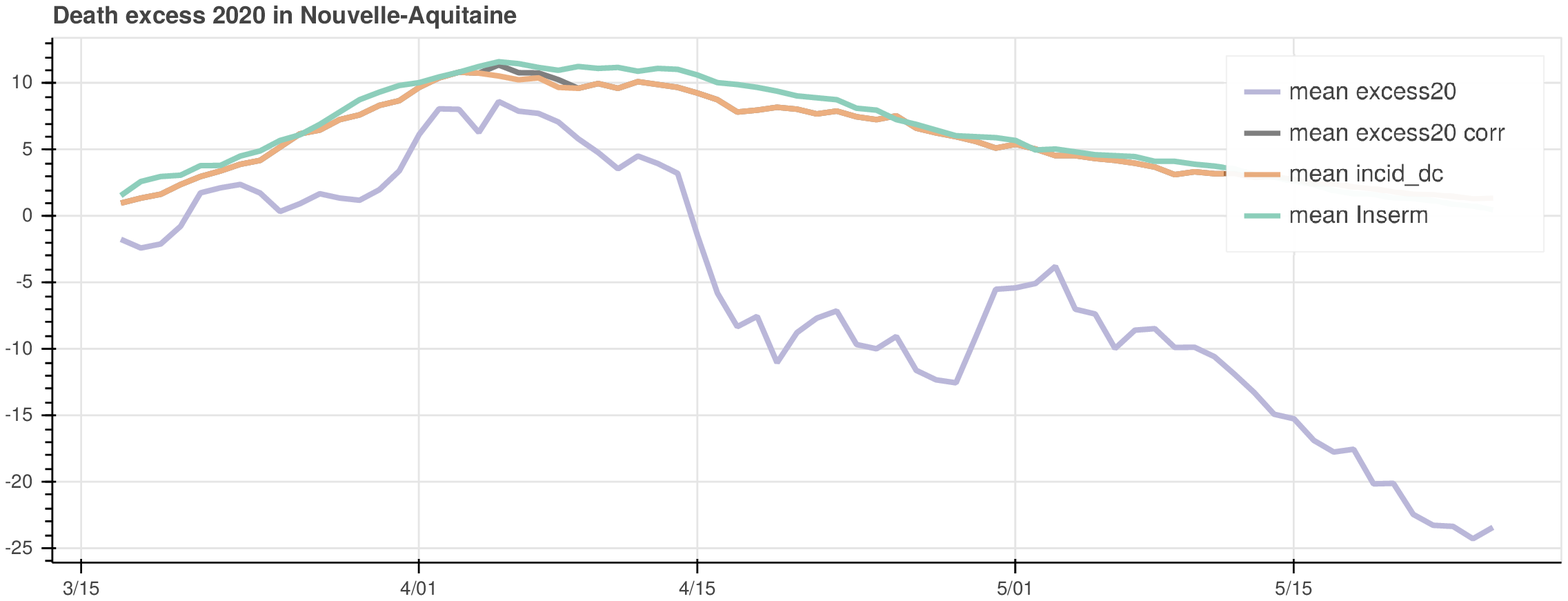}
 \caption{Excess of death in 2020 with respect to the average of 2018/2019, {\tt mean excess20}; deaths certified in hospitals {\tt mean incid\_dc} and deaths electronically certified to Inserm {\tt mean Inserm}, averaged over a 14-days-long window. Corrected average of excess {\tt mean excess20 corr}, in the identification interval March 17th - April 28th, in Normandie (28) and Nouvelle-Aquitaine (75) regions.} 
\end{figure}

\begin{figure}[H]
 \centering
\includegraphics[width=6.8cm]{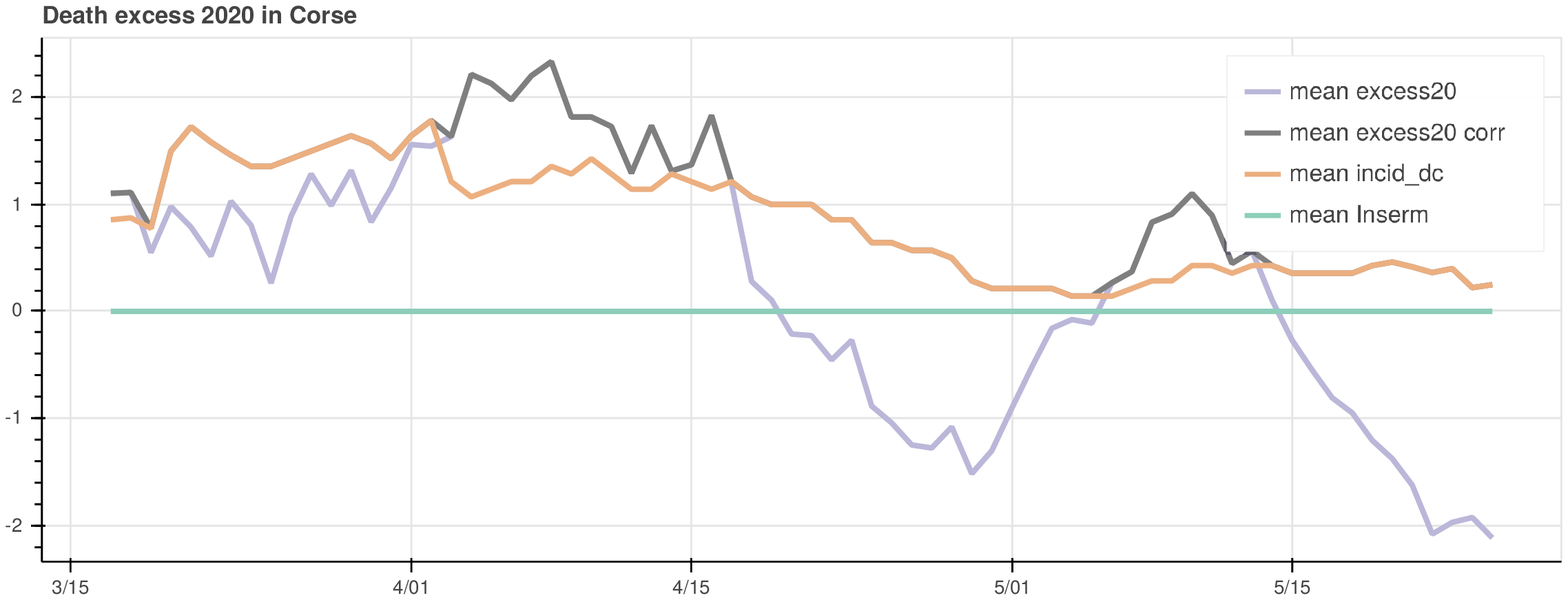}
 \caption{Excess of death in 2020 with respect to the average of 2018/2019, {\tt mean excess20}; deaths certified in hospitals {\tt mean incid\_dc} and deaths electronically certified to Inserm {\tt mean Inserm}, averaged over a 14-days-long window. Corrected average of excess {\tt mean excess20 corr}, in the identification interval March 17th - April 28th, in Corse (94) region.} 
\end{figure}

\newpage

\section{Validation on other data: results}\label{app:valid}

% Île-de-France (11) and Grand Est (44) regions
% Auvergne-Rhône-Alpes (84) and Hauts-de-France (32)
% Provence-Alpes-Côte d'Azur (93) and Bourgogne-Franche-Comté (27)
% Occitanie  (76) and Pays de la Loire (52)
% Centre-Val de Loire (24) and Bretagne (53)
% Normandie (28) and Nouvelle-Aquitaine (75)
% Corse (94) region
% 

\subsection{Provence-Alpes-Côte d'Azur}

 \vspace{-0.2cm}

\begin{figure}[H]
 \centering
\includegraphics[width=6.8cm]{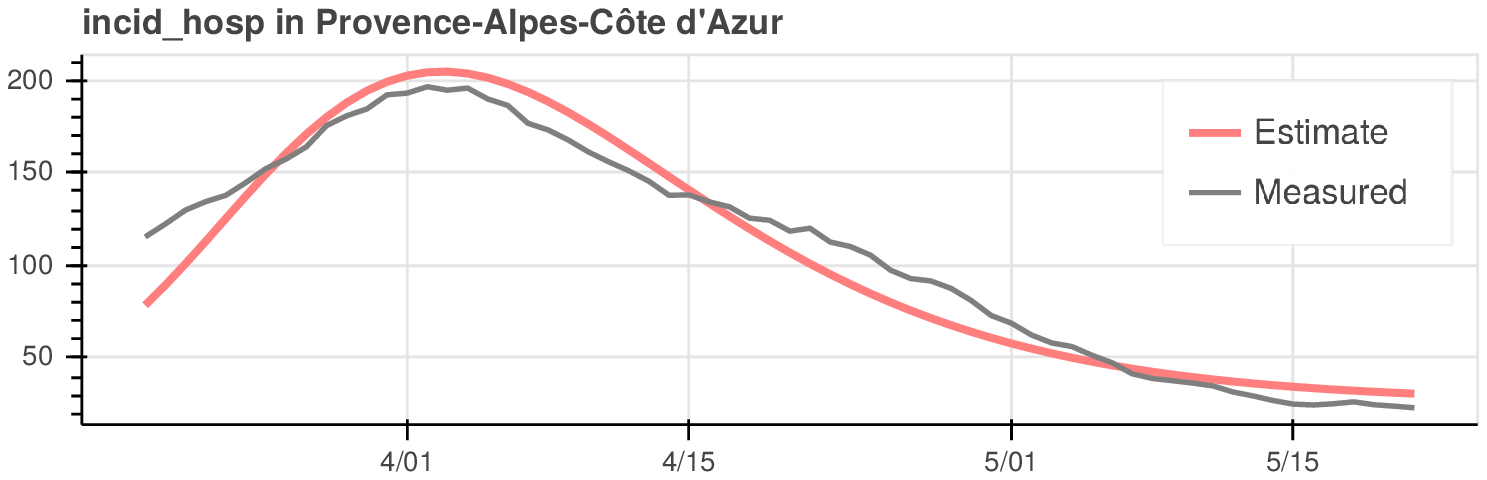}
\includegraphics[width=6.8cm]{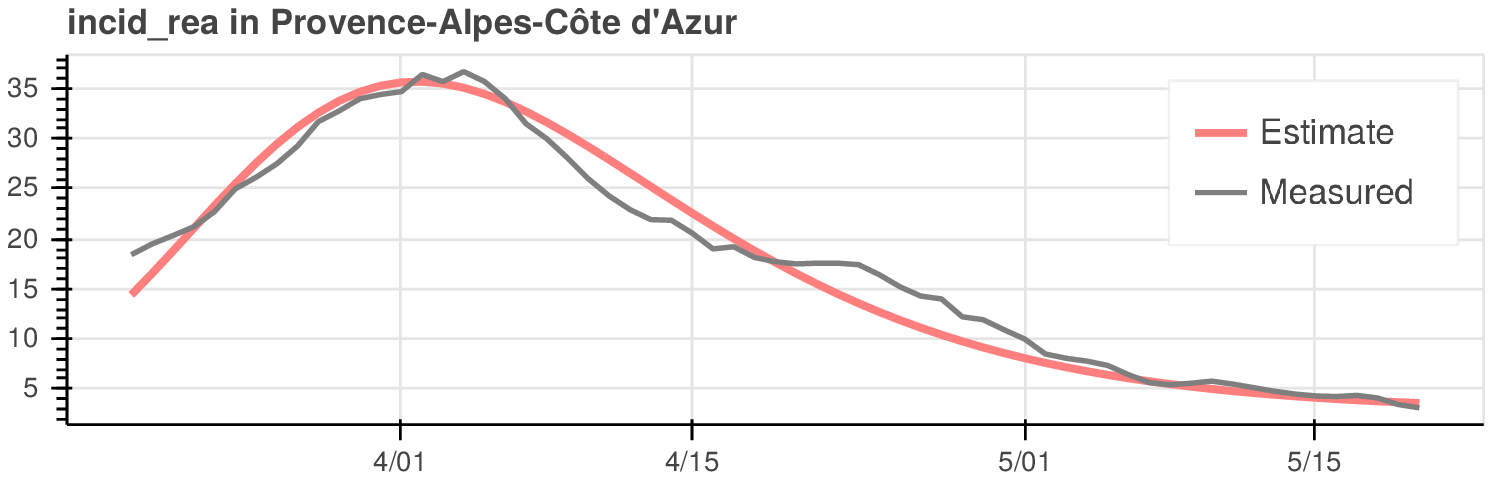}
 \vspace{-0.7cm}
 \caption{Comparison of time series {\tt incid\_hosp} and {\tt incid\_rea} averaged over a 14-days-long window, with the fitted outputs, for Provence-Alpes-Côte d'Azur (93) region.}
\end{figure}

 \vspace{-0.7cm}
\begin{figure}[H]
 \centering
\includegraphics[width=6.8cm]{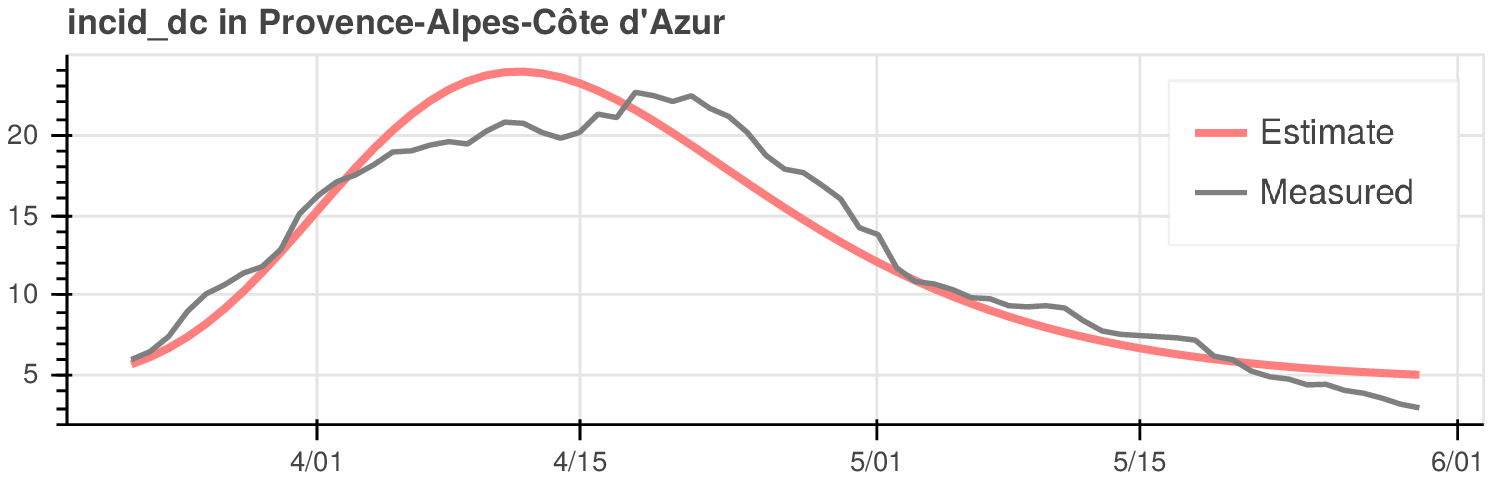}
\includegraphics[width=6.8cm]{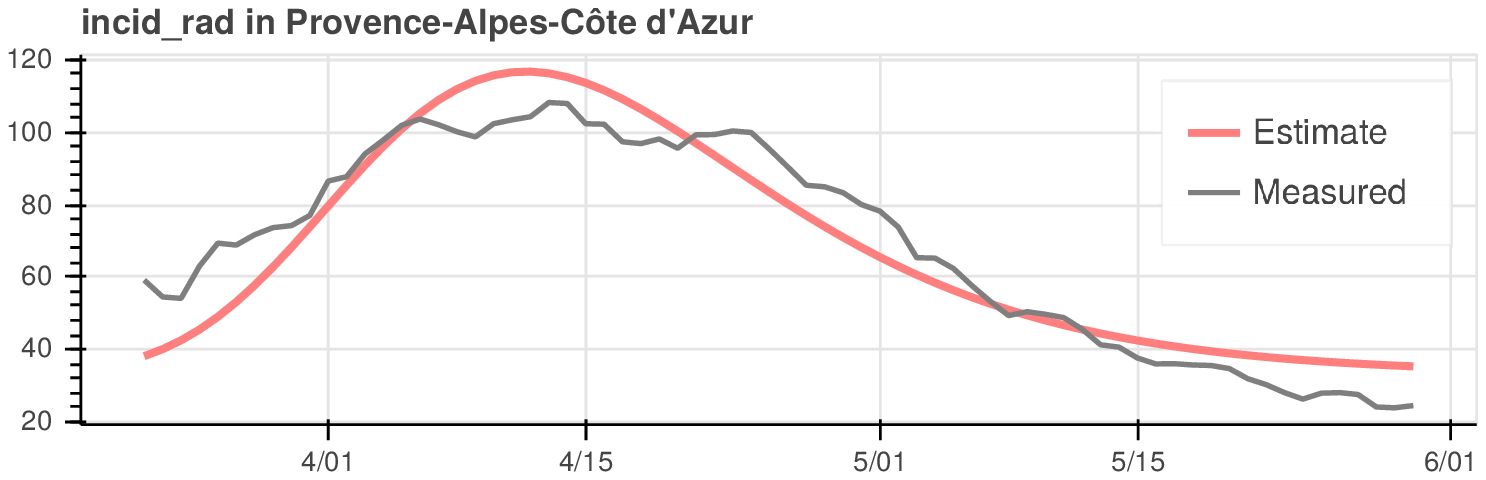}
 \vspace{-0.7cm}
 \caption{Comparison of time series {\tt incid\_dc} and {\tt incid\_rad} averaged over a 14-days-long window, with the fitted outputs, for Provence-Alpes-Côte d'Azur (93) region.}
\end{figure}

 \vspace{-0.7cm}
\begin{figure}[H]
 \centering
\includegraphics[width=6.8cm]{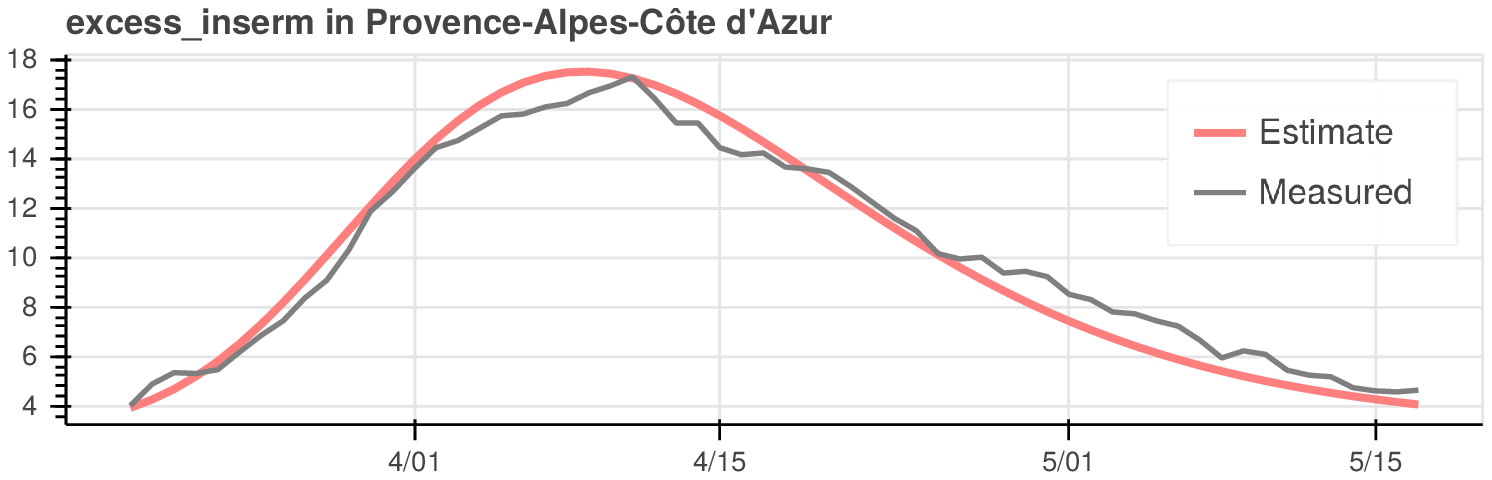}
\includegraphics[width=6.8cm]{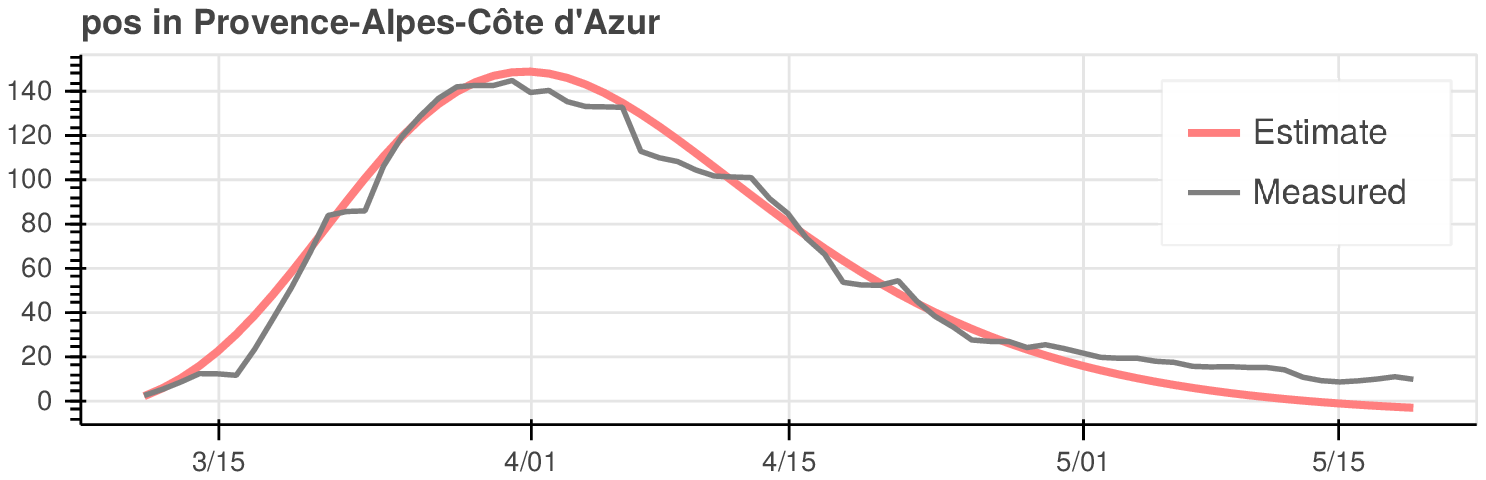}
 \vspace{-0.7cm}
 \caption{Comparison of time series {\tt incid\_inserm} and {\tt pos} averaged over a 14-days-long window, with the fitted outputs, for Provence-Alpes-Côte d'Azur (93) region.}
\end{figure}

 \vspace{-0.7cm}
\begin{figure}[H]
 \centering
\includegraphics[width=6.8cm]{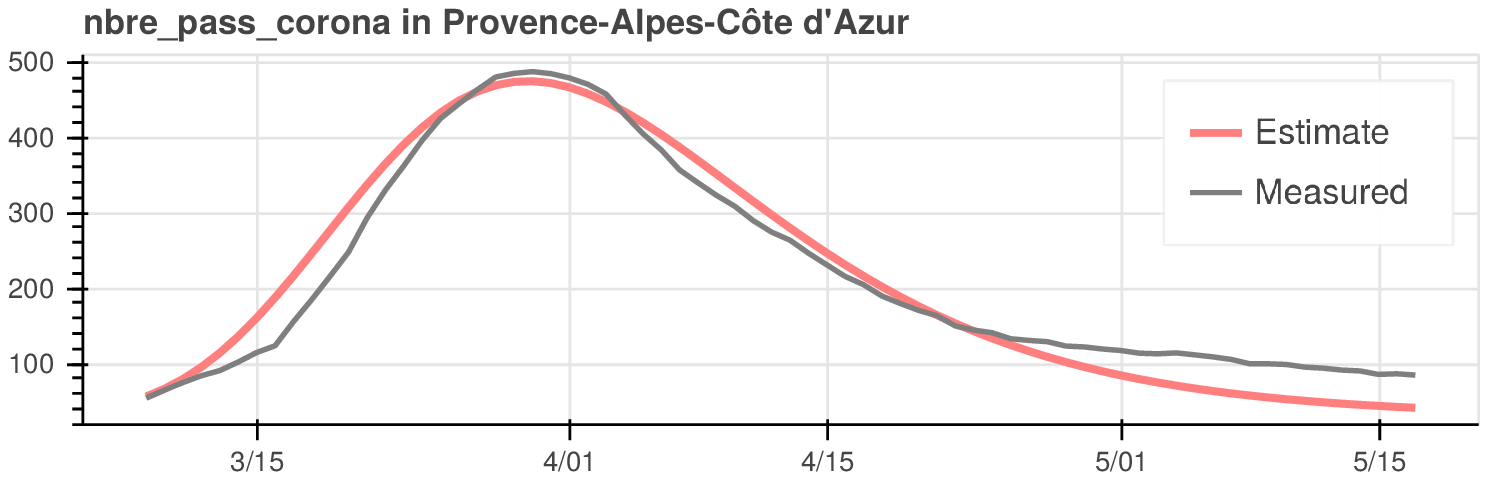}
\includegraphics[width=6.8cm]{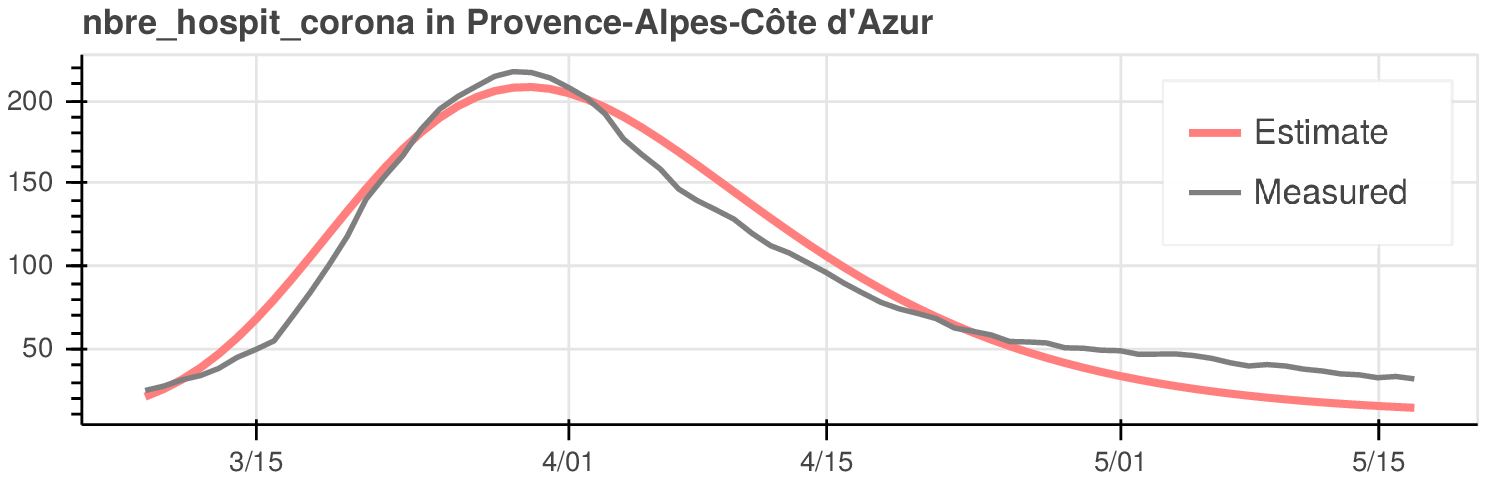}
 \vspace{-0.7cm}
 \caption{Comparison of time series {\tt nbre\_pass\_corona} and 
 {\tt nbre\_hospit\_corona} averaged over a 14-days-long window, with the fitted outputs, for Provence-Alpes-Côte d'Azur (93) region.}
\end{figure}
 \vspace{-0.7cm}
\begin{figure}[H]
 \centering
\includegraphics[width=6.8cm]{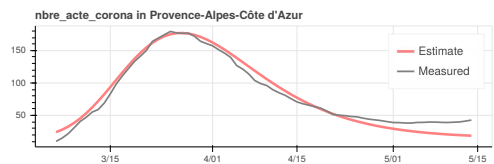}
 \vspace{-0.3cm}
 \caption{Comparison of time series {\tt nbre\_acte\_corona} averaged over a 14-days-long window, with the fitted output, for Provence-Alpes-Côte d'Azur (93) region.}
\end{figure}
% Île-de-France (11) and Grand Est (44) regions
% Auvergne-Rhône-Alpes (84) and Hauts-de-France (32)
% Provence-Alpes-Côte d'Azur (93) and Bourgogne-Franche-Comté (27)
% Occitanie  (76) and Pays de la Loire (52)
% Centre-Val de Loire (24) and Bretagne (53)
% Normandie (28) and Nouvelle-Aquitaine (75)
% Corse (94) region
% 

\subsection{Bourgogne-Franche-Comté}

 \vspace{-0.2cm}

\begin{figure}[H]
 \centering
\includegraphics[width=6.8cm]{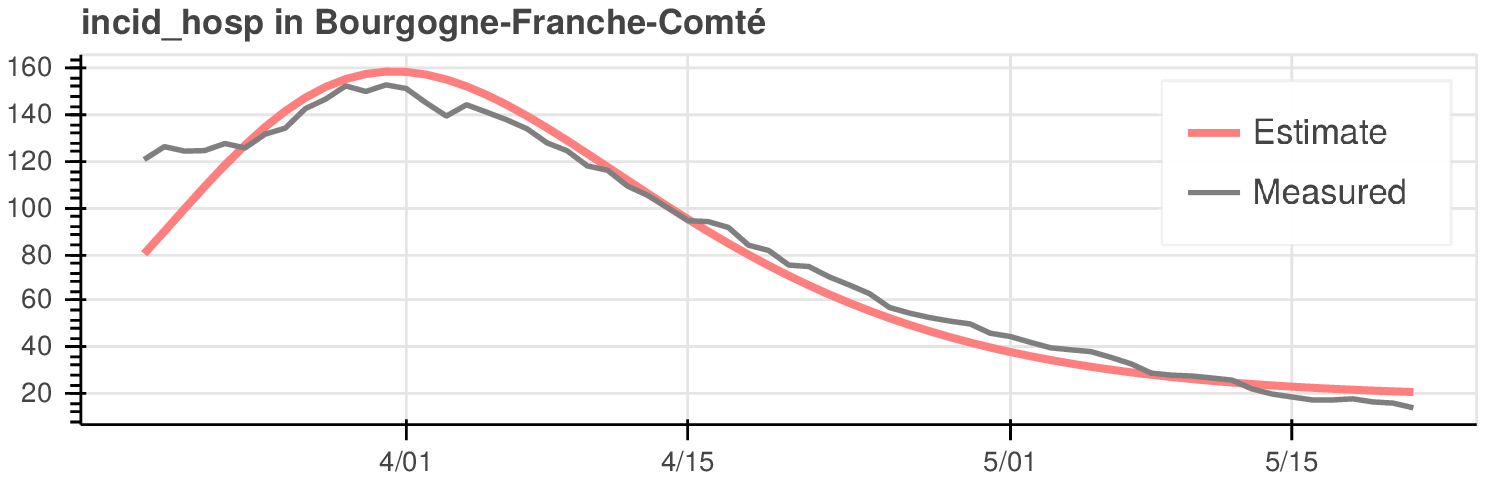}
\includegraphics[width=6.8cm]{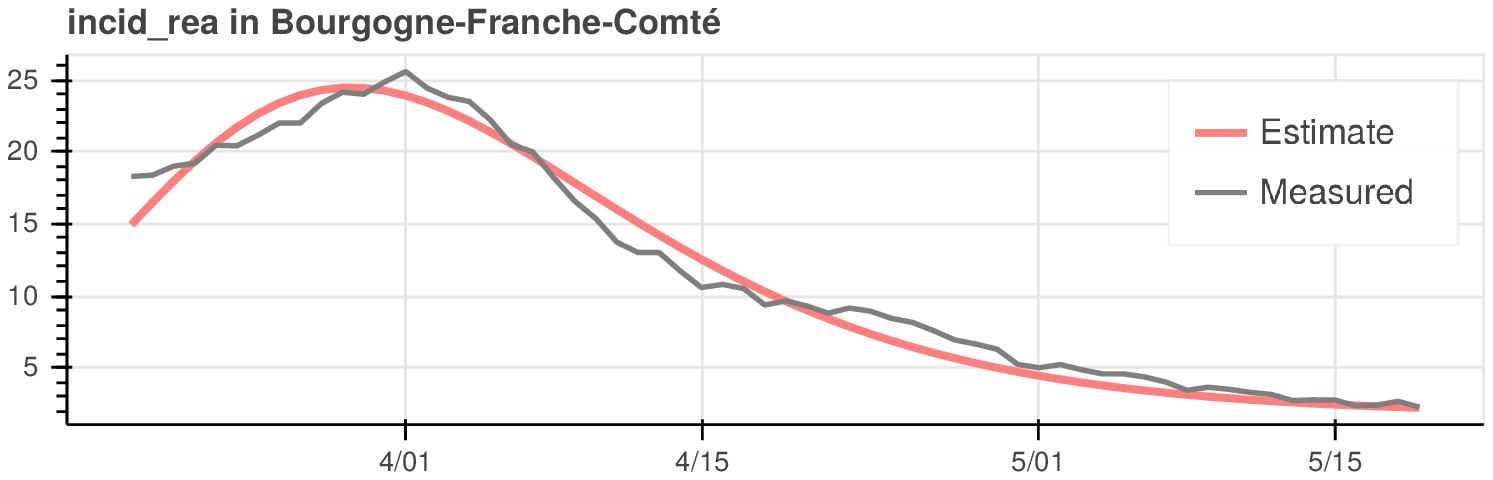}
  \vspace{-0.7cm}
\caption{Comparison of time series {\tt incid\_hosp} and {\tt incid\_rea} averaged over a 14-days-long window, with the fitted outputs, for Bourgogne-Franche-Comté (27) region.}
\end{figure}

\vspace{-0.7cm}
\begin{figure}[H]
 \centering
\includegraphics[width=6.8cm]{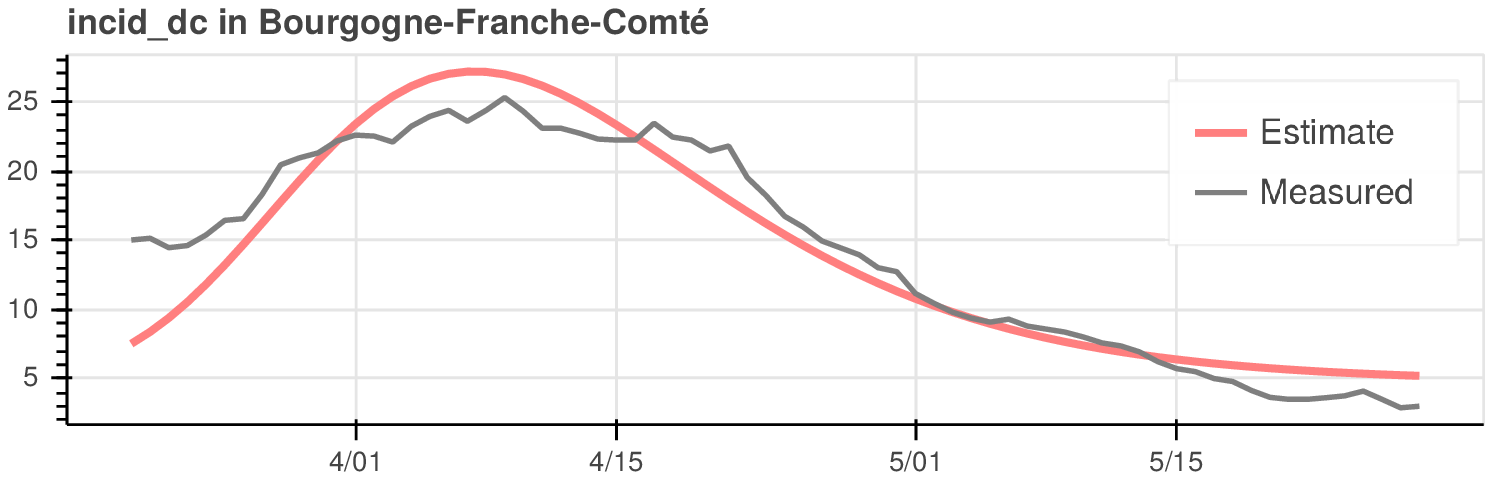}
\includegraphics[width=6.8cm]{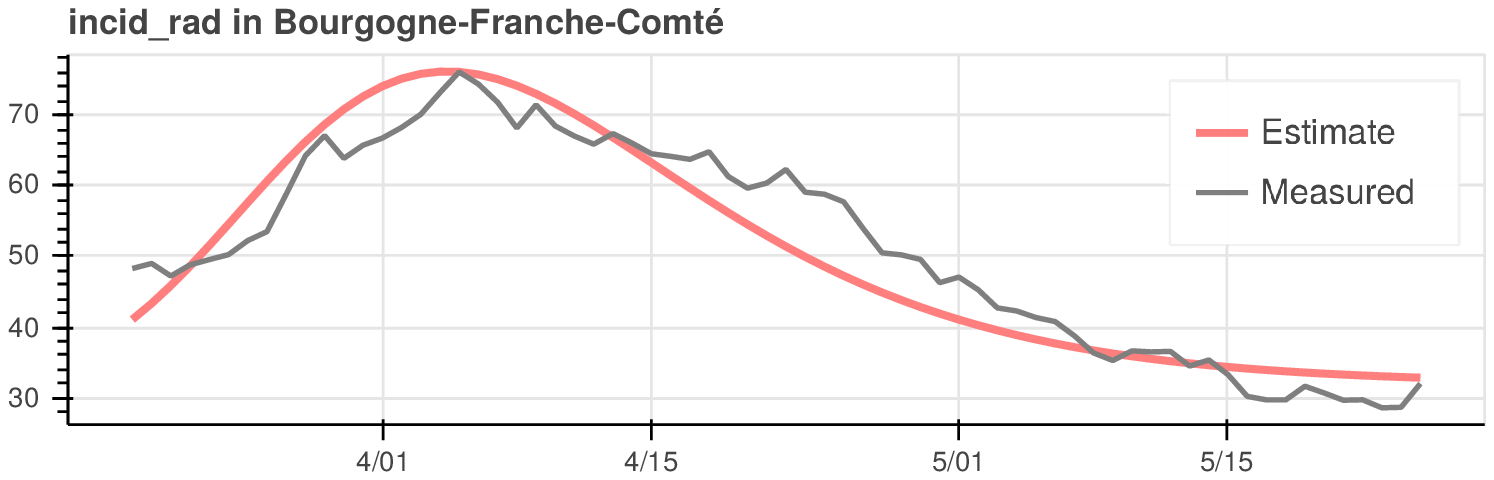}
\vspace{-0.7cm}
 \caption{Comparison of time series {\tt incid\_dc} and {\tt incid\_rad} averaged over a 14-days-long window, with the fitted outputs, for Bourgogne-Franche-Comté (27) region.}
\end{figure}

\vspace{-0.7cm}
\begin{figure}[H]
 \centering
\includegraphics[width=6.8cm]{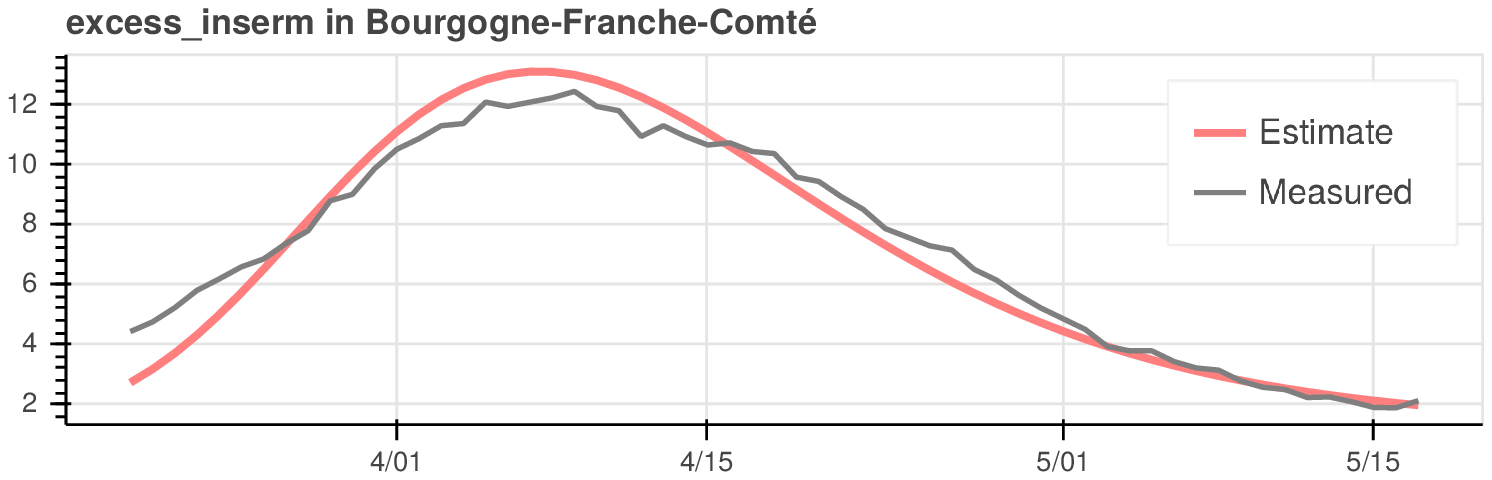}
\includegraphics[width=6.8cm]{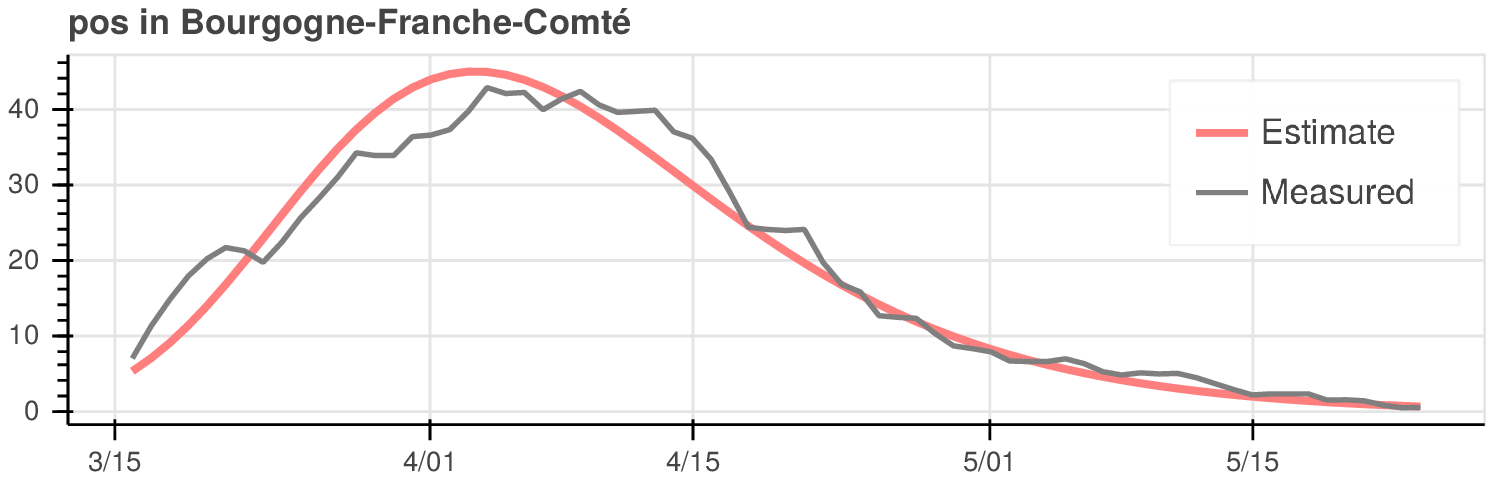}
\vspace{-0.7cm}
 \caption{Comparison of time series {\tt incid\_inserm} and {\tt pos} averaged over a 14-days-long window, with the fitted outputs, for Bourgogne-Franche-Comté (27) region.}
\end{figure}

\vspace{-0.7cm}
\begin{figure}[H]
 \centering
\includegraphics[width=6.8cm]{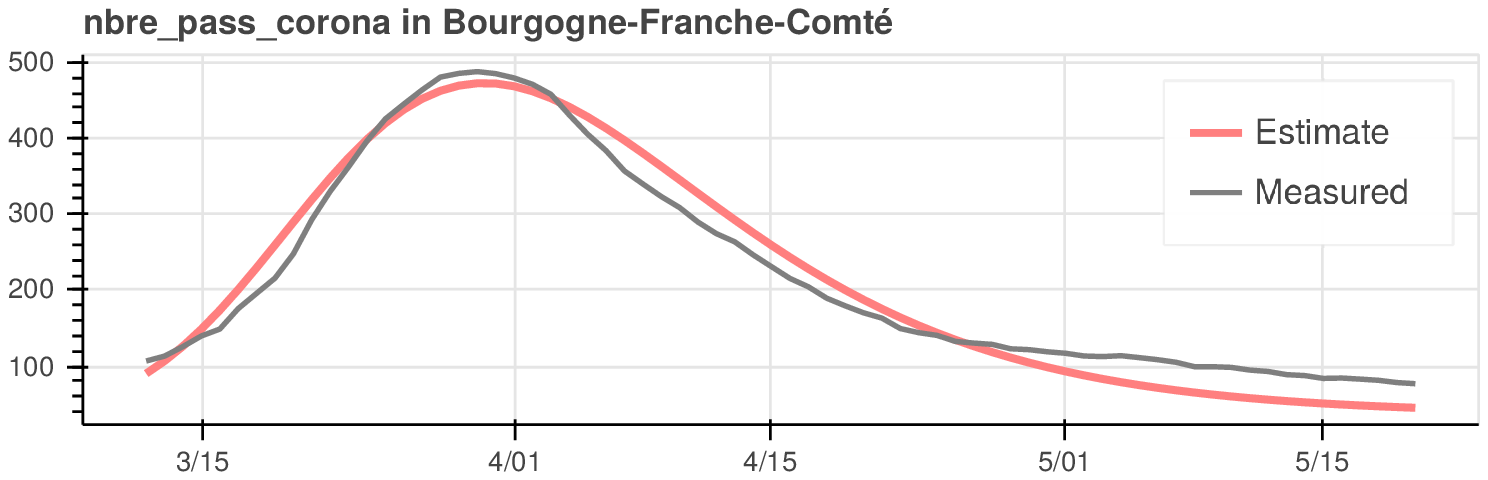}
\includegraphics[width=6.8cm]{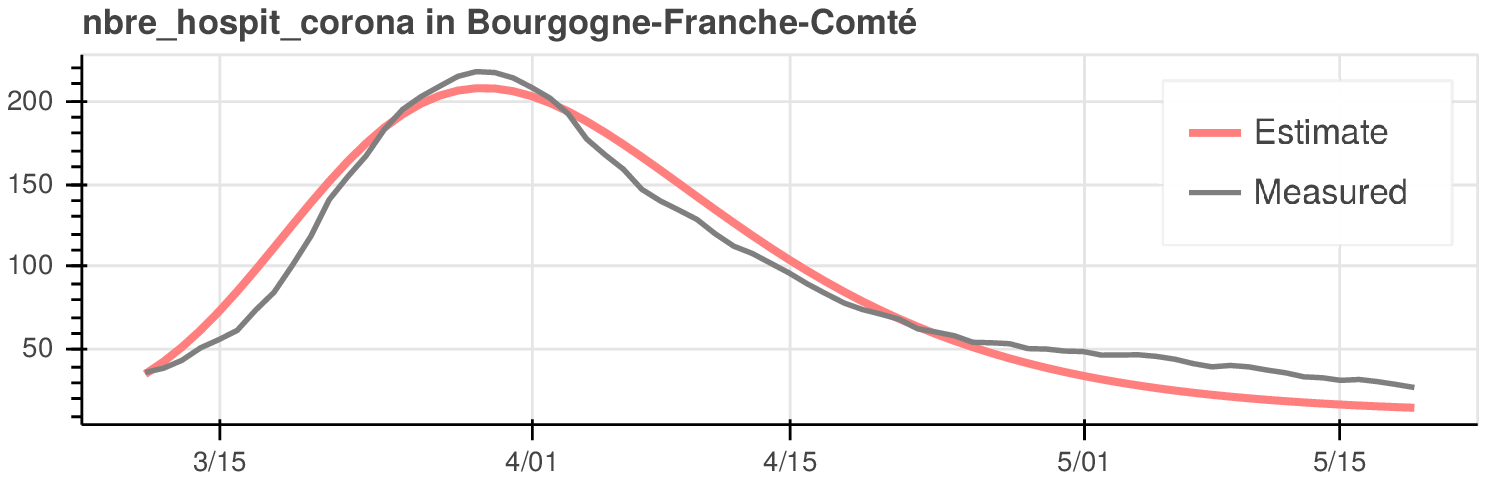}
\vspace{-0.7cm}
 \caption{Comparison of time series {\tt nbre\_pass\_corona} and 
 {\tt nbre\_hospit\_corona} averaged over a 14-days-long window, with the fitted outputs, for Bourgogne-Franche-Comté (27) region.}
\end{figure}
\vspace{-0.7cm}
\begin{figure}[H]
 \centering
\includegraphics[width=6.8cm]{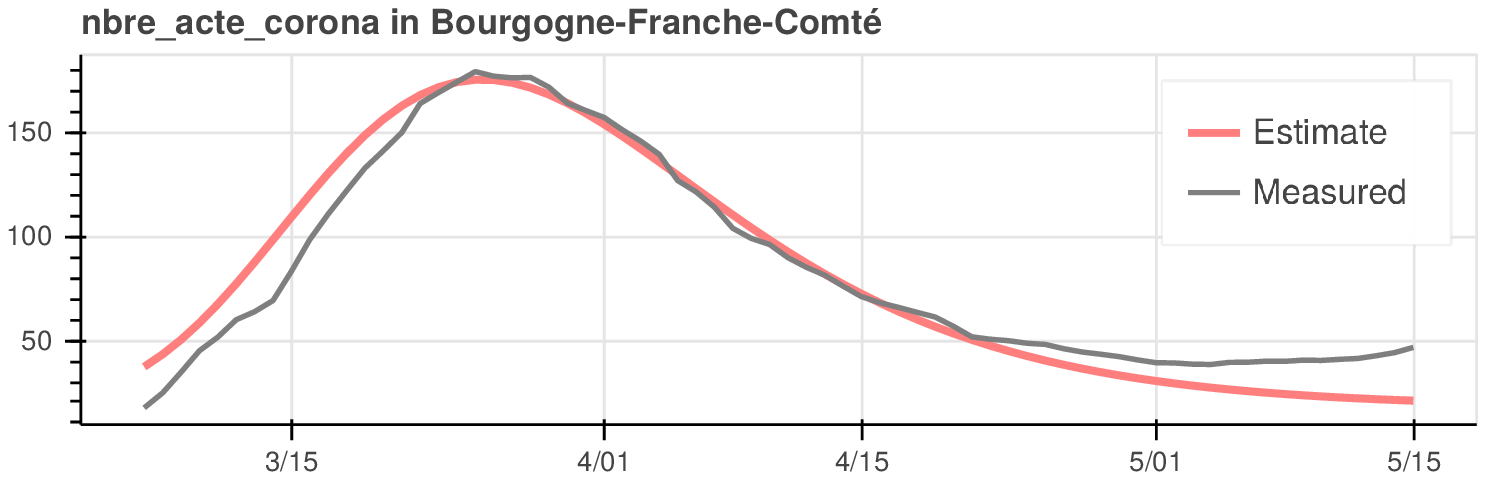}
\vspace{-0.3cm}
 \caption{Comparison of time series {\tt nbre\_acte\_corona} averaged over a 14-days-long window, with the fitted output, for Bourgogne-Franche-Comté (27) region.}
\end{figure}

% Île-de-France (11) and Grand Est (44) regions
% Auvergne-Rhône-Alpes (84) and Hauts-de-France (32)
% Provence-Alpes-Côte d'Azur (93) and Bourgogne-Franche-Comté (27)
% Occitanie  (76) and Pays de la Loire (52)
% Centre-Val de Loire (24) and Bretagne (53)
% Normandie (28) and Nouvelle-Aquitaine (75)
% Corse (94) region
% 

\subsection{Occitanie}

\vspace{-0.2cm}

\begin{figure}[H]
 \centering
\includegraphics[width=6.8cm]{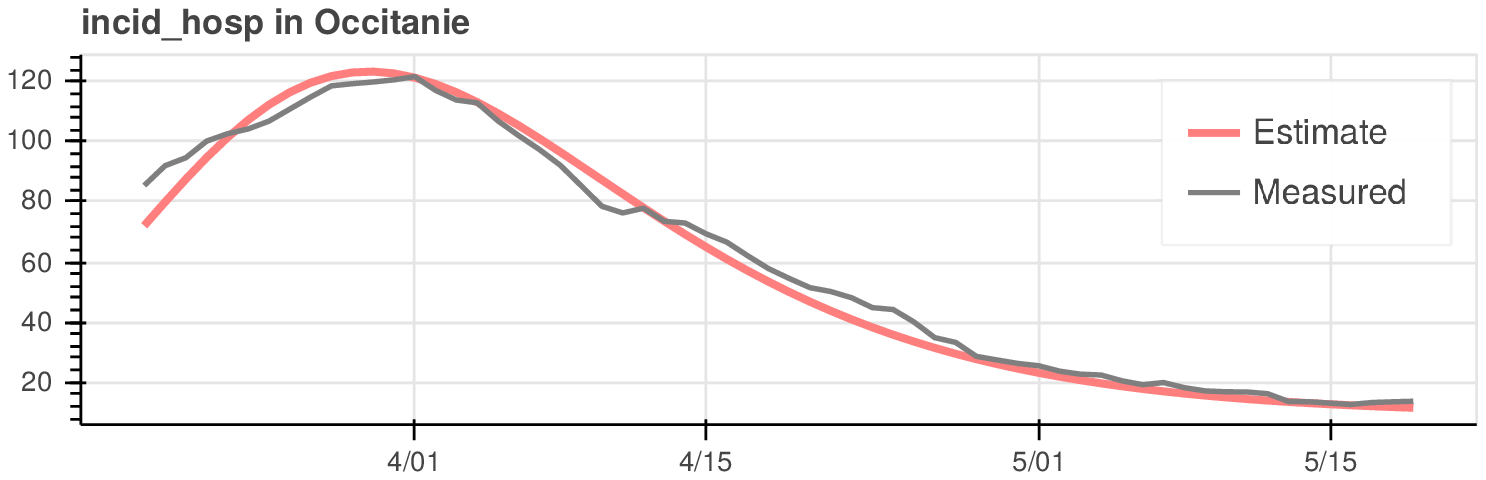}
\includegraphics[width=6.8cm]{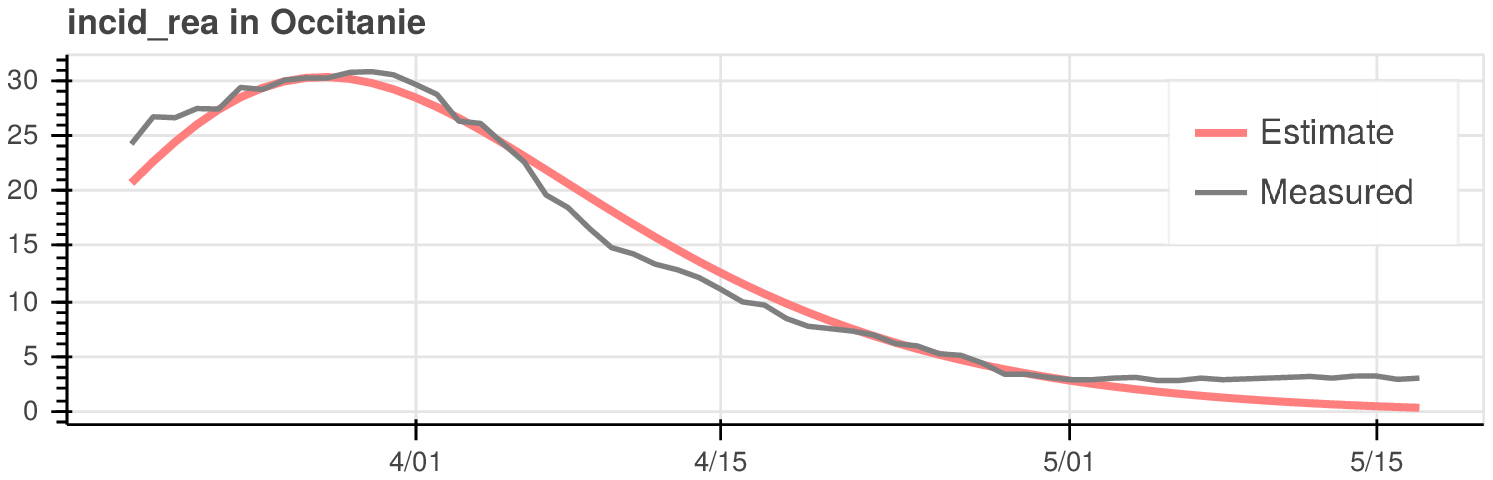}
\vspace{-0.7cm}
 \caption{Comparison of time series {\tt incid\_hosp} and {\tt incid\_rea} averaged over a 14-days-long window, with the fitted outputs, for Occitanie (76) region.}
\end{figure}

\vspace{-0.7cm}
\begin{figure}[H]
 \centering
\includegraphics[width=6.8cm]{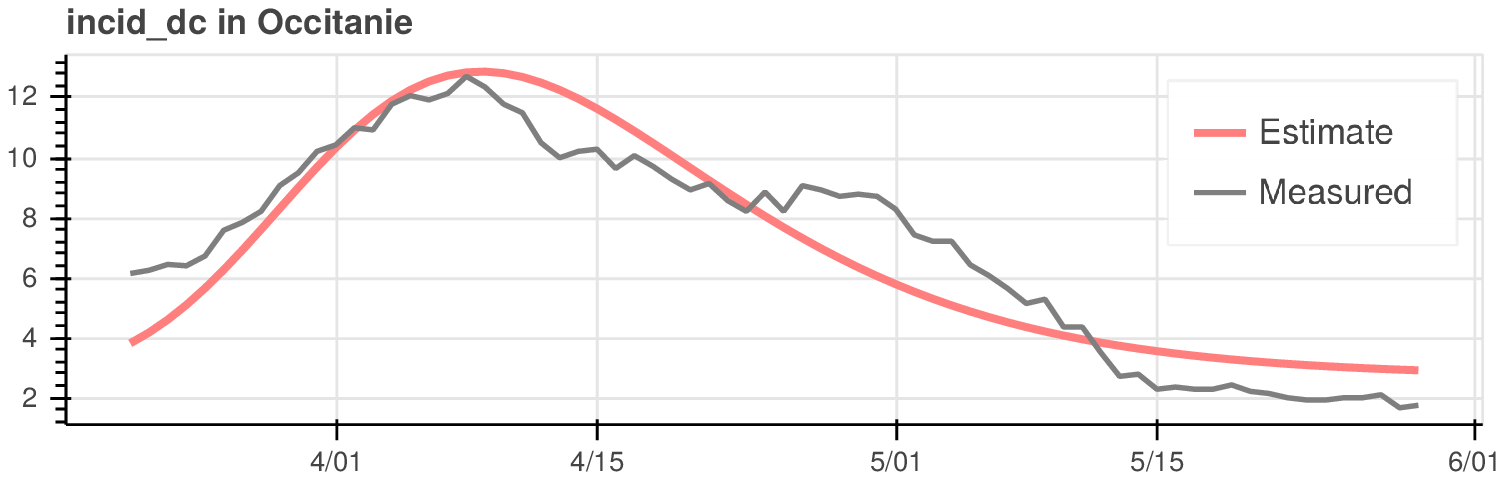}
\includegraphics[width=6.8cm]{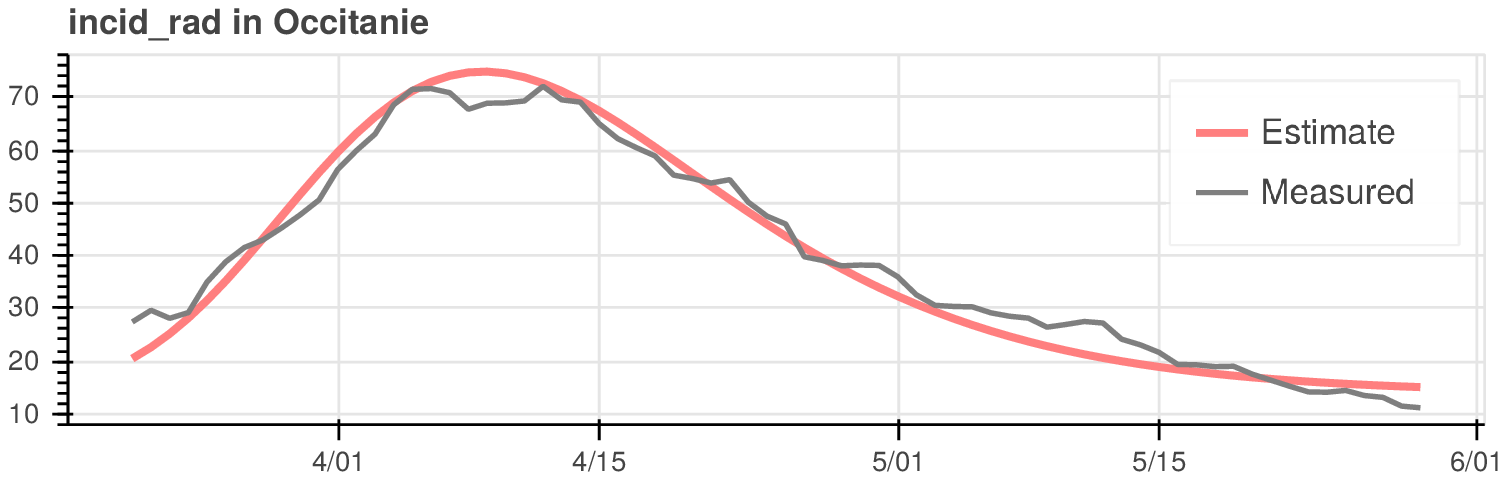}
\vspace{-0.7cm}
 \caption{Comparison of time series {\tt incid\_dc} and {\tt incid\_rad} averaged over a 14-days-long window, with the fitted outputs, for Occitanie (76) region.}
\end{figure}

\vspace{-0.7cm}
\begin{figure}[H]
 \centering
\includegraphics[width=6.8cm]{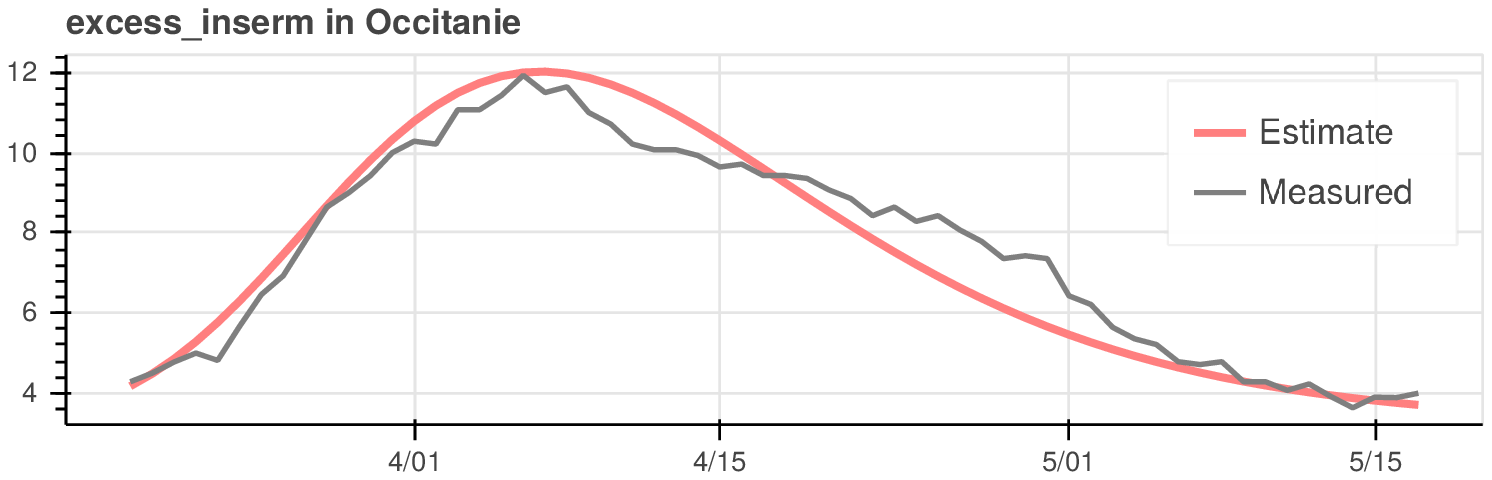}
\includegraphics[width=6.8cm]{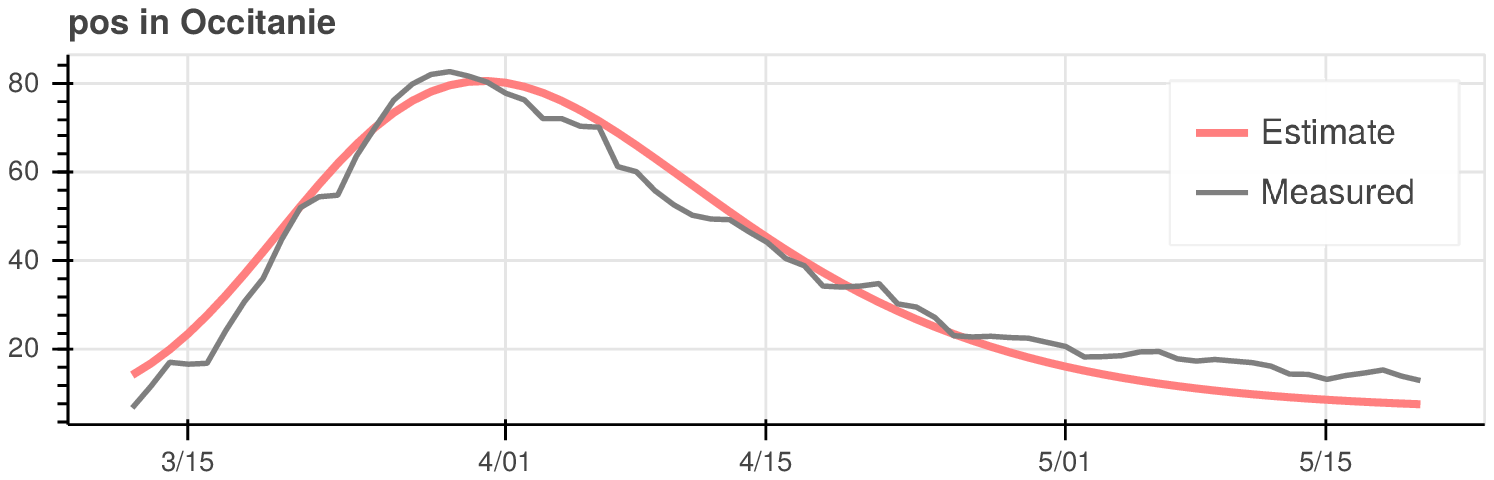}
\vspace{-0.7cm}
 \caption{Comparison of time series {\tt incid\_inserm} and {\tt pos} averaged over a 14-days-long window, with the fitted outputs, for Occitanie (76) region.}
\end{figure}

\vspace{-0.7cm}
\begin{figure}[H]
 \centering
\includegraphics[width=6.8cm]{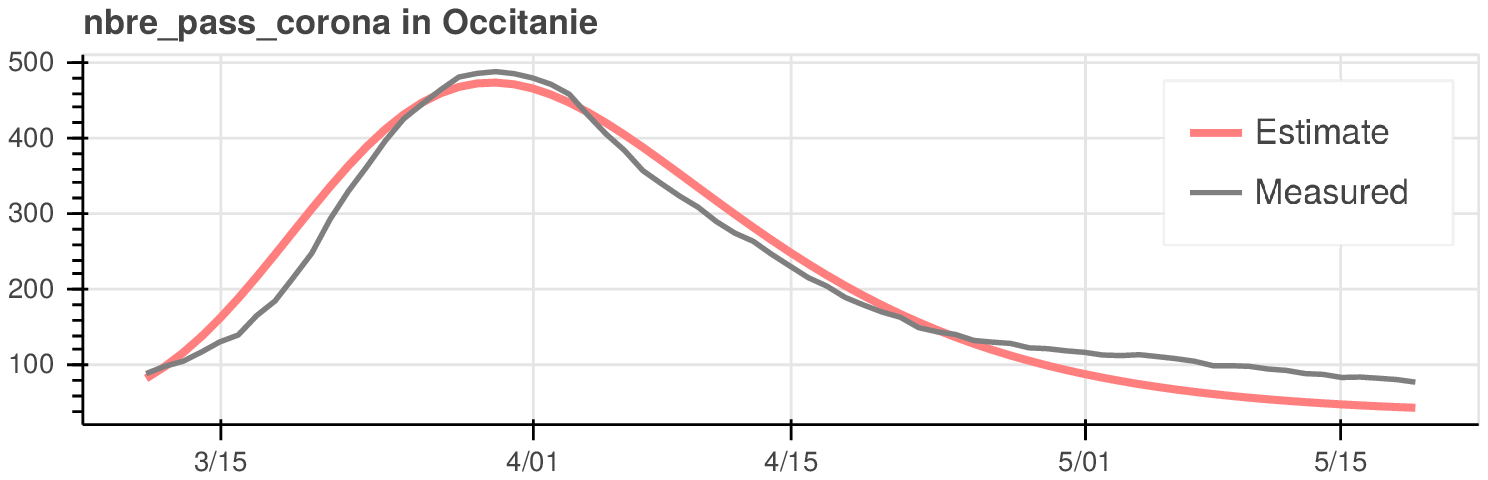}
\includegraphics[width=6.8cm]{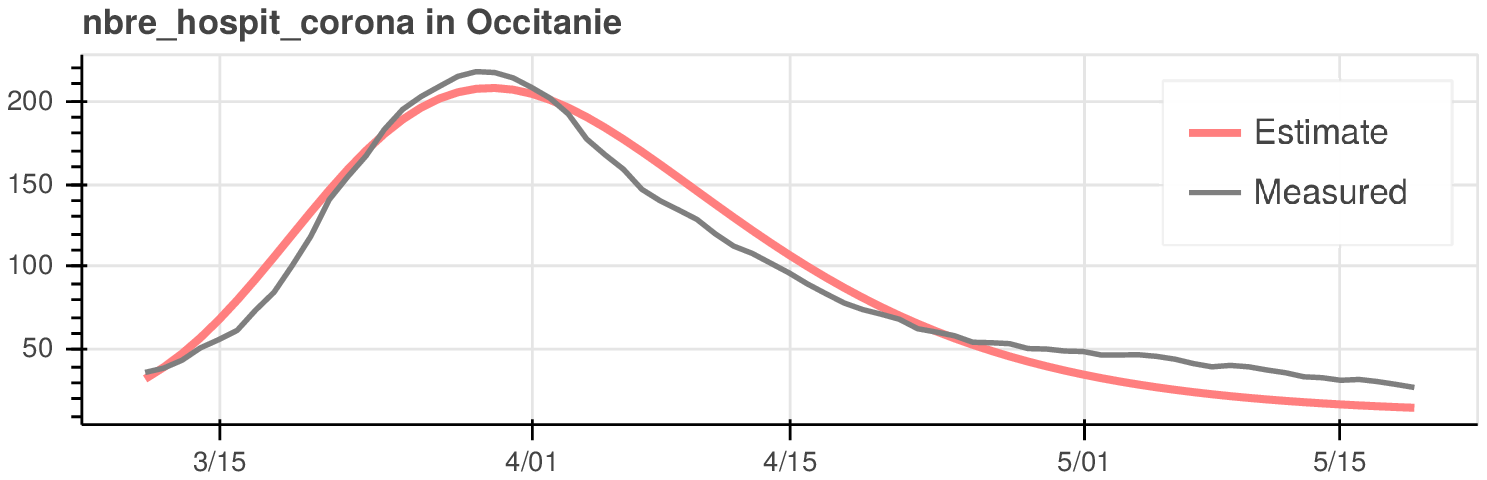}
\vspace{-0.7cm}
 \caption{Comparison of time series {\tt nbre\_pass\_corona} and 
 {\tt nbre\_hospit\_corona} averaged over a 14-days-long window, with the fitted outputs, for Occitanie (76) region.}
\end{figure}
\vspace{-0.7cm}
\begin{figure}[H]
 \centering
\includegraphics[width=6.8cm]{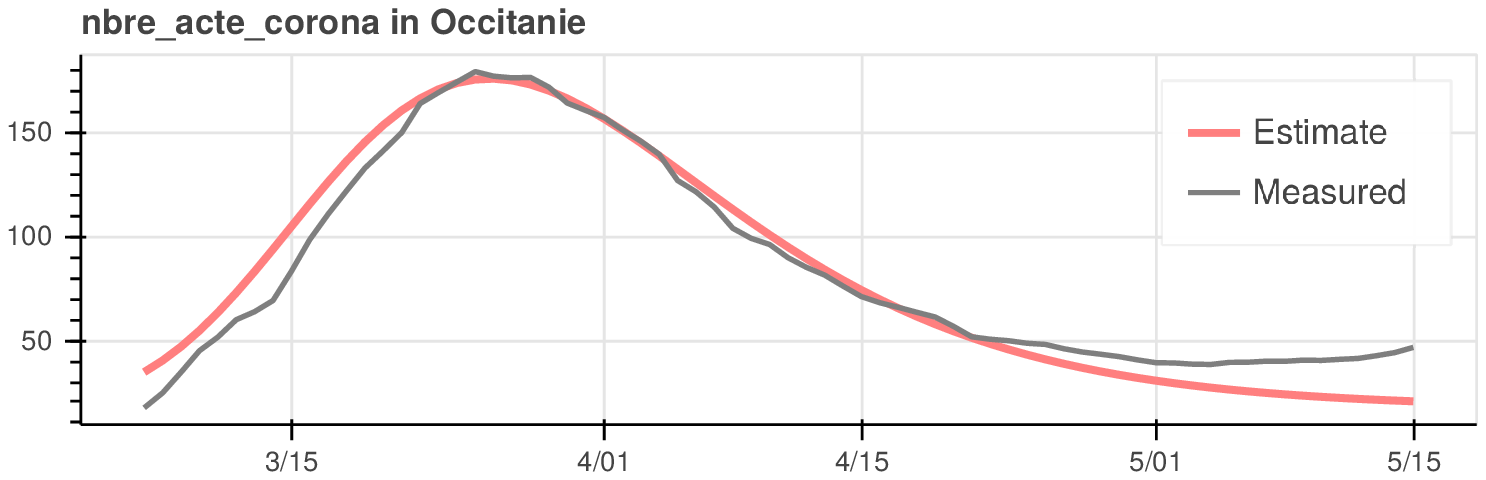}
\vspace{-0.3cm}
 \caption{Comparison of time series {\tt nbre\_acte\_corona} averaged over a 14-days-long window, with the fitted output, for Occitanie (76) region.}
\end{figure}
% Île-de-France (11) and Grand Est (44) regions
% Auvergne-Rhône-Alpes (84) and Hauts-de-France (32)
% Provence-Alpes-Côte d'Azur (93) and Bourgogne-Franche-Comté (27)
% Occitanie  (76) and Pays de la Loire (52)
% Centre-Val de Loire (24) and Bretagne (53)
% Normandie (28) and Nouvelle-Aquitaine (75)
% Corse (94) region
% 

\subsection{Pays de la Loire}

\vspace{-0.2cm}
\begin{figure}[H]
 \centering
\includegraphics[width=6.8cm]{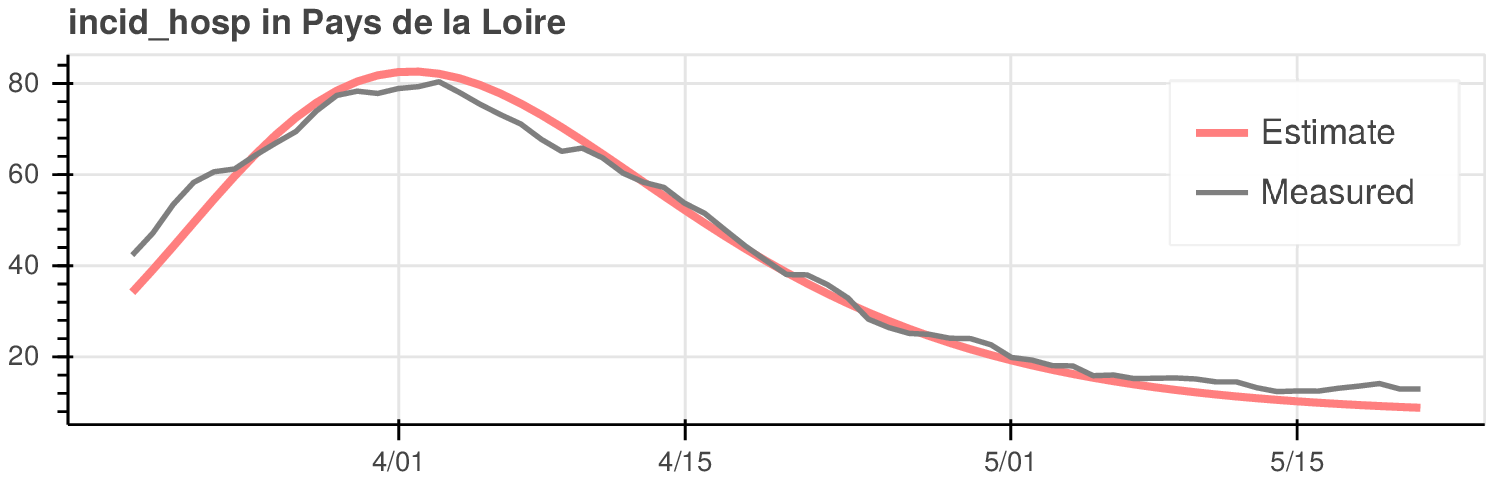}
\includegraphics[width=6.8cm]{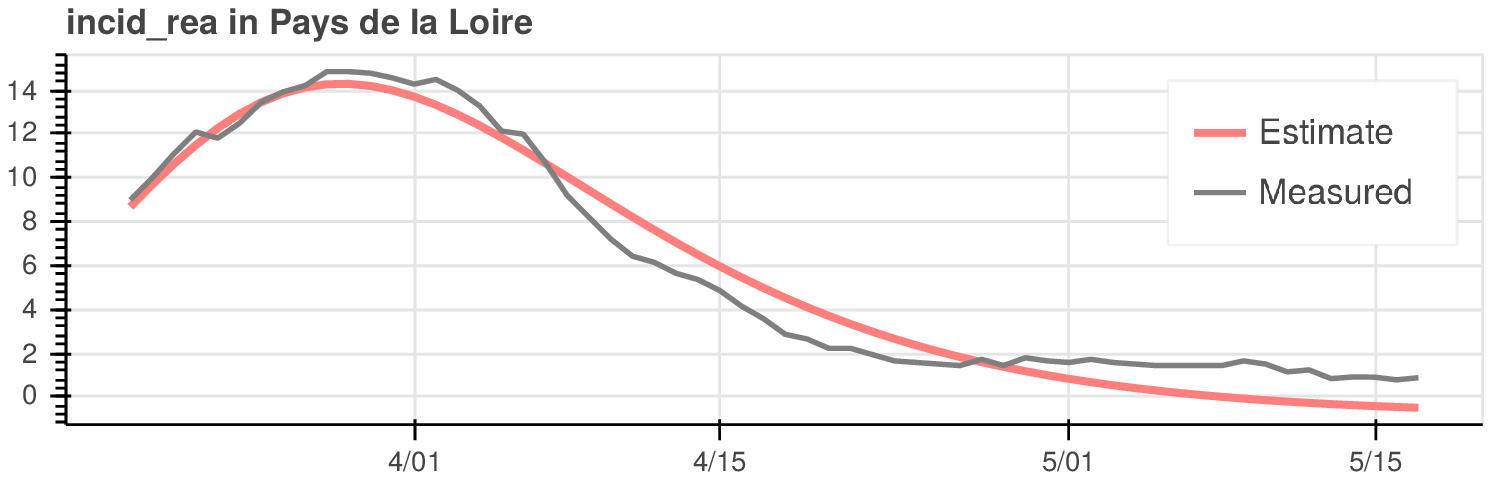}
\vspace{-0.7cm}
 \caption{Comparison of time series {\tt incid\_hosp} and {\tt incid\_rea} averaged over a 14-days-long window, with the fitted outputs, for Pays de la Loire (52) region.}
\end{figure}

\vspace{-0.7cm}
\begin{figure}[H]
 \centering
\includegraphics[width=6.8cm]{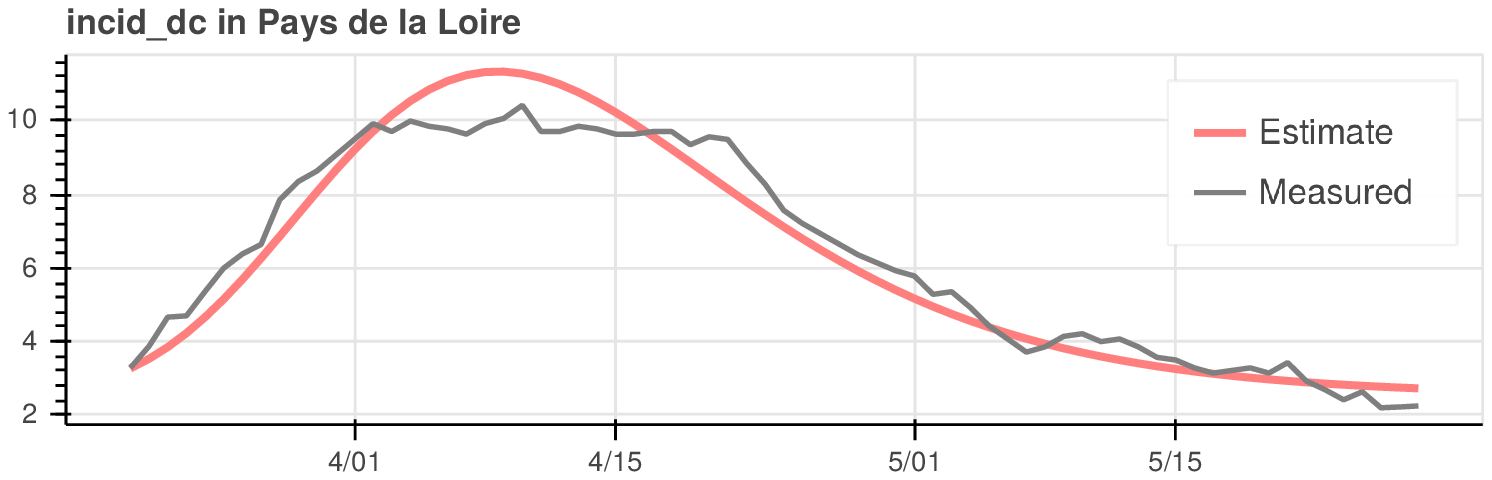}
\includegraphics[width=6.8cm]{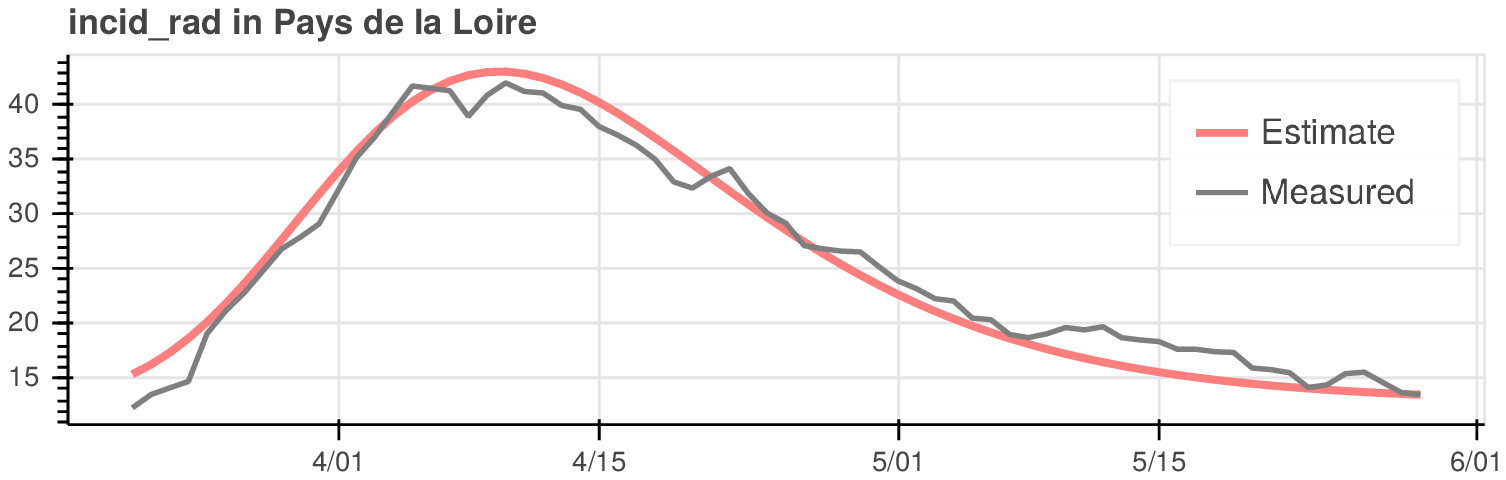}
\vspace{-0.7cm}
 \caption{Comparison of time series {\tt incid\_dc} and {\tt incid\_rad} averaged over a 14-days-long window, with the fitted outputs, for Pays de la Loire (52) region.}
\end{figure}

\vspace{-0.7cm}
\begin{figure}[H]
 \centering
\includegraphics[width=6.8cm]{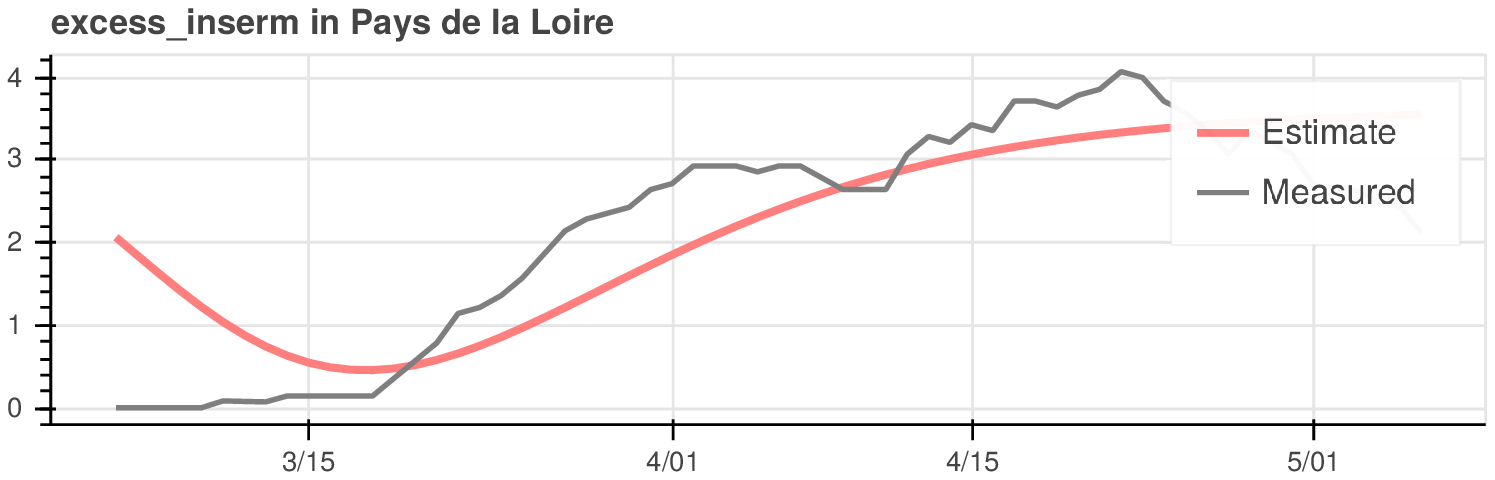}
\includegraphics[width=6.8cm]{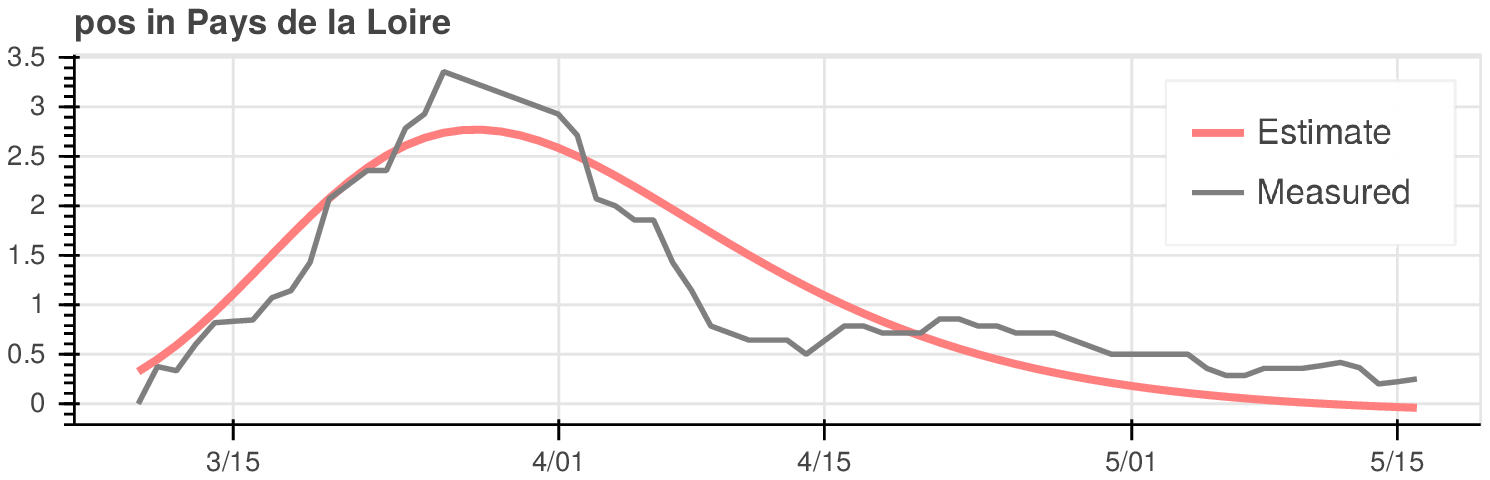}
\vspace{-0.7cm}
 \caption{Comparison of time series {\tt incid\_inserm} and {\tt pos} averaged over a 14-days-long window, with the fitted outputs, for Pays de la Loire (52) region.}
\end{figure}

\vspace{-0.7cm}
\begin{figure}[H]
 \centering
\includegraphics[width=6.8cm]{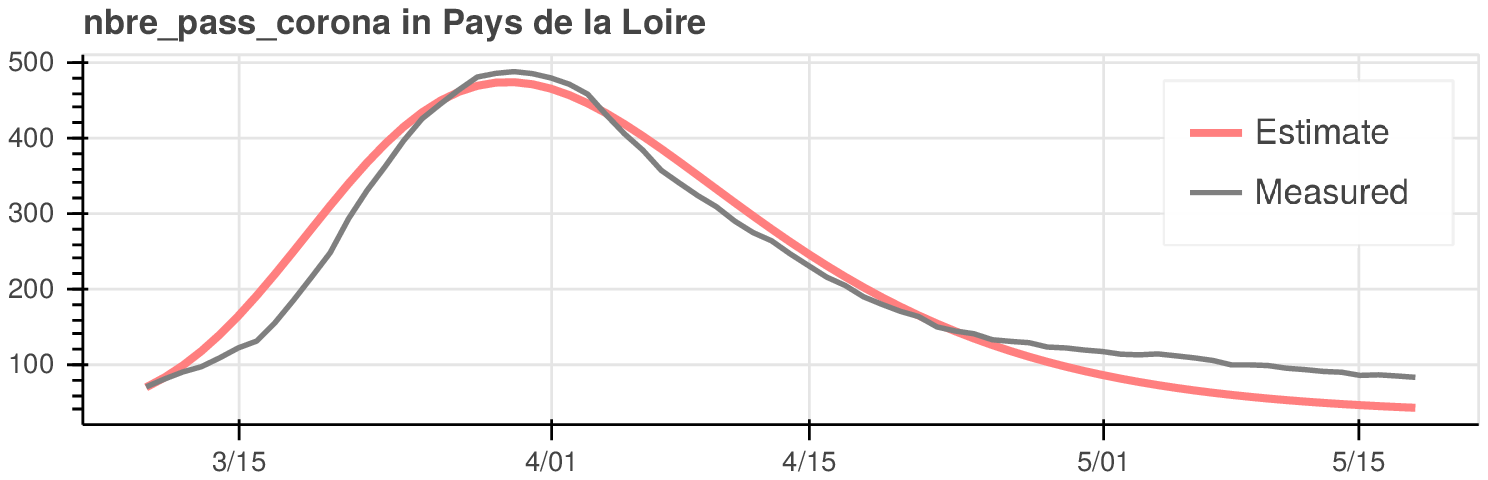}
\includegraphics[width=6.8cm]{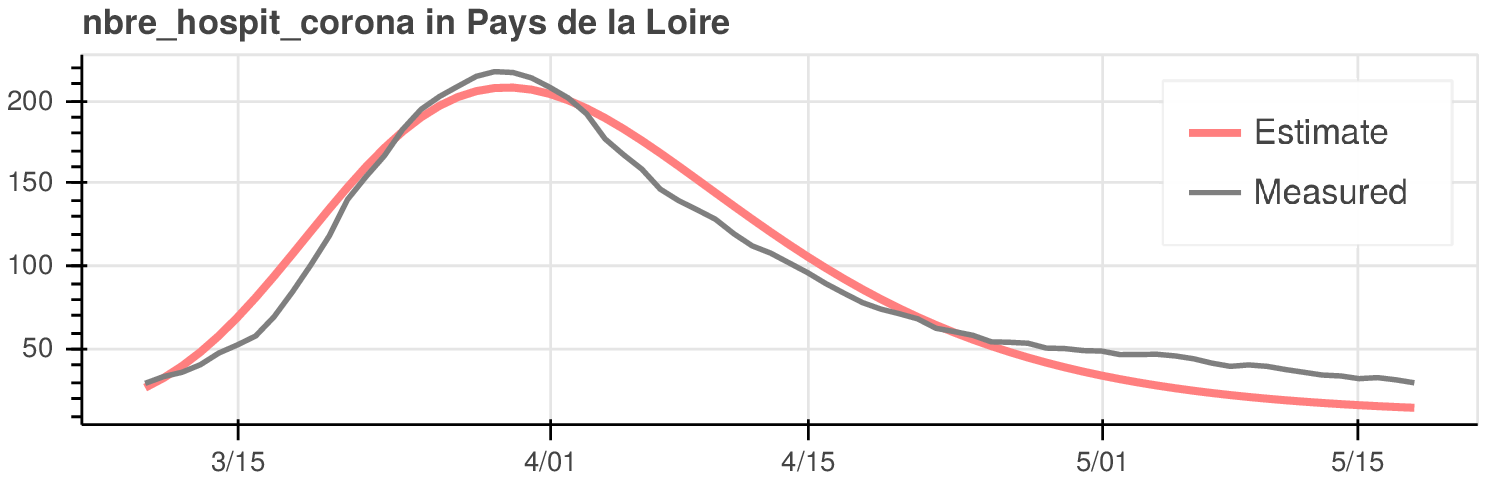}
\vspace{-0.7cm}
 \caption{Comparison of time series {\tt nbre\_pass\_corona} and 
 {\tt nbre\_hospit\_corona} averaged over a 14-days-long window, with the fitted outputs, for Pays de la Loire (52) region.}
\end{figure}
\vspace{-0.7cm}
\begin{figure}[H]
 \centering
\includegraphics[width=6.8cm]{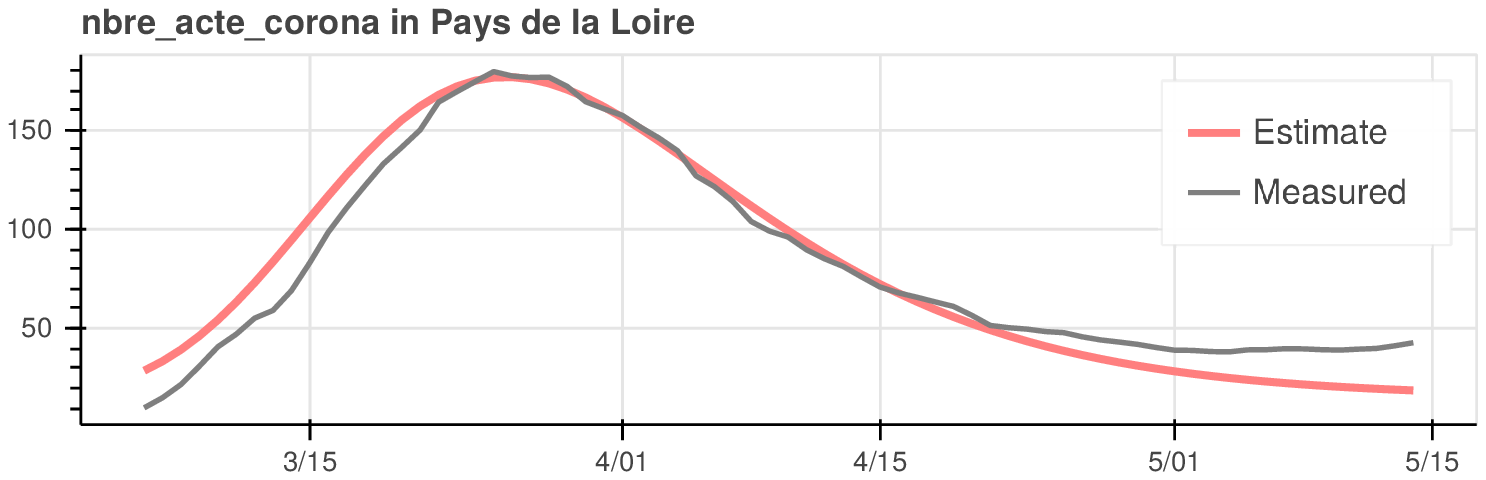}
\vspace{-0.3cm}
 \caption{Comparison of time series {\tt nbre\_acte\_corona} averaged over a 14-days-long window, with the fitted output, for Pays de la Loire (52) region.}
\end{figure}

\subsection{Centre-Val de Loire}

\vspace{-0.2cm}

\begin{figure}[H]
 \centering
\includegraphics[width=6.8cm]{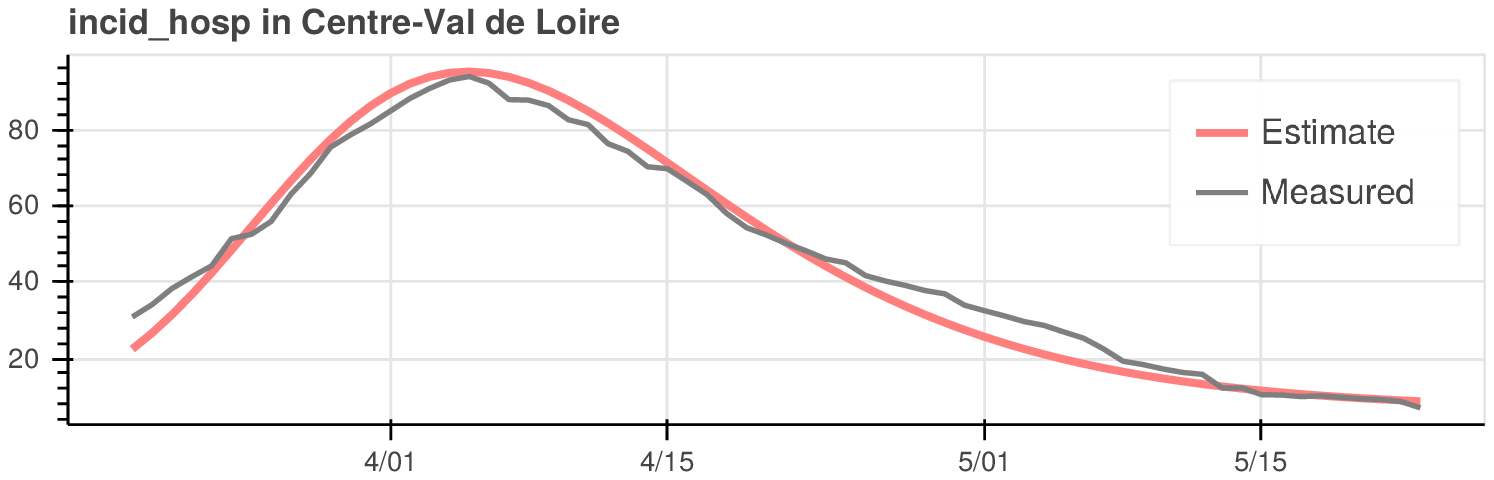}
\includegraphics[width=6.8cm]{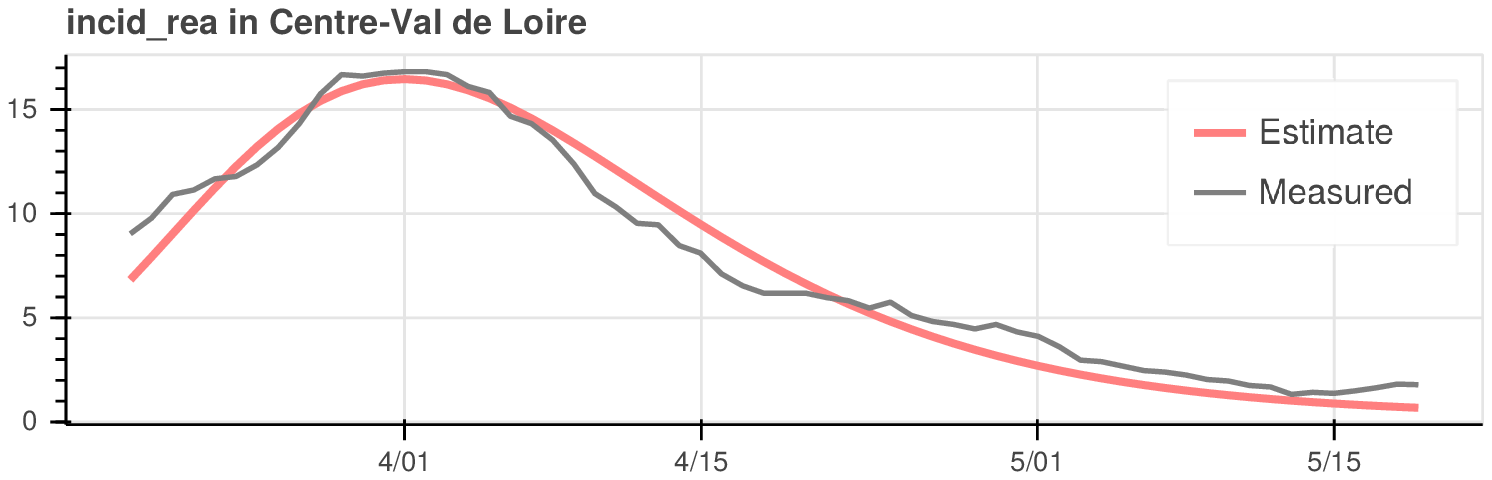}
\vspace{-0.7cm}
 \caption{Comparison of time series {\tt incid\_hosp} and {\tt incid\_rea} averaged over a 14-days-long window, with the fitted outputs, for Centre-Val de Loire (24) region.}
\end{figure}

\vspace{-0.7cm}
\begin{figure}[H]
 \centering
\includegraphics[width=6.8cm]{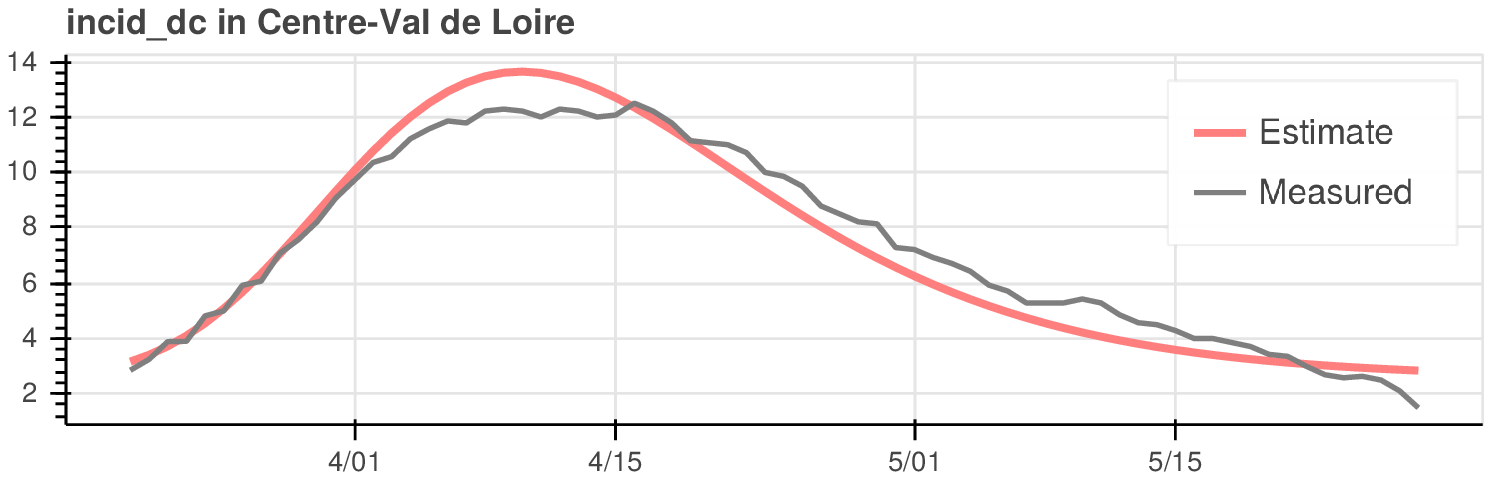}
\includegraphics[width=6.8cm]{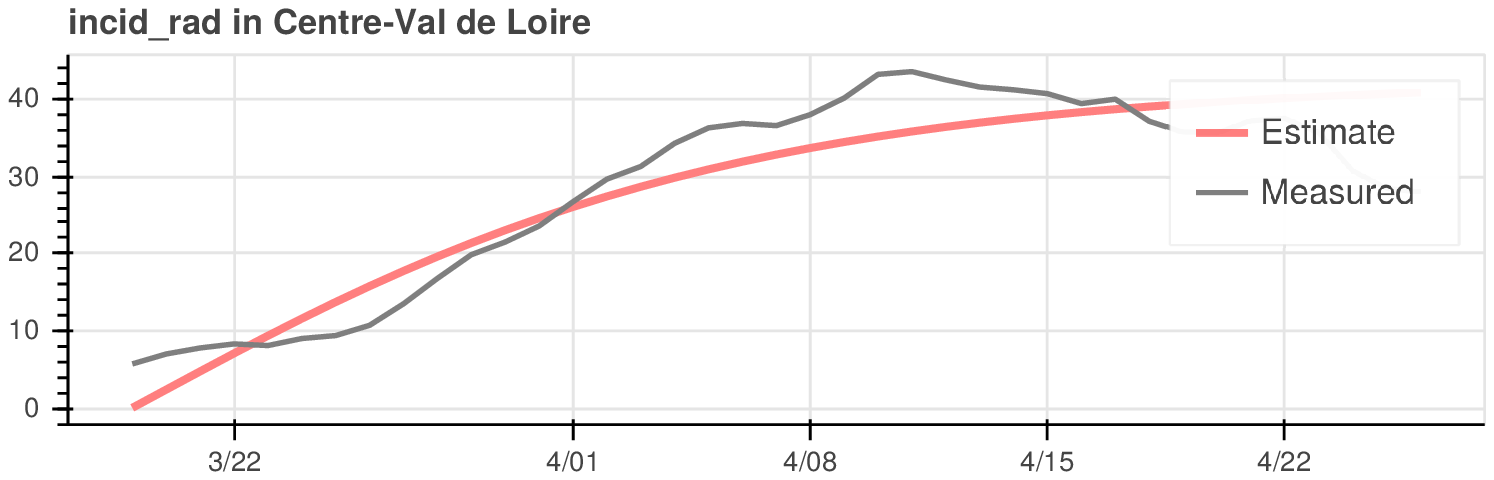}
\vspace{-0.7cm}
 \caption{Comparison of time series {\tt incid\_dc} and {\tt incid\_rad} averaged over a 14-days-long window, with the fitted outputs, for Centre-Val de Loire (24) region.}
\end{figure}

\vspace{-0.7cm}
\begin{figure}[H]
 \centering
\includegraphics[width=6.8cm]{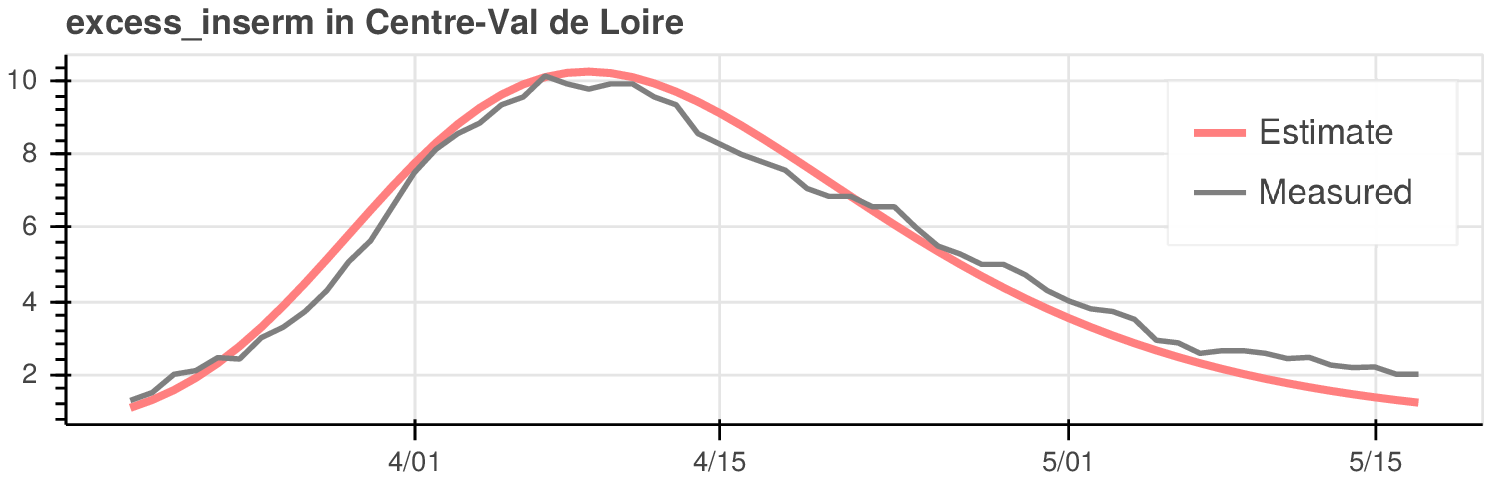}
\includegraphics[width=6.8cm]{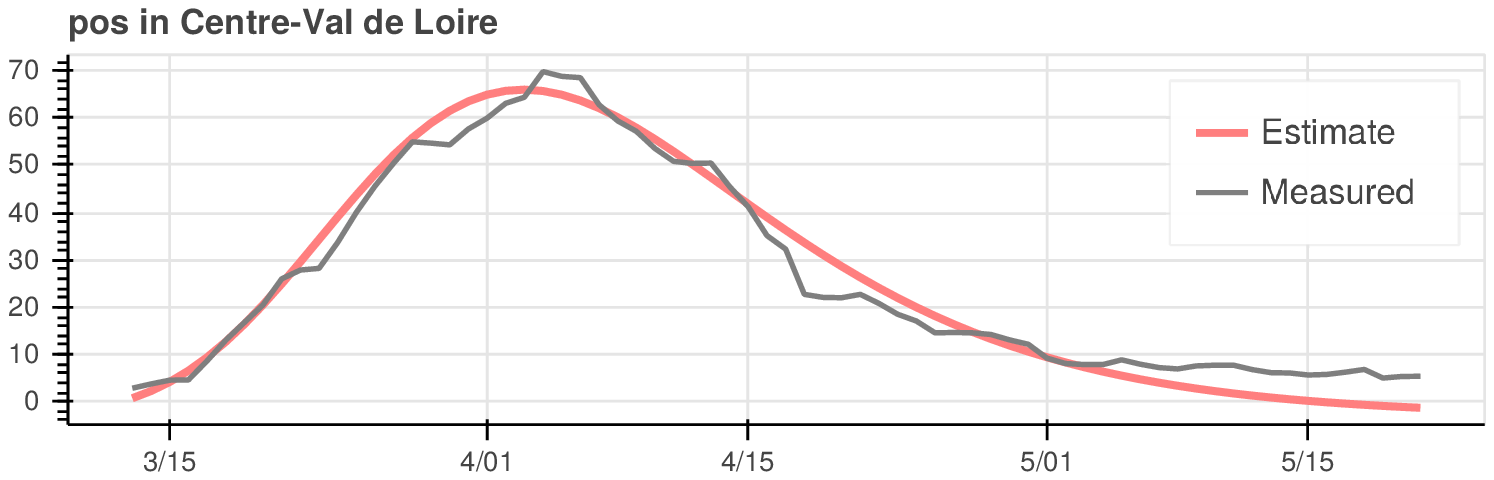}
\vspace{-0.7cm}
 \caption{Comparison of time series {\tt incid\_inserm} and {\tt pos} averaged over a 14-days-long window, with the fitted outputs, for Centre-Val de Loire (24) region.}
\end{figure}

\vspace{-0.7cm}
\begin{figure}[H]
 \centering
\includegraphics[width=6.8cm]{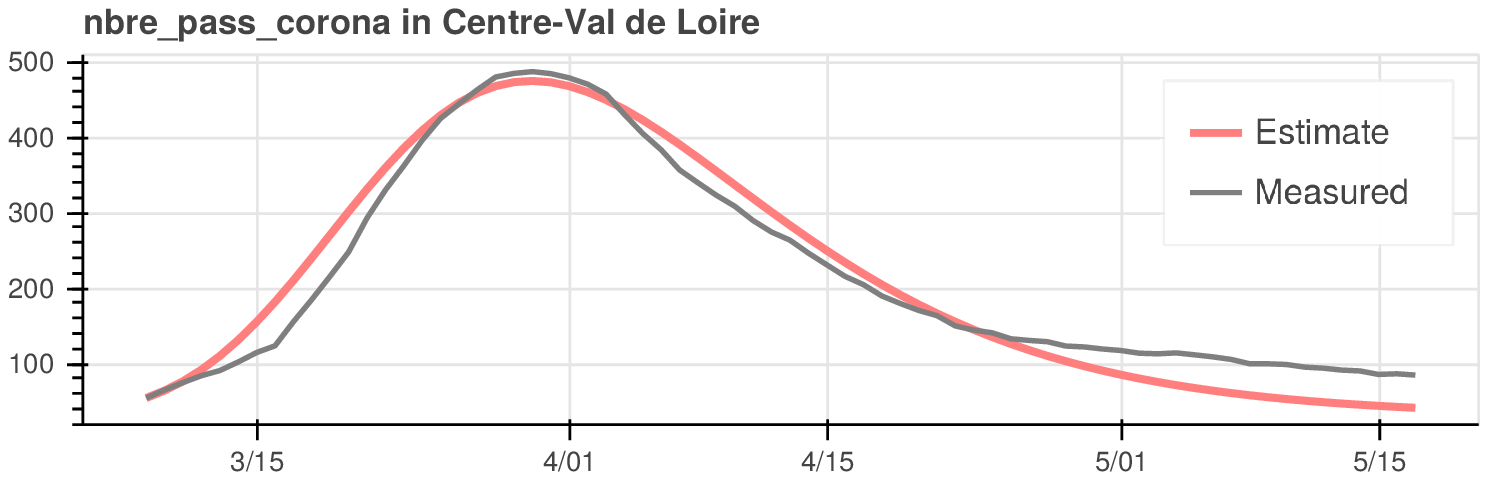}
\includegraphics[width=6.8cm]{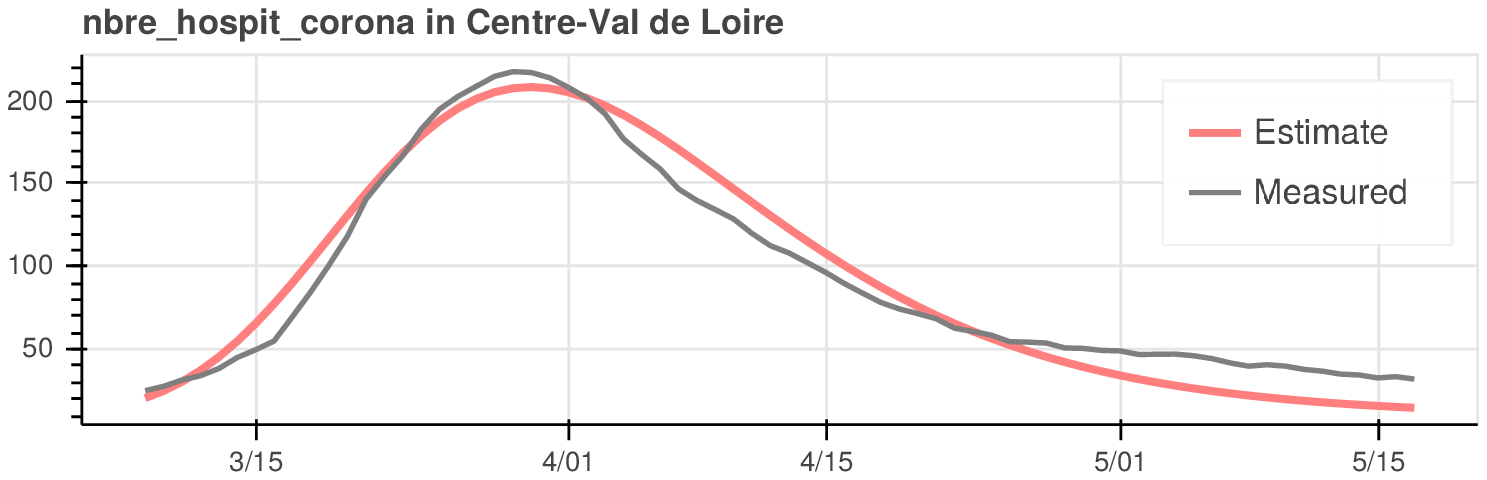}
\vspace{-0.7cm}
 \caption{Comparison of time series {\tt nbre\_pass\_corona} and 
 {\tt nbre\_hospit\_corona} averaged over a 14-days-long window, with the fitted outputs, for Centre-Val de Loire (24) region.}
\end{figure}
\vspace{-0.7cm}
\begin{figure}[H]
 \centering
\includegraphics[width=6.8cm]{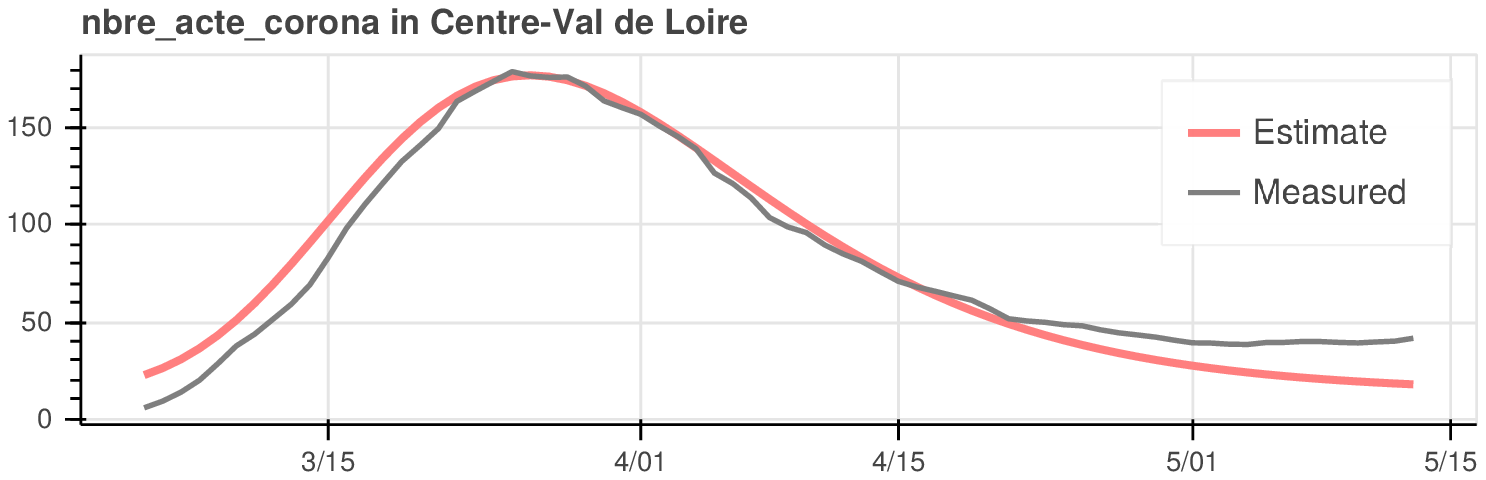}
\vspace{-0.3cm}
 \caption{Comparison of time series {\tt nbre\_acte\_corona} averaged over a 14-days-long window, with the fitted output, for Centre-Val de Loire (24) region.}
\end{figure}

% Île-de-France (11) and Grand Est (44) regions
% Auvergne-Rhône-Alpes (84) and Hauts-de-France (32)
% Provence-Alpes-Côte d'Azur (93) and Bourgogne-Franche-Comté (27)
% Occitanie  (76) and Pays de la Loire (52)
% Centre-Val de Loire (24) and Bretagne (53)
% Normandie (28) and Nouvelle-Aquitaine (75)
% Corse (94) region
% 

\subsection{Bretagne}

\vspace{-0.2cm}

\begin{figure}[H]
 \centering
\includegraphics[width=6.8cm]{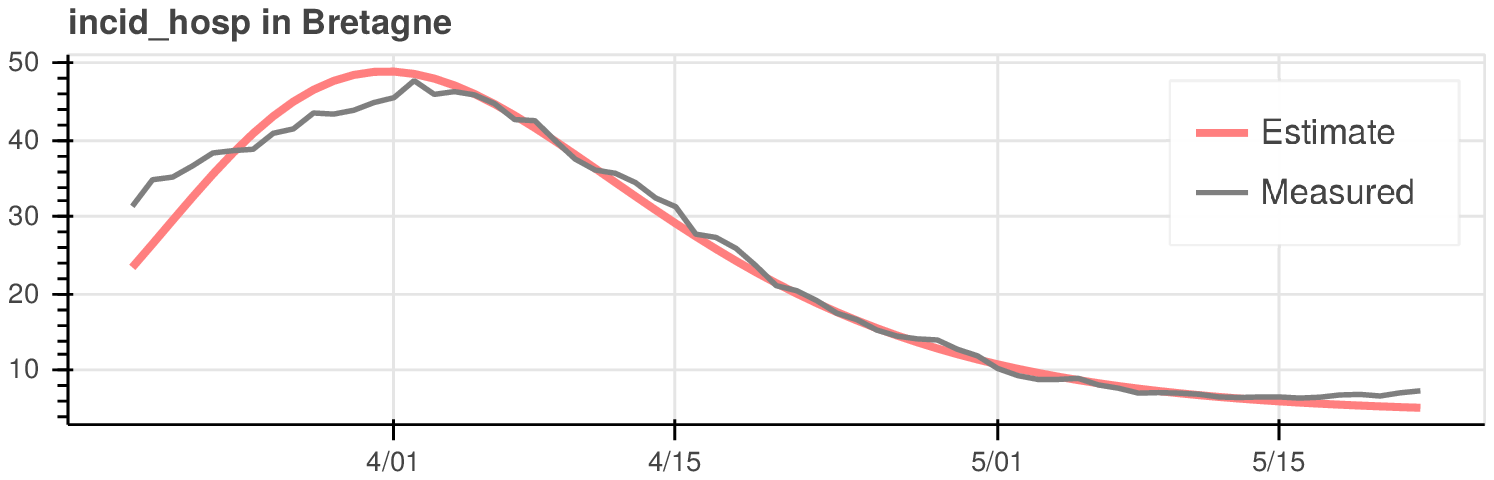}
\includegraphics[width=6.8cm]{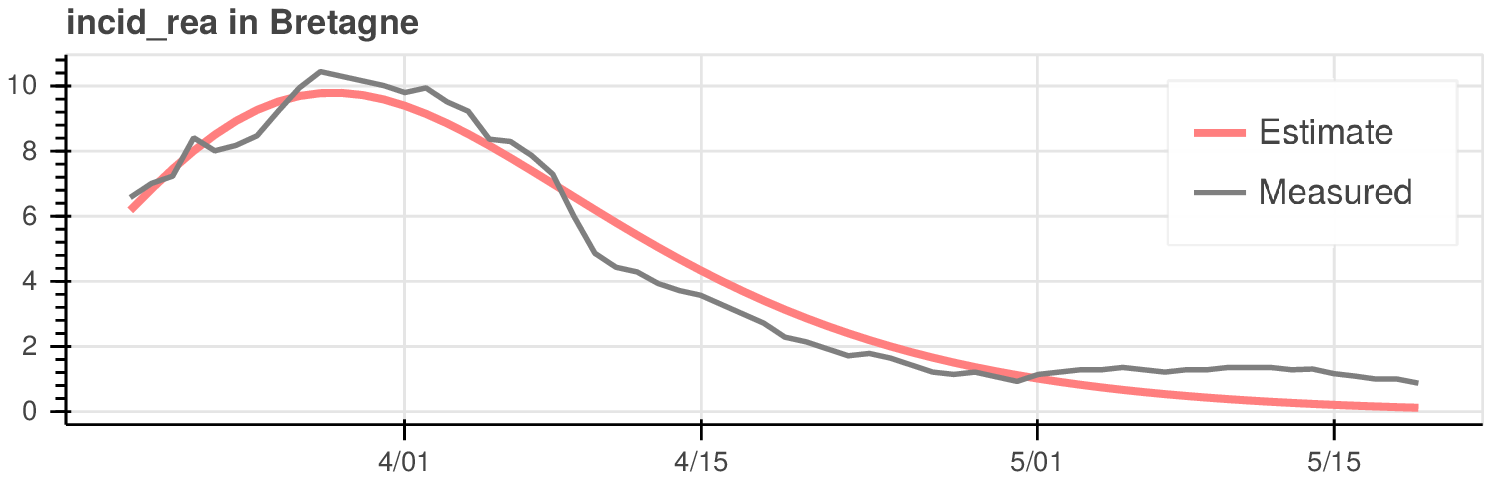}
\vspace{-0.7cm}
 \caption{Comparison of time series {\tt incid\_hosp} and {\tt incid\_rea} averaged over a 14-days-long window, with the fitted outputs, for Bretagne (53) region.}
\end{figure}

\vspace{-0.7cm}
\begin{figure}[H]
 \centering
\includegraphics[width=6.8cm]{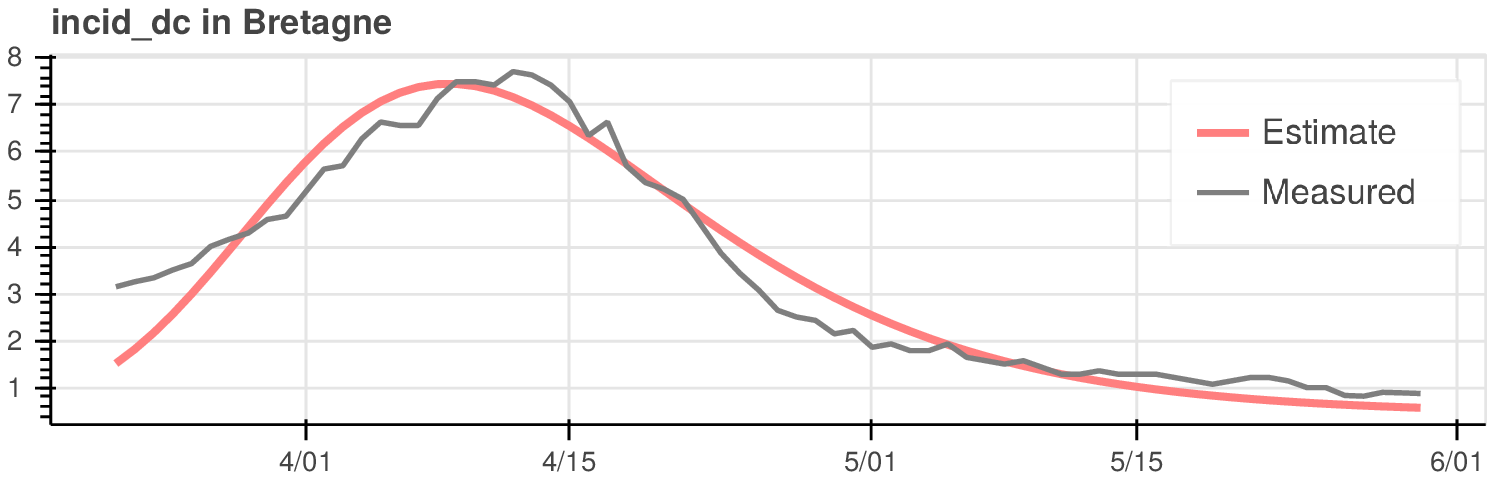}
\includegraphics[width=6.8cm]{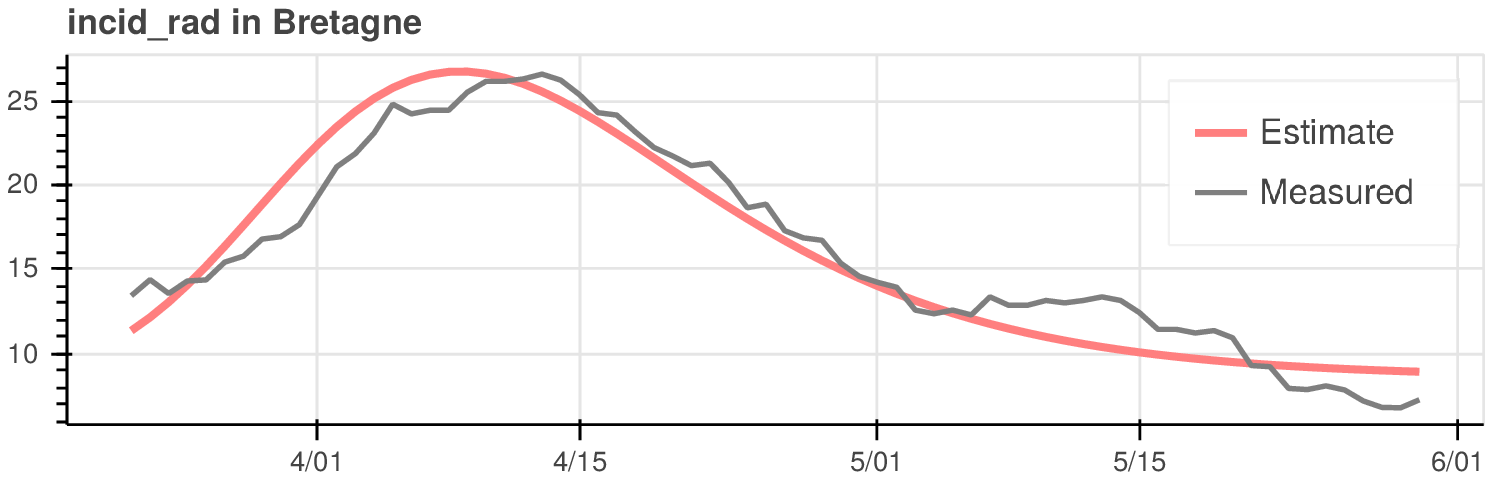}
\vspace{-0.7cm}
 \caption{Comparison of time series {\tt incid\_dc} and {\tt incid\_rad} averaged over a 14-days-long window, with the fitted outputs, for Bretagne (53) region.}
\end{figure}

\vspace{-0.7cm}
\begin{figure}[H]
 \centering
\includegraphics[width=6.8cm]{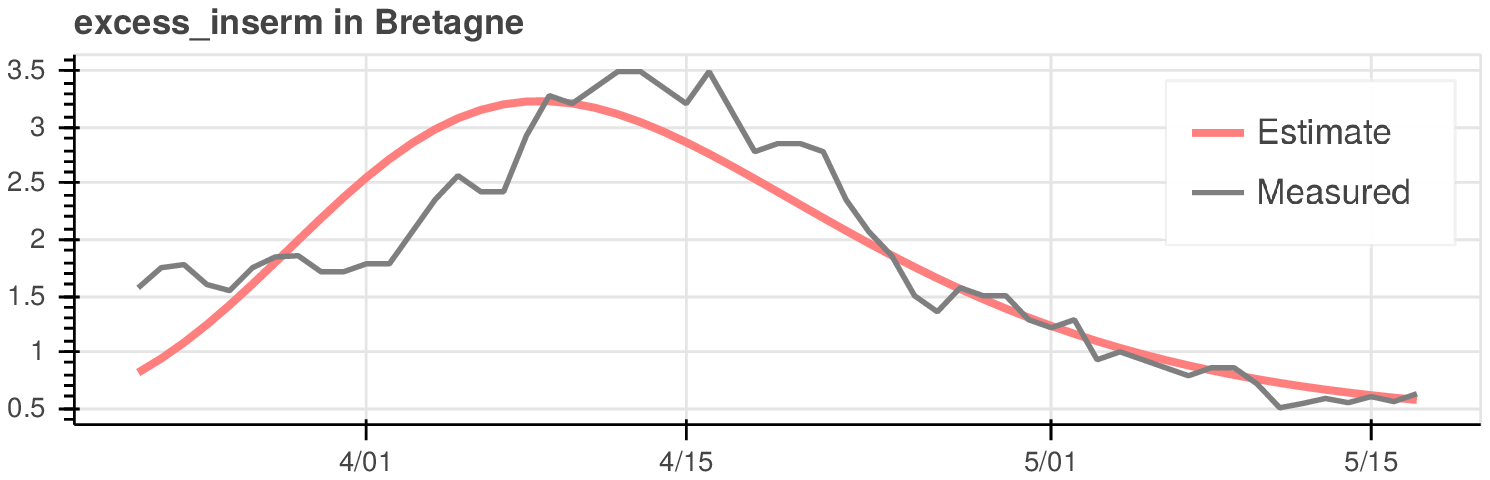}
\includegraphics[width=6.8cm]{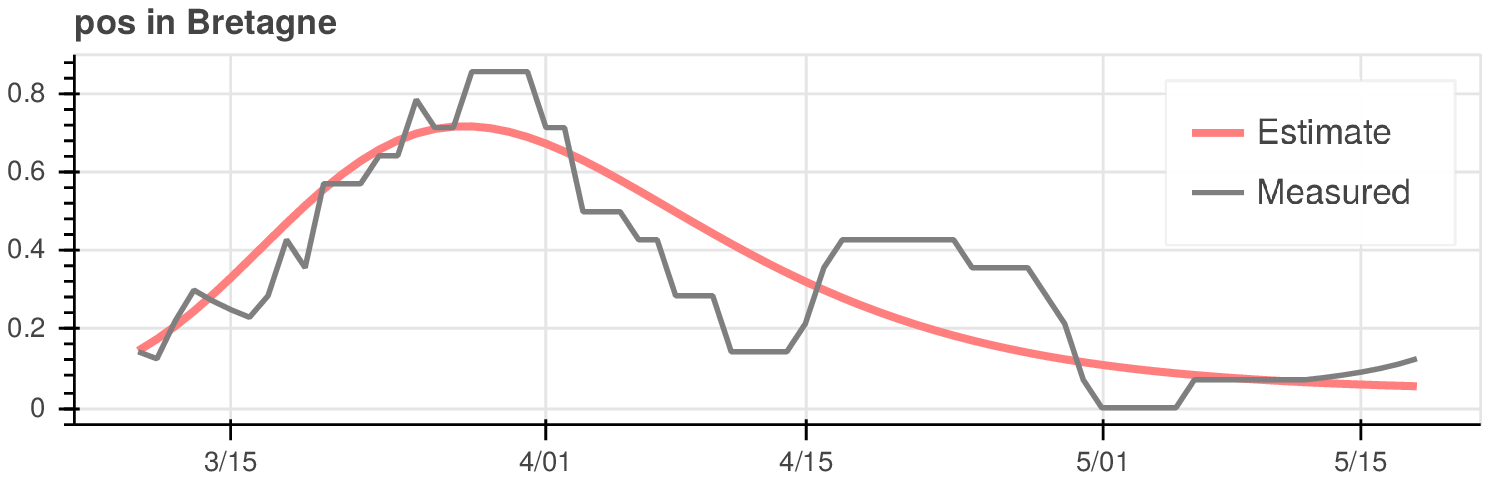}
\vspace{-0.7cm}
 \caption{Comparison of time series {\tt incid\_inserm} and {\tt pos} averaged over a 14-days-long window, with the fitted outputs, for Bretagne (53) region.}
\end{figure}

\vspace{-0.7cm}
\begin{figure}[H]
 \centering
\includegraphics[width=6.8cm]{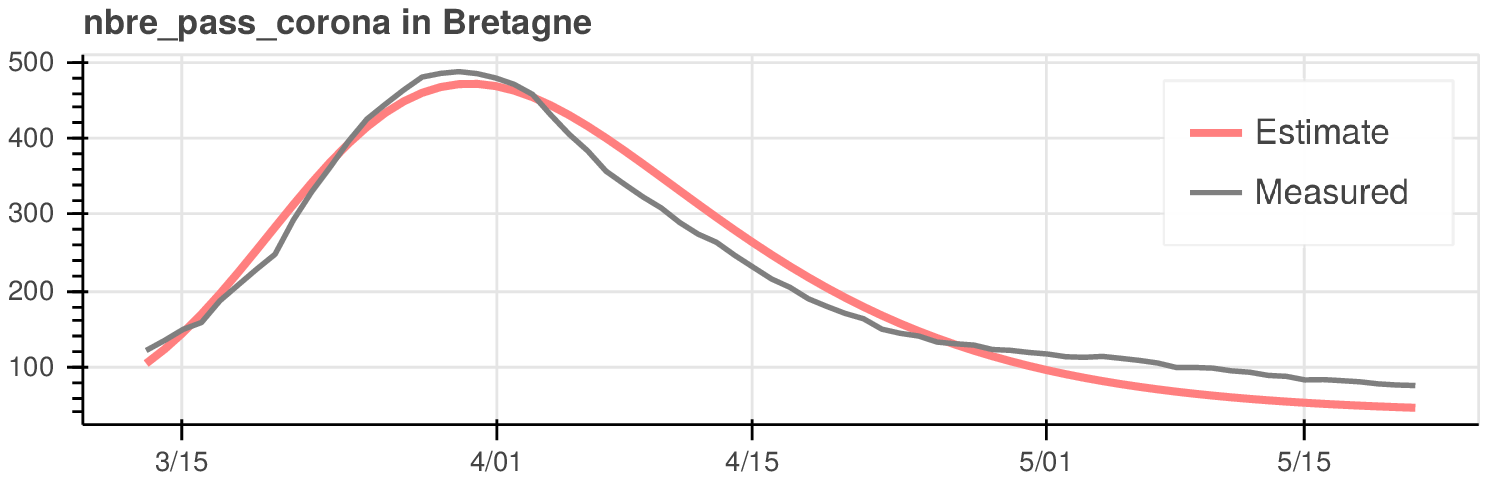}
\includegraphics[width=6.8cm]{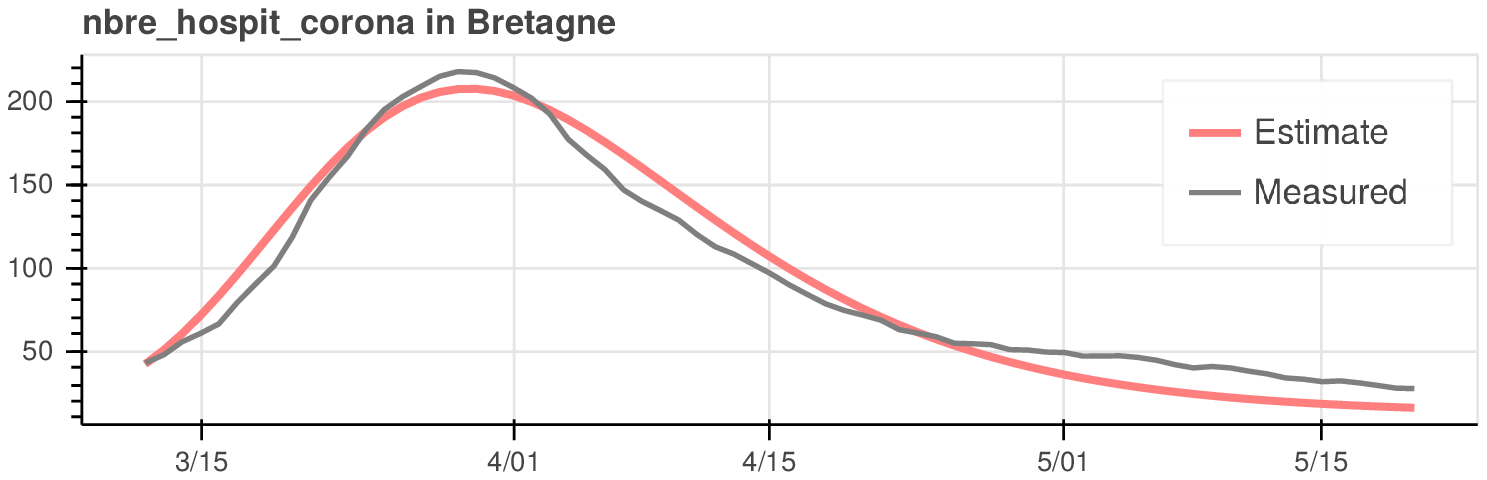}
\vspace{-0.7cm}
 \caption{Comparison of time series {\tt nbre\_pass\_corona} and 
 {\tt nbre\_hospit\_corona} averaged over a 14-days-long window, with the fitted outputs, for Bretagne (53) region.}
\end{figure}
\vspace{-0.7cm}
\begin{figure}[H]
 \centering
\includegraphics[width=6.8cm]{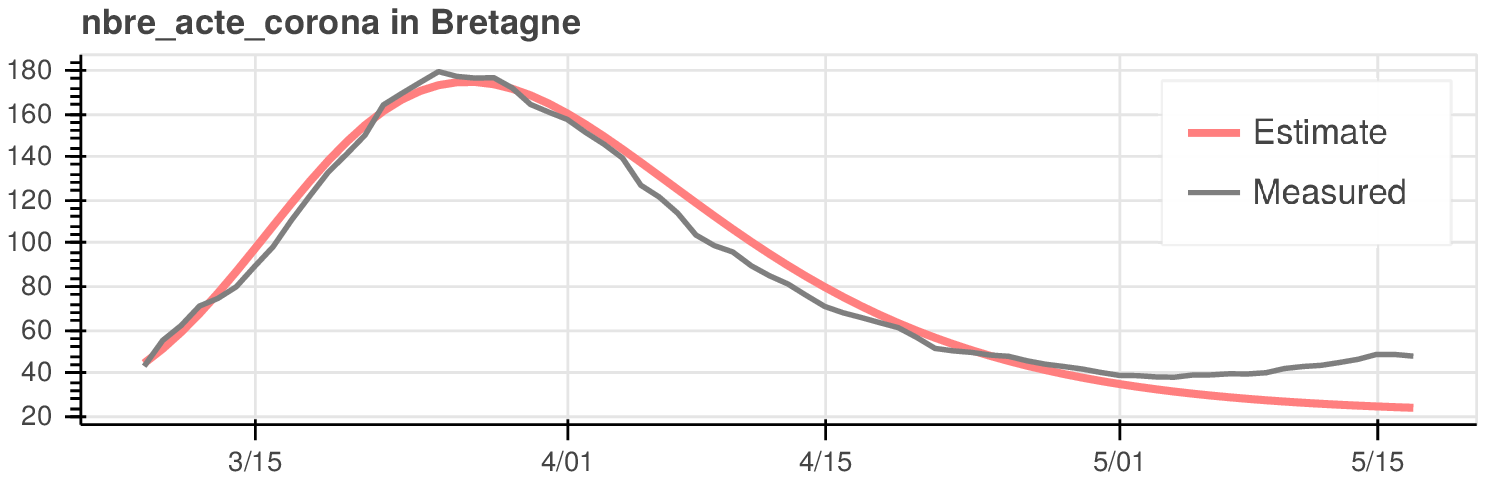}
\vspace{-0.3cm}
 \caption{Comparison of time series {\tt nbre\_acte\_corona} averaged over a 14-days-long window, with the fitted output, for Bretagne (53) region.}
\end{figure}
% Île-de-France (11) and Grand Est (44) regions
% Auvergne-Rhône-Alpes (84) and Hauts-de-France (32)
% Provence-Alpes-Côte d'Azur (93) and Bourgogne-Franche-Comté (27)
% Occitanie  (76) and Pays de la Loire (52)
% Centre-Val de Loire (24) and Bretagne (53)
% Normandie (28) and Nouvelle-Aquitaine (75)
% Corse (94) region
% 

\subsection{Normandie}

\vspace{-0.2cm}

\begin{figure}[H]
 \centering
\includegraphics[width=6.8cm]{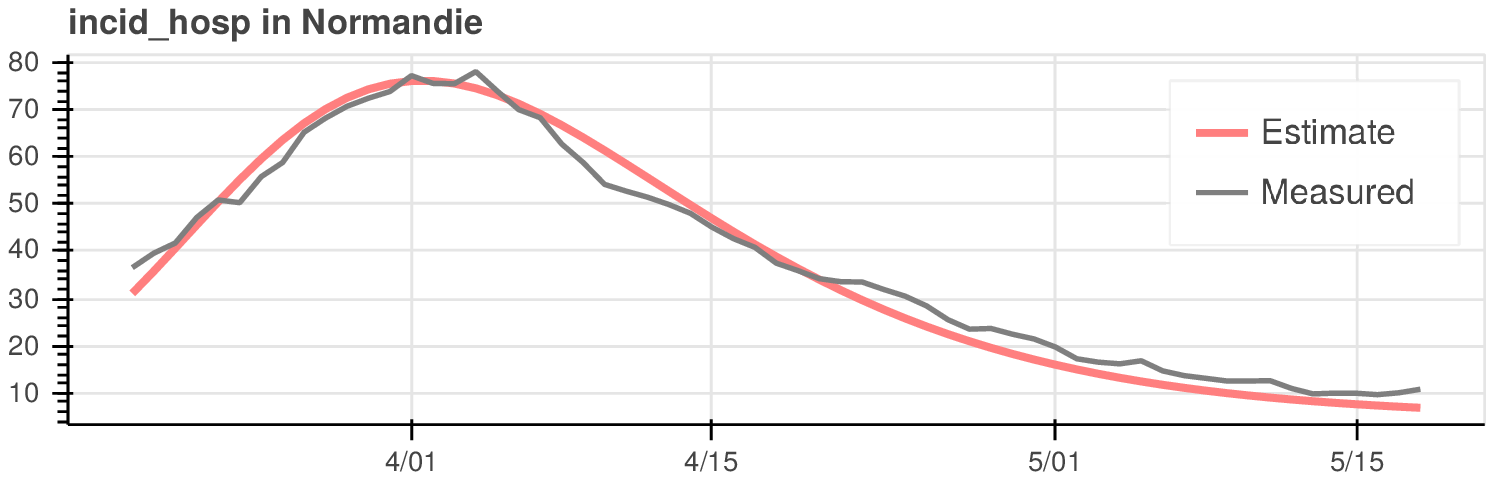}
\includegraphics[width=6.8cm]{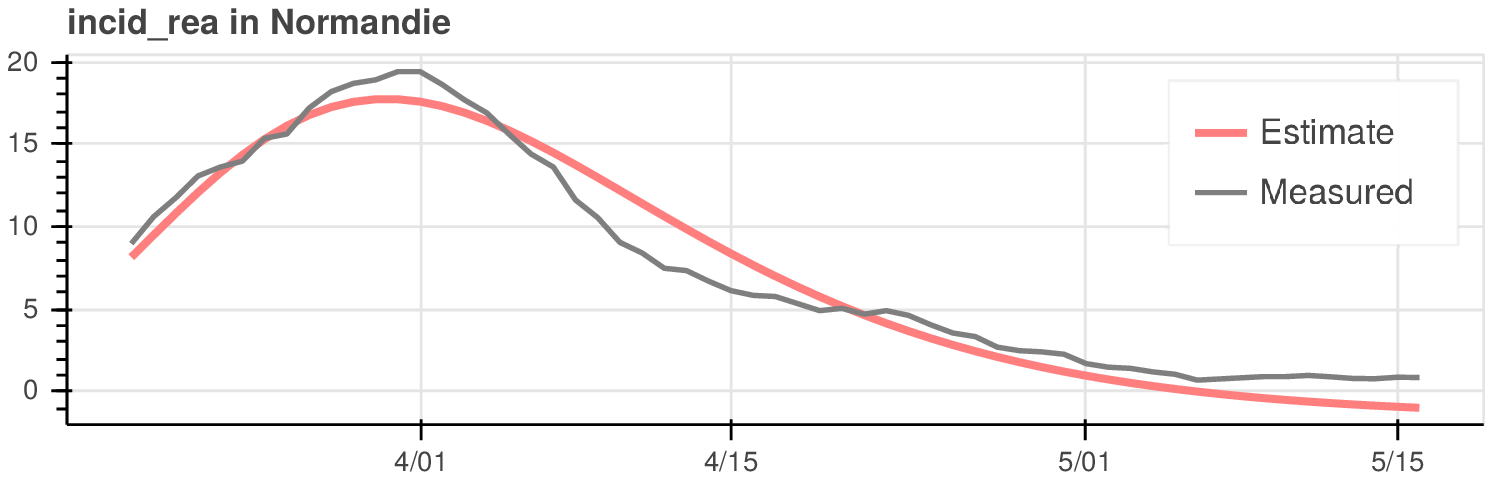}
\vspace{-0.7cm}
 \caption{Comparison of time series {\tt incid\_hosp} and {\tt incid\_rea} averaged over a 14-days-long window, with the fitted outputs, for Normandie (28) region.}
\end{figure}

\vspace{-0.7cm}
\begin{figure}[H]
 \centering
\includegraphics[width=6.8cm]{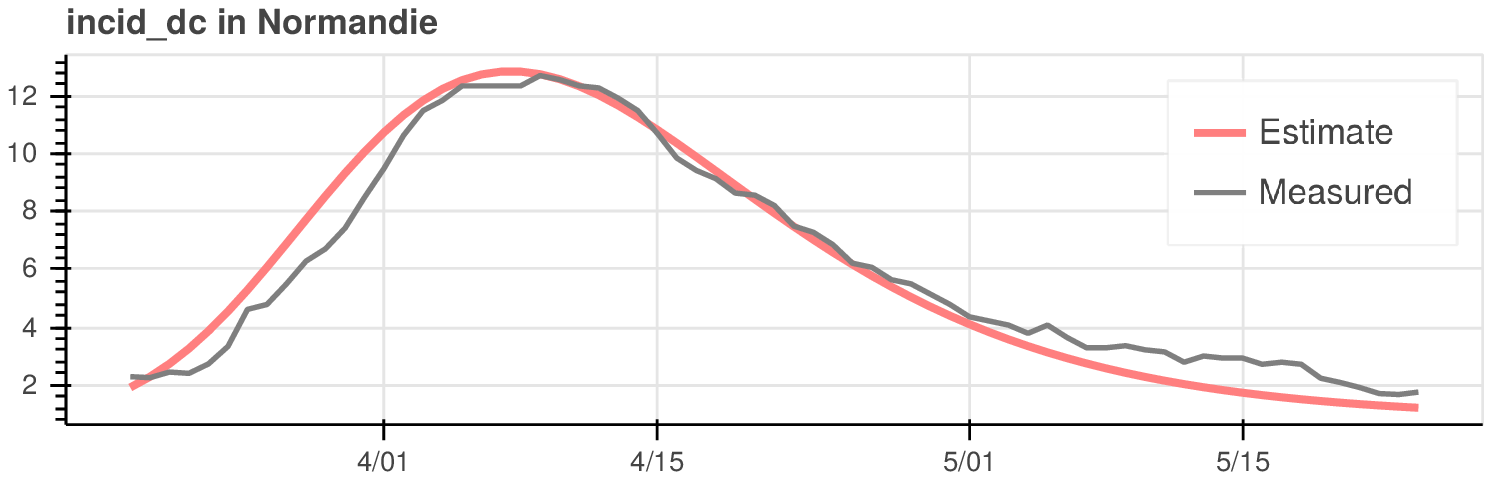}
\includegraphics[width=6.8cm]{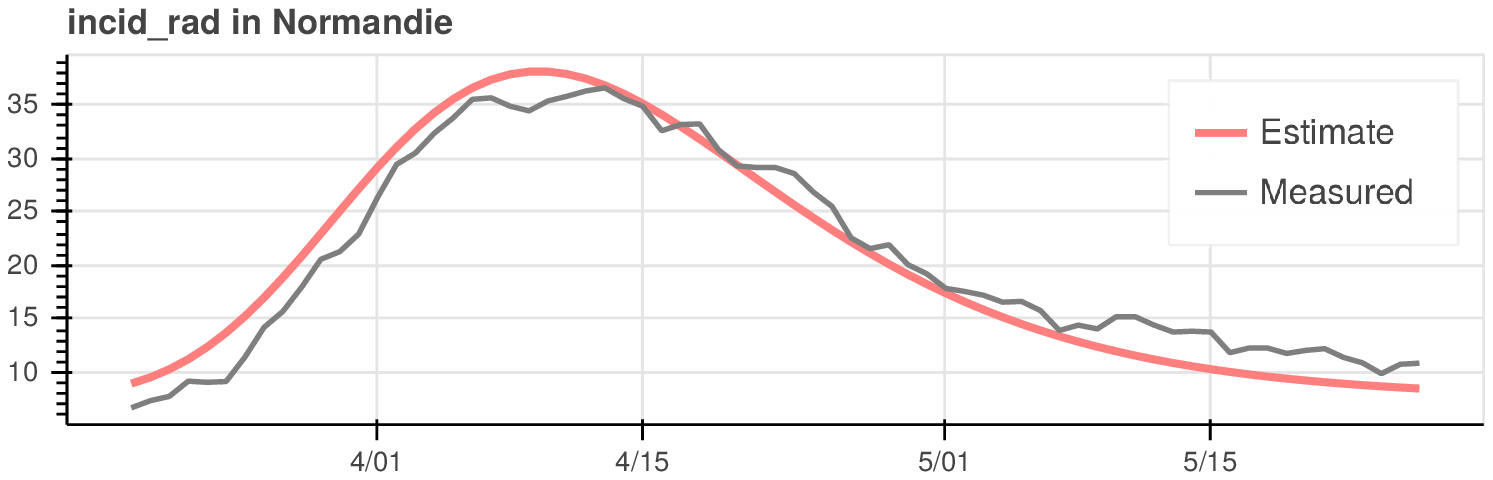}
\vspace{-0.7cm}
 \caption{Comparison of time series {\tt incid\_dc} and {\tt incid\_rad} averaged over a 14-days-long window, with the fitted outputs, for Normandie (28) region.}
\end{figure}

\vspace{-0.7cm}
\begin{figure}[H]
 \centering
\includegraphics[width=6.8cm]{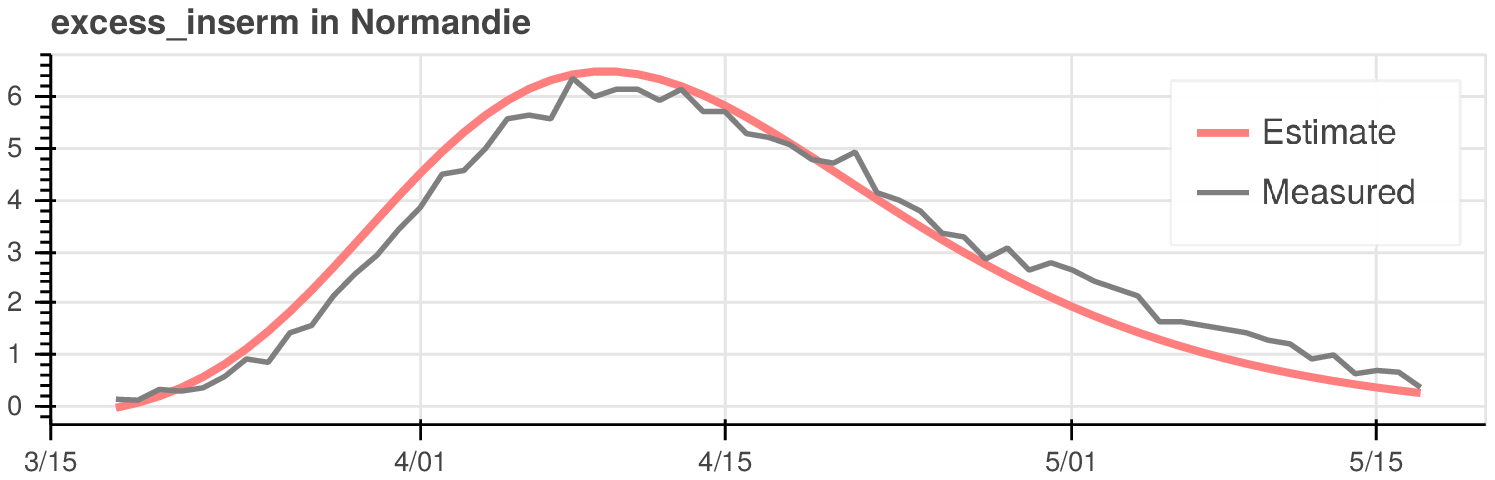}
\includegraphics[width=6.8cm]{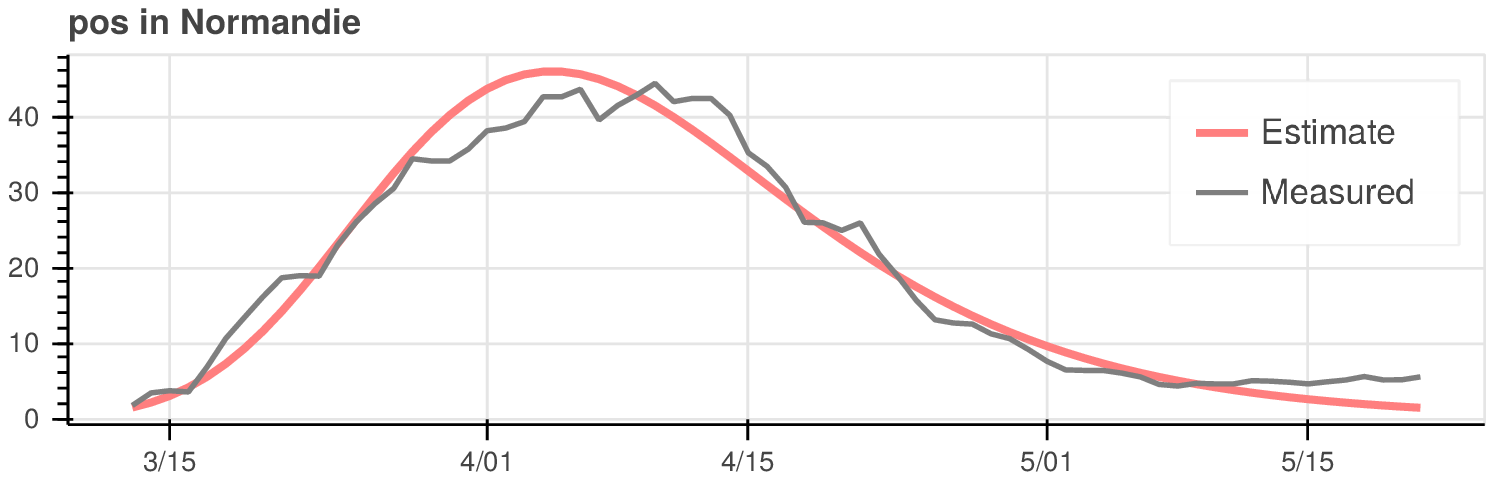}
\vspace{-0.7cm}
 \caption{Comparison of time series {\tt incid\_inserm} and {\tt pos} averaged over a 14-days-long window, with the fitted outputs, for Normandie (28) region.}
\end{figure}

\vspace{-0.7cm}
\begin{figure}[H]
 \centering
\includegraphics[width=6.8cm]{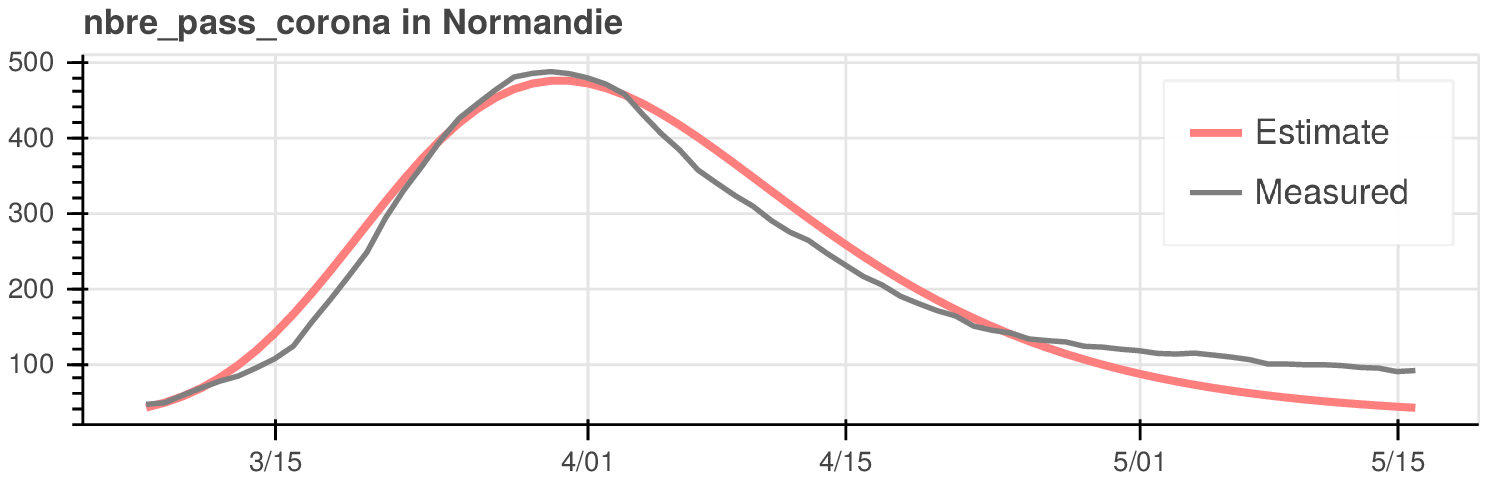}
\includegraphics[width=6.8cm]{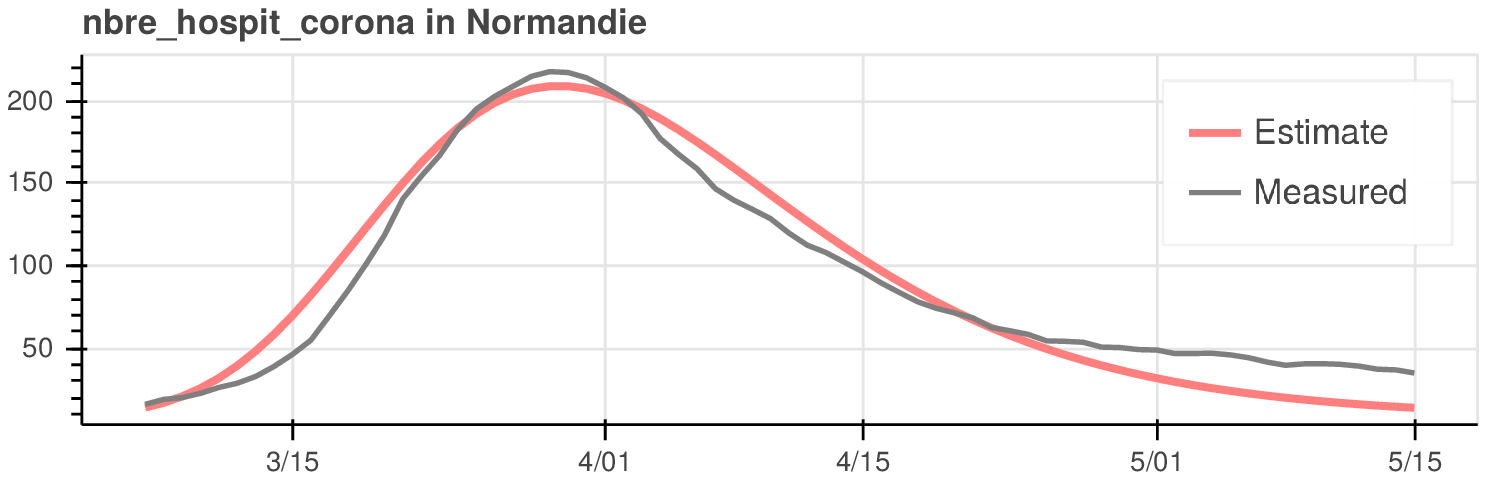}
\vspace{-0.7cm}
 \caption{Comparison of time series {\tt nbre\_pass\_corona} and 
 {\tt nbre\_hospit\_corona} averaged over a 14-days-long window, with the fitted outputs, for Normandie (28) region.}
\end{figure}
\vspace{-0.7cm}
\begin{figure}[H]
 \centering
\includegraphics[width=6.8cm]{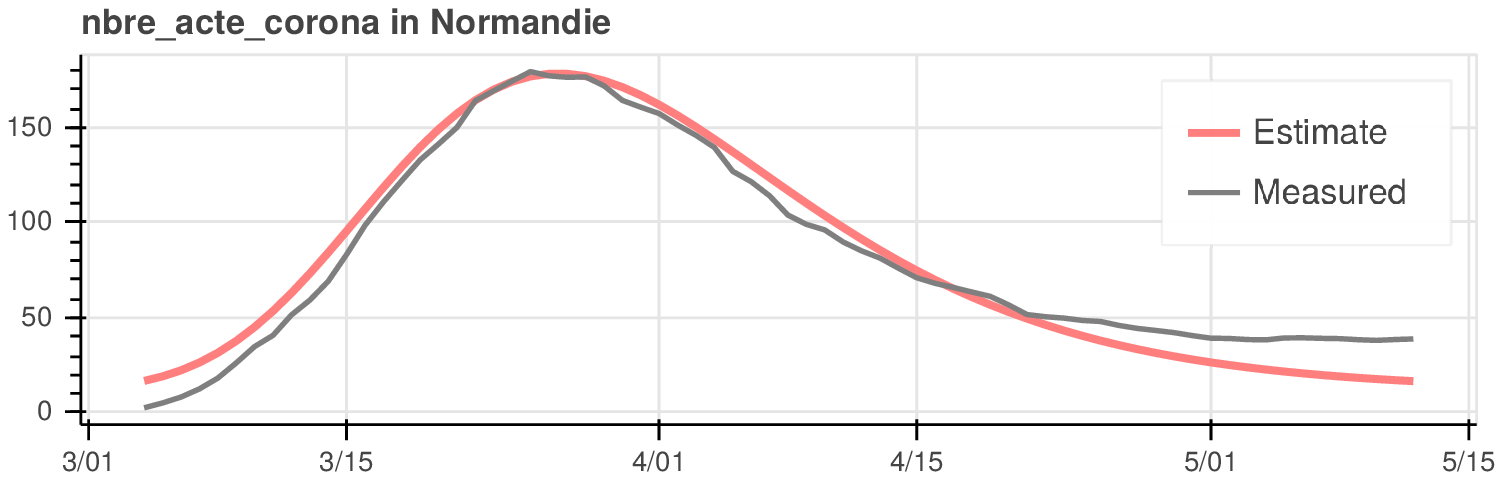}
\vspace{-0.3cm}
 \caption{Comparison of time series {\tt nbre\_acte\_corona} averaged over a 14-days-long window, with the fitted output, for Normandie (28) region.}
\end{figure}

\subsection{Nouvelle-Aquitaine}

\vspace{-0.2cm}

\begin{figure}[H]
 \centering
\includegraphics[width=6.8cm]{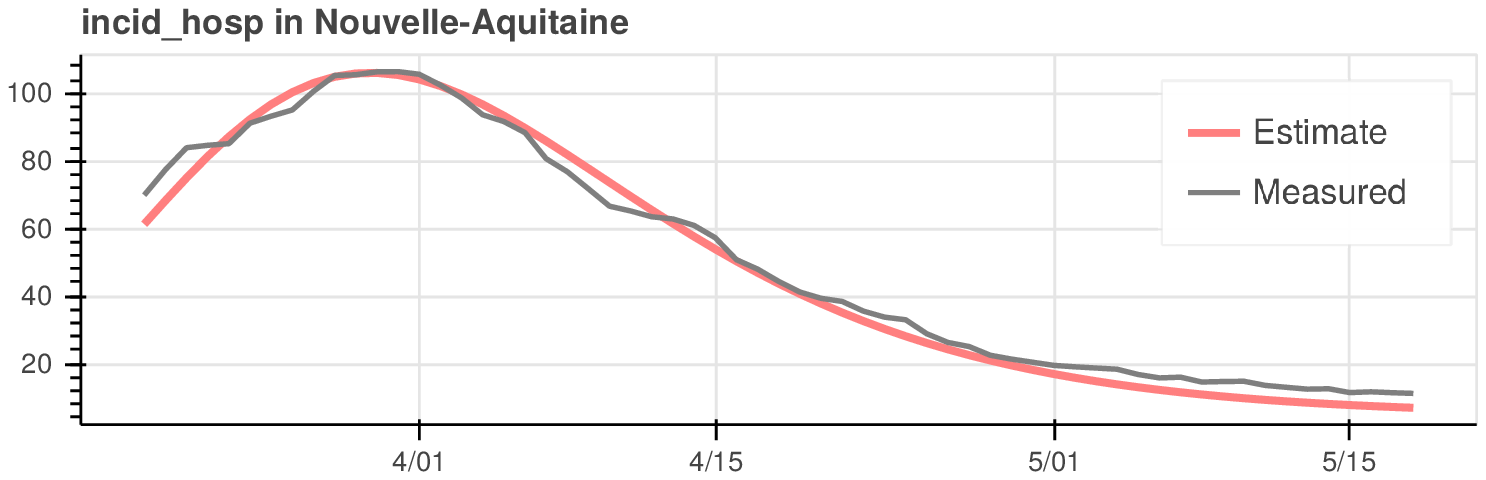}
\includegraphics[width=6.8cm]{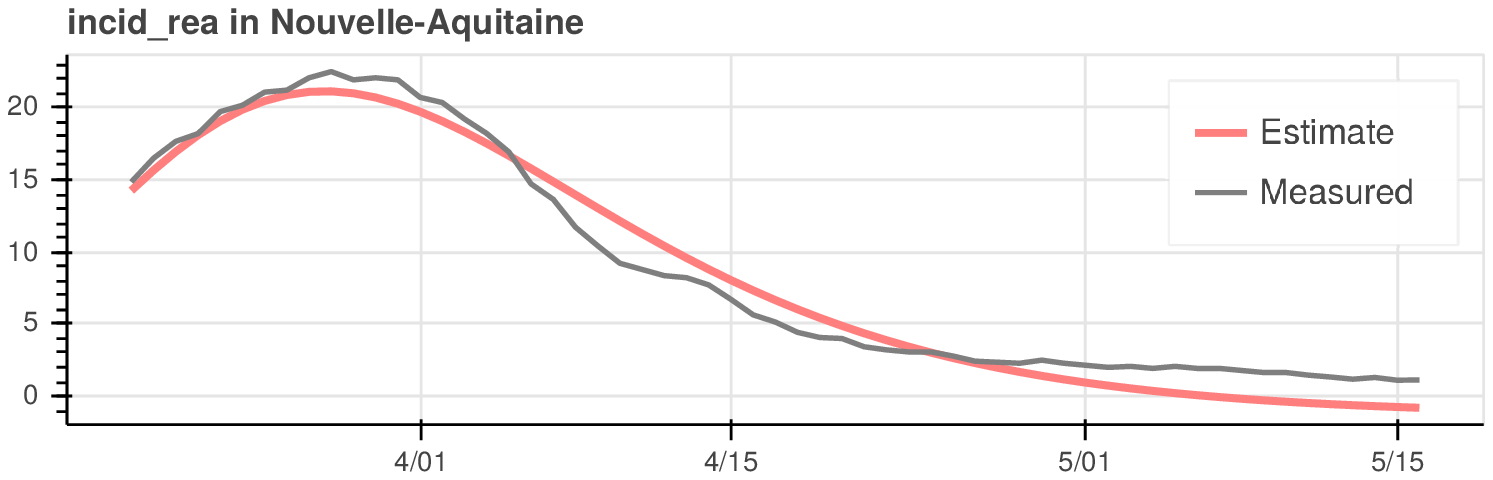}
\vspace{-0.7cm}
 \caption{Comparison of time series {\tt incid\_hosp} and {\tt incid\_rea} averaged over a 14-days-long window, with the fitted outputs, for Nouvelle-Aquitaine (75) region.}
\end{figure}

\vspace{-0.7cm}
\begin{figure}[H]
 \centering
\includegraphics[width=6.8cm]{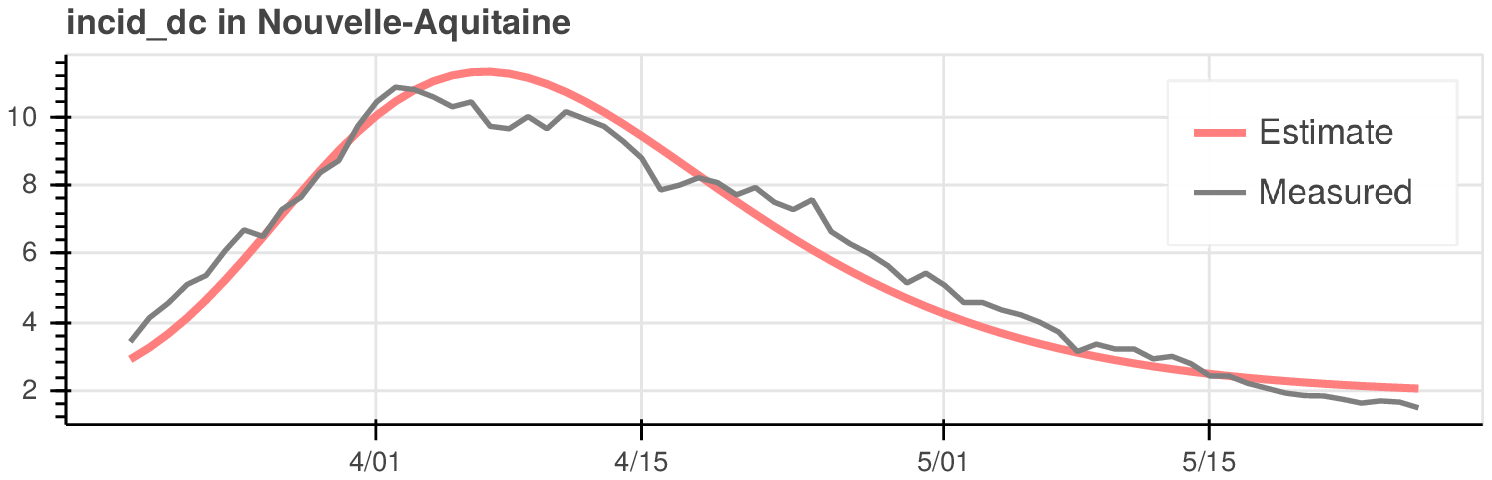}
\includegraphics[width=6.8cm]{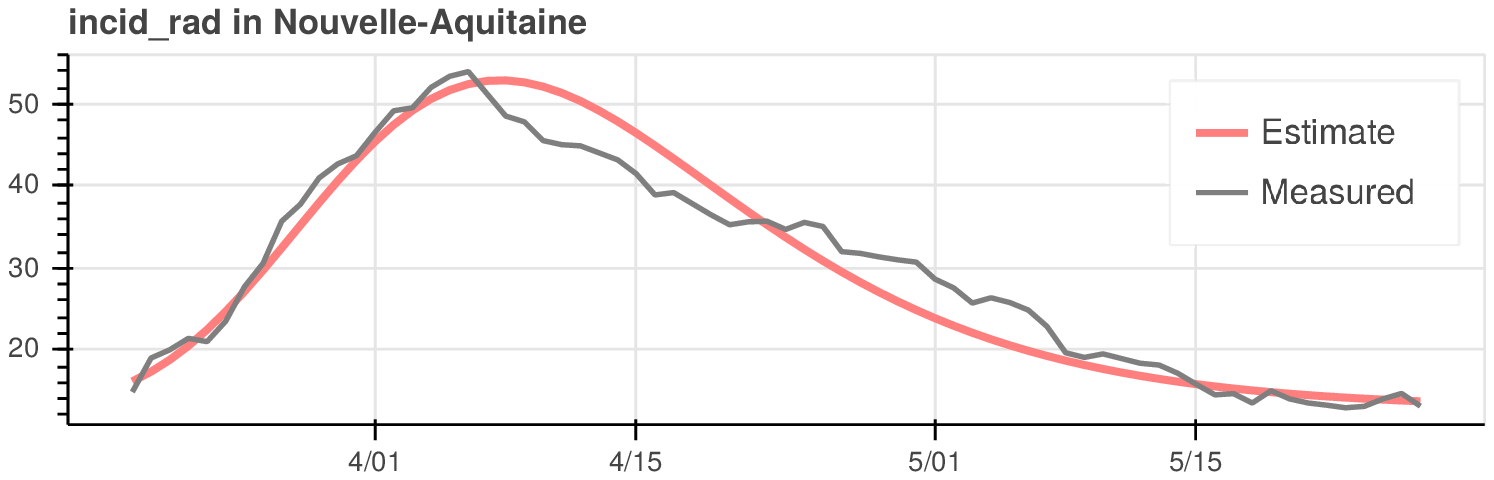}
\vspace{-0.7cm}
 \caption{Comparison of time series {\tt incid\_dc} and {\tt incid\_rad} averaged over a 14-days-long window, with the fitted outputs, for Nouvelle-Aquitaine (75) region.}
\end{figure}

\vspace{-0.7cm}
\begin{figure}[H]
 \centering
\includegraphics[width=6.8cm]{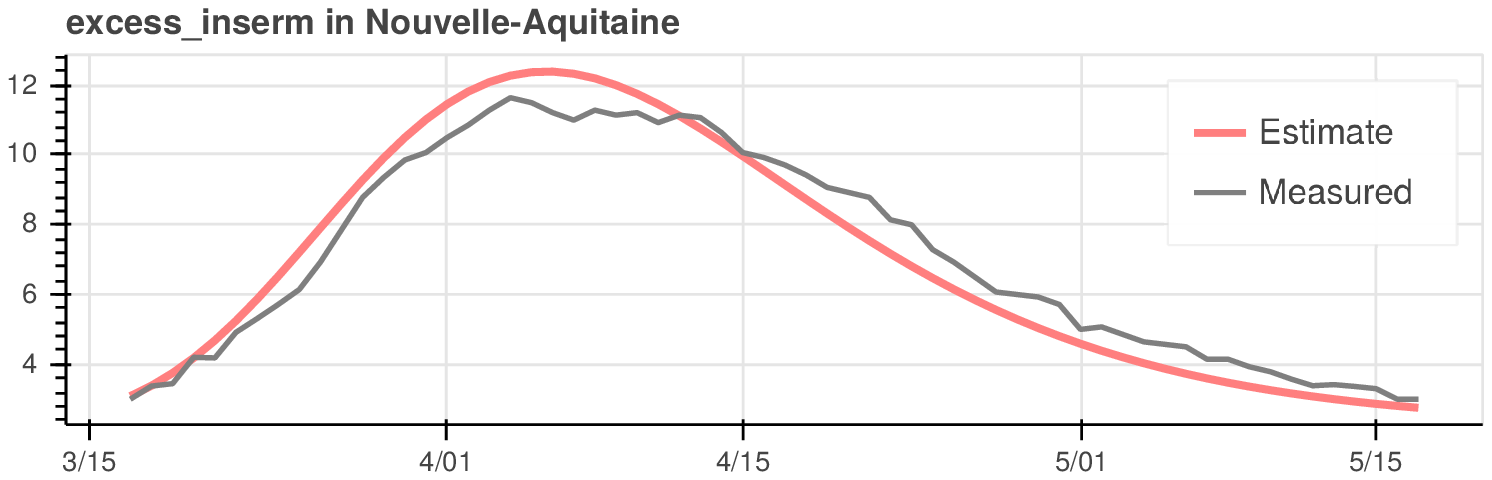}
\includegraphics[width=6.8cm]{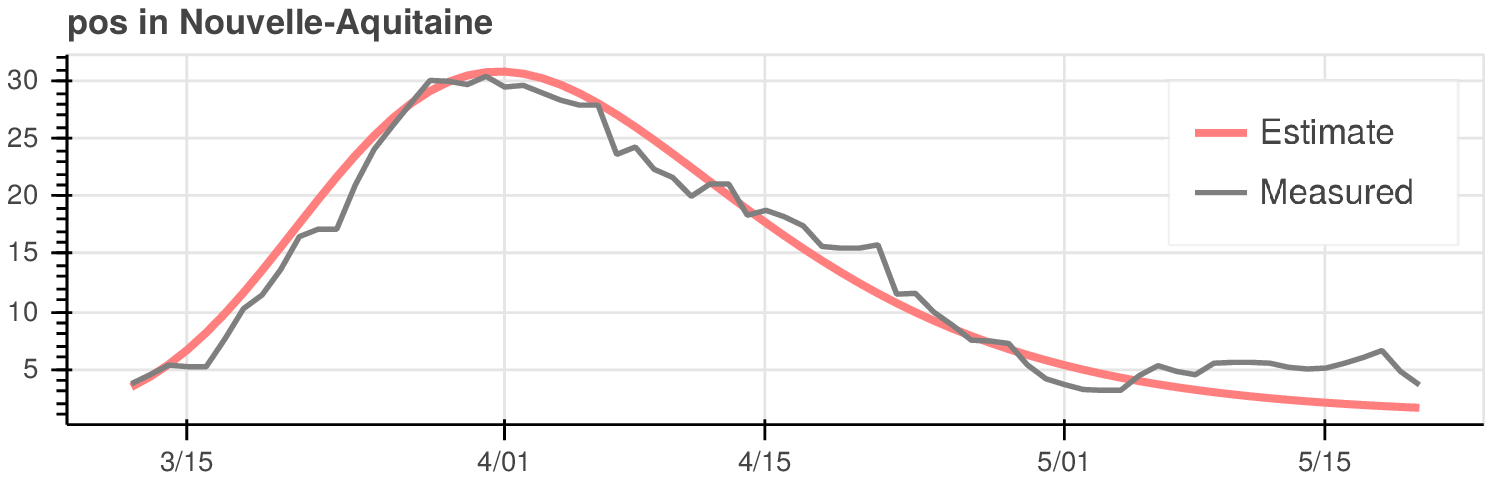}
\vspace{-0.7cm}
 \caption{Comparison of time series {\tt incid\_inserm} and {\tt pos} averaged over a 14-days-long window, with the fitted outputs, for Nouvelle-Aquitaine (75) region.}
\end{figure}

\vspace{-0.7cm}
\begin{figure}[H]
 \centering
\includegraphics[width=6.8cm]{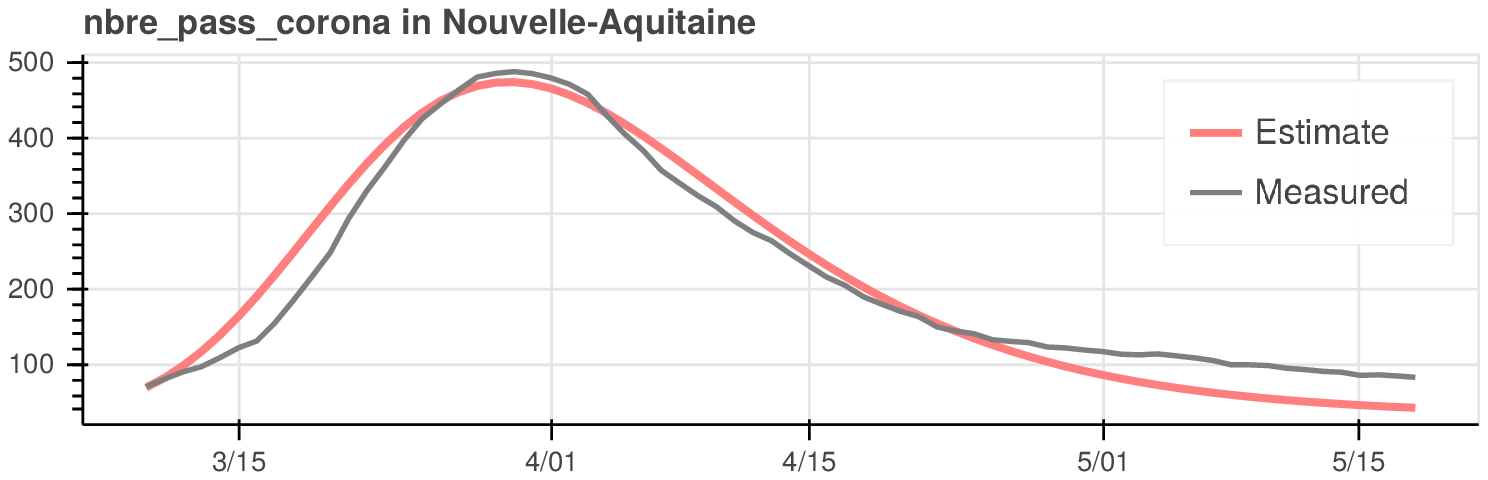}
\includegraphics[width=6.8cm]{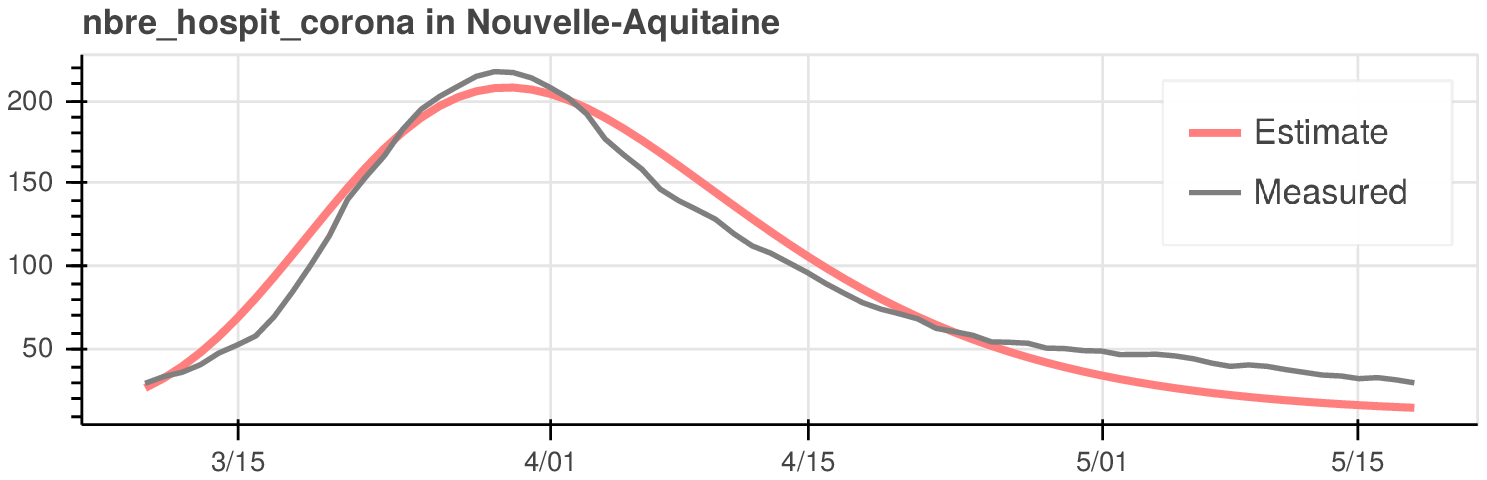}
\vspace{-0.7cm}
 \caption{Comparison of time series {\tt nbre\_pass\_corona} and 
 {\tt nbre\_hospit\_corona} averaged over a 14-days-long window, with the fitted outputs, for Nouvelle-Aquitaine (75) region.}
\end{figure}
\vspace{-0.7cm}
\begin{figure}[H]
 \centering
\includegraphics[width=6.8cm]{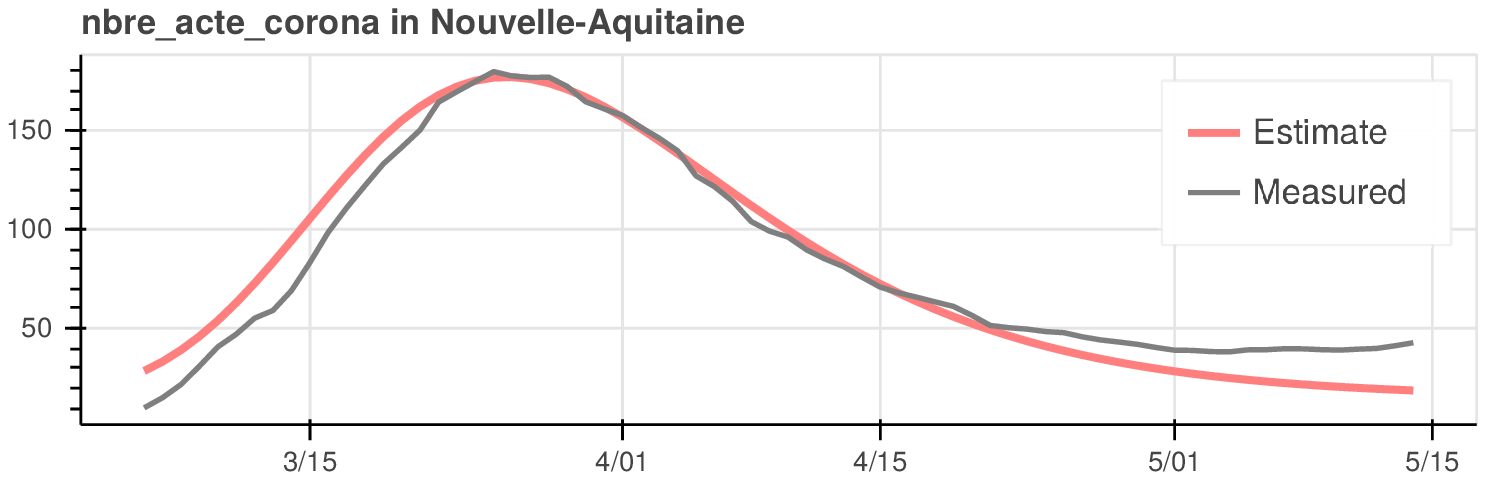}
\vspace{-0.3cm}
 \caption{Comparison of time series {\tt nbre\_acte\_corona} averaged over a 14-days-long window, with the fitted output, for Nouvelle-Aquitaine (75) region.}
\end{figure}

\subsection{Corse}

\vspace{-0.2cm}

\begin{figure}[H]
 \centering
\includegraphics[width=6.8cm]{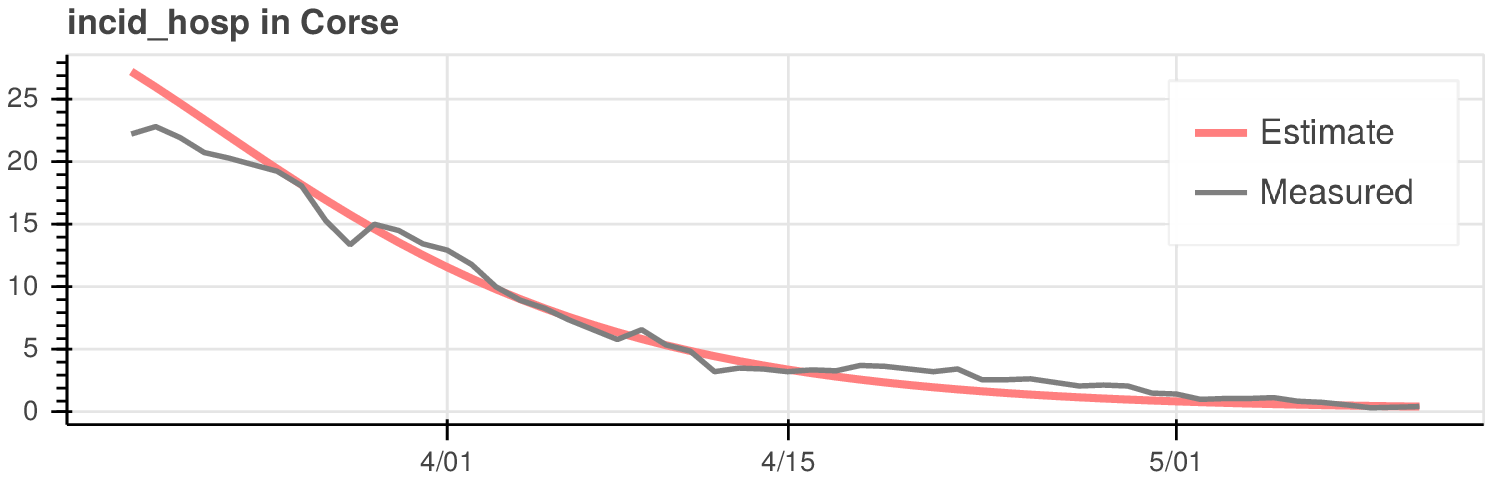}
\includegraphics[width=6.8cm]{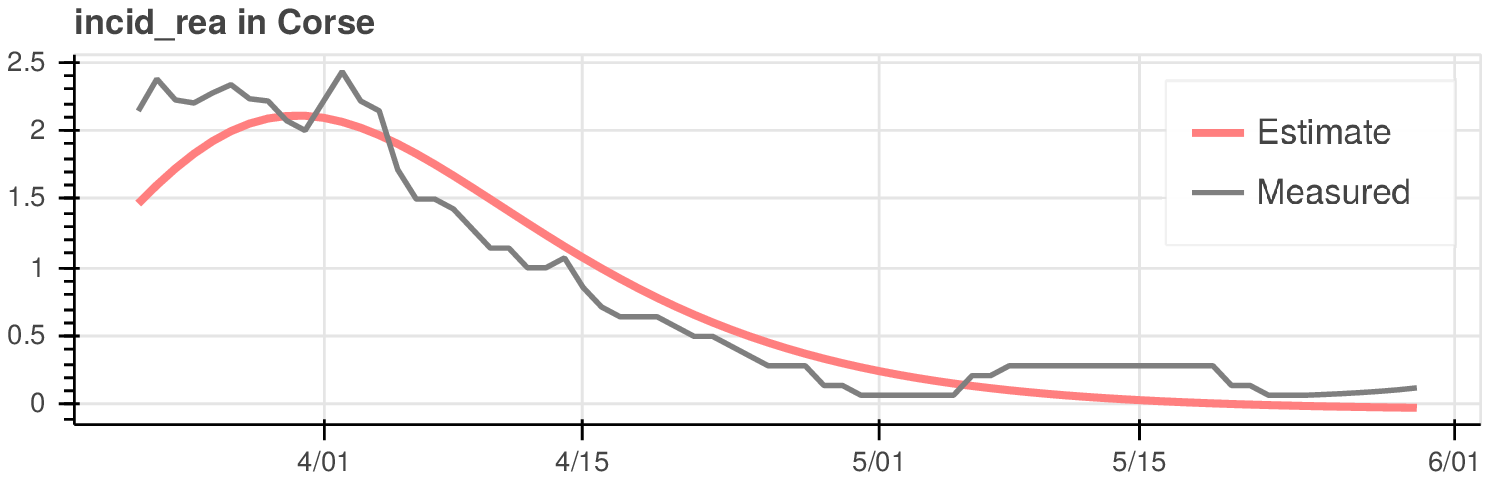}
\vspace{-0.7cm}
 \caption{Comparison of time series {\tt incid\_hosp} and {\tt incid\_rea} averaged over a 14-days-long window, with the fitted outputs, for Corse (94) region.}
\end{figure}

\vspace{-0.7cm}
\begin{figure}[H]
 \centering
\includegraphics[width=6.8cm]{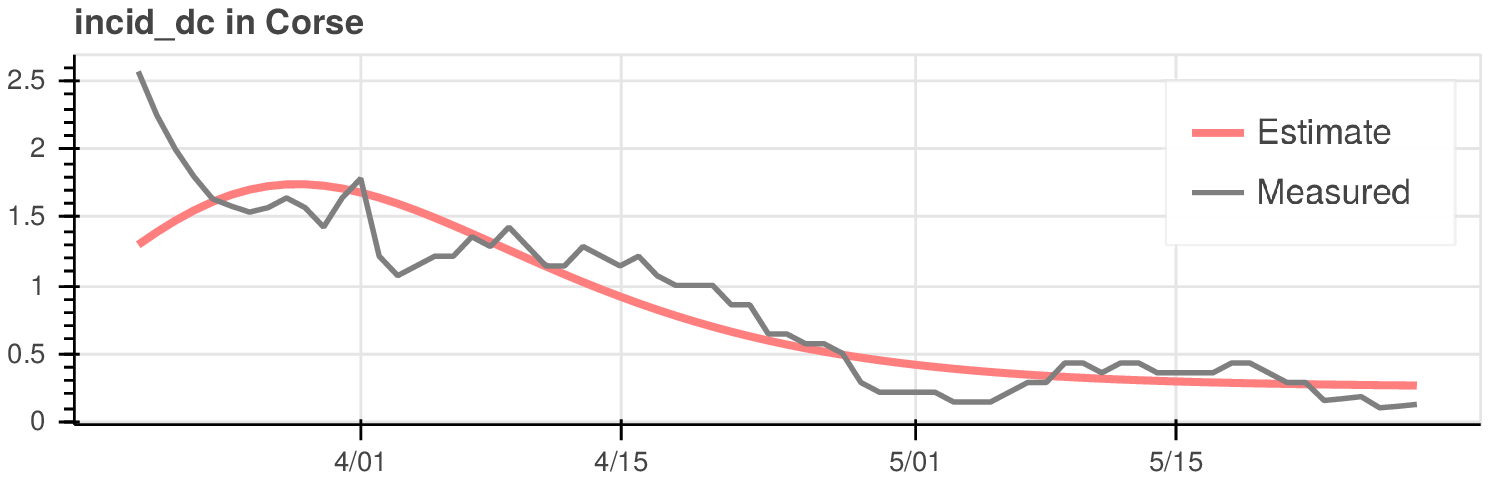}
\includegraphics[width=6.8cm]{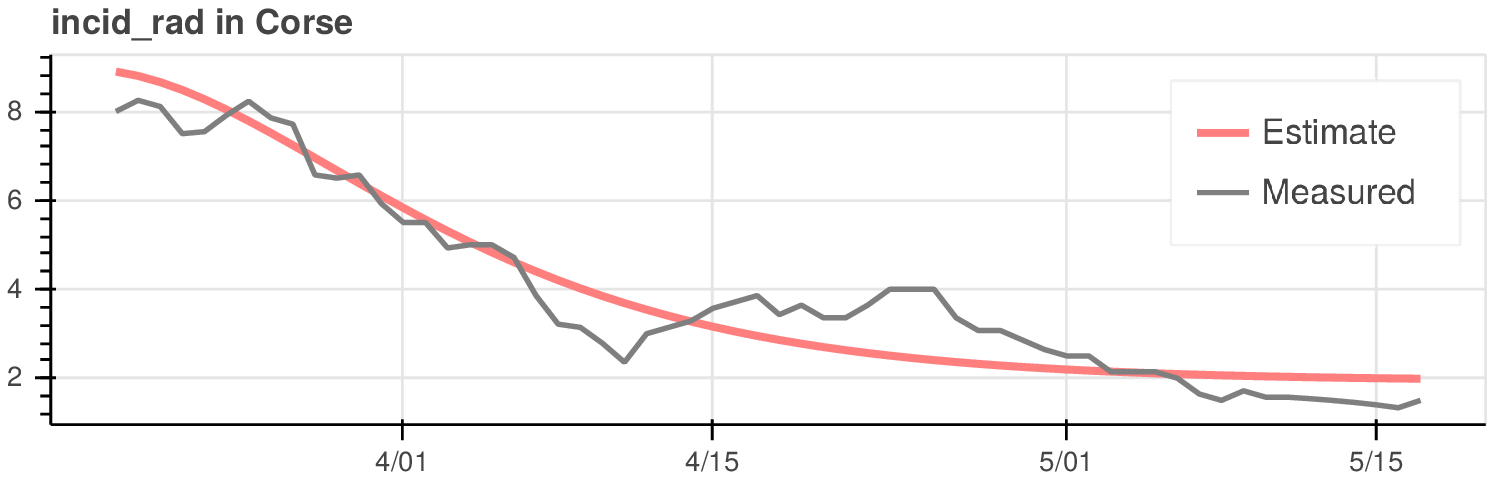}
\vspace{-0.7cm}
 \caption{Comparison of time series {\tt incid\_dc} and {\tt incid\_rad} averaged over a 14-days-long window, with the fitted outputs, for Corse (94) region.}
\end{figure}

\vspace{-0.7cm}
\begin{figure}[H]
 \centering
\includegraphics[width=6.8cm]{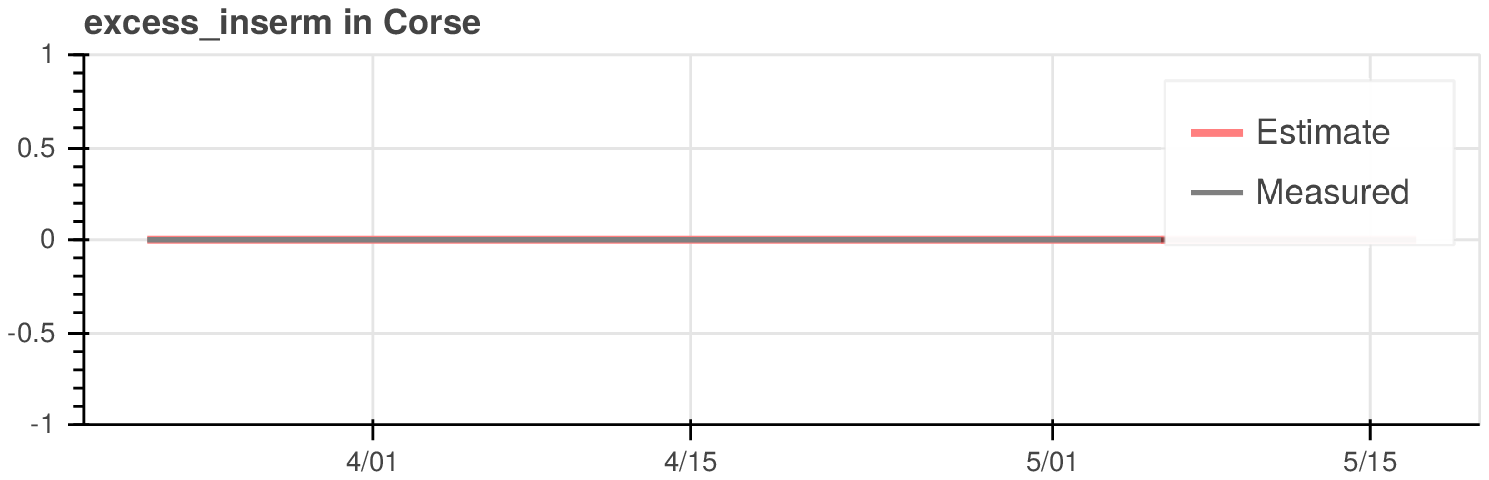}
\includegraphics[width=6.8cm]{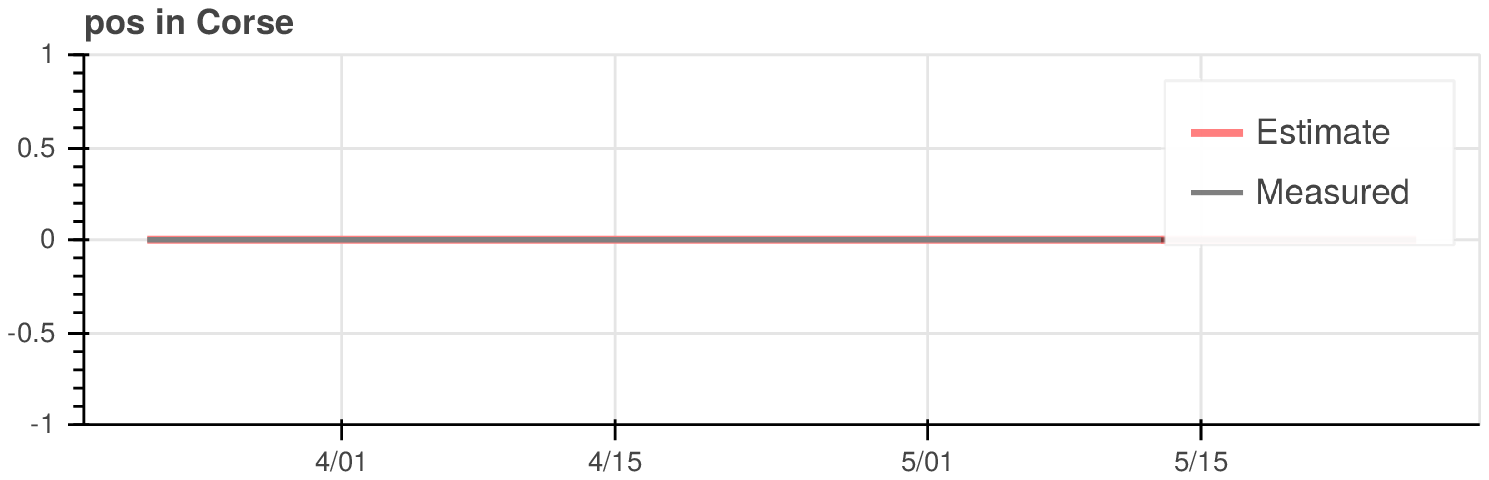}
\vspace{-0.7cm}
 \caption{Comparison of time series {\tt incid\_inserm} and {\tt pos} averaged over a 14-days-long window, with the fitted outputs, for Corse (94) region.}
\end{figure}

\vspace{-0.7cm}
\begin{figure}[H]
 \centering
\includegraphics[width=6.8cm]{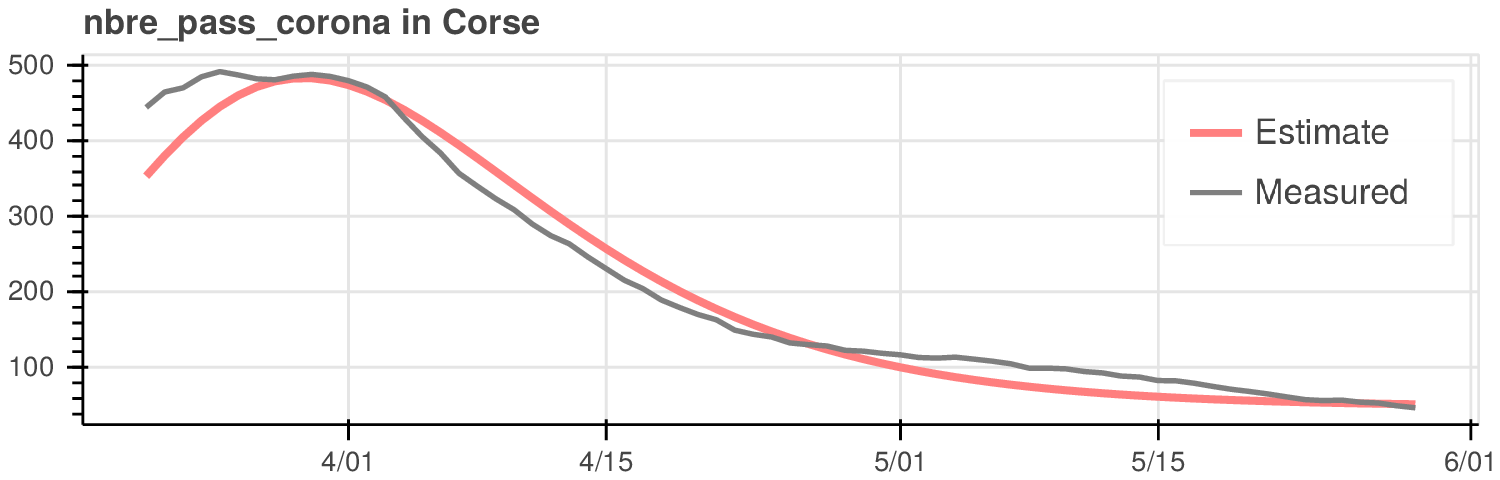}
\includegraphics[width=6.8cm]{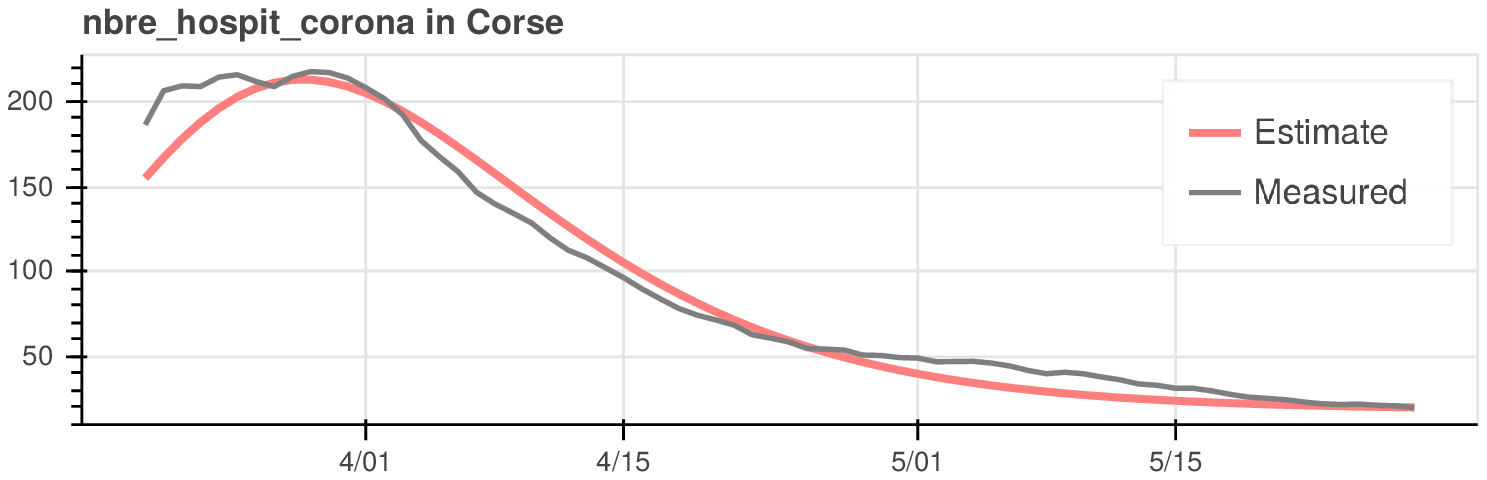}
\vspace{-0.7cm}
 \caption{Comparison of time series {\tt nbre\_pass\_corona} and 
 {\tt nbre\_hospit\_corona} averaged over a 14-days-long window, with the fitted outputs, for Corse (94) region.}
\end{figure}
\vspace{-0.7cm}
\begin{figure}[H]
 \centering
\includegraphics[width=6.8cm]{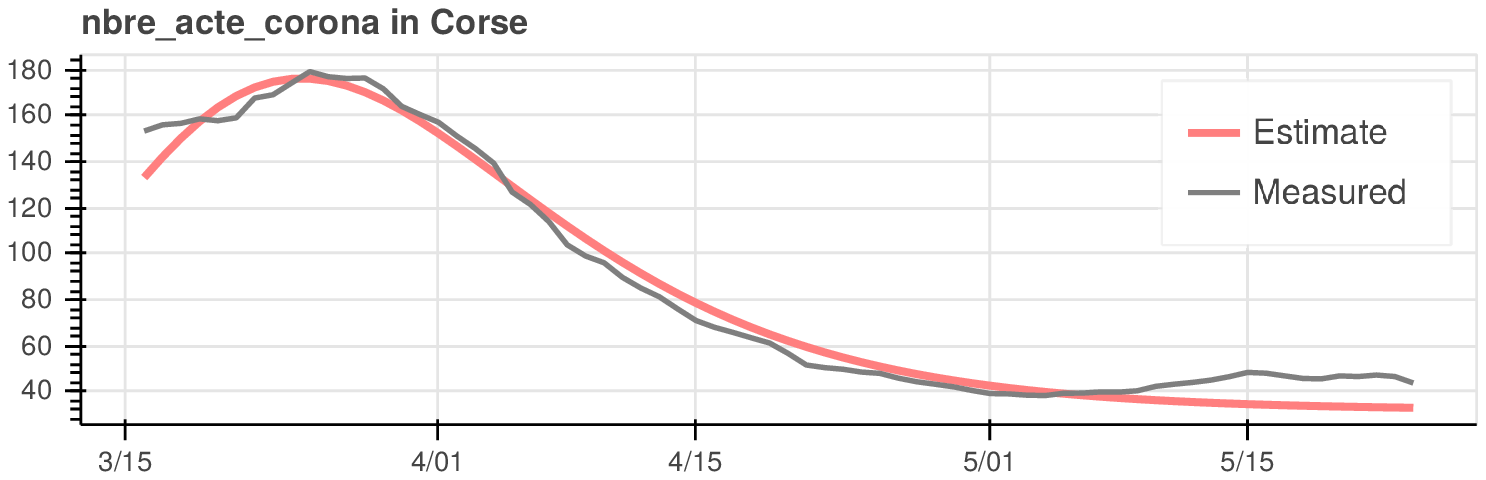}
\vspace{-0.3cm}
 \caption{Comparison of time series {\tt nbre\_acte\_corona} averaged over a 14-days-long window, with the fitted output, for Corse (94) region.}
\end{figure}

\end{document}